 \documentclass[final,3p,times]{elsarticle}

\usepackage{framed,multirow}
\usepackage[utf8x]{inputenc}

\usepackage{amssymb}
\usepackage{latexsym}
\usepackage{amsmath}
\usepackage{amsfonts}
\usepackage{physics}
\usepackage{enumitem}
\usepackage{caption}
\usepackage{tabu}
\usepackage{stackengine}
\usepackage{lineno}
\usepackage{graphicx}
\usepackage{floatrow}
\usepackage{hhline}
\usepackage{cellspace} 
\usepackage{ulem}  %barrer texte
\usepackage{cancel} % idem

\usepackage{diagbox}   % ADDED
\usepackage{hyperref}  % ADDED hypertext capabilities
\usepackage{pifont}    % ADDED for cross symbol in tables
\usepackage{threeparttable} % ADDED for legend in RK table

\usepackage{caption}    % ADDED
\usepackage{subcaption} % ADDED

\usepackage{fontawesome} % ADDED

\usepackage{layouts}
\usepackage{hhline}

\usepackage{url}
\usepackage[svgnames]{xcolor}

\xdefinecolor{dgreen}{named}{DarkOliveGreen}
\definecolor{newcolor}{rgb}{.8,.349,.1}

\usepackage{tikz}
\usetikzlibrary{patterns}
\usetikzlibrary{decorations}
\usetikzlibrary{decorations.pathmorphing}
\usetikzlibrary{decorations.pathreplacing}
\usetikzlibrary{decorations.shapes}
\usetikzlibrary{decorations.text}
\usetikzlibrary{decorations.markings}
\usetikzlibrary{decorations.fractals}
\usetikzlibrary{decorations.footprints}
\usetikzlibrary{arrows,calc}
\usetikzlibrary{shapes}

\usetikzlibrary{matrix}

\usepackage{lineno}   %% <- no mathlines option
\usepackage{amsmath}  %% <- after lineno
\usepackage{etoolbox} %% <- for \cspreto, \csappto
\newcommand*\linenomathpatch[1]{%
  \cspreto{#1}{\linenomath}%
  \cspreto{#1*}{\linenomath}%
  \csappto{end#1}{\endlinenomath}%
  \csappto{end#1*}{\endlinenomath}%
}
\linenomathpatch{equation}
\linenomathpatch{gather}
\linenomathpatch{multline}
\linenomathpatch{align}
\linenomathpatch{alignat}
\linenomathpatch{flalign}
\newcommand\thefont{\expandafter\string\the\font}

\usetikzlibrary{fit}

\newcommand{\linetriangleleft}{\raisebox{0.5pt}{\protect\tikz{\protect\node[draw,scale=0.32,regular polygon, regular polygon sides = 3, fill=none, color=black, line width=0.25mm, rotate=90](){};\protect\draw[color=black, line width=0.25mm] (-0.2,0)--(0.2,0);}}}
\newcommand{\dashedlinetriangleleft}{\raisebox{0.5pt}{\protect\tikz{\protect\node[draw,scale=0.32,regular polygon, regular polygon sides = 3, fill=none, color=black, line width=0.25mm, rotate=90](){};\protect\draw[color=black, line width=0.25mm, densely dashed] (-0.3,0)--(0.3,0);}}}

\newcommand{\linetriangleright}{\raisebox{0.5pt}{\protect\tikz{\protect\node[draw,scale=0.32,regular polygon, regular polygon sides = 3, fill=none, color=black, line width=0.25mm, rotate=-90](){};\protect\draw[color=black, line width=0.25mm] (-0.2,0)--(0.2,0);}}}
\newcommand{\dashedlinetriangleright}{\raisebox{0.5pt}{\protect\tikz{\protect\node[draw,scale=0.32,regular polygon, regular polygon sides = 3, fill=none, color=black, line width=0.25mm, rotate=-90](){};\protect\draw[color=black, line width=0.25mm, densely dashed] (-0.3,0)--(0.3,0);}}}

\newcommand{\blackdashedline}{\raisebox{0.5pt}{\protect\tikz[baseline=-0.5ex]{\protect\draw[color=black, line width=0.17mm, densely dashed] (-0.2,0)--(0.2,0);}}}
\newcommand{\blackdasheddottedline}{\raisebox{0.5pt}{\protect\tikz[baseline=-0.5ex]{\protect\draw[color=black, line width=0.25mm, dash dot] (-0.35,0)--(0.35,0);}}}
\newcommand{\blackdottedline}{\raisebox{0.5pt}{\protect\tikz[baseline=-0.5ex]{\protect\draw[color=black, line width=0.20mm, densely dotted] (-0.2,0)--(0.2,0);}}}

\newcommand{\blacklinefine}{\raisebox{0.5pt}{\protect\tikz[baseline=-0.5ex]{\protect\draw[color=black, line width=0.17mm] (-0.2,0)--(0.2,0);}}}

\newcommand{\bluelinetriangleright}{\raisebox{0.5pt}{\protect\tikz{\protect\node[draw,scale=0.32,regular polygon, regular polygon sides = 3, fill=none, color=blue, line width=0.25mm, rotate=-90](){};\protect\draw[color=blue, line width=0.25mm] (-0.2,0)--(0.2,0);}}}
\newcommand{\greenlinetriangleleft}{\raisebox{0.5pt}{\protect\tikz{\protect\node[draw,scale=0.32,regular polygon, regular polygon sides = 3, fill=none, color=black!60!green, line width=0.25mm, rotate=90](){};\protect\draw[color=black!60!green, line width=0.25mm] (-0.2,0)--(0.2,0);}}}
\newcommand{\redlinetriangleup}{\raisebox{0.5pt}{\protect\tikz{\protect\node[draw,scale=0.32,regular polygon, regular polygon sides = 3, fill=none, color=red, line width=0.25mm, rotate=0](){};\protect\draw[color=red, line width=0.25mm] (-0.2,0)--(0.2,0);}}}

\newcommand{\blacklineveryfine}{\raisebox{0.5pt}{\protect\tikz[baseline=-0.5ex]{\protect\draw[color=black, line width=0.12mm] (-0.2,0)--(0.2,0);}}}
\newcommand{\dashedlinecircle}{\raisebox{0.5pt}{\protect\tikz[baseline=-0.5ex]{\protect\node[draw,scale=0.4, circle, fill=none, color=black, line width=0.12mm, rotate=-90](){};\protect\draw[color=black, line width=0.12mm, densely dashed] (-0.3,0)--(0.3,0);}}}

\newcommand{\dottedlinecross}{\raisebox{0.5pt}{\protect\tikz[baseline=-0.5ex]{\protect\node[ scale=0.7, rotate=45] at (0, 0.){$\boldsymbol{\times}$};\protect\draw[color=black, line width=0.17mm, densely dotted] (-0.3,0.)--(0.3,0.);}}}

\newcommand{\dashedlinecross}{\raisebox{0.5pt}{\protect\tikz[baseline=-0.5ex]{\protect\node[ scale=0.7, rotate=45] at (0, 0.){$\boldsymbol{\times}$};\protect\draw[color=black, line width=0.25mm, densely dashed] (-0.3,0.)--(0.3,0.);}}}

\usepackage[OT1]{fontenc}
\usepackage{geometry}
\journal{Journal of Computational Physics}

\begin{document}

\begin{frontmatter}
	
\author[label1]{Gauthier Wissocq\corref{cor1}}
\ead{gaulthier.wissocq@univ-amu.fr}
\author[label1]{Pierre Sagaut}
\address[label1]{Aix Marseille Univ, CNRS. Centrale Marseille, M2P2 UMR 7340, 13451 Marseille, France}

\title{Hydrodynamic limits and numerical errors of isothermal lattice Boltzmann schemes}

\begin{abstract}

With the aim of better understanding the numerical properties of the lattice Boltzmann method (LBM), a general methodology is proposed to derive its hydrodynamic limits in the discrete setting. It relies on a Taylor expansion in the limit of low Knudsen numbers. With a single asymptotic analysis, two kinds of deviations with the Navier-Stokes (NS) equations are explicitly evidenced: consistency errors, inherited from the kinetic description of the LBM, and numerical errors attributed to its space and time discretization. The methodology is applied to the Bhatnagar-Gross-Krook (BGK), the regularized and the multiple relaxation time (MRT) collision models in the isothermal framework. Deviation terms are systematically confronted to linear analyses in order to validate their expressions, interpret them and provide explanations for their numerical properties. The low dissipation of the BGK model is then related to a particular pattern of its error terms in the Taylor expansion. Similarly, dissipation properties of the regularized and MRT models are explained by a phenomenon referred to as \textit{hyperviscous degeneracy}. The latter consists in an unexpected resurgence of high-order Knudsen effects induced by a large numerical pre-factor. It is at the origin of over-dissipation and severe instabilities in the low-viscosity regime.

\end{abstract}

\begin{keyword}
% MSC codes here, in the form: \MSC code \sep code
% or \MSC[2008] code \sep code (2000 is the default)
%\MSC 41A05\sep 41A10\sep 65D05\sep 65D17
% Keywords
% \KWD lattice Boltzmann \sepvon Neumann analysis
lattice Boltzmann \sep asymptotic analysis \sep numerical errors \sep hydrodynamic limits \sep BGK \sep regularization \sep MRT \sep TRT
\end{keyword}

\end{frontmatter}

\section{Introduction}
\label{sec:Intro}

Over the last two decades, the lattice Boltzmann method (LBM) has become a promising alternative to conventional methods in computational fluid dynamics (CFD)~\cite{GUO_Book_2013, HUANG_Book_2015, Kruger2017, SUCCI_Book_2018}. Inherited from lattice gas automata~\cite{Frisch1986}, it relies on a simplified statistical description of a gas inspired by the Boltzmann equation, encompassing a macroscopic flow behavior usually modelled by the Navier-Stokes (NS) equations~\cite{Boltzmann1872, CHAPMAN_Book_3rd_1970}. The great success of the LBM lies in the simplicity of its numerical scheme. It is based on a two-step Strang-like splitting method \cite{Dellar2013b}: first a local collision step designed to mimic the effects of inter-particle collisions, followed by a node-to-node streaming of discrete populations on a Cartesian grid. The ensuing method offers an easy handling of complex geometries~\cite{Touil2014} together with an efficient and simply parallelizable algorithm~\cite{Schornbaum2016}. However, in addition to a common athermal hypothesis required to preserve the narrow stencil of the method~\cite{Qian1992, SHAN2006}, standard lattice Boltzmann (LB) solvers suffer from a lack of robustness when the Mach number increases and in the inviscid limit~\cite{Lallemand2000, Siebert2008a}. This defect is often attributed to the presence of unstable non-hydrodynamic modes consequent of the kinetic fluid description~\cite{Lallemand2000, Dellar2002, Wissocq2019}.

Historical attempts to enhance the numerical stability of LB solvers turned towards the pursuit of a more sophisticated collision kernel than the notorious Bhatnagar-Gross-Krook (BGK) approximation~\cite{Bhatnagar1954}. 
The latter aims at relaxing distribution functions towards a discrete counterpart of the Maxwell-Boltzmann equilibrium with a single relaxation rate, eventually related to the kinematic viscosity. 
Noticing its stability issues, several authors proposed an extension of this approach using multiple relaxation times (MRT) in the aim of dissipating independently non-hydrodynamic modes, while preserving the correct macroscopic equations in the hydrodynamic limit~\cite{DHumieres1994, Lallemand2000, DHumieres2002}. 
Based on this idea, several models subsequently emerged, including the two-relaxation-time (TRT) scheme designed to reduce the large number of degrees of freedom allowed by the MRT~\cite{Ginzburg2005, Ginzburg2010}, or the so-called cascaded models, performing the collision step in the local flow frame with a view to reduce Galilean invariance errors~\cite{Geier2006, Dubois2015, DeRosis2017}. A second family of collision models, referred to as \textit{regularized models}, aims at filtering out the non-hydrodynamic content of the distribution functions before the collision step~\cite{Latt2006, Malaspinas2015, Mattila2017, Coreixas2017, Jacob2018}. They are built by analogy with a Chapman-Enskog expansion, showing that only first-order deviations from the equilibrium are required to recover the intended macroscopic behavior in the hydrodynamic limit. 
An equivalence can be drawn between some regularized and MRT models, indicating that they share similarities~\cite{Latt2007, Coreixas2019, Wissocq2020}. 
Another family of collision kernels is referred to as \textit{entropic models}. Their purpose is to restore a discrete counterpart of Boltzmann's $H$-theorem so as to increase the robustness of the collision step~\cite{Karlin1998, Boghosian2001, Ansumali2003, Karlin2014, Frapolli2015, ATIF_PRL_119_2017}. They might either enforce the existence of a Lyapunov functional by locally solving a minimization problem, or by ensuring the relaxation towards a numerical equilibrium defined as the maximum of a pseudo-entropy.
An important feature of these models is that they are not obtained performing an expansion of the continuous Maxwellian, but that they are built \textit{ab initio} enforcing a set of \textit{a priori} constraints.
In the same spirit, an enhanced robustness can be noted by simply changing the equilibrium distribution function of the BGK model to a numerical one~\cite{Latt2020}, or by increasing the number of equilibrium moments imposed by the standard polynomial equilibrium~\cite{Dellar2002, Coreixas2018, Wissocq2019, Wissocq2020}.

\begin{figure}[h!]
\begin{center}
\begin{tikzpicture}

%\begin{scope}[xshift=1cm,yshift=-1.2cm]
\draw [line width=0.3mm,,>=stealth, ->] (-4.3,0) -- (2,0);
\draw [line width=0.3mm,,>=stealth, ->, dashed] (-5.,-0.3) -- (-5.,-1.7);
\draw [line width=0.3mm,,>=stealth, ->] (2.7,-0.3) -- (3.4,-1.7);
\draw [line width=0.3mm,,>=stealth, ->] (5.2,-0.3) -- (4.7,-1.7);
\draw [line width=0.3mm,,>=stealth, ->, dashed] (-2.8,-2.2) -- (1.9,-2.2);
\draw (-0.5,0.3) node{$\Delta t, \Delta x \rightarrow 0$};
\draw (-0.5,-0.3) node{\cite{He1998, Dellar2013b}};
\draw (4.,-0.6) node{$\mathrm{Kn} \rightarrow 0$};
\draw (4.,-1.) node{\cite{CHAPMAN_Book_3rd_1970, SHAN2006, PAM2019}};
\draw (-5.8,-1) node{$\mathrm{Kn} \rightarrow 0$};
%\draw (-3.5,-1.2) node{AP?~\cite{Jin2012}};
\draw (-5,0.) node{LBM};
\draw (2.7,0.) node{DVBE};
\draw (5.7,0.) node{Boltzmann equation};
\draw (4.1,-2.) node{Continuous hydrodynamic};
\draw (4.1,-2.4) node{equations: Euler, NS};
\draw (-5,-2) node{Discretized};
\draw (-5,-2.4) node{hydrodynamic equations};
\draw (0,-1.9) node{$\Delta t, \Delta x \rightarrow 0$};
\draw (-0.5,-2.5) node{Convergence? (Lax-type~\cite{Lax1956})};
%\draw [line width=0.2mm, densely dotted] (-4.3,0.3) -- (-1.9,0.3);
%\draw [line width=0.2mm, densely dotted] (-4.3,0.3) -- (-4.3,-2.6);
%\draw [line width=0.2mm, densely dotted] (-4.3,-2.6) -- (3.6,-2.6);
%\draw [line width=0.2mm, densely dotted] (3.6,-2.6) -- (3.6,-1.4);
%\draw [line width=0.2mm, densely dotted] (3.6,-1.4) -- (-1.9,-1.4);
%\draw [line width=0.2mm, densely dotted] (-1.9,-1.4) -- (-1.9, 0.3);
%\end{scope}
\end{tikzpicture}
\caption{Illustration, inspired from~\cite{Jin2012}, of the asymptotic convergence of the LBM towards the Euler and NS equations, involving the time and space discretizations $\Delta t$ and $\Delta x$ and the Knudsen number $\mathrm{Kn}$. Dashed arrows are scrutinized in the present work.}
\label{fig:asymptotic_convergence_LBM}
\end{center}
\end{figure}
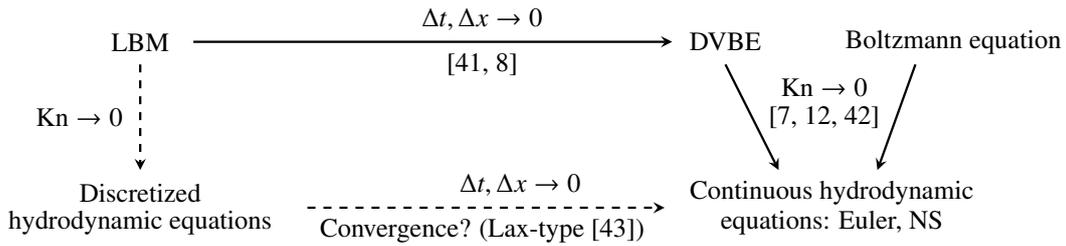

Most of the aforementioned collision models share a common feature: they have been built upon physical considerations inherited from the kinetic theory of gases (hydrodynamic variables, Chapman-Enskog expansion, $H$-theorem and positivity of distribution functions). However, the stability problem can be raised from a different angle, noticing that the LBM is nothing but a numerical scheme solving a counterpart of the Boltzmann equation on a velocity lattice, namely the discrete-velocity Boltzmann equation (DVBE)~\cite{He1998}. Based on a sufficiently large lattice, the DVBE is precisely designed to recover the Navier-Stokes equations as hydrodynamic limit~\cite{SHAN2006}, as illustrated in Fig.~\ref{fig:asymptotic_convergence_LBM}. Linear analyses of the DVBE indicate that its stability is ensured for a large range of Mach numbers in the limit of low Knudsen numbers~\cite{PAM2019}. Therefore, there is some evidence that the instabilities of the LBM are solely induced by the time and space discretization of the numerical algorithm. As such, they might be investigated using common tools of numerical analyses like von Neumann linearizations or Taylor expansions to obtain equivalent equations.

In light of this, the relationship between the LBM and an intended macroscopic behavior (Euler or NS equations) involves two asymptotic parameters as illustrated in Fig.~\ref{fig:asymptotic_convergence_LBM}: the discretization variables ($\Delta x$, $\Delta t$) indicating the numerical resolution of the DVBE, and the Knusen number $\mathrm{Kn}$, drawing a link between microscopic and macroscopic descriptions. The distinction between numerical and hydrodynamic expansions is the cornerstone of so-called \textit{asymptotic preserving} (AP) schemes, which are designed to preserve the asymptotic limits from the microscopic to the macroscopic models in the discrete settings~\cite{Jin2012}. This topic is surprisingly poorly addressed in the LBM community. Yet, even though $\Delta t$ and $\Delta x$ can be related through a (diffusive or acoustic) scaling, the coexistence of two independent smallness parameters ($\Delta t$ and $\mathrm{Kn}$) makes Taylor expansions challenging in LBM. For instance, performing a Taylor expansion in 
$\Delta t$ only yields an open system of macroscopic equations, which does not allow providing a clear physical interpretation to the numerical errors~\cite{Wissocq2019_these, Farag2021}. To overcome this issue, several authors assume an explicit relation between the discretization parameters and the Knudsen number~\cite{Junk2005, Caiazzo2009, Dong2010, Geier2015, Geier2018}. In other words, deviations from the equilibrium allegedly scale as the mesh size or the time step. A single asymptotic analysis is then sufficient to recover the Euler and NS equations as a \textit{consistency} study, rather than as a hydrodynamic limit~\cite{Caiazzo2009}. In other works, a similar off-equilibrium scaling is obtained under the hypothesis that the collision does not involve the time step~\cite{Holdych2004, Dubois2008, Silva2014, Otomo2017, Geier2017, Fucik2021}. However, $\mathrm{Kn}$ and $\Delta t$ are rigorously independent from each other, even though the spatial discretization might yield a maximal reachable Knudsen number~\cite{PAM2019}, which acts as a numerical switch between the microscopic and macroscopic representations in AP schemes~\cite{Jin2012}. Unfortunately, supposing a systematic link between the Knudsen number and the discretization parameters prevents the distinction between numerical errors and hydrodynamic consistency.

The aim of the present work is to propose an original Taylor expansion of the LBM based on the Knudsen number, without assuming any relationship between the deviations from the equilibrium and the discretization parameters. For given mesh size and time step, the proposed methodology offers the benefit to make explicitly appear both numerical errors and inconsistency with the desired physics, with a single asymptotic expansion. Such an analysis is also the opportunity to answer two questions, which are paramount for any AP scheme: 1) What are the hydrodynamic limits of the LBM in the discrete settings? 2) Are these limits stable and consistent (\textit{i.e.} convergent in the sense of Lax~\cite{Lax1956}) with regards to the Navier-Stokes equations? Every error terms will be systematically confronted with linear stability analyses in order to validate their expression, to interpret them, as well as to provide explanations for the numerical errors exhibited in a previous work~\cite{Wissocq2020}. In particular, the present study exhibits a phenomenon referred to as \textit{hyperviscous degeneracy}, consisting in a resurgence of unintended high-order Knudsen effects induced by a large numerical pre-factor.

The present article is structured as follows. In Sec.~\ref{sec:Methodology}, the general methodology adopted in this work is described. A generic variable change is proposed regardless the form of the scheme, so that it can be applied to any collision model. In Sec.~\ref{sec:BGK}, it is applied to the BGK collision model, providing a theoretical rationale for its low numerical dissipation. 
A general form of regularized collision models is investigated in Sec.~\ref{sec:regularized_models}. Two particular regularizations are then scrutinized: a complete shear stress reconstruction (Sec.~\ref{sec:Regul_FR}) and the more standard \textit{projected} (PR) and \textit{recursive} (RR) regularizations (Sec.~\ref{sec:Regul_PR_RR}). Finally, the analysis is extended to several forms of MRT models in Sec.~\ref{sec:MRT_models}. All the forthcoming work is conducted under the isothermal approximation in the context of an acoustic scaling, so as to investigate the asymptotic behavior of LBM with regards to the isothermal weakly compressible NS equations.

\section{Methodology}
\label{sec:Methodology}

\subsection{Isothermal Discrete Velocity Boltzmann Equation (DVBE)}

The aim of any LB scheme is to numerically solve a DVBE. In the present context, considering the BGK collision model~\cite{Bhatnagar1954} and including a general body-force term $\Psi_i$, it reads
\begin{align}
	\partial_t f_i + \xi_{i, \alpha} \partial_{\alpha} f_i = -\frac{1}{\tau} \left( f_i - f_i^{eq} \right) + \Psi_i, \qquad \forall i \in \llbracket 0, V-1 \rrbracket,
	\label{eq:dimensional_DVBE}
\end{align}
where $(f_i)_{i \in \llbracket 0, V-1 \rrbracket}$ is the set of distribution functions, considered as unknowns of the DVBE, $V$ is the number of velocities of the lattice composed of discrete velocities $(\boldsymbol{\xi_i})_{i \in \llbracket 0, V-1 \rrbracket}$, $\tau$ is the relaxation time of the BGK collision model and $(f_i^{eq})_{i \in \llbracket 0, V-1 \rrbracket}$ is the set of equilibrium distribution functions, yet to define. In the following, only one- and two-dimensional cases will be considered. In  one dimension, $\alpha=x$, while in two dimensions, an implicit summation is performed in Eq.~(\ref{eq:dimensional_DVBE}) on the index $\alpha \in \{x, y\}$. In the present context, and without loss of generality on the methodology, the equilibrium distribution function will be expressed as a Hermite polynomial expansion, as proposed by Grad~\cite{Grad1949} and Shan \textit{et al.}~\cite{SHAN2006}:
\begin{align}
	f_i^{eq} = \sum_{n=0}^N \frac{w_i}{n! c^{2n}} \boldsymbol{a}_{eq}^{(n)}:\boldsymbol{\mathcal{H}}_i^{(n)},
	\label{eq:Hermite_equilibrium_expression}
\end{align}
where $w_i$ are the gaussian weights of the lattice, $c$ is a reference velocity, $\boldsymbol{\mathcal{H}}_i^{(n)}$ are Hermite polynomials ($n$th-order tensors) defined by the Rodrigues' formula,
\begin{align}
	\mathcal{H}^{(n)}_{i, \alpha_1..\alpha_n} = \frac{(-c^2)^n}{w(\boldsymbol{\xi}_i)} \frac{\partial^n w}{\partial \xi_{\alpha_1}..\partial \xi_{\alpha_n}} (\boldsymbol{\xi}_i), \qquad w(\boldsymbol{\xi}) = \frac{1}{\left(2 \pi c^2\right)^{D/2}} \exp \left(- \frac{\xi^2}{2c^2} \right),
\end{align}
the symbol `:' stands for the Frobenius inner product, $D$ is the spatial dimension and $\boldsymbol{a}_{eq}^{(n)}$ are Hermite equilibrium coefficients. Furthermore, $N$ stands for the maximal order of equilibrium coefficients that can be imposed by Eq.~(\ref{eq:Hermite_equilibrium_expression}).  In order to recover the isothermal form of the Navier-Stokes equations as hydrodynamic limit of the DVBE, \textit{e.g.} thanks to a Chapman-Enskog expansion~\cite{CHAPMAN_Book_3rd_1970}, the first four coefficients have to be defined as
\begin{align}
	& a_{eq}^{(0)} = \rho, \qquad a_{eq, \alpha}^{(1)} = \rho u_\alpha, \qquad a_{eq, \alpha \beta}^{(2)} = \rho u_\alpha u_\beta + \rho c^2 (\theta-1) \delta_{\alpha \beta}, \nonumber \\
	& a_{eq, \alpha \beta \gamma}^{(3)} = \rho u_\alpha u_\beta u_\gamma + \rho c^2 (\theta-1)(u_\alpha \delta_{\beta \gamma} + u_\beta \delta_{\alpha \gamma} + u_\gamma \delta_{\alpha \beta}),
\end{align}
where $\delta$ is the Kronecker symbol, $\rho$ and $\boldsymbol{u}$ are respectively the macroscopic density and velocity defined as
\begin{align}
	\rho = \sum_i f_i, \qquad \rho u_\alpha = \sum_i \xi_{i, \alpha} f_i,
\end{align}
and $\theta$ is a dimensionless constant defined so that $c \sqrt{\theta}$ is the actual isothermal (Newtonian) sound speed~\cite{Dellar2008}. In a thermal framework, $\theta$ can be viewed as the ratio between the local temperature and a reference one. \newline

In the present work, two standard lattices are considered: the one-dimensional D1Q3 lattice and the two-dimensional D2Q9 one. Their discrete velocities and Gaussian weights read~\cite{Qian1992}:
\begin{align}
	& \mathrm{D1Q3}: \boldsymbol{\xi}_x = [0, 1, -1]\, c/c_s, \quad w_i = [2/3, 1/6, 1/6], \\
	& \mathrm{D2Q9}: \boldsymbol{\xi}_x = [0, 1, 1, 0, -1, -1 ,-1, 0, 1]\, c/c_s, \quad \boldsymbol{\xi}_y = [0, 0, 1, 1, 1, 0 ,-1, -1, -1]\, c/c_s, \nonumber \\
	& \qquad \qquad w_i = [2/3, 1/9, 1/36, 1/9, 1/36, 1/9, 1/36, 1/9, 1/36],
\end{align}
where $c_s=1/\sqrt{3}$ is called the lattice constant. Note that with both of these lattices, the quadrature order is not sufficiently large to correctly impose third-order Hermite equilibrium coefficients. The polynomial expansion of Eq.~(\ref{eq:Hermite_equilibrium_expression}) has to be truncated at $N=2$, leading to a weakly compressible assumption, where the isothermal NS equations are affected by a non-Galilean-invariant error in Mach number~\cite{QIAN_EPL_21_1993, SHAN2006}. With the D2Q9 lattice, partial third-order ($a^{(3)}_{xxy}$ and $a^{(3)}_{xyy}$) and fourth-order ($a^{(3)}_{xxyy}$) expansions can be considered, without, however, increasing the validity range of the macroscopic equations~\cite{Wissocq2019_these}. The latter equilibria will be denoted $N=3^*$ and $N=4^*$ in the following. \newline

The body-force term $\Psi_i$ can precisely be included to address this issue. In a previous work by Prasianakis and Karlin, this error is corrected by modifying the first-order moment of the DVBE, yielding a correct momentum equation~\cite{Prasianakis2007}. However, this affects the computation of the macroscopic velocity, leading to an implicit problem to solve. In the present context, it is decided to adopt the strategy of Feng \textit{et al.}, which only affects the second-order moment of the DVBE~\cite{Feng2019}. With the D1Q3 lattice and the D2Q9 one with $N=3^*, 4^*$, this correction reads
\begin{align}
	\Psi_i = -\frac{w_i}{2c^4} \mathcal{H}_{i,\alpha \alpha} \,\partial_\alpha a_{eq, \alpha \alpha \alpha}^{(3)}.
	\label{eq:correction_D1Q3}
\end{align}
With the D2Q9 lattice with $N=2$, the correction term has to fix the \textit{complete} absence of third-order moments in $f_i^{eq}$ and reads
\begin{align}
	\Psi_i = -\frac{w_i}{2c^4} \mathcal{H}_{i,\alpha \beta} \,\partial_\gamma a_{eq, \alpha \beta \gamma}^{(3)},
	\label{eq:correction_D2Q9_N2}
\end{align}
where implicit summations are performed on $\alpha$, $\beta$ and $\gamma$. In the following, models for which $\Psi_i$ intends to fix incorrect  third-order moments in $f_i^{eq}$ will be referred to as \textit{corrected} models.

\subsection{Dimensionless DVBE}
\label{sec:dimensionless_DVBE}

The first step of the present methodology is to obtain a dimensionless form of the DVBE~(\ref{eq:dimensional_DVBE}) in order to make characteristic non-dimensional numbers appear. This is done by considering a characteristic length $l_p$, defined as the length scale of a perturbation of a quantity $X$ as
\begin{align}
	l_p \sim \frac{|X|}{|\boldsymbol{\nabla{X}}|},
\end{align}
where $\boldsymbol{\nabla}$ stands for the gradient operator. In the context of the linear analyses of Sec.~\ref{sec:validation_LSA}, it can simply be reduced to $l_p=1/k$, where $k$ is the wavenumber of a considered perturbation. For the sake of consistency with the linear analyses, the characteristic length will be referred to as $1/k$ in the following. Furthermore, considering $c/c_s$ as a characteristic velocity, a characteristic time can be defined as $c_s/(kc)$. This leads to the following dimensionless variables:
\begin{align}
	t^* = \frac{kct}{c_s},\qquad x^*=kx, \qquad y^*=ky.
\end{align}
With this non-dimensionalization, the DVBE can be re-written as
\begin{align}
	\partial_{t^*}f_i + e_{i,\alpha} \partial_{\alpha^*} f_i = -\frac{c_s}{k \tau c} \left( f_i - f_i^{eq} \right) + \Psi_i^*,
	\label{eq:DVBE_dimensionless_1}
\end{align}
where $\boldsymbol{e_i}= \boldsymbol{\xi_i}\, c_s/c$ and $\Psi_i^* = \Psi_i\, c_s/(kc)$. Note that with these definitions and given the values of $\boldsymbol{\xi_i}$ for the D1Q3 and D2Q9 lattices, the dimensionless discrete velocities $\boldsymbol{e_i}$ are integers, which make them consistent with the dimensionless quantities of a LB solver. Finally, the relaxation coefficient appearing in Eq.~(\ref{eq:DVBE_dimensionless_1}) can be directly related to the Knudsen number $\mathrm{Kn}$. The latter is indeed defined as the ratio between the mean free path of particles and a characteristic length. Considering $1/k$ as characteristic length and defining the mean free path as $\tau c/c_s$ following~\cite{PAM2019}, the Knudsen number can be defined in the present formalism as
\begin{align}
	\mathrm{Kn} = \frac{k \tau c}{c_s}.
\end{align}
The dimensionless DVBE finally reads
\begin{align}
	\partial_{t^*}f_i + e_{i,\alpha} \partial_{\alpha^*} f_i &= -\frac{1}{\mathrm{Kn}} \left( f_i - f_i^{eq} \right) + \Psi_i^* = -\frac{1}{\mathrm{Kn}} \left( f_i - f_i^{eq} - \mathrm{Kn} \, \Psi_i^* \right),
	\label{eq:DVBE_dimensionless_2}
\end{align}
where the second equality can be very useful to interpret any body-force term as a change of equilibrium distribution function, which was, for instance, recently proposed by Saadat \textit{et al.}~\cite{Saadat2021}. This dimensionless equation will be considered as the target any LB solver intends to solve.

\subsection{Taylor expansion of the dimensionless LB scheme}
\label{sec:Taylor_expansion_Kn}

In a very general way, a LB scheme can be decomposed into two steps acting on a set of discrete distribution functions $(g_i)_{i \in \llbracket 0, V-1 \rrbracket}$ (which are different than the $f_i$ of the DVBE): a streaming step and a collision step. Since the collision step depends on the adopted model, a LB scheme can be systematically identified as a streaming step as
\begin{align}
	g_i(\boldsymbol{x} + \boldsymbol{\xi}_{i} \Delta t, t+\Delta t) = g_i^c(\boldsymbol{x}, t),
	\label{eq:LBM_streaming}
\end{align}
where $\Delta t$ is the time-step and $g_i^c$ are the post-collision distribution functions, which can always be written as explicit functions of $(g_i)_{i \in \llbracket 0, V-1 \rrbracket}$. Note that in presence of a body force term $\Psi_i$, it can be included under a discrete form (eventually evaluating spatial gradients with finite differences) in the collision step which makes Eq.~(\ref{eq:LBM_streaming}) general. Following the same characteristic length, time and velocity as adopted in Sec.~\ref{sec:dimensionless_DVBE}, a dimensionless LB scheme reads
\begin{align}
	g_i \left(\boldsymbol{x}^* + \boldsymbol{e_i} \, \mathrm{Kn} \frac{\Delta t}{\tau}, t^* + \mathrm{Kn} \frac{\Delta t}{\tau} \right) = g_i^c(\boldsymbol{x}^*, t^*).
	\label{eq:LBM_dimensionless}
\end{align}
In agreement with Fig.~\ref{fig:asymptotic_convergence_LBM}, two dimensionless parameters appear:
\begin{enumerate}
	\item the Knudsen number $\mathrm{Kn}$, which is the only parameter driving the flow physics since it is the only one appearing in the DVBE,
	\item the ratio $\Delta t/\tau$, which is a purely numerical quantity representing the effects of the space and time discretization.
\end{enumerate}
In particular, taking the limit $\Delta t/\tau \rightarrow 0$ should yield the DVBE providing that the LB scheme is consistent with it. In order to obtain the equivalent partial differential equations of the LB scheme, it seems tempting to perform a Taylor expansion of it for low values of $\Delta t/\tau$, by analogy with numerical analyses of common discrete schemes . This is for example the strategy adopted in~\cite{Wissocq2019_these, Farag2021}. As discussed in Sec.~\ref{sec:Intro}, many Taylor expansions of the litterature assume a relationship between $\Delta t/\tau$ and $\mathrm{Kn}$. In the present work, no such assumption is performed and an original Taylor expansion in the physical parameter $\mathrm{Kn}$ rather than the numerical parameter $\Delta t/\tau$ is preferred. This is motivated by several observations:
\begin{itemize}
	\item In many applications of the LBM for simulations of air flows, \textit{e.g.} in the aeronautical field, the common order of magnitude of $\tau/\Delta t$ is $[10^{-7}-10^{-3}]$~\cite{Kruger2017}. The numerical parameter $\Delta t/\tau$ is then much greater than one in these cases, which makes a Taylor expansion for $\Delta t/\tau \rightarrow 0$ not convenient for the quantification of numerical errors.
	\item A Taylor expansion in $\Delta t/\tau$ is not sufficient to interpret any numerical error of the LBM as a deviation from the macroscopic equations, \textit{e.g.} the Navier-Stokes (NS) equations. It would only exhibit numerical errors with respect to the DVBE, which is not a closed set of equations involving macroscopic variables ($\rho$ and $u_\alpha$) only. In any case, a supplementary investigation of the hydrodynamic limit of the equations would be mandatory~\cite{Gendre2018}.
	\item Since the Knudsen number is directly related to the flow physics, a Taylor expansion in $\mathrm{Kn}$ can evidence the effect of numerical errors at each physical scale (Euler, Navier-Stokes, Burnett,...). Furthermore, it will be shown in Sec.~\ref{sec:macroscopic_equations} that a Taylor expansion in $\mathrm{Kn}$ is sufficient to derive a closed set of macroscopic equations in a given hydrodynamic limit.
	\item In addition to remaining small for physical reasons, the constraint $\mathrm{Kn} \ll 1$ is ensured for numerical reasons with the LB scheme. This is a consequence of the Nyquist-Shannon theorem~\cite{Nyquist1928, Shannon1949}, stipulating that no spatial fluctuations can be discretized with less than two points per wavelength. In other words, the characteristic length $1/k$ is limited by the mesh size $\Delta x$, so that $k \leq \pi/\Delta x$ for linear perturbations. Together with the relationship relating $\Delta x$ with the time step $\Delta t$ for an acoustic scaling~\cite{Kruger2017} ($c=c_s \sqrt{\theta} \Delta x/\Delta t$), this leads to 
	\begin{align}
	    \mathrm{Kn} \leq \mathrm{Kn}_s = \pi \sqrt{\theta} \frac{\tau}{\Delta t},
	\end{align}
	which ensures that the Knudsen number remains small in common simulations of air flows. The dimensionless number $\mathrm{Kn}_s$ has been referred to as the Knudsen-Shannon number by Masset and Wissocq~\cite{PAM2019}. This property is central for the numerical switch of AP schemes between microscopic and macroscopic descriptions~\cite{Jin2012}.
\end{itemize}
A Taylor expansion of Eq.~(\ref{eq:LBM_dimensionless}) for $\mathrm{Kn} \rightarrow 0$ yields
\begin{align}
	\sum_{n \geq 0} \frac{\mathrm{Kn}^n}{n!} \left( \frac{\Delta t}{\tau} \right)^n D_i^n g_i = g_i^c,
	\label{eq:LBM_Taylor_1}
\end{align}
where $D_i$ is the operator defined as $D_i = \partial_{t^*} + e_{i, \alpha} \partial_{\alpha^*}$ and $D_i^n$ stands for $n$ successive applications of this operator. To draw a parallel with the DVBE, Eq.~(\ref{eq:LBM_Taylor_1}) can then be re-written as
\begin{align}
	D_i g_i = - \frac{\tau}{\Delta t} \frac{1}{\mathrm{Kn}} \left( g_i - g_i^c \right) - \sum_{n\geq 2} \frac{\mathrm{Kn}^{n-1}}{n!} \left(\frac{\Delta t}{\tau} \right)^{n-1} D_i^n g_i.
	\label{eq:LBM_Taylor_2}
\end{align}

\subsection{Generic variable change}

Noticing now that the first right-hand-side term of Eq.~(\ref{eq:LBM_Taylor_2}) is very close to the collision term of the DVBE~(\ref{eq:DVBE_dimensionless_2}), the following generic variable change can be assumed:
\begin{align}
	 g_i - g_i^c = \frac{\Delta t}{\tau} \left(f_i - f_i^{eq} - \mathrm{Kn}\, \Psi_i^* \right).
	 \label{eq:variable_change}
\end{align}
This variable change strongly depends on the collision model and it will be shown in Sec.~\ref{sec:BGK} that, in the case of the BGK collision, it reduces to the well-known relation obtained by He \textit{et al.}~\cite{He1998} in their \textit{a priori} derivation of the BGK-LB scheme. However, when different collision models are adopted, other variable changes actually occur, which will we be shown for both regularized models in Sec.~\ref{sec:regularized_models} and MRT models in Sec.~\ref{sec:MRT_models}. Whatever the collision model, one has to check that the variable change is admissible, meaning that it does not affect the computation of macroscopic variables:
\begin{align}
	\sum_i g_i = \sum_i f_i = \rho, \qquad \sum_i e_{i, \alpha} g_i = \sum_i e_{i, \alpha} f_i = \rho u_\alpha^* = \rho u_\alpha \frac{c_s}{c},
\end{align}
which ensures that the variable change is completely transparent on the macroscopic quantities of interest.

In the following, the Taylor expansion in $\mathrm{Kn}$ will be re-written on the new variable $f_i$ for several collision models, leading to the generic equation:
\begin{align}
	D_i f_i = -\frac{1}{\mathrm{Kn}}\left(f_i - f_i^{eq} - \mathrm{Kn}\, \Psi_i^* \right) + \sum_{n\geq 1} \mathrm{Kn}^n \, E_i^{(n)},
	\label{eq:DVBE_Errors}
\end{align}
where $E_i^{(n)}$ are the deviations from the DVBE at each order in Knudsen number, which can be explicitly expressed as functions of $\Delta t/\tau$ and spatial derivatives of the macroscopic quantities of interest. In presence of a body-force term, $E_i^{(n)}$ can also account for the numerical errors induced by their numerical discretization (\textit{e.g.} centered or upwind finite differences). In particular, it will be shown that in the linear approximation, $E_i^{(n)}$ involves $(n+1)$th-order spatial derivatives only. %Also note that focusing on the explicit expressions of $E_i^{(n)}$ will provide strong conclusions on two important numerical aspects of the LB scheme:
%\begin{itemize}
%	\item looking at the behavior of $E_i^{(n)}$ when $\Delta t/\tau \rightarrow 0$ highlights the \textit{consistency} of the LB scheme with a desired physics (at a given Knudsen order),
%	\item looking at $(\Delta t/\tau)$-dependent terms evidences the numerical errors induced by the time and space discretization of the LB scheme.
%\end{itemize}

The $*$ exponent will be dropped in the rest of the article for the time $t$, position $\boldsymbol{x}$ and velocity $\boldsymbol{u}$.

\subsection{Macroscopic equations at leading Knudsen orders}
\label{sec:macroscopic_equations}

Once the deviation terms from the DVBE $E_i^{(n)}$ are found at given Knudsen orders for the considered collision model, the next step of the analysis is to evidence their impact on the macroscopic evolution equations. After computing the moments of Eq.~(\ref{eq:DVBE_Errors}) and following the procedure detailed in~\ref{app:macroscopic_equations}, the following equations are obtained:
\begin{align}
	& \partial_t \rho + \partial_\alpha (\rho u_\alpha) = \mathrm{Kn} E_\rho^{(1)} + \mathrm{Kn}^2 E_\rho^{(2)} + \mathrm{Kn}^3 E_\rho^{(3)} + O(\mathrm{Kn}^4),
	\label{eq:mass_errors} \\
	& \partial_t (\rho u_\alpha) + \partial_\beta (\rho u_\alpha u_\beta) + c_s^2 \theta \partial_\alpha \rho = \mathrm{Kn} \left( \rho c_s^2 \theta \partial_\beta S_{\alpha \beta} \right) + \mathrm{Kn}E_{\rho u_\alpha}^{(1)} + \mathrm{Kn}^2 E_{\rho u_\alpha}^{(2)} + \mathrm{Kn}^3 E_{\rho u_\alpha}^{(3)} + O(\mathrm{Kn}^4),
	\label{eq:momentum_errors}
\end{align} 
with $S_{\alpha \beta} = \partial_\alpha u_\beta + \partial_\beta u_\alpha$. This closed set of equations is obtained as a hydrodynamic limit (in the sense $\mathrm{Kn} \rightarrow 0$) without performing the time-derivative expansion in smallness parameter proposed by Chapman and Enskog~\cite{CHAPMAN_Book_3rd_1970}. Hence, the Taylor expansion in Knudsen number is sufficient to exhibit the hydrodynamic limits of Eq.~(\ref{eq:DVBE_Errors}). In a certain manner, the procedure adopted here is similar to the Taylor expansion method proposed by Dubois~\cite{Dubois2008}, but with different notations yielding more straight physical interpretations thanks to the explicit appearance of $\mathrm{Kn}$.

Defining a dynamic viscosity $\mu=\mathrm{Kn} \rho c_s^2 \theta$, Eqs.~(\ref{eq:mass_errors})-(\ref{eq:momentum_errors}) can be identified as the isothermal form of the NS equations including a bulk viscosity, as previously shown by Dellar~\cite{Dellar2001}. As expected by a Chapman-Enskog expansion, viscous effects explicitly appear at first-order in Knudsen number. These equations are affected by the deviation terms $E_{\rho}^{(n)}$ and $E_{\rho u_\alpha}^{(n)}$. Like the $E_i^{(n)}$ terms of Eq.~(\ref{eq:DVBE_Errors}), they can be explicitly expressed as derivatives of macroscopic quantities, and \textit{a priori} depend on the numerical parameter $\Delta t/\tau$. Moreover, they can be systematically decomposed into two contributions:
\begin{itemize}
	\item Looking at the behavior of $E_\rho^{(n)}$ and $E_{\rho u_\alpha}^{(n)}$ when $\Delta t/\tau \rightarrow 0$ highlights the \textit{consistency} of the LB scheme with a desired physics (at a given Knudsen order).
	\item Looking at $(\Delta t/\tau)$-dependent terms evidences the numerical errors induced by the time and space discretization of the LB scheme.
\end{itemize}
For instance, regarding the first-order deviations $E_\rho^{(1)}-E_{\rho u_\alpha}^{(1)}$, a consistency error is expected for non-corrected models with the D1Q3 and D2Q9 lattices, which is the well known Mach error in the shear stress tensor. Moreover, no first-order numerical error is expected for second-order accurate schemes, like the BGK one~\cite{Dubois2008}. Since these terms appear at the same level (first-order in Kn) as the viscous effects, they can be directly interpreted as an additional viscosity arising in the macroscopic equations. A second-order accuracy is then mandatory if one wants to recover the Navier-Stokes behavior in the asymptotic limit. Any first-order error would result in an incorrect viscosity, depending on the numerical model under consideration. In the linear approximation, it will be shown that these terms only involve second-order gradients.

With a similar reasoning, it can be shown that:
\begin{itemize}
	\item second-order deviations $E_\rho^{(2)}-E_{\rho u_\alpha}^{(2)}$, involving third-order gradients only in the linear approximation, correspond to the leading \textit{dispersion} errors,
	\item third-order deviations $E_\rho^{(3)}-E_{\rho u_\alpha}^{(3)}$, involving fourth-order gradients only in the linear approximation, can be interpreted as a \textit{numerical hyperviscosity} in the macroscopic equations, yielding a \textit{dissipation} error.
\end{itemize}
In the following, the deviations terms $E_\rho^{(n)}$ and $E_{\rho u_\alpha}^{(n)}$ will be computed for several models up to the third-order assuming a linear approximation, which is sufficient to exhibit their effective viscosity and hyperviscosity. For this purpose and given the complexity of the problems under consideration, the computer algebra system Maxima~\cite{maxima} is used following the methodology proposed in \ref{app:macroscopic_equations}.

\subsection{Hyperviscous degeneracy}
\label{sec:hyperviscous_degeneracy}

In the forthcoming investigations of several collision models, it will be shown that the $E_i^{(1)}$ term has no macroscopic contribution in most of the LB schemes of interest, meaning that they do not include any numerical viscosity. In order to explain their stability and dissipation properties, it is therefore required to focus on the $E_i^{(3)}$ term, related to hyperviscous effects, which is the purpose of the present section.

\begin{figure}
    \centering
    \begin{subfigure}[b]{0.48\textwidth}
    \centering
    \includegraphics[scale=0.9]{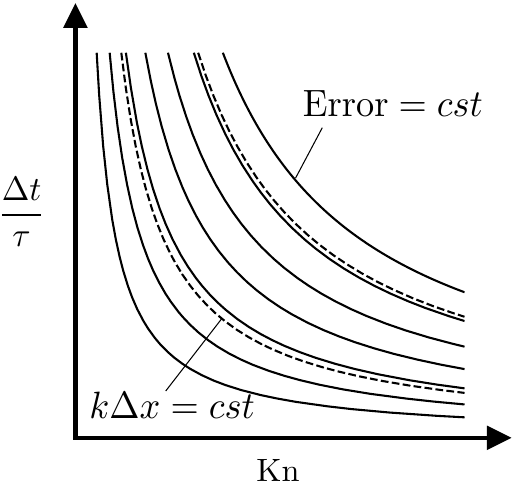}
    \caption{No degeneracy: Error$=O \left( \mathrm{Kn}^2 \left( \frac{\Delta t}{\tau} \right)^2 \right)$}
    %\label{.}
    \end{subfigure}
    \hfill
    \begin{subfigure}[b]{0.48\textwidth}
    \centering
    \includegraphics[scale=0.9]{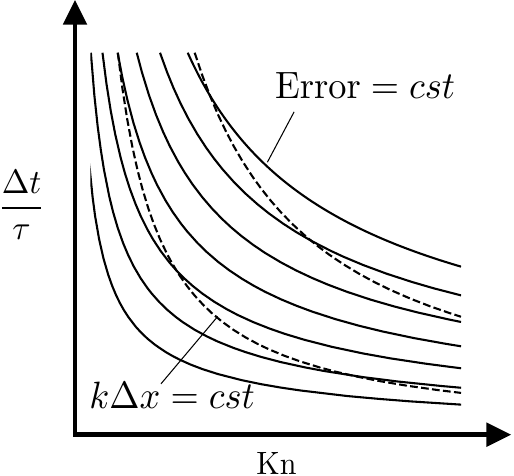}
    \caption{Degeneracy: Error$=O \left( \mathrm{Kn}^2 \left( \frac{\Delta t}{\tau} \right)^3 \right)$}
    %\label{.}
    \end{subfigure}
    \caption{Schematic plot of the link between numerical ($\Delta t/\tau$-related) and asymptotic ($\mathrm{Kn}$-related) errors for a discretized spatial phenomena described by a constant value of $k \Delta x = \mathrm{Kn} \, \Delta t/\tau$. \label{fig:Kn_dt}}
\end{figure}

A particular quantity of interest is the order of magnitude of the ratio between the numerical hyperviscosity and the expected viscous effects, expressed as:
\begin{align}
	\Gamma = \frac{\mathrm{Kn}^3 E_i^{(3)}}{\mathrm{Kn}} = \mathrm{Kn}^2 E_i^{(3)}.
	\label{eq:ratio_visco_hypervisco}
\end{align}
In order to obtain a more explicit expression of this quantity, note that the definition of the Knudsen number yields
\begin{align}
	\mathrm{Kn} = k \Delta x \frac{\tau}{\Delta t}, 
\end{align}
where $\Delta x$ is the mesh size, linked with the time step by an acoustic scaling $c=c_s \sqrt{\theta} \Delta x/\Delta t$~\cite{Kruger2017}. $k\Delta x$ is directly related to the number of points used to discretize the characteristic length $l_p=1/k$. Especially, for a plane monochromatic wave, the number of points per wavelength is $N_{ppw} = 2\pi/(k\Delta x)$. Furthermore, it will be shown that, in every analysis performed in the present article, the third-order deviation in $\mathrm{Kn}$ largely depends on the numerical parameter $\Delta t/\tau$, so that
\begin{align}
	E_i^{(3)} \sim \left(\frac{\Delta t}{\tau}\right)^n \qquad \mathrm{when} \qquad \frac{\tau}{\Delta t} \ll 1,
\end{align} 
where $n$, the maximal power of $(\Delta t/\tau)$ present in $E_i^{(3)}$, depends on the collision model under investigation. Hence, Eq.~(\ref{eq:ratio_visco_hypervisco}) becomes
\begin{align}
	\Gamma \sim (k \Delta x)^2 \left( \frac{\Delta t}{\tau} \right)^{n-2}.
\end{align}
For a given finite value of $k\Delta x$, \textit{i.e.} for a given discretized phenomenon, three cases can now be distinguished:
\begin{enumerate}
	\item $n < 2$: hyperviscous effects induced by $E_i^{(3)}$ are negligible compared to the expected viscous effects when the dimensionless relaxation time is low ($\tau/\Delta t \ll 1$),
	\item $n=2$: the ratio $\Gamma$ between hyperviscous and viscous effects is not affected by the dimensionless relaxation time $\tau/\Delta t$,
	\item $n>2$: hyperviscous effects induced by $E_i^{(3)}$ become very large compared to the expected viscous effects when $\tau/\Delta t \ll 1$.
\end{enumerate}
The third case is referred to as a \textit{hyperviscous degeneracy} in the rest of the article. It consists in a resurgence, unexpected from a physical point of view, of high-order Knudsen effects caused by a large numerical pre-factor ($E_i^{(3)}$ in this case). %Since the latter has a purely numerical origin, the macroscopic effects of the degeneracy on a solution are hardly predictable \textit{a priori}, this is why particular investigations are required for each collision model. 
This phenomenon is illustrated in Fig.~\ref{fig:Kn_dt}, where typical iso-contours of errors are represented in solid lines. For a given discretized spatial phenomena (represented by constant values of $k \Delta x$ in dashed lines), the error is independent of $\Delta t/\tau$ in the non-degenerated case, but only depends on the resolution $k \Delta x$. On the contrary, it increases as $\Delta t/\tau$ increases in the degenerated one.

In Sec.~\ref{sec:BGK}, it will be shown that the BGK model does not suffer from such a discrepancy. However, regularized models are subject to this phenomenon as shown in Sec.~\ref{sec:regularized_models}. %This leads to either severe instabilities or over-dissipation when the dimensionless relaxation time $\tau/\Delta t$ decreases, as evidenced in a previous work thanks to linear stability analyses (LSA)~\cite{Wissocq2020}. 
In Sec.~\ref{sec:MRT_models}, it will be shown that any MRT model with constant relaxation times can \textit{a priori} be subject to a degeneracy.

\subsection{Numerical validation: linear analyses}
\label{sec:validation_LSA}

Once the errors terms of a given LB collision model have been found, they can be systematically validated by performing a linear analysis of the mass and momentum equations~(\ref{eq:mass_errors})-(\ref{eq:momentum_errors}) and comparing it with an analysis of the LB scheme of Eq.~(\ref{eq:LBM_streaming}). As a reminder, linear (von Neumann) analyses are a powerful tool to exhibit the numerical properties of a linearized scheme~\cite{VonNeumann1950, sengupta2007}, which has been largely used to analyse the properties of the LBM~\cite{Sterling1996, Worthing1997, Lallemand2000, Marie2009, xu2011, xu2012, Cleon2014, Wissocq2019, Hosseini2019a, Coreixas2020, Wissocq2020}.

%In the present context, to perform analyses of Eqs.~(\ref{eq:mass_errors})-(\ref{eq:momentum_errors}), the linearized deviation terms $E_\Phi^{(n)}$ can be written under the following general form:
%\begin{align}
%	E_\Phi^{(n)} = A_{\Phi,\alpha_1..\alpha_{n+1}}^{(n+1)} \frac{\partial^{n+1} \rho}{\partial \alpha_1 .. \partial \alpha_{n+1}} + B_{\Phi,\alpha_1..\alpha_{n+1}}^{(n+1)} \frac{\partial^{n+1} (\rho u_x)}{\partial {\alpha_1} .. \partial {\alpha_{n+1}}} + C_{\Phi,\alpha_1..\alpha_{n+1}}^{(n+1)} \frac{\partial^{n+1} (\rho u_y)}{\partial {\alpha_1} .. \partial {\alpha_{n+1}}}, \qquad \Phi \in \{\rho, \rho u_\alpha \},
%	\label{eq:deviation_terms_linearized}
%\end{align} 
%where implicit summations are performed on indices $ \alpha_1,..,\alpha_{n+1} \in \{x,y\} $ and $\boldsymbol{A}_\Phi^{(n)}$, $\boldsymbol{B}_\Phi^{(n)}$ and $\boldsymbol{C}_\Phi^{(n)}$ are constant coefficients involving the mean base (space- and time-averaged) density and velocity. In 1D, every index is equal to $x$ and $\boldsymbol{C}_\Phi^{(n)}=0$ so that, for instance, the first-order deviation of the mass equation $E_\rho^{(1)}$ can be written as
%\begin{align}
%	E_\rho^{(1)} = A_{\rho,xx}^{(2)} \frac{\partial^2 \rho}{\partial x^2} + B_{\rho,xx}^{(2)} \frac{\partial^2 (\rho u_x)}{\partial x^2}.
%\end{align}
Applied to a system like Eqs.~(\ref{eq:mass_errors})-(\ref{eq:momentum_errors}), the purpose of the von Neumann analysis is to seek linear solutions as
\begin{align}
	\Phi(\boldsymbol{x}, t) &= \overline{\Phi} + \Phi'(\boldsymbol{x}, t), \qquad \Phi' \ll \overline{\Phi}, \\
	&= \overline{\Phi} + \hat{\Phi} \exp\left( \frac{\mathrm{i}}{k \Delta x} (\boldsymbol{k}\cdot \boldsymbol{x} \Delta x - \omega t \Delta t) \right),
	\label{eq:plane_monochromatic_wave}
\end{align}
for $\Phi \in \{\rho, \rho u_\alpha \}$, where $\overline{\Phi}$ are mean base quantities, constant in space and time and $\Phi'$ are small fluctuations with regards to $\overline{\Phi}$. In the von Neumann formalism, they are sough as plane monochromatic waves, involving the (complex) amplitudes $\hat{\Phi}$, the complex unity $\mathrm{i}$ defined such that $\mathrm{i}^2=-1$, the real wavenumber vector $\boldsymbol{k}=(k_x, k_y)$ and the complex pulsation $\omega$. Note that in Eq.~(\ref{eq:plane_monochromatic_wave}), time and space variables $\boldsymbol{x}$ and $t$ are dimensionless using the non-dimensionalization of Sec.~\ref{sec:dimensionless_DVBE}, which explains the presence of the factor $1/(k\Delta x)$ in the equation. Moreover, $k$, related to the characteristic length of Sec.~\ref{sec:dimensionless_DVBE}, is defined here as the norm of the wavenumber vector. Using the Nyquist-Shannon sampling theorem~\cite{Nyquist1928, Shannon1949}, the wavenumber vector is restricted to $k_x, k_y \leq \pi/\Delta x$. For this reason, the set of dimensionless parameters $(k_x \Delta x, k_y \Delta x, \tau/\Delta t)$ will be adopted in the linear analyses, instead of the Knudsen number which is a combination of them.

A linearization of Eqs.~(\ref{eq:mass_errors})-(\ref{eq:momentum_errors}) followed by the injection of plane monochromatic waves yield an eigenvalue problem 
\begin{align}
    \omega^* \mathbf{U} = \mathbf{M}^{(p)} \mathbf{U} + O \left({k^*}^{p+2} \right).
    \label{eq:LSA_macros_Errors}
\end{align}
where $\omega^*=\omega \Delta t$, $\mathbf{U}=[\rho', (\rho u_\alpha)']^T$, $\mathbf{M}$ is a matrix of size $(D+1)$ and $p$ is the order of truncation of the deviation terms. The explicit expression of this matrix, depending on $\boldsymbol{k}$, $\Delta x$, $\tau$, $\Delta t$ and the deviation terms, is provided in~\ref{app:matrices_LSA}. 
The complex pulsation $\omega$ can then be found as an eigenvalue of the matrix $\mathbf{M}$.

The eigenvalues of the linear system (\ref{eq:LSA_macros_Errors}) will be systematically compared to that of the linearized LB scheme, which reads in the general form
\begin{align}
    e^{-\mathrm{i} \omega^*_{\mathrm{LBM}}} \mathbf{F} = \mathbf{M}^{\mathrm{LBM}} \mathbf{F},
	\label{eq:LSA_LBM}
\end{align}
where the matrix $\mathbf{M}^{\mathrm{LBM}}$ depends on the adopted collision model. The corresponding matrices are recalled in \ref{app:matrices_LSA} for the collision models under consideration in the present work.

\begin{figure}
    \centering
    \includegraphics[scale=0.9]{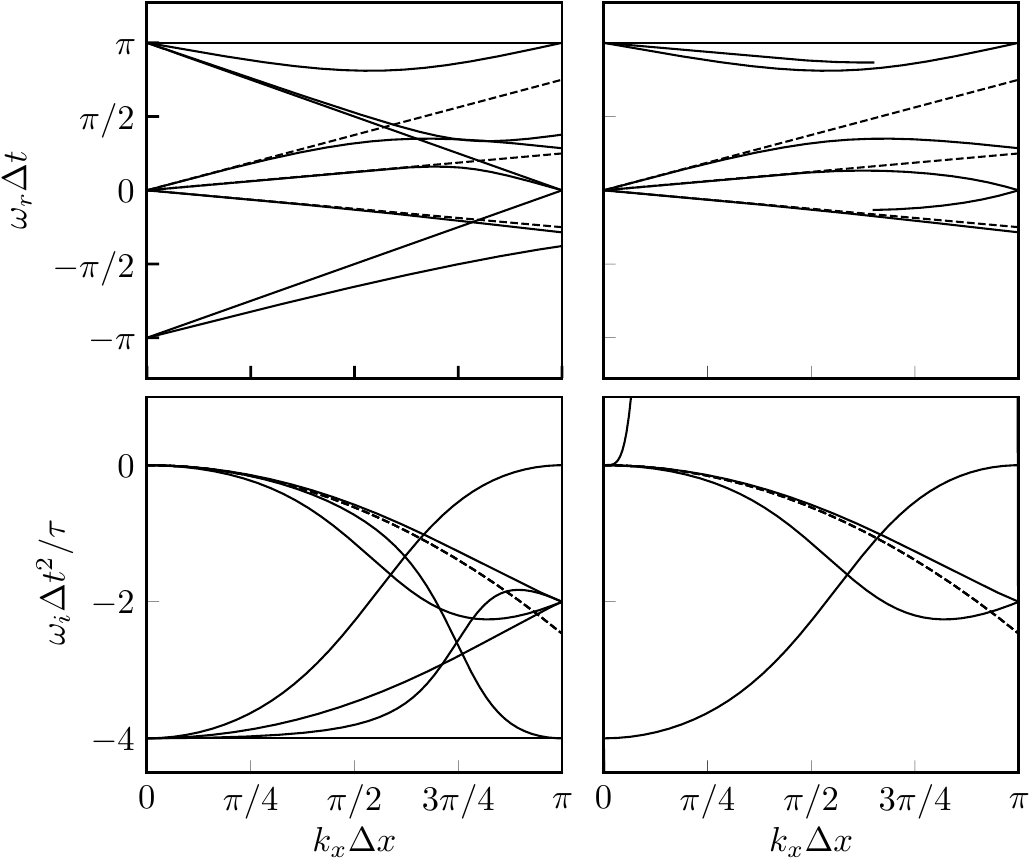}
    \caption{Propagation (top) and dissipation (bottom) curves of the linear analyses performed on the isothermal NS equations (\blackdashedline) and the LB scheme with the non-corrected D2Q9 lattice and $N^*=4$ (\blacklinefine). A horizontal mean flow is considered with $\overline{u_x}=0.5 c_s\sqrt{\theta}$, $\theta=0.75$ and $\tau/\Delta t=10^{-5}$. Left: BGK, right: PR collision models.} 
    \label{fig:LSA_NS_BGK_PR}
\end{figure}

Such analyses are illustrated in Fig.~\ref{fig:LSA_NS_BGK_PR}, where the eigenvalues of both systems (\ref{eq:LSA_macros_Errors})-(\ref{eq:LSA_LBM}) are displayed for $k_x \Delta x \in [0, \pi]$, $k_y=0$. In the case of Eq.~(\ref{eq:LSA_macros_Errors}), no deviation term is considered here, meaning that the isothermal NS equations are actually studied. Regarding Eq.~(\ref{eq:LSA_LBM}), two models are displayed for the non corrected D2Q9 lattice with $N=4^*$: the BGK and the projected regularization (PR). %The parameters of these analyses are: $\tau/\Delta t=10^{-5}$, $\theta=0.75$, $\overline{u_x}= \mathrm{Ma}\, c_s \sqrt{\theta}$ with $\mathrm{Ma}=0.5$ and $\overline{u_y}=0$, as usually adopted as representative of common LB simulations of air flow. 
In-depth discussion of such analyses can be found for instance in~\cite{Wissocq2020}. Here, one can focus on two observations:
\begin{enumerate}
	\item While the three eigenvalues of the isothermal NS equations tend towards zero when $k_x \Delta x \rightarrow 0$ (\textit{i.e.} in the well-resolved limit), this is not the case of all the eigenvalues of the LBM. With the BGK (resp. PR) model, 6 modes (resp. 3 modes) tend towards $\pi$ in this limit. The corresponding modes will be referred to as \textit{non-hydrodynamic} (NH) modes.
	\item The analysis of the PR model evidences an instability ($\omega_i/\tau >0$) of a hydrodynamic mode, even in the low-wavenumber limit. Further analyses allow identifying this unstable mode as a shear wave~\cite{Wissocq2020}. This instability is not present with the BGK collision model, for which the dissipation of the hydrodynamic modes is close to that of the NS equations.  
\end{enumerate}
Note that because of modal interactions occurring in the LB scheme, the identification of NH modes as continuous functions $\omega^*=f(k^*)$ is not straightforward~\cite{Wissocq2019}. However, since the present work focuses on Taylor expansions for $\mathrm{Kn} \rightarrow 0$, it is sufficient to adopt here the above definition of NH modes. The particular case of these modes is discussed in the next section. The fact that the BGK model induces a very low numerical dissipation will be theoretically explained in Sec.~\ref{sec:BGK}, and the instability of the PR model will be explained by a hyperviscous degeneracy in Sec.~\ref{sec:Regul_PR_RR}.

\subsection{Particular case of non-hydrodynamic modes}
\label{sec:non_hydrodynamic_modes}

The validity of the methodology proposed in the present work is actually ensured provided that distribution functions $g_i$ are sufficiently continuous so that a Taylor expansion in $\mathrm{Kn}$ can be rigorously performed. One necessary condition is then
\begin{align}
	g_i \left( \boldsymbol{x}+ \boldsymbol{e_i}\mathrm{Kn} \frac{\Delta t}{\tau}, t+\mathrm{Kn} \frac{\Delta t}{\tau} \right) = g_i^c (\boldsymbol{x}, t) \rightarrow g_i (\boldsymbol{x}, t) \qquad \mathrm{when} \qquad \mathrm{Kn} \rightarrow 0.
	\label{eq:continuity_condition}
\end{align}
It is shown in the present section that this condition is not necessarily ensured, which further explains the behavior of non-hydrodynamic modes evidenced in Fig.~\ref{fig:LSA_NS_BGK_PR}. 
For this purpose, let us consider the non-corrected BGK collision model~\cite{Bhatnagar1954}, for which
\begin{align}
	g_i^c = g_i -\frac{\Delta t}{\tau + \Delta t/2} \left( g_i - f_i^{eq} \right).
\end{align}
With this model, the continuity condition (\ref{eq:continuity_condition}) for a given ratio $\Delta t/\tau>0$ is equivalent to
\begin{align}
	g_i(\boldsymbol{x}, t) \rightarrow f_i^{eq}(\boldsymbol{x},t) \qquad \mathrm{when} \qquad \mathrm{Kn} \rightarrow 0.
\end{align}
Let us now consider a given set of non-zero distribution functions $g_i$ verifying
\begin{align}
	\sum_i g_i = 0, \qquad \sum_i e_{i, \alpha} g_i = 0.
	\label{eq:null_macros}
\end{align}
By virtue of the rank-nullity theorem, the possible set of distribution functions verifying this condition is not reduced to zero: it is of dimension 6 with the D2Q9 lattice and of dimension 1 with the D1Q3 lattice. Given that zeroth- and first-order moments are both null with this choice, one has $f_i^{eq}(g_j)=0$ so that the BGK-LB scheme reads
\begin{align}
	g_i \left(\boldsymbol{x} + \boldsymbol{e_i} \, \mathrm{Kn} \frac{\Delta t}{\tau}, t+ \mathrm{Kn}\frac{\Delta t}{\tau} \right) = \frac{\tau - \Delta t/2}{\tau + \Delta t/2} \, g_i(\boldsymbol{x}, t),
	\label{eq:over_relaxation}
\end{align}
which does not tend towards $g_i(\boldsymbol{x},t)$ when $\mathrm{Kn} \rightarrow 0$. As a consequence, the Taylor expansion in Knudsen number cannot be performed on distributions functions verifying Eq.~(\ref{eq:null_macros}). Looking now more carefully at Eq.~(\ref{eq:over_relaxation}), the linear modes $\hat{g}_i$ ensuring this equation are those for which $\tau \neq \Delta t/2$ and
\begin{align}
	e^{- \mathrm{i} \omega^*_{\mathrm{LBM}}} = \frac{\tau - \Delta t/2}{\tau + \Delta t/2},
\end{align}
when $\mathrm{Kn}$ tends towards zero. This leads to two cases:
\begin{align}
	&\tau/\Delta t < 1/2: \qquad \Re\left(\omega_{\mathrm{LBM}}^*\right) = \pi,\qquad \Im\left(\omega_{\mathrm{LBM}}^*\right) = -\ln \left( \frac{\Delta t/2-\tau}{\tau+\Delta t/2} \right), \nonumber \\
	&\tau/\Delta t > 1/2:\qquad \Re\left(\omega_{\mathrm{LBM}}^*\right) = 0,\qquad \Im\left(\omega_{\mathrm{LBM}}^*\right) = -\ln \left( \frac{\tau - \Delta t/2}{\tau+\Delta t/2} \right). \nonumber
\end{align}
The first case corresponds to \textit{over-relaxed} modes~\cite{Kruger2017, Dellar2003}, whose amplitude is reversed at each iteration due to a real dimensionless pulsation close to $\pi$. They exactly match the non-hydrodynamic modes evidenced in the linear analyses of Sec.~\ref{sec:validation_LSA}. This spurious numerical behavior can be observed in practice, and may be very harmful for low-Reynolds and under-resolved simulations~\cite{Astoul2020}. One of the aim of more sophisticated collision models is precisely to damp, or even filter out, these modes. The second case corresponds to \textit{under-relaxed} modes~\cite{Kruger2017}. \newline

As a consequence, non-hydrodynamic modes are \textit{discontinuous} modes. The Taylor expansion method in Knudsen number, as proposed in this work, cannot be used to predict their numerical behavior. The discontinuity of the deviation terms involved in these modes could even be simply guessed by looking at the linear analyses of Fig.~\ref{fig:LSA_NS_BGK_PR}: due to the phase-shift of $\pi$ encountered by these modes, the deviation terms are necessarily discontinuous in $\mathrm{Kn}$. Note that this discontinuity is not in disagreement with a consistency of the LB scheme with the DVBE, which requires an investigation of the limit $\tau/\Delta t \rightarrow 0$ as shown in Fig.~\ref{fig:asymptotic_convergence_LBM}.

Hence, the rest of the article will focus on hydrodynamic modes only, more precisely on the ability of the present methodology to predict and quantify their dispersion and dissipation properties. It is therefore sufficient to investigate the deviation terms of the macroscopic equations (\ref{eq:mass_errors})-(\ref{eq:momentum_errors}), for which only hydrodynamic modes are considered, contrary to the DVBE (\ref{eq:DVBE_Errors}). Every analysis will have to be validated \textit{a posteriori}, which will be performed thanks to the linear analyses introduced in Sec.~\ref{sec:validation_LSA}.

\subsection{Summary of the methodology}

The methodology described in this section can be applied to any LB collision model. It is summarized as follows:
\begin{enumerate}
	\item Non-dimensionalization of the LB scheme, so that two dimensionless numbers appear: the Knudsen number $\mathrm{Kn}$ and the numerical parameter $\Delta t/\tau$.
	\item Taylor expansion in $\mathrm{Kn}$, involving $\Delta t/\tau$-dependent pre-factors.
	\item A variable change $f_i = \psi(g_i)$ is then performed to mimic the DVBE, in particular its collision term. The admissibility of this variable change ($\sum_i f_i = \sum_i g_i$ and $\sum_i e_{i, \alpha} f_i = \sum_i e_{i, \alpha} g_i$) has to be checked. Deviation terms from the DVBE can then be exhibited, as shown in Eq.~(\ref{eq:DVBE_Errors}). %The latter include potential consistency errors with the DVBE and numerical errors induced by the discretization.
	\item A closed set of macroscopic equations can then be obtained at a desired Knudsen order, in the linear approximation, following~\ref{app:macroscopic_equations}. They can be identified as the isothermal form of NS equation with deviation terms. The latter include: (1) consistency errors with the NS equations, (2) $\Delta t/\tau$-dependent numerical errors.
	\item Validation of the macroscopic deviation terms using the linear analyses introduced in Sec.~\ref{sec:validation_LSA}. 
\end{enumerate}

In all the analyses, a potential hyperviscous degeneracy will be scrutinized by looking at the numerical pre-factors of third-order deviations, as discussed in Sec.~\ref{sec:hyperviscous_degeneracy}.

\section{Application to the BGK collision model}
\label{sec:BGK}

This section aims at applying the Taylor expansion in $\mathrm{Kn}$ to the common BGK collision model, so as to quantify and validate the impacts of numerical errors on the intended physics. Given the poor practical interest of the BGK model, the main interests of this section are rather pedagogical, since it provides a simple application of the above methodology. Moreover, the resulting deviation terms will be used as a reference when comparing other schemes. Notably, it will provide some explanations to the very low numerical dissipation of the BGK collision model compared to other ones, previously observed in the literature~\cite{Marie2009, Wissocq2020}.

Both D1Q3 and D2Q9 lattices will be considered in this section including different Hermite-based equilibria, with and without correction terms. In the former case, the effect of their discretization using finite differences, with upwind and centered schemes, will be investigated.

\subsection{General Taylor expansion of the BGK model}

The general form of post-collision distribution functions involved in the BGK collision model reads:
\begin{align}
	g_i^c = g_i - \frac{\Delta t}{\tau + \Delta t/2} \left( g_i - f_i^{eq} - \mathrm{Kn} \Psi_i^d \right),
\end{align}
where $\Psi_i^d$ is a general discretized form of the body-force term. It can be either zero in the case of non-corrected LBM, or equal to the $\Psi_i$ terms of Eqs.~(\ref{eq:correction_D1Q3})-(\ref{eq:correction_D2Q9_N2}) with possible deviations induced by the finite-difference estimation of gradients. The latter will be further detailed for each case under consideration.

After performing the Taylor expansion of Sec.~\ref{sec:Taylor_expansion_Kn}, the variable change suggested by Eq.~(\ref{eq:variable_change}) reads
\begin{align}
	\frac{\Delta t}{\tau + \Delta t/2} \left( g_i - f_i^{eq} - \mathrm{Kn} \Psi_i^d \right) = \frac{\Delta t}{\tau} \left(f_i - f_i^{eq} - \mathrm{Kn}\, \Psi_i^d \right),
\end{align}
which leads to
\begin{align}
	g_i = f_i + \frac{\Delta t}{2 \tau} \left( f_i - f_i^{eq} - \mathrm{Kn} \Psi_i^d \right).
\end{align}
This change of variable is in line with what can be found in the literature, especially in an \textit{a priori} construction of the BGK-LB scheme~\cite{He1998}. Moreover, since the correction term of Feng \textit{et al.}~\cite{Feng2019} involves a second-order Hermite moment only (if not null), its zeroth- and first-order moments vanish, which leads to
\begin{align}
	\sum_i g_i = \sum_i f_i = \rho,\qquad \sum_i {e_{i,\alpha}} g_i = \sum_i {e_{i,\alpha}}f_i = \rho {u}_\alpha,
\end{align}
so that the variable change is admissible. The following equivalent partial differential equation is then obtained on $f_i$:
\begin{align}
	D_i f_i = -\frac{1}{\mathrm{Kn}} f_i^{(1)} - \sum_{n\geq 2} \frac{\mathrm{Kn}^{n-1}}{n!} \left( \frac{\Delta t}{\tau} \right) ^{n-1} D_i^n f_i - \sum_{n\geq 1} \frac{\mathrm{Kn}^{n-1}}{2n!} \left( \frac{\Delta t}{\tau} \right)^n D_i^n f_i^{(1)},
	\label{eq:Taylor_BGK_1}
\end{align}
where $f_i^{(1)}$ is defined as
\begin{align}
	f_i^{(1)} = f_i - f_i^{eq} - \mathrm{Kn} \Psi_i^d.
	\label{eq:def_fi1}
\end{align}
In order to obtain a more convenient form of the deviation terms, one then needs to express the derivatives of $f_i^{(1)}$ appearing in Eq.~(\ref{eq:Taylor_BGK_1}) as function of $f_i$ only. For this purpose, Eq.~(\ref{eq:Taylor_BGK_1}) can be written as:
\begin{align}
	f_i^{(1)} = -\sum_{n \geq 1} \frac{\mathrm{Kn}^n}{n!} \left( \frac{\Delta t}{\tau} \right)^{n-1} D_i^n f_i - \sum_{n \geq 1} \frac{\mathrm{Kn}^n}{2n!} \left( \frac{\Delta t}{\tau} \right)^n D_i^n f_i^{(1)}.
	\label{eq:Taylor_BGK_2}
\end{align}
Let us now seek a series expansion of $f_i^{(1)}$ under the following form:
\begin{align}
	f_i^{(1)} = \sum_{n \geq 1} \frac{c_n}{n!} \mathrm{Kn}^n \left( \frac{\Delta t}{\tau} \right)^{n-1} D_i^n f_i.
	\label{eq:f1_series}
\end{align}
Injecting this expression in Eq.~(\ref{eq:Taylor_BGK_2}) yields
\begin{align}
	\sum_{n \geq 1} \frac{c_n}{n!} \mathrm{Kn}^n \left( \frac{\Delta t}{\tau} \right)^{n-1} D_i^n f_i &= - \sum_{n \geq 1} \frac{\mathrm{Kn}^n}{n!} \left( \frac{\Delta t}{\tau} \right)^{n-1} D_i^n f_i - \frac{1}{2} \sum_{n\geq 1} \sum_{m \geq 1} \frac{c_m}{n! m!} \mathrm{Kn}^{n+m} \left( \frac{\Delta t}{\tau} \right)^{n+m-1} D_i^{n+m} f_i \nonumber \\
	&= - \sum_{n \geq 1} \frac{\mathrm{Kn}^n}{n!} \left( \frac{\Delta t}{\tau} \right)^{n-1} D_i^n f_i - \frac{1}{2} \sum_{n \geq 2}\mathrm{Kn}^n \left( \frac{\Delta t}{\tau} \right)^{n-1} \left( \sum_{k=1}^{n-1} \frac{c_k}{k! (n-k)!} \right) D_i^n f_i. \nonumber
\end{align}
This leads to the following recursive relation on the coefficients $c_n$:
\begin{align}
c_1 = -1, \qquad \forall n \geq 2,\ c_n = -1 - \frac{1}{2} \sum_{k=1}^{n-1} \binom{n}{k}c_k,
\end{align}
which can also be written as
\begin{align}
	c_0=2,\qquad \forall n\geq 1, \ c_n &= - \sum_{k=0}^{n}\binom{n}{k}c_k.
\end{align}
In \ref{app:Genocchi_Bernoulli}, it is shown that these coefficients are related to both the Genocchi numbers $G_n$ and the Bernoulli numbers $B^-_n$ as
\begin{align}
	\forall n \geq 0,\ c_n=2\frac{G_{n+1}}{n+1}=4 \left(1-2^{n+1} \right)\frac{B_{n+1}^-}{n+1}.
\end{align}
This relation is helpful since the Bernoulli numbers are well documented in the literature. The interested reader may refer to~\cite{Arfken, Comtet, abramowitz-stegun, nist-handbook} for more informations on Genocchi and Bernoulli numbers. 
Some relations between Taylor expansions of the LB scheme and the Bernoulli polynomials have already been noticed in the literature~\cite{Dong2010}, but no demonstration could be provided to the authors' knowledge. The first ten coefficients of this expansion are detailed in Table~\ref{tab:cn_coefficients}. An important feature is that every odd Bernoulli number is null, so that only odd coefficients remain in Eq.~(\ref{eq:f1_series}). 

\begin{table}[h]
\begin{center}
\begin{tabular}{|c||c|c|c|c|c|c|c|c|c|c|c|}
  \hline
  $n$ & 0 & 1 & 2 & 3 & 4 & 5 & 6 & 7 & 8 & 9 & 10 \\
  \hline
  $B_n^-$ & - & -1/2 & 1/6 & 0 & -1/30 & 0 & 1/42 & 0 & -1/30 & 0 & 5/66 \\
  $G_n$ & - & 1 & -1 & 0 & 1 & 0 & -3 & 0 & 17 & 0 & -155 \\
  $c_n$ & 2 & -1 & 0 & 1/2 & 0 & -1 & 0 & 17/4 & 0 & -31 & 0    \\
  \hline
\end{tabular}
\end{center}
\caption{Expression of the ten firsts Bernoulli numbers $B^-_n$, Genocchi numbers $G_n$ and the $c_n$ coefficients involved in the series expansion of $f_i^{(1)}$.}
\label{tab:cn_coefficients}
\end{table}

Finally, injecting these expressions into Eq.~(\ref{eq:f1_series}) and after some manipulations, a general form of the Taylor expansion is obtained for the BGK collision model
\begin{align}
	D_i f_i = -\frac{1}{\mathrm{Kn}} \left( f_i - f_i^{eq} \right) + \Psi_i^d + \frac{\mathrm{Kn}^2}{12} \left( \frac{\Delta t}{\tau} \right)^2 D_i^3 f_i - \frac{\mathrm{Kn}^4}{5!} \left( \frac{\Delta t}{\tau} \right)^4 D_i^5 f_i + \sum_{p \geq 3} \frac{c_{2p+1}}{(2p+1)!}  \mathrm{Kn}^{2p} \left( \frac{\Delta t}{\tau} \right)^{2p} D_i^{2p+1} f_i.
	\label{eq:Taylor_BGK_3}
\end{align}
This equation differs from the dimensionless DVBE of Eq.~(\ref{eq:DVBE_dimensionless_2}) by the discretization errors in $\Psi_i^d$ and the deviation terms in $O(\mathrm{Kn}^2)$. Note that no first-order term in $\mathrm{Kn}$ appear, which is a direct consequence of the second-order accuracy of the BGK-LB scheme~\cite{He1998, Dubois2008, Dellar2013b}. Following the macroscopic interpretation of Sec.~\ref{sec:macroscopic_equations}, it means that the BGK-LBM induces no numerical viscosity.

In order to obtain an explicit expression of the deviation terms $E_i^{(n)}$ involved in Eq.~(\ref{eq:Taylor_BGK_3}), further development of the temporal and spatial derivatives included in $D_i^n f_i$ is required. This will be the purpose of Sec.~\ref{sec:BGK_errors_linearized} in the linear approximation. But before that, one can focus at the expression as it in order to explain the low-dissipation properties of the BGK collision model.

\subsection{Low dissipation of the BGK collision model}
\label{sec:low_dissipation_BGK}

An important feature of the BGK collision model directly appears in Eq.~(\ref{eq:Taylor_BGK_3}): when written under this form, only even orders in Knudsen number appear in the expansion. This is a direct consequence of the relation found between the $c_n$ coefficients and the Bernoulli numbers. 
Let us recall here that even and odd Knudsen orders are directly related to dispersion and dissipation errors, respectively. 
Hence, no dissipation error appears in the above relation. 
As a matter of fact, a dissipation error is implicitly present in the $D_i^{2p} f_i$ terms of the even order errors. A first insight into these dissipative terms is provided in this section in absence of body-force term ($\Psi_i^d=0$). For instance, let us focus on the second-order term of Eq.~(\ref{eq:Taylor_BGK_3}). Using the fact that $f_i=f_i^{eq} - \mathrm{Kn}\, D_i f_i^{eq}  + O(\mathrm{Kn}^2)$, one has
\begin{align}
	\frac{\mathrm{Kn}^2}{12} \left( \frac{\Delta t}{\tau} \right)^2 D_i^3 f_i = \frac{\mathrm{Kn}^2}{12} \left( \frac{\Delta t}{\tau} \right)^2 D_i^3 f_i^{eq} - \frac{\mathrm{Kn}^3}{12} \left( \frac{\Delta t}{\tau} \right)^2 D_i^4 f_i^{eq} + O(\mathrm{Kn}^4),
	\label{eq:Taylor_BGK_4}
\end{align}
so that a third-order error, related to the numerical hyperviscosity, actually arises. The latter involves the parameter $\Delta t/\tau$ at a power $n=2$. Following Sec.~\ref{sec:hyperviscous_degeneracy}, it means that the numerical hyperviscosity remains bounded and is not affected by the value of the dimensionless relaxation time. Hence, no hyperviscous degeneracy occurs. Given the parity of the deviation terms, this fundamental property of the BGK collision model may be extended to any error term of odd order, so that the numerical dissipation does not depend on the value of the dimensionless relaxation time. It is at the origin of the very low-dissipation properties of the model, despite its second-order accuracy, especially in comparison with other solvers~\cite{Marie2009}.

\subsection{Error terms in the linear approximation}
\label{sec:BGK_errors_linearized}

Starting from Eq.~(\ref{eq:Taylor_BGK_3}), the purpose of the present section is to obtain an explicit expression of the deviation terms appearing in Eq.~(\ref{eq:DVBE_Errors}) involving spatial derivatives of macroscopic quantities only. This is a prerequisite for obtaining the explicit deviations from the macroscopic equations, as proposed in~\ref{app:macroscopic_equations}. The focus is put here on the first-, second- and third-order terms, and a linear assumption will be assumed so as to simplify the expressions. \newline

First, the discretized body-force term $\Psi_i^d$ of Eq.~(\ref{eq:Taylor_BGK_3}) has to be considered. A general Taylor expansion in $\mathrm{Kn}$ is proposed for this term in~\ref{app:discretized_correction_terms} in the case of second-order centered (DCO2) and first-order upwind (DUO1) finite differences. 
Note that, while no first-order error is explicitly present in Eq.~(\ref{eq:Taylor_BGK_3}), such a term appears in the Taylor expansion of the body-force term with an upwind discretization. \newline

Then, a second-order estimation of the $D_i^3 f_i$ terms appearing in Eq.~(\ref{eq:Taylor_BGK_3}) is required. It starts by writing
\begin{align}
	D_i^3 f_i = D_i^3 f_i^{eq} - \mathrm{Kn}\, D_i^4 f_i^{eq} + \mathrm{Kn} \, D_i^3 \Psi_i + O(\mathrm{Kn}^2),
\end{align}
where the fact that $\Psi_i^d = \Psi_i + O(\mathrm{Kn})$ has been used. The $D_i$ terms can then be related to time and space derivatives as
\begin{align}
	& D_i^3  = \partial^3_{ttt} + 3 e_{i, \alpha} \partial^3_{tt\alpha} + 3 e_{i, \alpha} e_{i, \beta} \partial^3_{t\alpha \beta} + e_{i, \alpha} e_{i, \beta} e_{i, \gamma} \partial^3_{\alpha \beta \gamma}, \\
    & D_i^4 = \partial^4_{tttt} + 4 e_{i, \alpha} \partial^4_{ttt\alpha} + 6 e_{i, \alpha} e_{i, \beta} \partial^4_{tt\alpha \beta} + 4 e_{i, \alpha} e_{i, \beta} e_{i, \gamma} \partial^4_{t\alpha \beta \gamma} + e_{i, \alpha} e_{i, \beta} e_{i, \gamma} e_{i, \delta} \partial^4_{\alpha \beta \gamma \delta}.
\end{align}
These terms can be related to derivatives of macroscopic quantities only since, for any function $\Phi(\rho, \boldsymbol{u})$, the following chain rule can be applied:
\begin{align}
	\partial^n_{\alpha_1..\alpha_n} \Phi = \frac{\partial \Phi}{\partial \rho} \partial^n_{\alpha_1..\alpha_n} \rho + \frac{\partial \Phi}{\partial (\rho u_\beta)} \partial^n_{\alpha_1..\alpha_n} (\rho u_\beta),
	\label{eq:chain_rule}
\end{align}
for any combination of indices $\alpha_i \in \{t, x, y\}$. This chain rule can be applied to both $f_i^{eq}$ and $\Psi_i$, which are explicit functions of $\rho$ and $\boldsymbol{u}$. Note that the linear assumption is applied at this step since $\partial \Phi/\partial \rho$ and $\partial \Phi/\partial (\rho u_\beta)$ are assumed constant as the pre-factors of time/space derivatives. One can then dispose of time-derivatives of macroscopic quantities using the conservation equations at the desired Knudsen order, obtained by computing the zeroth- and first-order moments of Eq.~(\ref{eq:Taylor_BGK_3}):
\begin{align}
	\partial_t \rho = - \partial_{\gamma} (\rho u_\gamma) + O(\mathrm{Kn}^2), \qquad \partial_{t} (\rho u_\alpha) = -\partial_{\beta} \Pi_{\alpha \beta} + O(\mathrm{Kn}^2), 
\end{align}
where the fact that $\sum_i \Psi^d_i=\sum_i e_{i, \alpha} \Psi_i^d = 0$ has been used, and where
\begin{align}
	\Pi_{\alpha \beta} = \Pi^{eq}_{\alpha \beta} - \mathrm{Kn} \left( \partial_t \Pi^{eq}_{\alpha \beta} + \partial_\gamma Q^{eq}_{\alpha \beta \gamma} \right) + \mathrm{Kn} \sum_i e_{i, \alpha} e_{i, \beta} \Psi_i + O(\mathrm{Kn}^2).
\end{align}
Time and space derivatives of $\boldsymbol{\Pi}^{eq}$ and $\boldsymbol{Q}^{eq}$ can finally be disposed thanks to the chain rule of Eq.~(\ref{eq:chain_rule}). All these relations form a closed set of equations allowing an expression of $D_i^3 f_i$ involving spatial derivatives of macroscopic quantity only, at first-order in $\mathrm{Kn}$. For this purpose, and given the complexity of the problems, the simplification of these equations is performed using the computer algebra system Maxima~\cite{maxima}. \newline

Finally, the explicit expressions of the $E_i^{(n)}$ terms being known, the macroscospic equations (\ref{eq:mass_errors})-(\ref{eq:momentum_errors}) are obtained following the general procedure detailed in~\ref{app:macroscopic_equations}. Next sections provide the resulting deviation terms $E_\Phi^{(n)}$ obtained by these computations.

\subsection{D1Q3 lattice}
\label{sec:BGK_deviations_D1Q3}

With the D1Q3 lattice, the following macroscopic equations are recovered:
\begin{align}
	& \partial_t \rho + \partial_x(\rho u_x) = \mathrm{Kn} \, E_\rho^{(1)} + \mathrm{Kn}^2 \, E_\rho^{(2)} + \mathrm{Kn}^3 \, E_\rho^{(3)} + O(\mathrm{Kn}^4), \label{eq:D1Q3_mass} \\
	& \partial_t (\rho u_x) + \partial_x (\rho u_x^2) + c_s^2 \theta \, \partial_x \rho = \mathrm{Kn} \left( 2\rho c_s^2 \theta \, \partial^2_{xx} u_x \right) + \mathrm{Kn}\, E_{\rho u_x}^{(1)} + \mathrm{Kn}^2 \, E_{\rho u_x}^{(2)} + \mathrm{Kn}^3 \, E_{\rho u_x}^{(3)} + O(\mathrm{Kn}^4), \label{eq:D1Q3_momentum}
\end{align}
where the macroscopic deviation terms $E_\rho^{(n)}$ and $E_{\rho u_x}^{(n)}$ are provided below,  involving consistency errors $C_i$ and numerical errors $N_i$, whose explicit expressions are detailed in \ref{app:deviation_terms}.

\subsubsection{Absence of correction}

The mass and momentum equations (\ref{eq:D1Q3_mass})-(\ref{eq:D1Q3_momentum}) are obtained with
\begin{align}
    & E^{(1)}_\rho = 0, \qquad E^{(1)}_{\rho u_x} = -\frac{\partial^2 a^{(3)}_{eq, xxx}}{\partial x^2}, \nonumber \\
    & E^{(2)}_\rho = \left( \frac{\Delta t}{\tau} \right)^2 \frac{1}{12} \left( -\frac{\partial^3 a^{(3)}_{eq, xxx}}{\partial x^3} + 2\rho c_s² \theta \frac{\partial^3 u_x}{\partial x^3} \right), \qquad  E^{(2)}_{\rho u_x} = C_1 + \left( \frac{\Delta t}{\tau} \right)^2 N_1, \nonumber \\  
    & E^{(3)}_\rho = \left( \frac{\Delta t}{\tau} \right)^2 \frac{1}{12} \partial_{x} C_1, \qquad E^{(3)}_{\rho u_x} = C_2 + \left( \frac{\Delta t}{\tau} \right)^2 N_2. \label{eq:deviation_terms_D1Q3_noCorr}
\end{align}
The only first-order deviation from the isothermal NS equations is a modeling error related to $a^{(3)}_{eq, xxx}$ in the equation of momentum, which is the well-known Galilean invariance error induced by the lattice closure~\cite{QIAN_EPL_21_1993, SHAN2006}. Similar consistency errors denoted with $C_i$, \textit{i.e.} the remaining deviations terms when $\Delta t/\tau \rightarrow 0$, appear at second- and third-orders in $\mathrm{Kn}$, but they may be neglected at the NS level. The only numerical errors arising at $\mathrm{Kn}^2$ (dispersion error) and $\mathrm{Kn}^3$ (numerical hyperviscosity) are proportional to $(\Delta t/\tau)^2$, which is in agreement with the second-order accuracy of the BGK-LBM. As mentioned in Sec.~\ref{sec:low_dissipation_BGK}, no hyperviscous degeneracy occurs in this case. One can finally note that the numerical errors affecting the mass equation at a given order $n$ (in Knudsen number) are directly related to the consistency error of the momentum equation at order $n-1$. It therefore seems that the numerical errors in the mass conservation originate from consistency errors multiplied by a numerical factor. 

\subsubsection{Analytically computed correction}

Including now a correction term $\Psi_i$ without considering its discretization errors,  Eqs.~(\ref{eq:D1Q3_mass})-(\ref{eq:D1Q3_momentum}) are recovered with
\begin{align}
    & E^{(1)}_\rho = 0, \qquad E^{(1)}_{\rho u_x} = 0, \nonumber \\
    &  E^{(2)}_\rho = \left( \frac{\Delta t}{\tau} \right)^2 \frac{1}{12} \left( -\frac{\partial^3 a^{(3)}_{eq, xxx}}{\partial x^3} + 2\rho c_s² \theta \frac{\partial^3 u_x}{\partial x^3} \right), \qquad
     E^{(2)}_{\rho u_x} = C_3 + \left( \frac{\Delta t}{\tau} \right)^2 N_1, \nonumber \\
     & E^{(3)}_\rho = \left( \frac{\Delta t}{\tau} \right)^2 \frac{1}{12} \partial_{x} C_3, \qquad E^{(3)}_{\rho u_x} = C_4 + \left( \frac{\Delta t}{\tau} \right)^2 N_3. \label{eq:deviation_terms_D1Q3_CorrAnalytics}
\end{align}
A direct consequence of the correction is that there is no more first-order deviation from the isothermal NS equations. First- and second-order deviation terms only differ from the non-corrected case by the consistency terms, while the numerical errors remain the same. On the contrary, the third-order numerical error (numerical hyperviscosity) is affected by the correction.
Furthermore, even though the first-order consistency error is corrected by the introduction of $\Psi_i$, the second-order numerical error appearing in the mass conservation equation is surprisingly still related to the deviation term $a^{(3)}_{eq, xxx}$ induced by the lattice closure. Except this observation, the numerical error $E_\rho^{(3)}$ can still be related to the consistency error of $E_{\rho u_x}^{(2)}$.

\subsubsection{Correction with the DCO2 scheme}

Including now a correction with a DCO2 discretization, Eqs.~(\ref{eq:D1Q3_mass})-(\ref{eq:D1Q3_momentum}) are recovered with the exactly same deviation terms as with the analytically computed correction, except for the third-order term in the momentum equation:
\begin{align}
    & E^{(3)}_{\rho u_x} = C_4 + \left(\frac{\Delta t}{\tau} \right)^2 \left( N_3 + \frac{1}{6} \frac{\partial^4 a^{(3)}_{eq, xxx}}{\partial x^4} \right).
\end{align}
Naturally, the discretization of $\Psi_i$ only impacts the numerical error and does not affect the consistency of the scheme.

\subsubsection{Correction with the DUO1 scheme}

When the correction term is discretized with a DUO1 scheme, the same deviations terms as with the analytically computed correction, except for
\begin{align}
    & E^{(2)}_{\rho u_x} = C_3 + \left( \frac{\Delta t}{\tau} \right)^2 N_1 - \left( \frac{\Delta t}{2\tau} \right) \mathrm{sgn}(u_x) \frac{\partial^3 a^{(3)}_{eq, xxx}}{\partial x^3}, \qquad E^{(3)}_{\rho u_x} = C_4 +  \left( \frac{\Delta t}{\tau} \right)^2 \left( N_3 + \frac{1}{6} \frac{\partial^4 a^{(3)}_{eq, xxx}}{\partial x^4} \right) + \left( \frac{\Delta t}{2\tau} \right) \mathrm{sgn}(u_x) N_4.
\end{align}
It only differs from the DCO2 case by numerical error terms in $\Delta t/\tau$, which is a direct consequence of the first-order accuracy of the upwind gradients. Note that, despite this first-order accuracy in $\Delta t$, the mass and momentum equations are recovered without error at the NS level, \textit{i.e.} without error in $O(\mathrm{Kn})$. 
This is a consequence of the fact that zeroth- and first-order moments of $\Psi_i^d$ are null. Then, the discretization error only affects the macroscopic equations through the moments cascade, so that its contribution is at the order $O(\mathrm{Kn}^2)$. As a consequence, no numerical viscosity originates from the choice of discretization of the correction term.

\subsection{D2Q9 lattice}

% With the D2Q9 lattice, the following macroscopic equations are obtained:
% \begin{align}
% 	& \partial_t \rho + \partial_x(\rho u_x) + \partial_y(\rho u_y) = \mathrm{Kn} \, E_\rho^{(1)} + \mathrm{Kn}^2 \, E_\rho^{(2)} + \mathrm{Kn}^3 \, E_\rho^{(3)} + O(\mathrm{Kn}^4), \label{eq:D2Q9_mass} \\
% 	& \partial_t (\rho u_x) + \partial_x (\rho u_x^2) + \partial_y(\rho u_x u_y) + c_s^2 \theta \, \partial_x \rho = \mathrm{Kn} \left( \rho c_s^2 \theta (\partial_x S_{xx} + \partial_y S_{xy}) \right) + \mathrm{Kn}\, E_{\rho u_x}^{(1)} + \mathrm{Kn}^2 \, E_{\rho u_x}^{(2)} + \mathrm{Kn}^3 \, E_{\rho u_x}^{(3)} + O(\mathrm{Kn}^4), \label{eq:D2Q9_momentum_x} \\
% 	& \partial_t (\rho u_y) + \partial_x (\rho u_x u_y) + \partial_y (\rho u_y^2) + c_s^2 \theta \, \partial_y \rho = \mathrm{Kn} \left(\rho c_s^2 \theta ( \partial_x S_{xy} + \partial_y S_{yy} ) \right) + \mathrm{Kn}\, E_{\rho u_y}^{(1)} + \mathrm{Kn}^2 \, E_{\rho u_y}^{(2)} + \mathrm{Kn}^3 \, E_{\rho u_y}^{(3)} + O(\mathrm{Kn}^4), \label{eq:D2Q9_momentum_y}
% \end{align}
% with $S_{\alpha \beta}=\partial_\alpha u_\beta + \partial_\beta u_\alpha$. 

With the D2Q9 lattice, the macroscopic equations (\ref{eq:mass_errors})-(\ref{eq:momentum_errors}) are obtained with the deviation terms $E_\Phi^{(n)}$ detailed in \ref{app:deviations_BGK_D2Q9}. 
For the sake of simplicity, only the x-derivatives are provided in the deviation terms $E_{\rho u_x}^{(2)}$, $E_{\rho u_y}^{(2)}$, $E_\rho^{(3)}$, $E_{\rho u_x}^{(3)}$ and $E_{\rho u_y}^{(3)}$. The expressions of the other terms ($E_\rho^{(1)}$, $E_{\rho u_x}^{(1)}$, $E_{\rho u_x}^{(2)}$ and $E_\rho^{(2)}$) are much simpler, they are provided in their complete form. \newline

From the expressions of deviation terms, the main observations are:
\begin{itemize}
    \item the second-order numerical error in the mass equation ($E_\rho^{(2)}$) is always related to the cubic defect $a^{(3)}_{eq, \alpha \beta \gamma}$,
    \item adding a correction term $\Psi_i$ does not affect its value,
    \item on the contrary, enriching the equilibrium with third-order moments reduces this error,
    \item the DCO2 and DUO1 discretizations of $\Psi_i$ only affect the numerical errors of the momentum equation at $O(\mathrm{Kn}^2)$: they do not add any numerical viscosity,
    %\item when $x$-derivatives only are considered, enriching the equilibrium only affects the deviation terms of the $y$-momentum equation,
    \item the $N=3^*$- and $N=4^*$-equilibria yield similar deviation terms when $x$-derivatives only are considered.
\end{itemize}

\subsection{Validation with linear analyses}
\label{sec:BGK_validation}

The deviation terms provided in the previous section can now be validated by linear analyses. This is done here by comparing the eigenvalue problem of the macroscopic equations including deviation terms, given by Eq.~(\ref{eq:LSA_macros_Errors}), with that of the BGK-LB scheme, given by Eq.~(\ref{eq:LSA_LBM}). If the truncated system of Eq.~(\ref{eq:LSA_macros_Errors}) correctly takes into consideration the deviation terms of the macroscopic equations up to the order $p$ , then one necessarily has
\begin{align}
    \omega^* = \omega^*_{\mathrm{LBM}} + O \left({k^*}^{p+2} \right),
\end{align}
for any eigenvalue associated to a physical wave: acoustics or shear. In the following, the latter are identified as the ones whose real part is the closest to that of the expected physical waves:
\begin{align}
    \Re \left(\omega^*_{\mathrm{shear}} \right) = k^*_\alpha \overline{u_\alpha}, \qquad \Re \left(\omega^*_{\mathrm{ac+}} \right) = k^*_\alpha \overline{u_\alpha} + k^* c_s \sqrt{\theta}, \qquad \Re \left(\omega^*_{\mathrm{ac-}} \right) = k^*_\alpha \overline{u_\alpha} - k^* c_s \sqrt{\theta},
\end{align}
$\overline{\boldsymbol{u}}$ being the mean flow velocity of the linear analysis. For each of these waves, the validity of the deviation terms up to an order $p$ is finally assessed by checking that
\begin{align}
    \frac{\Delta \omega^*}{k^*} = \frac{|\omega^* - \omega^*_{\mathrm{LBM}}|}{k^*} = O \left({k^*}^{p+1} \right) = O \left( \frac{1}{N_{ppw}^{p+1}} \right),
    \label{eq:relation_DeltaOmega_Nppw}
\end{align}
where $N_{ppw}=2\pi/k^*$ is the number of points per wavelength of the monochromatic plane wave considered by the linear analysis. \newline

\begin{figure}
    \centering
    \includegraphics[scale=1]{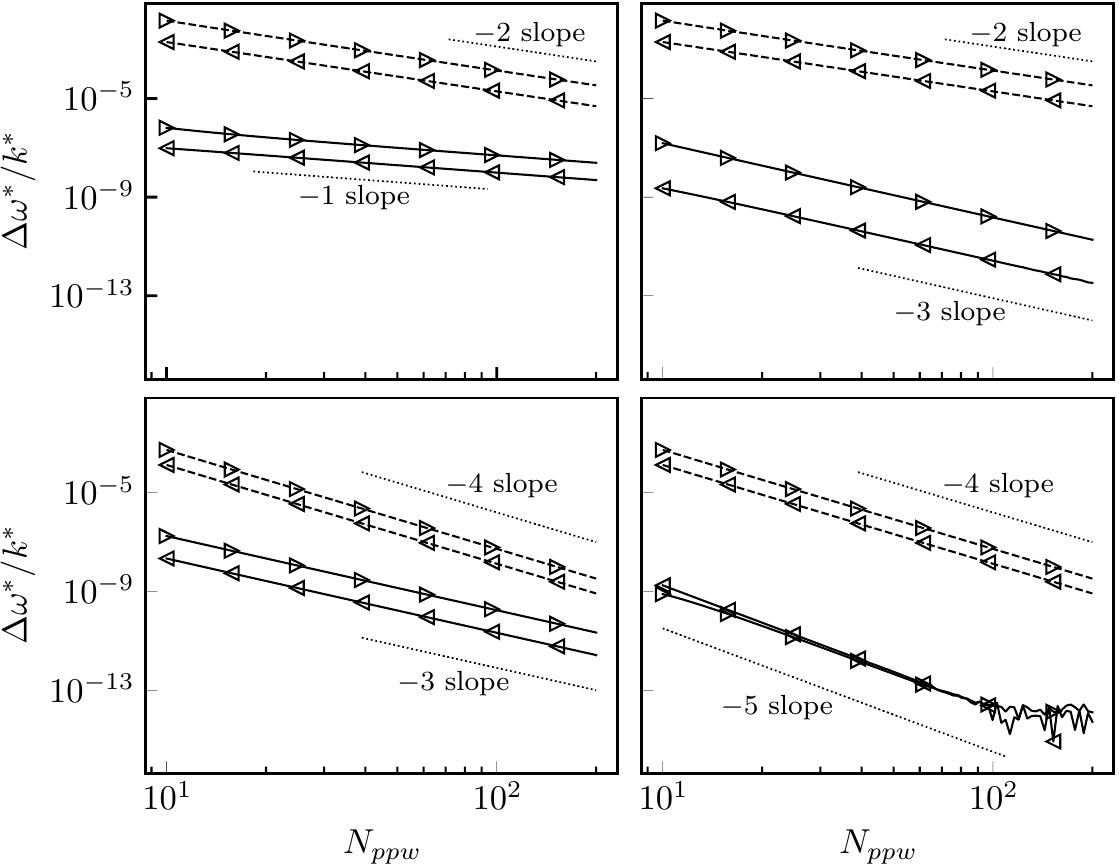}
    \caption{Relative error between the eigenvalues of the BGK-LBM and that of the isothermal NS equations including deviation terms up to an order $p$ provided in Eq.~(\ref{eq:deviation_terms_D1Q3_noCorr}), for the non corrected D1Q3 lattice with $\overline{u}=0.5 \mathrm{c_s} \sqrt{\theta}$, $\theta=0.75$ and $\tau/\Delta t=10^{-5}$. Top-left: $p=0$ (no deviation terms are considered), top-right: $p=1$, bottom-left: $p=2$, bottom-right: $p=3$. \dashedlinetriangleright: propagation error of the downstream acoustic eigenvalue, \dashedlinetriangleleft: propagation error of the upstream acoustic eigenvalue, \linetriangleright: dissipation error of the downstream acoustic eigenvalue, \linetriangleleft: dissipation error of the upstream acoustic eigenvalue.}
    \label{fig:BGK_D1Q3_noCorr_LSA_accuracy_allp}
\end{figure}

Real and imaginary parts of $\Delta \omega^*/k^*$ are displayed in Fig.~\ref{fig:BGK_D1Q3_noCorr_LSA_accuracy_allp} in log-scale for the two physical (acoustic) waves of the non-corrected D1Q3 lattice, as function of $N_{ppw}$. Four cases are considered: without considering any deviation term ($p=0$), and with deviation terms truncated up to the order $p=1$, $p=2$ and $p=3$, as provided by Eq.~(\ref{eq:deviation_terms_D1Q3_noCorr}). The mean flow velocity is arbitrarily prescribed as $\overline{u}=0.5 \mathrm{c_s} \sqrt{\theta}$, the constant $\theta$ is set as $\theta=0.8$ to avoid the particular case $\theta=1$ and the dimensionless relaxation time is set to a commonly encountered value in aeronautic simulations: $\tau/\Delta t=10^{-5}$. In any case, the adopted parameters do not impact the following conclusions:
\begin{itemize}
    \item As expected by Eq.~(\ref{eq:relation_DeltaOmega_Nppw}), an asymptotic behavior of the remaining error terms is exhibited by the slope of the logarithmic curve for each physical eigenvalue.
    \item Real parts of the eigenvalues, highlighting dispersion errors, are associated to even error orders, while imaginary parts, highlighting dissipation errors, are associated to odd error orders. This is a consequence to the fact that linear dispersion (resp. dissipation) is related to odd (resp. even) spatial derivatives.
    \item When considering the deviations from the isothermal NS equations up to an order $p$, the predominant error is well recovered as $O(1/{k^*}^{p+1})$, as predicted by Eq.~(\ref{eq:relation_DeltaOmega_Nppw}).
    \item For $p=0$ (without including deviation terms), the $-1$ slope exhibits the first-order deviation in Knudsen number, \textit{i.e.} in viscosity, caused by the lattice closure defect $a^{(3)}_{eq, xxx}$.
    \item With the BGK collision model, the dissipation error is always smaller than the dispersion one, in agreement with the conclusions of previous work~\cite{Marie2009}.
\end{itemize}
More importantly, the $-4$ slope recovered in the case $p=3$ allows validating all the deviation terms obtained in the linear approximation up to the third-order in $\mathrm{Kn}$, provided in Eq.~(\ref{eq:deviation_terms_D1Q3_noCorr}).

\begin{figure}
    \centering
    \includegraphics[scale=1]{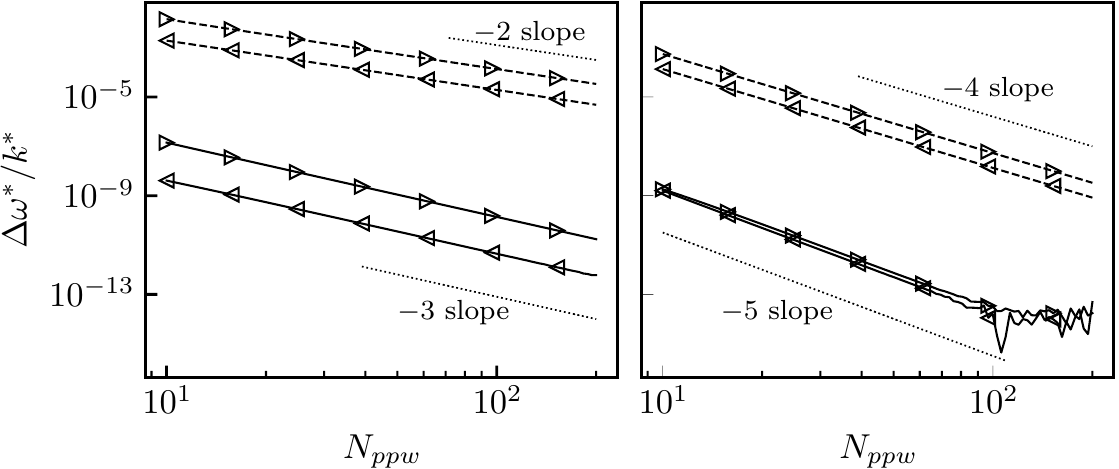}
    \caption{Relative error between the eigenvalues of the BGK-LBM and that of the isothermal NS equations including deviation terms up to an order $p$ provided in Eq.~(\ref{eq:deviation_terms_D1Q3_CorrAnalytics}), for the analytically corrected D1Q3 lattice with $\overline{u}=0.5 \mathrm{c_s} \sqrt{\theta}$, $\theta=0.75$ and $\tau/\Delta t=10^{-5}$. Left: $p=0$ (no deviation terms are considered), right: $p=3$. \dashedlinetriangleright: propagation error of the downstream acoustic eigenvalue, \dashedlinetriangleleft: propagation error of the upstream acoustic eigenvalue, \linetriangleright: dissipation error of the downstream acoustic eigenvalue, \linetriangleleft: dissipation error of the upstream acoustic eigenvalue. }
    \label{fig:BGK_D1Q3_CorrAnalyticss_LSA_accuracy_p0p3}
\end{figure}

Similarly, remaining errors $\Delta \omega^*/k^*$ are displayed for the analytically corrected D1Q3 model in Fig.~\ref{fig:BGK_D1Q3_CorrAnalyticss_LSA_accuracy_p0p3} for $p=0$ and $p=3$. Two observations can be made from these curves. First, without including any deviation term in the isothermal NS equations ($p=0$), a $-2$ slope is obtained for the remaining error, exhibiting an accuracy in $O({k^*}^2)$, or equivalently $O(\mathrm{Kn}^2)$. This indicates that the lattice defect $a^{(3)}_{eq, xxx}$ is well corrected thanks to the body-force term $\Psi_i$, so that no consistency error in viscosity occurs. Secondly, the $-4$ slope exhibited in the case $p=3$ allows validating all the deviation terms provided in the linear approximation in Eq.~(\ref{eq:deviation_terms_D1Q3_CorrAnalytics}).

A similar work performed on all the deviation terms provided for the BGK-D1Q3 LB scheme in Sec.~\ref{sec:BGK_deviations_D1Q3}, as well as for the D2Q9 lattice in \ref{app:deviations_BGK_D2Q9} further including a shear eigenvalue, allows validating them in the linear approximation. For the sake of compactness, these results are not provided here, the logarithmic curves of $\Delta \omega^*/k^*$ being very close to that of Figs.~\ref{fig:BGK_D1Q3_noCorr_LSA_accuracy_allp}-\ref{fig:BGK_D1Q3_CorrAnalyticss_LSA_accuracy_p0p3}. 

\subsection{Summary of the numerical properties of the BGK collision model}

In this section, the Taylor expansion in Knudsen number has been successfully applied to the BGK collision model for the D1Q3 and D2Q9 lattices, with and without body-force correction term. Even though the aim of such a work is mainly pedagogical, given the poor practical interest of the BGK collision model, one of its interesting features has been exhibited. Thanks to a close relationship linking the numerical errors of this model with the Bernoulli numbers, whose even terms are null, the numerical viscosity remains bounded and does not degenerate when the dimensionless relaxation time $\tau/\Delta t$ is decreased. This property turns out to be fundamental to explain the very low dissipation of the BGK collision compared to other models such as regularized ones or some MRT models, which will be investigated in the forthcoming sections.

Looking at the explicit expression of deviation terms with regards to the isothermal NS equations also underlines other interesting properties of the BGK collision model. The consistency errors exhibited in the present work are not problematic (except for the Mach error in the shear stress tensor): they are a simple consequence to the fact that a LB scheme does not intend to solve the NS equations, but a discrete counterpart of the Boltzmann equation. However, numerical errors exhibited in the mass equation are closely related to some consistency errors of the momentum equations. This observation might explain how enforcing the consistency of the LBM with the NS equations can improve the numerical behavior of the scheme, as observed in a previous work~\cite{Farag2021}. 

Hence, with the BGK collision model, the numerical error in the mass equation involves the lattice closure defect $a_{eq,\alpha \beta \gamma}^{(3)}$.
Interestingly, when the equilibrium distribution function is improved by recovering more equilibrium moments, as proposed with the D2Q9 lattice with $N=3^*$ and $N=4^*$ in \ref{app:deviations_BGK_D2Q9}, the numerical error in the mass equation can be reduced, as a consequence of the improved consistency of the momentum equation. On the contrary, when the defect $a_{eq, \alpha \beta \gamma}^{(3)}$ is explicitly corrected thanks to a body-force term $\Psi_i$, the mass equation remains related to it even though the consistency of the momentum equation is recovered up to the first-order in $\mathrm{Kn}$. The consequences of such a surprising observation may be the purpose of future work.

The main limitations of the BGK collision model remain its strong instability issues. The latter are caused by mode couplings arising in the linear approximation between hydrodynamic and non-hydrodynamic waves~\cite{Wissocq2019}, whose behavior cannot be investigated by the present Taylor expansion (\textit{cf.} Sec.~\ref{sec:non_hydrodynamic_modes}). For this reason, more sophisticated collision models are usually preferred in the aim of filtering out non-hydrodynamic modes. Among them, some regularized and MRT models are investigated in the next sections.

\section{Regularized LBM: general case}
\label{sec:regularized_models}

\subsection{Taylor expansion}

In the general case and including a correction term $\Psi_i$, the regularization step affects the post-collision distribution functions as~\cite{Latt2006, Malaspinas2015, Coreixas2017, Jacob2018, Renard2020}:
\begin{align}
    g_i^c = f_i^{eq} + \left( 1-\frac{\Delta t}{\tau + \Delta t/2} \right) g_i^{(1)} + \frac{\Delta t}{2\tau} \mathrm{Kn} \Psi_i^d,
    \label{eq:Regularization_general}
\end{align}
where the non-dimensionalization of Sec.~\ref{sec:dimensionless_DVBE} has been applied, and $g_i^{(1)}$ is a recomputed (regularized) off-equilibrium part, whose explicit expression depends on the model under consideration. Note that the BGK collision model can also be written under this form with $g_i^{(1)} = g_i - f_i^{eq} + \mathrm{Kn} \Delta t/(2 \tau) \Psi_i^d = (\tau+\Delta t/2)/\tau\, (f_i-f_i^{eq})$. After performing the Taylor expansion of Sec.~\ref{sec:Taylor_expansion_Kn}, the variable change suggested by Eq.~(\ref{eq:variable_change}) reads
\begin{align}
    g_i = h_i + \left( \frac{\Delta t}{\tau} - 1 \right) f_i^{(1)}, 
    \label{eq:Regul_variable_change}
\end{align}
with 
\begin{align}
    & h_i = f_i + \frac{\tau-\Delta t/2}{\tau + \Delta t/2} \, g_i^{(1)} + \left( \frac{\Delta t}{2\tau} - 1 \right) \mathrm{Kn} \Psi_i^d,\qquad f_i^{(1)} = f_i - f_i^{eq} - \mathrm{Kn} \Psi_i^d.
\end{align}
A necessary and sufficient condition for the admissibility of this variable change is: $ \sum_i g_i^{(1)} = \sum_i e_{i, \alpha} g_i^{(1)} = 0$, which is the case of all the regularization steps under consideration in this work. The following equivalent partial differential equation is then obtained:
\begin{align}
    \sum_{n \geq 1} \frac{\mathrm{Kn}^{n-1}}{n!} \left( \frac{\Delta t}{\tau} \right)^{n-1} D_i^n h_i + \left( \frac{\Delta t}{\tau} -1 \right) \sum_{n \geq 1} \frac{\mathrm{Kn}^{n-1}}{n!} \left( \frac{\Delta t}{\tau} \right)^{n-1} D_i^n f_i^{(1)} = -\frac{1}{\mathrm{Kn}} f_i^{(1)}.
\end{align}
The latter can be re-written as
\begin{align}
    f_i^{(1)} = - \sum_{n \geq 1} \frac{\mathrm{Kn}^{n}}{n!} \left( \frac{\Delta t}{\tau} \right)^{n-1} D_i^n h_i + \left( \frac{\tau}{\Delta t} -1 \right) \sum_{n \geq 1} \frac{\mathrm{Kn}^{n}}{n!} \left( \frac{\Delta t}{\tau} \right)^{n} D_i^n f_i^{(1)}.
    \label{eq:Regul_Taylor_1}
\end{align}
By analogy with the BGK model of Sec.~\ref{sec:BGK}, let us seek for a series expansion of $f_i^{(1)}$ under the following form:
\begin{align}
    f_i^{(1)} = \sum_{n \geq 1} \frac{d_n}{n!} \mathrm{Kn}^n \left( \frac{\Delta t}{\tau} \right)^{n-1} D_i^n h_i. 
    \label{eq:Regul_series_expansion_fi1}
\end{align}
Injecting this expression in Eq.~(\ref{eq:Regul_Taylor_1}) yields
\begin{align}
    \sum_{n \geq 1} \frac{d_n}{n!} \mathrm{Kn}^n \left( \frac{\Delta t}{\tau} \right)^{n-1} D_i^n h_i &= - \sum_{n \geq 1} \frac{\mathrm{Kn}^{n}}{n!} \left( \frac{\Delta t}{\tau} \right)^{n-1} D_i^n h_i + \left( \frac{\tau}{\Delta t} -1 \right) \sum_{n \geq 1} \sum_{m \geq 1} \frac{d_m}{n! m!} \mathrm{Kn}^{n+m} \left( \frac{\Delta t}{\tau} \right)^{n+m-1} D_i^{n+m} h_i \\
    &= - \sum_{n \geq 1} \frac{\mathrm{Kn}^{n}}{n!} \left( \frac{\Delta t}{\tau} \right)^{n-1} D_i^n h_i + \left( \frac{\tau}{\Delta t} -1 \right) \sum_{n \geq 2} \frac{\mathrm{Kn}^n}{n!} \left( \frac{\Delta t}{\tau} \right)^{n-1} \left( \sum_{k=1}^{n-1} \binom{n}{k} d_k \right) D_i^n h_i,
\end{align}
where a Cauchy product is used to obtain the second equation. This leads to the following recurrence relation on the coefficients $d_n$:
\begin{align}
    d_1 = -1, \qquad \forall n \geq 2,\ d_n = -1 + \left( \frac{\tau}{\Delta t} -1 \right) \sum_{k=1}^{n-1} \binom{n}{k} d_k.
    \label{eq:reccurrence_dn}
\end{align}
Contrary to the recurrence relation obtained in the BGK-case, the numerical parameter $\tau/\Delta t$ arises here, precluding the possibility to find an explicit relation between the $d_i$ and a set of well-documented numbers such as the Bernoulli ones. 
Anyway, the first $d_n$ coefficients can be easily computed using Eq.~(\ref{eq:reccurrence_dn}), leading to
\begin{align}
    d_2 = 1-2\frac{\tau}{\Delta t}, \quad d_3 = -1 + 6 \frac{\tau}{\Delta t} - 6 \left( \frac{\tau}{\Delta t} \right)^2, \quad d_4 = 1-14 \frac{\tau}{\Delta t} + 36 \left( \frac{\tau}{\Delta t} \right)^2 - 24 \left( \frac{\tau}{\Delta t} \right)^3.
\end{align}
Similar coefficients were found by Holdych \textit{et al.}~\cite{Holdych2004} in their analysis of the BGK scheme, and were related to the polylogarithm function. Dong \textit{et al.}~\cite{Dong2010} also noticed a similarity with the firsts Bernoulli polynomials. Finally, one should note that with every regularization step considered in the present article, the distribution functions are reconstructed such that $g_i^{(1)}=O(\mathrm{Kn})$. This allows defining a zeroth-order terms as
\begin{align}
    \hat{g}_i^{(1)} = \frac{1}{\mathrm{Kn}} \left( \frac{\tau}{\tau+\Delta t/2} \right) g_i^{(1)},
    \label{eq:Regul_def_hat_gi1}
\end{align}
where the ratio $\tau/(\tau+\Delta t/2)$ simply appears for convenience. Together with this definition, injecting the $d_n$ coefficients into Eq.~(\ref{eq:Regul_series_expansion_fi1}) and collecting every terms in $\mathrm{Kn}$ up to the third-order yields
\begin{align}
    D_i f_i &= -\frac{1}{\mathrm{Kn}} f_i^{(1)} + \mathrm{Kn} \left( \frac{\Delta t}{2 \tau} -1 \right) \left[ D_i^2 f_i + D_i \left( \hat{g}_i^{(1)} - \Psi_i^d  \right) \right] + \mathrm{Kn}^2 \left[ \left(-1 + \frac{\Delta t}{\tau} - \frac{1}{6} \left( \frac{\Delta t}{\tau} \right)^2 \right) D_i^3 f_i - \left( 1-\frac{\Delta t}{2 \tau} \right)^2 D_i^2 \left( \hat{g}_i^{(1)} - \Psi_i^d  \right) \right] \nonumber \\
    & + \mathrm{Kn}^3 \left[ \left( -1 + \frac{3 \Delta t}{2 \tau} - \frac{7}{12} \left( \frac{\Delta t}{\tau} \right)^2 + \frac{1}{24} \left( \frac{\Delta t}{\tau} \right)^3 \right)D_i^4 f_i + \left(-1 + \frac{\Delta t}{\tau} - \frac{1}{6} \left( \frac{\Delta t}{\tau} \right)^2 \right) \left( 1-\frac{\Delta t}{2 \tau} \right) D_i^3 \left( \hat{g}_i^{(1)} - \Psi_i^d  \right) \right] + O(\mathrm{Kn}^4).
    \label{eq:Regul_Taylor_2}
\end{align}
This equation will be the starting point for the forthcoming analyses of regularized models, especially to explicitly obtain the deviation terms $E_i^{(n)}$ and $E_\Phi^{(n)}$. Before going further, one can notice that whatever the way regularized distribution functions are computed, the Euler equations are recovered provided that the zeroth- and first-order moments of  $g_i^{(1)}$ are null. In particular, even with $g_i^{(1)}=0$, the Euler equations can be recovered in the hydrodynamic limit. However, a first-order error in $\mathrm{Kn}$ can possibly lead to an unintended numerical viscosity, which will be particularly investigated.

\subsection{Degenerating hyperviscosity with regularized models}

Looking at Eq.~(\ref{eq:Regul_Taylor_2}) also allows a first estimation of the numerical hyperviscosity induced by the regularization, especially its ratio with the expected physical viscosity $\Gamma=\mathrm{Kn}^2 \, E_i^{(3)}$. Neglecting the discretization errors of $\Psi_i$ and keeping only the higher order in $\Delta t/\tau$ yields
\begin{align}
    \Gamma \sim \frac{\mathrm{Kn}^2}{24} \left( \frac{\Delta t}{\tau} \right)^3 D_i^3 \left[ D_i f_i + 2 \left( \hat{g}_i^{(1)} - \Psi_i \right) \right] = \frac{(k \Delta x)^2}{24} \frac{\Delta t}{\tau} D_i^3 \left[ D_i f_i + 2 \left( \hat{g}_i^{(1)} - \Psi_i \right) \right].
    \label{eq:Regul_hyperviscous_degeneracy}
\end{align}
In the second equality, this ratio depends on $\Delta t/\tau$, as a consequence of the $(\Delta t/\tau)^3$ terms arising in Eq.~(\ref{eq:Regul_Taylor_2}). Hence, it is shown that for a given discretized spatial phenomenon characterized by a given value of $k \Delta x$, the unexpected effects of the numerical hyperviscosity can become predominant over the physical viscosity when $\tau/\Delta t$ is small. Since one usually has $\tau/\Delta t < 10^{-3}$, the numerical hyperviscosity is then likely to degenerate. Looking at Eq.~(\ref{eq:Regul_hyperviscous_degeneracy}), this actually occurs for physical phenomena for which
\begin{align}
    k \Delta x > \sqrt{\frac{\tau}{\Delta t}},
\end{align}
which might remain very small in practice compared to $1$. This means that even well-resolved phenomena can be largely affected by the numerical hyperviscosity. Since the latter is \textit{a priori} not controlled and strongly depends on the model under investigation, it can either lead to (1) over-dissipation or (2) instability. This can explain the unexpected dissipation properties of regularized models exhibited in a previous work~\cite{Wissocq2020}. \newline

Note that with the BGK collision model, one has $\hat{g}_i^{(1)} = (f_i-f_i^{eq})/\mathrm{Kn}$, so that
\begin{align}
    D_i f_i + 2 \left( \hat{g}_i^{(1)} - \Psi_i^d \right) = D_i f_i + \frac{2}{\mathrm{Kn}} f_i^{(1)} = -D_i f_i + O(\mathrm{Kn}).
\end{align}
Injecting this expression into Eq.~(\ref{eq:Regul_Taylor_2}) allows, after some manipulation of the equations, recovering the Taylor expansion of the BGK collision model of Eq.~(\ref{eq:Taylor_BGK_3}), where no hyperviscous degeneracy occurs.\newline

In order to exhibit the specific hyperviscous effects of the regularization, it is then necessary to focus on each model, which is the purpose of the following sections. 

%Finally, a similar degeneracy as the hyperviscous one is unfortunately susceptible to occur at any order in $\mathrm{Kn}$ with regularized models. This can be shown noticing that the leading term in the coefficients $d_n$ is
%\begin{align}
%    d_n = (-1)^n + O(\tau/\Delta t).
%\end{align}
%Hence, at any order $n$ in Knudsen number, the predominant term in $\Delta t/\tau$ in the numerical error is
%\begin{align}
%    \mathrm{Kn}^n \frac{(-1)^{n+1}}{(n+1)!} \left( \frac{\Delta t}{\tau} \right)^n D_i^n \left[ D_i f_i + \left( %\frac{n+1}{2} \right) \left( \hat{g}_i^{(1)} - \Psi_i^d \right) \right]
%\end{align}
% Je ne rentre pas dans les details de la degenerescence a tous les ordres, c'est trop compliqe a detailler ici et l'interet n'est pas enorme. Cela aura sa place dans un autre papier qui introduira les termes correctifs.

\section{Regularization with a shear stress tensor reconstruction}
\label{sec:Regul_FR}

In this section, the regularized model under consideration is referred to as a ``shear stress tensor reconstruction'', where $g_i^{(1)}$ is computed using the knowledge of macroscopic quantities $\rho$ and $u_\alpha$ only. It corresponds to the HRR collision model of Jacob \textit{et al.}~\cite{Jacob2018} in the particular case $\sigma=0$. The main interest of this model is that it is the only regularization procedure that can be applied with the D1Q3 lattice. It is therefore first studied here to obtain reasonable expressions of deviation terms in 1D. This model is based on the relationship obtained by a Chapman-Enskog expansion~\cite{CHAPMAN_Book_3rd_1970}, relating the off-equilibrium populations with the shear stress tensor. With the non-dimensionalization of Sec.~\ref{sec:dimensionless_DVBE} it reads
\begin{align}
    g_i^{(1)} = \mathrm{Kn} \left( 1 + \frac{\Delta t}{2 \tau} \right) \hat{g}_i^{(1)} \qquad \mathrm{with} \qquad  \hat{g}_i^{(1)} = -\frac{w_i}{c_s^2} \left( e_{i, \alpha} e_{i, \beta} - c_s^2 \delta_{\alpha \beta} \right) \rho \theta \left( \partial_\alpha u_\beta + \partial_\beta u_\alpha \right).
    \label{eq:Regul_AR_scheme}
\end{align}

Like the correction term $\Psi_i$, this off-equilibrium reconstruction involves space gradients which are, in practice, discretized by finite differences. The effects of a DCO2 and a DUO1 schemes will be particularly investigated by comparison with the analytical gradients of Eq.~(\ref{eq:Regul_AR_scheme}) in the following.

\subsection{First-order error term with the D1Q3 lattice}
\label{sec:Regul_AR_First-order_error}

Let us focus on the first-order error term in $\mathrm{Kn}$ of Eq.~(\ref{eq:Regul_Taylor_2}), which is expected to be a source of numerical viscosity in the model. It can be re-written as
\begin{align}
    E_i^{(1)} = \left( \frac{\Delta t}{2\tau} -1 \right) D_i \left( D_if_i + \hat{g}_i^{(1)} - \Psi_i \right).
\end{align}
The computation of this error term requires the computation of $D_i f_i$, which reads
\begin{align}
    D_i f_i = D_i f_i^{eq} + O(\mathrm{Kn}) &= \partial_t f_i^{eq} + e_{i,x} \partial_x f_i^{eq} + O(\mathrm{Kn}) \nonumber \\
    &= \frac{\partial f_i^{eq}}{\partial \rho} \partial_t \rho + \frac{\partial f_i^{eq}}{\partial (\rho u_x)} \partial_t (\rho u_x) + e_{i,x} \left( \frac{\partial f_i^{eq}}{\partial \rho} \partial_x \rho + \frac{\partial f_i^{eq}}{\partial (\rho u_x)} \partial_x (\rho u_x) \right) + O(\mathrm{Kn}) \nonumber \\
    &= -\frac{\partial f_i^{eq}}{\partial \rho} \partial_x (\rho u_x) - \frac{\partial f_i^{eq}}{\partial (\rho u_x)} \partial_x \Pi^{eq}_{xx} + e_{i,x} \left( \frac{\partial f_i^{eq}}{\partial \rho} \partial_x \rho + \frac{\partial f_i^{eq}}{\partial (\rho u_x)} \partial_x (\rho u_x) \right) + O(\mathrm{Kn}),
    \label{eq:Regul_FR_Difi}
\end{align}
where the fact that mass and momentum equations are recovered at first-order in $\mathrm{Kn}$ has been used. With the D1Q3 lattice, the equilibrium distribution can be written as
\begin{align}
    f_i^{eq} = w_i \left[ \rho + \frac{\rho u_x e_{i,x}}{c_s^2} + \frac{(\rho u_x)^2 e_{i,x}^2}{2 \rho c_s^4} - \frac{(\rho u_x)^2}{2 \rho c_s^2} + \rho (\theta-1) \frac{e_{i,x}^2-c_s^2}{2 c_s^2} \right], 
\end{align}
so that
\begin{align}
    \frac{\partial f_i^{eq}}{\partial \rho} = w_i \left[ 1 - \frac{u_x e_{i,x}}{2 c_s^4} + \frac{u_x^2}{2c_s^2} + (\theta-1) \frac{e_{i,x}^2 - c_s^2}{2c_s^2} \right], \qquad \frac{\partial f_i^{eq}}{\partial (\rho u_x)} = w_i \left[ \frac{e_{i,x}}{c_s^2} + \frac{u_x e_{i,x}^2}{c_s^4} - \frac{u_x}{c_s^2} \right].
\end{align}
Furthermore, 
\begin{align}
    \partial_x \Pi^{eq}_{xx} = \partial_x (\rho u_x^2 + \rho c_s^2 \theta) = (u_x^2 + c_s^2 \theta) \partial_x \rho + 2 \rho u_x \partial_x u_x.
\end{align}
Injecting these expressions in Eq.~(\ref{eq:Regul_FR_Difi}) yields after simplification
\begin{align}
    D_i f_i = \frac{w_i}{2c_s^2} (e_{i,x}^2 - c_s^2) \left[ \left( -\frac{u_x^3}{c_s^2} + 3u_x(1-\theta) \right) \partial_x \rho + \left( - \frac{3 \rho u_x^2}{c_s^2} + \rho(3-\theta) \right) \partial_x u_x \right] + O(\mathrm{Kn}),
\end{align}
so that
\begin{align}
    D_i f_i + \hat{g}_i^{(1)} = - \frac{w_i}{2c_s^4} \left( e_{i,x}^2 - c_s^2 \right) \partial_x \left(\rho u_x^3 + 3\rho c_s^2 (\theta-1) u_x \right) + O(\mathrm{Kn}).
\end{align}
Surprisingly, the right-hand-side term of this equation is exactly equal to the body-force term $\Psi_i$ of Eq.~(\ref{eq:correction_D1Q3}). As a consequence, using the fact that $\Psi_i^d = \Psi_i + O(\mathrm{Kn})$, one has in presence of correction
\begin{align}
    D_i f_i + \hat{g}_i^{(1)} - \Psi_i^d = O(\mathrm{Kn}).
    \label{eq:Regul_FR_numerical_viscosity_cancelled}
\end{align}
The first-order error in $\mathrm{Kn}$ of Eq.~(\ref{eq:Regul_Taylor_2}) then completely vanishes in presence of correction. Without any correction, a first-order error is expected, leading to undesired numerical viscosity, which can be cured thanks to the body-force term $\Psi_i$. It is therefore surprising that, while the correction term is designed to address a consistency error, it can be used to reduce a numerical error in the present case. This point can be interpreted as follows: the first-order error can be cancelled provided that the consistency of the viscous stress implicitly modelled by the LBM, and that explicitly imposed by $g_i^{(1)}$, is ensured.

This important conclusion should be kept in mind when using a regularized model based on this reconstruction, such as the HRR model~\cite{Jacob2018}. It has been demonstrated here with the D1Q3 lattice. It is shown in Sec.~\ref{sec:Regul_FR_D2Q9_deviations} that it remains true on the macroscopic equations of the D2Q9 lattice. This phenomenon explains why the correction had to be used in previous work based on the HRR collision model, even for low and moderate Mach numbers~\cite{Astoul2020, Astoul2020b}. 

\subsection{Corrected D1Q3 lattice: numerical properties of the degenerating hyperviscosity}
\label{sec:Regul_AR_degenerating_hyperviscosity}

The focus is now put on the corrected D1Q3 lattice, for which the numerical viscosity is cancelled. For the sake of simplicity, the numerical errors in $\Psi_i$ are first neglected. The numerical dissipation is here mainly caused by the degenerating hyperviscosity characterized by Eq.~(\ref{eq:Regul_hyperviscous_degeneracy}). Using the conclusions of Eq.~(\ref{eq:Regul_FR_numerical_viscosity_cancelled}), it can be re-written as
\begin{align}
    \Gamma \sim \frac{\mathrm{Kn}^2}{24} \left( \frac{\Delta t}{\tau} \right)^3 D_i^3 \left( \hat{g}_i^{(1)} - \Psi_i \right).
\end{align}
Studying the function $\hat{g}_i^{(1)} - \Psi_i$, which is explicitly known, should provide interesting properties on the degenerating hyperviscosity, \textit{a fortiori} on the dissipation and stability properties of the model. It reads
\begin{align}
    \hat{g}_i^{(1)} - \Psi_i = \frac{w_i}{2c_s^4} \left( e_{i,x}^2 - c_s^2 \right) \left[ \left( u_x^3 + 3c_s^2 (\theta-1) u_x \right) \partial_x \rho + \left( 3u_x^2 + c_s^2 (\theta-3) \right) \rho \partial_x u_x \right].
    \label{eq:Regul_g-Psi}
\end{align}
In order to exhibit how this error can affect the dissipation of the acoustics, it is now possible to re-express it as function of the Riemann invariants, which are the characteristic quantities of a given physical phenomenon. For isothermal 1D cases, the two Riemann invariants $R_+$ and $R_-$ are defined such that~\cite{Toro2009}
\begin{align}
    \rho \partial_x R_+ = \rho \partial_x u_x + c_s \sqrt{\theta} \partial_x \rho, \qquad \rho \partial_x R_- = \rho \partial_x u_x - c_s \sqrt{\theta} \partial_x \rho,
\end{align}
where $R_+$ is the characteristic variable of the downstream acoustic wave, and $R_-$ of the upstream one. Hence,
\begin{align}
    \partial_x u_x = \frac{1}{2} \left( \partial_x R_+ + \partial_x R_- \right), \qquad \partial_x \rho = \frac{\rho}{2 c_s\sqrt{\theta}} \left( \partial_x R_+ - \partial_x R_- \right).
\end{align}
Injecting them in Eq.~(\ref{eq:Regul_g-Psi}) yields after simplification
%\begin{align}
%    \hat{g}_i^{(1)} - \Psi_i = \frac{w_i}{4c_s^4} \left( e_{i,x}^2 - c_s^2 \right) \rho \Bigg[ & \left( \frac{u_x^3}{c_s \sqrt{\theta}} + \frac{3 c_s (\theta-1)u_x}{\sqrt{\theta}} + 3u_x^2 + c_s^2(\theta-3) \right) \partial_x R_+ \nonumber \\
%    & + \left( \frac{-u_x^3}{c_s \sqrt{\theta}} - \frac{3 c_s (\theta-1)u_x}{\sqrt{\theta}} + 3u_x^2 + c_s^2(\theta-3) \right) \partial_x R_- \Bigg].
%\end{align}
%It can be recast as 
\begin{align}
    \hat{g}_i^{(1)} - \Psi_i = \frac{w_i}{4c_s^2} \left( e_{i,x}^2 - c_s^2 \right) \rho \theta \Big[ P_+(\mathrm{Ma})\, \partial_x R_+ + P_-(\mathrm{Ma})\, \partial_x R_- \Big],
\end{align}
where $\mathrm{Ma}=u_x/(c_s \sqrt{\theta})$ is the Mach number and
\begin{align}
    & P_+(\mathrm{Ma}) =  \mathrm{Ma}^3 + 3 \mathrm{Ma}^2 + 3 \mathrm{Ma} \left( 1-\frac{1}{\theta} \right) + 1 - \frac{3}{\theta}, \qquad P_-(\mathrm{Ma}) = P_+(-\mathrm{Ma}).
\end{align}
This polynomial can be factorized as
\begin{align}
    P_+(\mathrm{Ma}) = \left( \mathrm{Ma} +1 \right) \left( \mathrm{Ma}- \left( \frac{1}{c_s \sqrt{\theta}}-1 \right) \right) \left( \mathrm{Ma}+ \left( \frac{1}{c_s \sqrt{\theta}}+1 \right) \right),
\end{align}
which makes its three real roots clearly appear. For the downstream acoustic wave, $\mathrm{Ma}=1/(c_s \sqrt{\theta})-1$ is the only positive root of $P_+$, which can be re-written as a Courant-Friedrich-Lewy condition~\cite{Courant1967},
\begin{align}
    \mathrm{CFL} = 1, \qquad \mathrm{with} \qquad \mathrm{CFL} = u_x + c_s \sqrt{\theta} = c_s \sqrt{\theta} (\mathrm{Ma}+1).
\end{align}
Regarding the upstream acoustic wave, $P_-$ has three positive roots: $\mathrm{Ma}=1$, $\mathrm{Ma}=1+1/(c_s \sqrt{\theta})$ and $\mathrm{Ma}=1-1/(c_s \sqrt{\theta})$, which is positive if $\theta > 3$.

As a consequence, the sign of the degenerating hyperviscosity is expected to theoretically change on a given wave for each of these roots, switching from a stabilizing error (over-dissipation) to a destabilizing one (negative dissipation). It can be summarized as follows:
\begin{itemize}
    \item for the downstream acoustic wave, the hyperviscous error switches at $\mathrm{CFL}=1$,
    \item for the upstream acoustic wave, it switches at $\mathrm{Ma}=1$, at $\mathrm{Ma}=1+1/(c_s \sqrt{\theta})$ and at $\mathrm{Ma}=1-1/(c_s \sqrt{\theta})$ when $\theta >3$.
\end{itemize}

This demonstration is in complete agreement with the stability limits of the HRR collision model observed by linear analyses in a previous work~\cite{Renard2020b}, namely $\mathrm{Ma}<1$ and $\mathrm{CFL}<1$. Note that similar limitations could even be recovered in a fully compressible framework, coupling the LB scheme with an additional energy equation.

In the following, this demonstration will be verified by explicitly computing the deviation terms of the D1Q3 lattice. Such conclusions will also be extended to the D2Q9 lattice in Sec.~\ref{sec:Regul_FR_D2Q9_deviations} and \ref{app:deviations_AR_D2Q9}.

\subsection{Deviation terms of the D1Q3 lattice in the linear approximation}
\label{sec:Regul_FR_deviation_terms}

Starting from Eq.~(\ref{eq:Regul_Taylor_2}), an explicit expression of the deviation terms appearing in Eq.~(\ref{eq:DVBE_Errors}) involving spatial derivatives of macroscopic quantities has to be obtained. This is a prerequisite for obtaining the explicit deviations from the macroscopic equations, following \ref{app:macroscopic_equations}.

To this extent, an estimation of $D^2_i f_i$, $D_i^3 f_i$ and $D_i^4 f_i$ is required with respectively a second-, first- and zeroth-order accuracy in $\mathrm{Kn}$. The same goes for $D_i (\hat{g}_i^{(1)} - \Psi_i^d)$, $D_i^2 (\hat{g}_i^{(1)} - \Psi_i^d)$ and $D_i^3 (\hat{g}_i^{(1)} - \Psi_i^d)$. For this purpose, the computer algebra system Maxima~\cite{maxima} is used. Of considerable complexity, the procedure employed for these computations is not provided in details here. It is basically based on the following relations:
\begin{itemize}
    \item the Taylor expansion of $\Psi_i^d$ proposed in \ref{app:discretized_correction_terms}, which makes $\Psi_i$ explicitly appear,
    \item a Taylor expansion of the gradient terms involved in $\hat{g}_i^{(1)}$ depending on the discretization scheme (DCO2 or DUO1), similarly as in \ref{app:discretized_correction_terms},
    \item a second-order estimation of $f_i$ as
\begin{align}
     f_i = f_i^{eq} + \mathrm{Kn} \left( \Psi_i - D_i f_i^{eq} \right) + \mathrm{Kn}^2 \left( D_i^2 f_i^{eq} - D_i \Psi_i + E_i^{(1)} \right) + O(\mathrm{Kn}^3),
 \end{align}
 on which the derivation operator $D_i^n$ can be applied,
    \item the chain rule of Eq.~(\ref{eq:chain_rule}), that can be applied to $f_i^{eq}$, $\Psi_i$ and $\hat{g}_i^{(1)}$,
    \item the macroscopic equations of \ref{app:macroscopic_equations} in order to transform any time-derivative to space-derivative.
\end{itemize}
Once the deviation terms $E_i^{(n)}$ obtained, their macroscopic counterpart $E_\Phi^{(n)}$ can be computed following the general procedure of \ref{app:macroscopic_equations}. The isothermal NS equations can finally be recovered as in Eqs.~(\ref{eq:mass_errors})-(\ref{eq:momentum_errors}) with the deviation terms detailed below. All the consistency errors $C_i$ and numerical terms $N_i$ are provided in  \ref{app:deviation_terms}.

\subsubsection{Non corrected D1Q3, analytical reconstruction of $g_i^{(1)}$}

In the non-corrected case and neglecting the discretization errors in $g_i^{(1)}$, the mass and momentum equations (\ref{eq:mass_errors})-(\ref{eq:momentum_errors}) are recovered with
\begin{align}
    & E_\rho^{(1)} = 0, \qquad  E^{(1)}_{\rho u_x} = -\left( \frac{\Delta t}{2 \tau} \right) \partial^2_{xx} a^{(3)}_{eq, xxx}, \nonumber \\
    & E^{(2)}_\rho = \left( \frac{\Delta t}{\tau} \right)^2 \frac{1}{12} \left( -\frac{\partial^3 a^{(3)}_{eq, xxx}}{\partial x^3} + 2\rho c_s² \theta \frac{\partial^3 u_x}{\partial x^3} \right), \qquad  E^{(2)}_{\rho u_x} = \left( \frac{\Delta t}{\tau} \right) \frac{C_3}{2} + \left( \frac{\Delta t}{\tau} \right)^2 N_{11}, \nonumber \\
    & E^{(3)}_\rho = \frac{1}{12} \left( \frac{\Delta t}{\tau} \right)^2 \partial_x C_3 + \left( \frac{\Delta t}{\tau} \right)^3 N_{12}, \qquad E^{(3)}_{\rho u_x} =  \left( \frac{\Delta t}{\tau} \right) N_{13} + \left( \frac{\Delta t }{\tau} \right)^2 N_{14} + \left( \frac{\Delta t}{\tau} \right)^3 N_{16}.
\end{align}

It is remarkable that there is no more consistency error with the isothermal NS equations in this model. In particular, even the well-known Mach error is removed thanks to the explicit computation of the off-equilibrium distribution functions based on a correct viscous stress tensor. Following some previous work~\cite{Farag2021}, this indicates that the shear stress imposed \textit{via} the reconstruction is the actual shear stress in the limit $\Delta t \rightarrow 0$. The only remaining terms are numerical errors which, contrary to the BGK model, include:
\begin{itemize}
    \item a $\Delta t/\tau$-dependent term in $E_{\rho u_x}^{(1)}$, indicating the numerical viscosity highlighted in Sec.~\ref{sec:Regul_AR_First-order_error},
    \item a $(\Delta t/\tau)^3$-term in $E_\rho^{(3)}$ and $E_{\rho u_x}^{(3)}$, indicating the degenerating hyperviscosity of Sec.~\ref{sec:Regul_AR_degenerating_hyperviscosity}.
\end{itemize}
Also note that the numerical viscosity is directly related to the lattice defect $a^{(3)}_{eq, xxx}$. The ratio between this numerical viscosity and the physically expected one being of the order $\Delta t/\tau$, this error is likely to have a very strong impact on the dissipation and the stability properties when the dimensionless relaxation time is small. For this reason, the correction term $\Psi_i$, which is expected to address this issue (\textit{cf.} Sec.~\ref{sec:Regul_AR_First-order_error}), is introduced in all the forthcoming analyses.

\subsubsection{Analytically corrected D1Q3, analytical reconstruction of $g_i^{(1)}$}

Introducing a correction term and neglecting the discretization errors in $g_i^{(1)}$ and $\Psi_i$, the mass and momentum equations (\ref{eq:mass_errors})-(\ref{eq:momentum_errors}) are recovered with

\begin{align}
    & E^{(1)}_\rho = 0, \qquad E^{(1)}_{\rho u_x} = 0, \nonumber \\
    & E^{(2)}_\rho = \left( \frac{\Delta t}{\tau} \right)^2 \frac{1}{12} \left( -\frac{\partial^3 a^{(3)}_{eq, xxx}}{\partial x^3} + 2\rho c_s² \theta \frac{\partial^3 u_x}{\partial x^3} \right), \qquad  E^{(2)}_{\rho u_x} = \left( \frac{\Delta t}{2\tau} \right) C_{3} + \left( \frac{\Delta t}{\tau} \right)^2 N_1, \nonumber \\
    & E^{(3)}_\rho = \frac{1}{12} \left( \frac{\Delta t}{\tau} \right)^2 \partial_x C_{3}, \qquad E^{(3)}_{\rho u_x} = \left( \frac{\Delta t}{\tau} \right) N_{13} + \left( \frac{\Delta t}{\tau} \right)^2 N_{17} + \left( \frac{\Delta t}{\tau} \right)^3 N_{18}.
    \label{eq:Regul_errors_D1Q3_AR_fneqAnalytical_PsiAnalytical}
\end{align}

As predicted, the numerical viscosity is cancelled by the introduction of the correction term $\Psi_i$. An interesting feature of this term appears here: while it aims at correcting the first-order consistency error with the BGK collision model, it can be now used to correct a purely numerical error.

\subsubsection{Correction with a DCO2 scheme, reconstruction of $g_i^{(1)}$ with a DCO2 scheme}

Compared to Eq.~(\ref{eq:Regul_errors_D1Q3_AR_fneqAnalytical_PsiAnalytical}), only the hyperviscous term is affected in the momentum equation as
\begin{align}
   & E^{(3)}_{\rho u_x} = \left( \frac{\Delta t}{\tau} \right) N_{13} + \left( \frac{\Delta t}{\tau} \right)^2 \left( N_{17} + \frac{\rho \theta}{9} \frac{\partial^4 u_x}{\partial x^4} \right) + \left( \frac{\Delta t}{\tau} \right)^3 \left( N_{18} + \frac{1}{12}\frac{\partial^4 a^{(3)}_{eq, xxx}}{\partial x^4} - \frac{\rho \theta}{18} \frac{\partial^4 u_x}{\partial x^4} \right).
\end{align}
%As a consequence, the $\mathrm{Ma}=1$ and $\mathrm{CFL}=1$ limitations exhibited in Sec.~\ref{sec:Regul_AR_degenerating_hyperviscosity}, do not apply for this model, as will be shown in Sec.~\ref{sec:SSTR_LSA_hypernu_degen}.

\subsubsection{Correction with a DCO2 scheme, reconstruction of $g_i^{(1)}$ with a DUO1 scheme}

Compared to Eq.~(\ref{eq:Regul_errors_D1Q3_AR_fneqAnalytical_PsiAnalytical}), only the numerical errors in the momentum equation are affected as
\begin{align}
    & E^{(2)}_{\rho u_x} = \left( \frac{\Delta t}{\tau} \right) \left( \frac{C_3}{2} - \mathrm{sgn}(u_x) \rho c_s^2 \theta \frac{\partial^3 u_x}{\partial x^3} \right) + \left( \frac{\Delta t}{\tau} \right)^2 \left( N_1 + \mathrm{sgn}(u_x) \frac{\rho \theta}{6} \frac{\partial^3 u_x}{\partial x^3} \right), \nonumber \\
   & E^{(3)}_{\rho u_x} = \left( \frac{\Delta t}{\tau} \right) N_{13} + \left( \frac{\Delta t}{\tau} \right)^2 \left( N_{17} + \frac{\rho \theta}{9} \frac{\partial^4 u_x}{\partial x^4} - \mathrm{sgn}(u_x) \frac{\partial_x C_3}{4} \right) + \left( \frac{\Delta t}{\tau} \right)^3 \left( N_{18} + \frac{1}{12}\frac{\partial^4 a^{(3)}_{eq, xxx}}{\partial x^4} - \frac{\rho \theta}{18} \frac{\partial^4 u_x}{\partial x^4} + \mathrm{sgn}(u_x) \frac{\partial_x C_3}{8} \right).
\end{align}

Note that even if a first-order error in $\Delta t/\tau$ is introduced in the computation of $g_i^{(1)}$, it does not lead to numerical viscosity.

%\subsubsection*{Correction with a DUO1 scheme, analytical reconstruction of $g_i^{(1)}$:}

\subsubsection{Correction with a DUO1 scheme, reconstruction of $g_i^{(1)}$ with a DCO2 scheme}

Compared to Eq.~(\ref{eq:Regul_errors_D1Q3_AR_fneqAnalytical_PsiAnalytical}), only the numerical errors in the momentum equation are affected as
\begin{align}
    & E^{(2)}_{\rho u_x} = \left( \frac{\Delta t}{2\tau} \right) C_{3} + \left( \frac{\Delta t}{\tau} \right)^2 \left( N_1 - \frac{\mathrm{sgn}(u_x)}{4} \frac{\partial^3 a^{(3)}_{eq, xxx}}{\partial x^3} \right), \nonumber \\
   & E^{(3)}_{\rho u_x} = \left( \frac{\Delta t}{\tau} \right) N_{13} + \left( \frac{\Delta t}{\tau} \right)^2 \left( N_{17} + \frac{\rho \theta}{9} \frac{\partial^4 u_x}{\partial x^4} \right) + \left( \frac{\Delta t}{\tau} \right)^3 \left( N_{18} + \frac{1}{12}\frac{\partial^4 a^{(3)}_{eq, xxx}}{\partial x^4} - \frac{\rho \theta}{18} \frac{\partial^4 u_x}{\partial x^4} + \mathrm{sgn}(u_x) \frac{N_4}{8} \right).
\end{align}

\subsubsection{Correction with a DUO1 scheme, reconstruction of $g_i^{(1)}$ with a DUO1 scheme}

Compared to Eq.~(\ref{eq:Regul_errors_D1Q3_AR_fneqAnalytical_PsiAnalytical}), only the numerical errors in the momentum equation are affected as
\begin{align}
    & E^{(2)}_{\rho u_x} = \left( \frac{\Delta t}{\tau} \right) \left(\frac{C_3}{2} - \mathrm{sgn}(u_x) \rho c_s^2 \theta \frac{\partial^3 u_x}{\partial x^3} \right) + \left( \frac{\Delta t}{\tau} \right)^2 \left( N_1 - \frac{\mathrm{sgn}(u_x)}{4} \frac{\partial^3 a^{(3)}_{eq, xxx}}{\partial x^3} + \mathrm{sgn}(u_x) \frac{\rho \theta}{6} \frac{\partial^3 u_x}{\partial x^3} \right), \nonumber \\
   & E^{(3)}_{\rho u_x} = \left( \frac{\Delta t}{\tau} \right) N_{13} + \left( \frac{\Delta t}{\tau} \right)^2 \left( N_{17} + \frac{\rho \theta}{9} \frac{\partial^4 u_x}{\partial x^4} - \mathrm{sgn}(u_x) \frac{\partial_x C_3}{4} \right) + \left( \frac{\Delta t}{\tau} \right)^3 \left( N_{18} + \frac{1}{12}\frac{\partial^4 a^{(3)}_{eq, xxx}}{\partial x^4} - \frac{\rho \theta}{18} \frac{\partial^4 u_x}{\partial x^4} + \frac{\mathrm{sgn}(u_x)}{8} \left( N_4 + \partial_x C_3 \right) \right).
\end{align}

\subsection{Deviation terms of the D2Q9 lattice in the linear approximation}
\label{sec:Regul_FR_D2Q9_deviations}

A similar work is performed on the D2Q9 lattice, for which deviation terms with the isothermal NS equations are provided in \ref{app:deviations_AR_D2Q9} in the linear approximation. Only the second-order equilibrium ($N=2$) is considered, noticing that changing the order of the Hermite-based equilibrium has few impact on the general conclusions. Similar observations as with the D1Q3 lattice can be drawn: (1) the shear stress tensor reconstruction with $g_i^{(1)}$ allows recovering a perfect consistency with the NS equation in every case, (2) in absence of the body-force term correction $\Psi_i$, a numerical viscosity arises, (3) the latter can be cancelled by introducing the correction, eventually discretized, (4) in any case, a degenerating hyperviscosity occurs. 

\subsection{Linear analyses: impact of the numerical errors on the dissipation}

All the expressions of error terms have been systematically validated by linear analyses, in the same way as with the BGK collision model of Sec.~\ref{sec:BGK_validation}. Looking at the remaining errors $\Delta \omega^*/k^*$ between the eigenvalues of the LB scheme and that of the equivalent macroscopic equations allows for confirming the validity of the terms at every order in $\mathrm{Kn}$. In this section, only the effect of error terms on the numerical dissipation is discussed.

\begin{figure}[ht]
    \centering
    \includegraphics[scale=1]{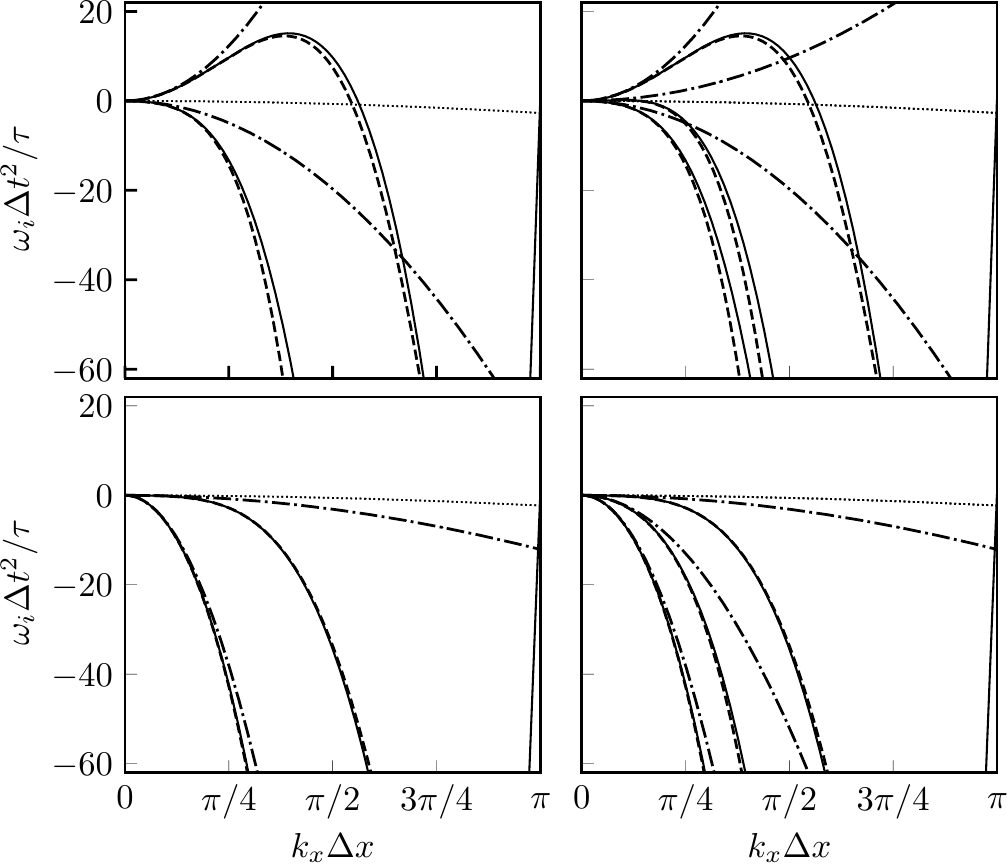}
    \caption{Dissipation curves of the non-corrected D1Q3 lattice (left) and D2Q9 lattice with $N=2$ (right) equipped with the complete regularization with an analytical computation of $g_i^{(1)}$. A horizontal mean flow is considered with $\overline{u_x}=0.5 c_s\sqrt{\theta}$ and $\tau/\Delta t=10^{-3}$. Top: $\theta=0.82$, bottom: $\theta=0.7$. \blacklinefine: LB scheme, \blackdottedline: isothermal NS equations, \blackdasheddottedline: isothermal NS with numerical viscosity, \blackdashedline: isothermal NS with numerical viscosity and hyperviscosity. } 
    \label{fig:FR_numerical_viscosity}
\end{figure}

Fig.~\ref{fig:FR_numerical_viscosity} displays the dissipation curves of the non corrected D1Q3 and D2Q9 ($N=2$) lattices with the shear stress tensor reconstruction. The parameters are arbitrarily chosen for the sake of readability at a horizontal Mach number $\mathrm{Ma}=0.5$, $\tau/\Delta t=10^{-3}$ and two different values of $\theta$: $\theta=0.82$ and $\theta=0.7$. The analyses of the LB schemes are compared to those of the isothermal NS equations without deviation terms ($p=0$) and including the first-order deviation ($p=1$) and up to the third-order one ($p=3$). 
The last two cases emphasize the effects of the numerical viscosity and hyperviscosity, respectively. 

In agreement with previous analyses, this particular regularization yields only two (resp. three) physical waves in 1D (resp. 2D), meaning that all the non-hydrodynamic modes are filtered out of a simulation~\cite{Astoul2020, Wissocq2020}. 
With $\theta=0.82$, the first-order error is responsible for a destabilization of one acoustic wave with the D1Q3 lattice, as well as the additional shear wave of the D2Q9 lattice. 
These instabilities even occur in the well-resolved limit ($k_x \ll \Delta x$) and their order of magnitude can be very large compared to the isothermal NS equations. For all the waves, the numerical hyperviscosity has a stabilizing effect, which yet does not cancel the instability caused by the numerical viscosity for low wavenumbers. This competition between viscous and hyperviscous effects provides explanations to the instabilities observed in previous linear analyses of regularized models~\cite{Wissocq2020}. 
The bottom part of Fig.~\ref{fig:FR_numerical_viscosity} ($\theta=0.7$) highlights the strong impact of $\theta$ on the behavior of the numerical error. In this particular case, both numerical viscosity and hyperviscosity induce an over-dissipation, leading to a stable model in the $x$-direction. Note that this does not allow drawing any conclusion on the linear stability of the scheme in other directions. Because of this large and unpredictable numerical viscosity, non-corrected models will no longer be discussed in the following, especially since this error can be simply addressed by introducing the body-force term $\Psi_i$. \newline

\begin{figure}[ht]
    \centering
    \includegraphics[scale=1]{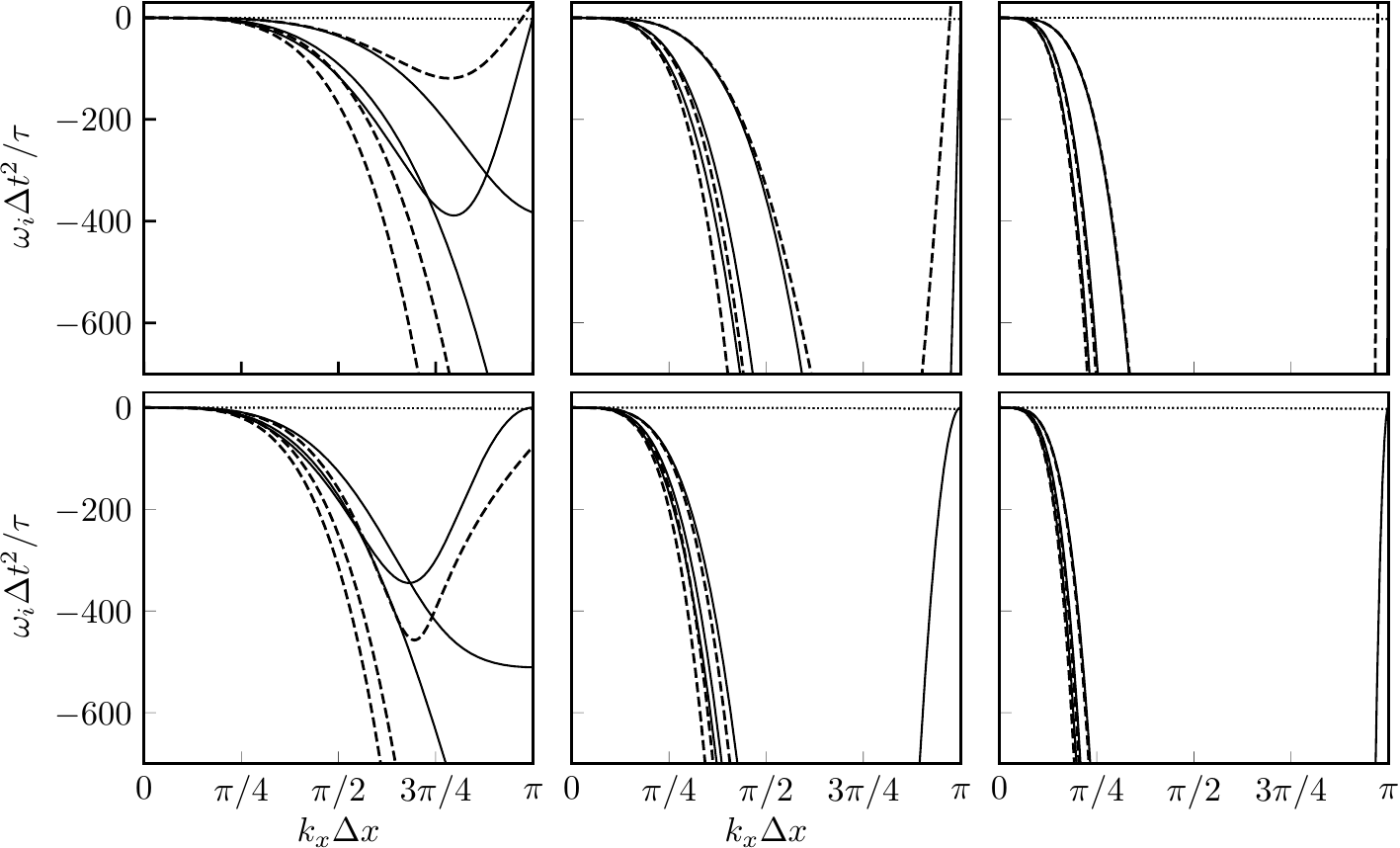}
    \caption{Dissipation curves of the corrected D2Q9 lattice equipped with the complete regularization. A horizontal mean flow is considered with $\overline{u_x}=0.5 c_s\sqrt{\theta}$, $\theta=0.8$ and three values of $\tau/\Delta t$: $10^{-3}$ (left), $10^{-4}$ (middle), $10^{-5}$ (right). Top: analytical computation of $\Psi_i$ and $g_i^{(1)}$, bottom: a DCO2 scheme is used for their space-derivatives. \blacklinefine: LB scheme, \blackdottedline: isothermal NS equations, \blackdashedline: isothermal NS with numerical hyperviscosity.} 
    \label{fig:FR_hyperviscosity}
\end{figure}

Fig.~\ref{fig:FR_hyperviscosity} exhibits the dissipation properties of the corrected D2Q9 model with $N=2$ for a horizontal mean flow at $\mathrm{Ma}=0.5$, $\theta=0.8$ and three values of the dimensionless relaxation time: $\tau/\Delta t=10^{-3}$, $\tau/\Delta t=10^{-4}$ and $\tau/\Delta t=10^{-5}$. 
The effect of the DCO2 discretization of $\Psi_i$ and $g_i^{(1)}$ is also investigated. In any case and for the three expected physical waves, no more numerical viscosity is observed and the hyperviscosity induces a large over-dissipation compared to the isothermal NS equations. In addition to a qualitative validation of the error terms, the analyses performed at different values of $\tau/\Delta t$ exhibit the numerical origin of the error. The over-dissipation is all the more important as the dimensionless relaxation time is reduced, which further confirms the error in $\Delta t/\tau$ predicted by the hyperviscous degeneracy. 
Furthermore, the bottom part of Fig.~\ref{fig:FR_hyperviscosity} points out the impact of the discretization of $\Psi_i$ and $g_i^{(1)}$ with a DCO2 scheme. Compared to their analytical computation, the numerical dissipation is further increased, but to a much lesser extent than that caused by the shear stress reconstruction. The over-dissipative character of this model, which was attributed to errors in the finite-difference estimation of the viscous stress tensor by previous authors~\cite{Jacob2018}, is therefore largely dominated by the degenerating hyperviscosity caused by the regularization procedure itself. This point is of paramount importance since it is precisely this limitation that led to the introduction of the $\sigma$ parameter in the HRR collision model, hybridizing it with a recursive regularization.

\subsection{Impact of the degenerating hyperviscosity on the linear stability}
\label{sec:SSTR_LSA_hypernu_degen}

Up to now, no clear discretization effect of $\Psi_i$ and $g_i^{(1)}$ could be exhibited, which yet clearly affects the numerical error in hyperviscosity. 
In this section, the focus is put on its degenerating part, which may provide important conclusions on the choice of the discretization scheme. This is done here by considering the effect of the degenerating hyperviscosity only on the dissipation/amplification properties of the physical waves. For this purpose, the Euler equations are simply considered, onto which the $(\Delta/\tau)^3$-related terms of $E^{(3)}_{\rho u_ \alpha}$, referred to as $H_\alpha$, are superimposed:
\begin{align}
    & \partial_t \rho + \partial_\alpha (\rho u_\alpha) = 0,\qquad \partial_t (\rho u_\alpha) + \partial_x (\rho u_\alpha u_\beta) + c_s^2 \theta \partial_\alpha \rho = H_\alpha.
    \label{eq:Regul_AR_D1Q3_degen_hypernu_only}
\end{align}
For instance, with the corrected D1Q3 lattice with analytical derivatives of Eq.~(\ref{eq:Regul_errors_D1Q3_AR_fneqAnalytical_PsiAnalytical}), one has $H_x=N_{18}$. One interest of such a system is that it does not involve the dimensionless relaxation time anymore, which corresponds to the behavior of the complete Eq.~(\ref{eq:Regul_errors_D1Q3_AR_fneqAnalytical_PsiAnalytical}) in the limit of infinitely small values of $\tau/\Delta t$. Then, a linearization followed by a diagonalization of this reduced system, as proposed in \ref{app:matrices_LSA}, leads to two acoustic eigenvalues. Studying the sign of their imaginary part $\omega_i$ in the low wavenumber limit provides information on the stabilizing or destabilizing effect of the degenerating hyperviscosity. \newline

\begin{figure}[ht]
    \centering
    \includegraphics[scale=1]{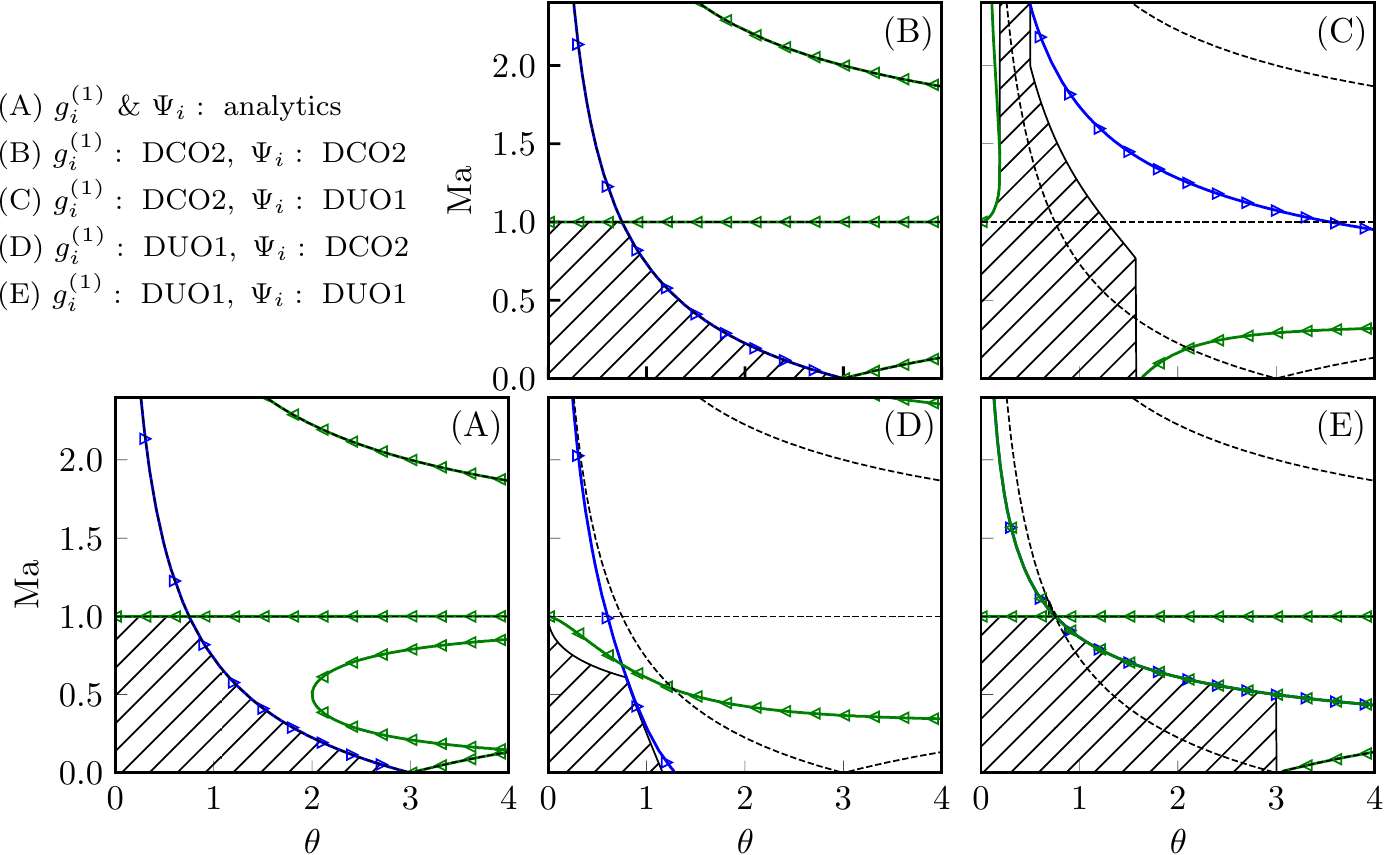}
    \caption{Impact of the degenerating hyperviscosity on the linear stability of the corrected D1Q3 lattice equipped with the complete regularization. Contours of a null imaginary eigenvalue $\omega_i$ of Eq.~(\ref{eq:Regul_AR_D1Q3_degen_hypernu_only}) are displayed as: downstream acoustics (\bluelinetriangleright), upstream acoustics (\greenlinetriangleleft). Dash lines represent the theoretical curves of Eq.~(\ref{eq:Regul_AR_theory_Ma_theta}). Hatched area: linear stability region of the LB scheme at $\tau/\Delta t=10^{-5}$.} 
    \label{fig:D1Q3_AR_degen_hypernu}
\end{figure}

Fig.~\ref{fig:D1Q3_AR_degen_hypernu} displays the nullity curves of the two imaginary eigenvalues $\omega_i$ of the system (\ref{eq:Regul_AR_D1Q3_degen_hypernu_only}) applied to the corrected D1Q3 lattice, for different values of the Mach number $\mathrm{Ma}=u_x/(c_s \sqrt{\theta})$ and $\theta$. Fives cases are considered depending on the way space gradients in $g_i^{(1)}$ and $\Psi_i$ are discretized. These curves are compared with the theoretical values obtained in Sec.~\ref{sec:Regul_AR_degenerating_hyperviscosity}:
\begin{align}
    \mathrm{Ma} = 1, \qquad \mathrm{Ma} = \frac{1}{c_s \sqrt{\theta}}-1, \qquad \mathrm{Ma} = \frac{1}{c_s \sqrt{\theta}}+1, \qquad \mathrm{Ma} = 1- \frac{1}{c_s \sqrt{\theta}}.
    \label{eq:Regul_AR_theory_Ma_theta}
\end{align}
Note that the second equation corresponds to the common condition $\mathrm{CFL}=1$.

Fig.~\ref{fig:D1Q3_AR_degen_hypernu} also displays hatched areas representing the linear stability regions obtained by performing a linear stability analysis of the isothermal LB scheme for $\tau/\Delta t=10^{-5}$. In this case, stability is ensured provided that, for any value of $k_x \Delta x \in [-\pi, \pi]$ every eigenvalue $\omega$ is such that $\omega_i < 0$. The following observations can be made:
\begin{itemize}
    \item Linear stability regions are always located below the annulment curves (\bluelinetriangleright) and (\greenlinetriangleleft). In most of the cases, the instability is even encountered when crossing these curves. This confirms that the degenerating hyperviscosity is at the origin of the linear instability.
    \item The theoretical estimations of Sec.~\ref{sec:Regul_AR_degenerating_hyperviscosity} (\ref{eq:Regul_AR_theory_Ma_theta}) fit remarkably well the actual sign inversion of the degenerating hyperviscosity for the two acoustic waves when $g_i^{(1)}$ and $\Psi_i$ are both analytically computed or estimated by a DCO2 discretization. However, using the DUO1 scheme in the estimation of $\Psi_i$ or $g_i^{(1)}$, Eq.~(\ref{eq:Regul_AR_theory_Ma_theta}) cannot be used to predict its behavior. 
    \item With a DCO2 discretization, the stability limits are: $\mathrm{Ma}=1$ and $\mathrm{CFL}=1$, which is clearly caused by a sign inversion in the degenerating hyperviscosity. This is in complete agreement with previous observations~\cite{Renard2020b}. 
    \item Evaluating $\Psi_i$ with a DUO1 scheme and $g_i^{(1)}$ with a DCO2 one has an interesting effect on the stability: the theoretical limitations $\mathrm{Ma}=1$ and $\mathrm{CFL}=1$ can be overcome. This is done thanks to the modification of the hyperviscous error, which is now stabilizing for larger values of $(\mathrm{Ma}, \theta)$. This point is in agreement with an improved stability of the HRR collision model when the body-force term is evaluated by an upwind scheme, already observed in the literature~\cite{Renard2020, Renard2020b, Guo2020}.
\end{itemize}

\begin{figure}[h!]
    \begin{minipage}[b]{\linewidth}
    \centering
    \includegraphics[scale=1.]{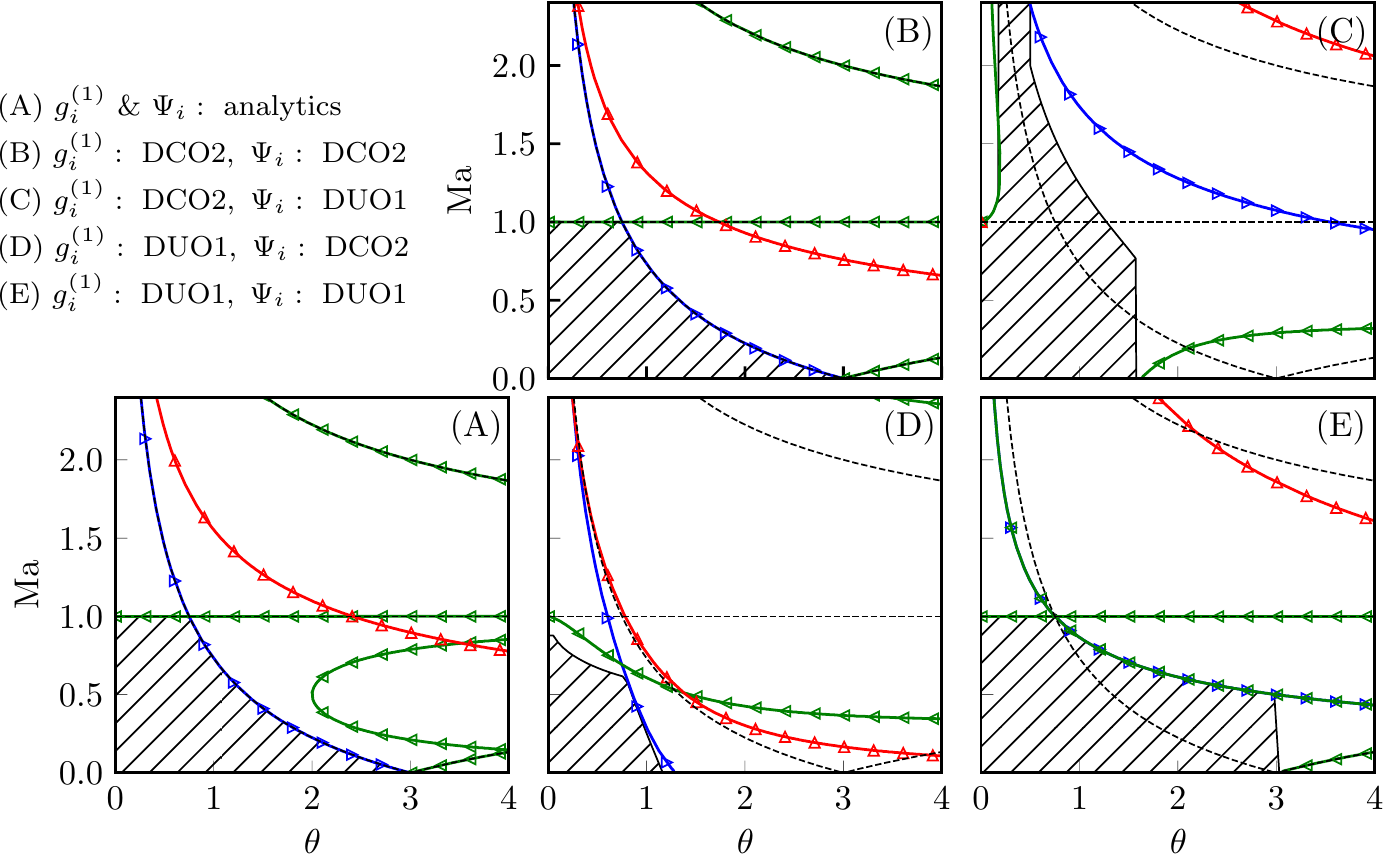}
    \subcaption{$N=2$}
    \end{minipage}\\ \vspace{5mm}
    \begin{minipage}[b]{\linewidth}
    \centering
    \includegraphics[scale=1.]{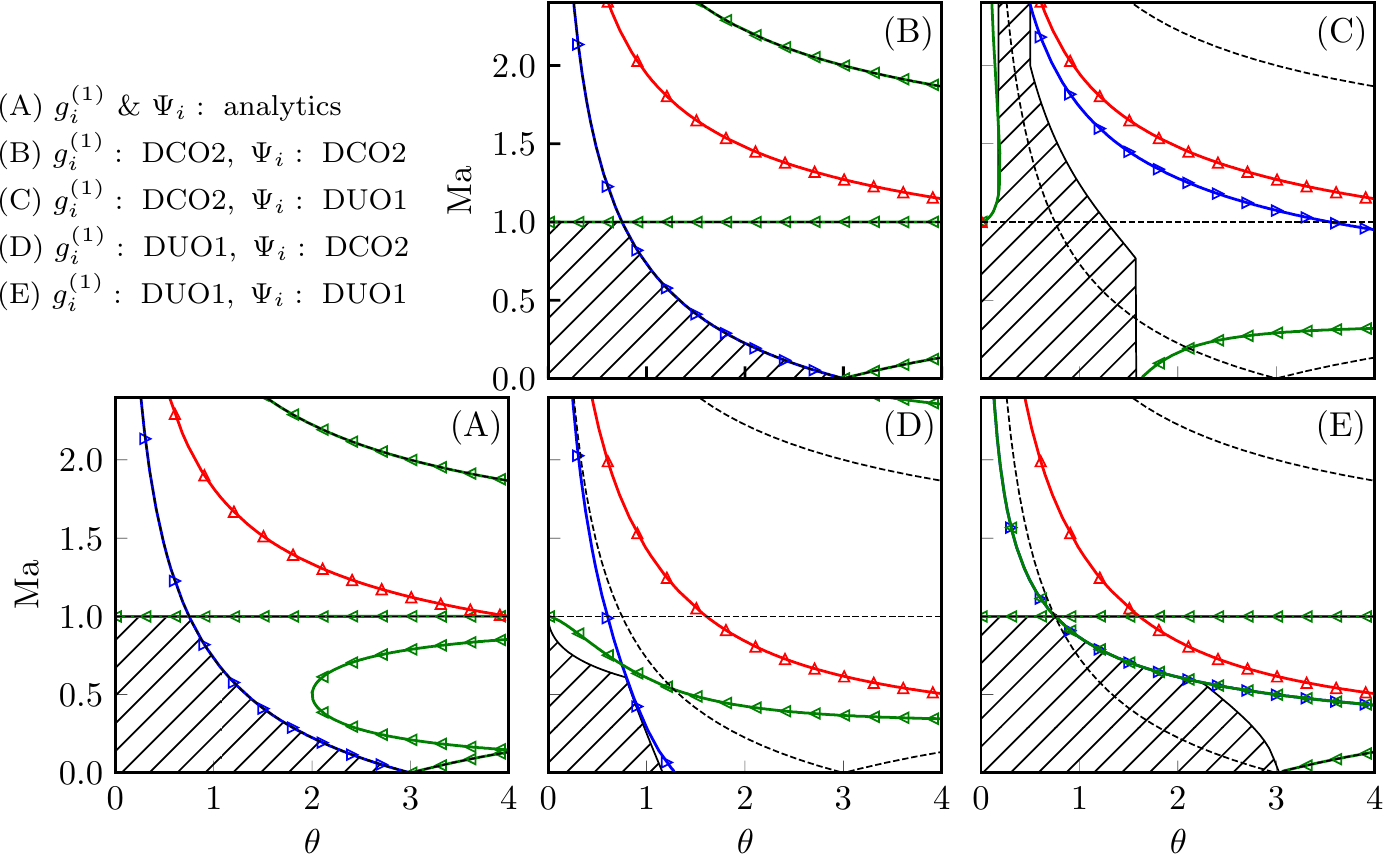}
    \subcaption{$N=3^*$, $N=4^*$}
    \end{minipage}
    \caption{Impact of the degenerating hyperviscosity on the linear stability of the corrected D2Q9 lattice equipped with the complete regularization, with a Hermite-based equilibrium at order $N$, considering the $x$-direction only. Contours of a null imaginary eigenvalue $\omega_i$ of Eq.~(\ref{eq:Regul_AR_D1Q3_degen_hypernu_only}) are displayed as: downstream acoustics (\bluelinetriangleright), upstream acoustics (\greenlinetriangleleft), shear wave (\redlinetriangleup). Dash lines represent the theoretical curves of Eq.~(\ref{eq:Regul_AR_theory_Ma_theta}). Hatched area: linear stability region of the LB scheme at $\tau/\Delta t=10^{-5}$.} 
    \label{fig:D2Q9_AR_degen_hypernu_xOnly}
\end{figure}

Similar studies are then performed on the D2Q9 lattice, considering the three possible Hermite-based equilibria. Fig.~\ref{fig:D2Q9_AR_degen_hypernu_xOnly} displays the nullity curves of the degenerating hyperviscosity of the three physical eigenvalues, as well as the linear stability region, considering the $x$-direction only (for the gradients and the mean flow). Regarding the acoustics, the curves are identical to the D1Q3 lattice of Fig.~\ref{fig:D1Q3_AR_degen_hypernu}. Only the appearance of a shear related eigenvalue can be noticed, whose behavior is slightly affected by the order of the equilibrium distribution function. In any case, the stability of acoustic waves remains the more constraining when looking at the $x$-direction only, and similar conclusions as with the D1Q3 lattice can be drawn. %Especially, the limitations of the analytical and DCO2 computations of $g_i^{(1)}$ and $f_i^{(1)}$ are still: $\mathrm{Ma}<1$ and $\mathrm{CFL}<1$. 
Also note that no difference are exhibited when comparing the $N=3^*$ and $N=4^*$ cases in the $x$-direction only. \newline

\begin{figure}[h!]
    \begin{minipage}[b]{\linewidth}
    \centering
    \includegraphics[scale=1.]{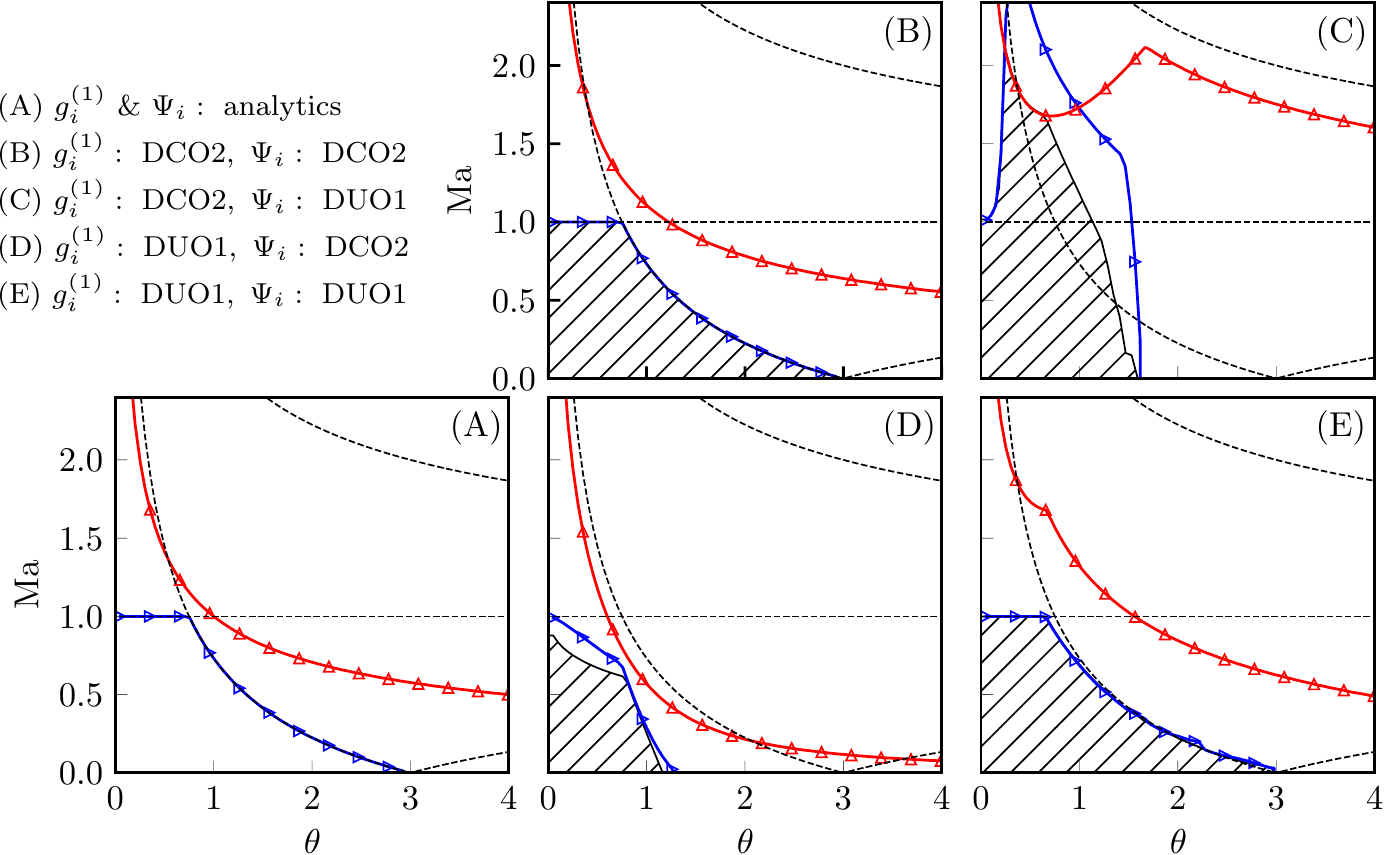}
    \subcaption{$N=2$}
    \end{minipage}\\ \vspace{5mm}
    \begin{minipage}[b]{\linewidth}
    \centering
    \includegraphics[scale=1.]{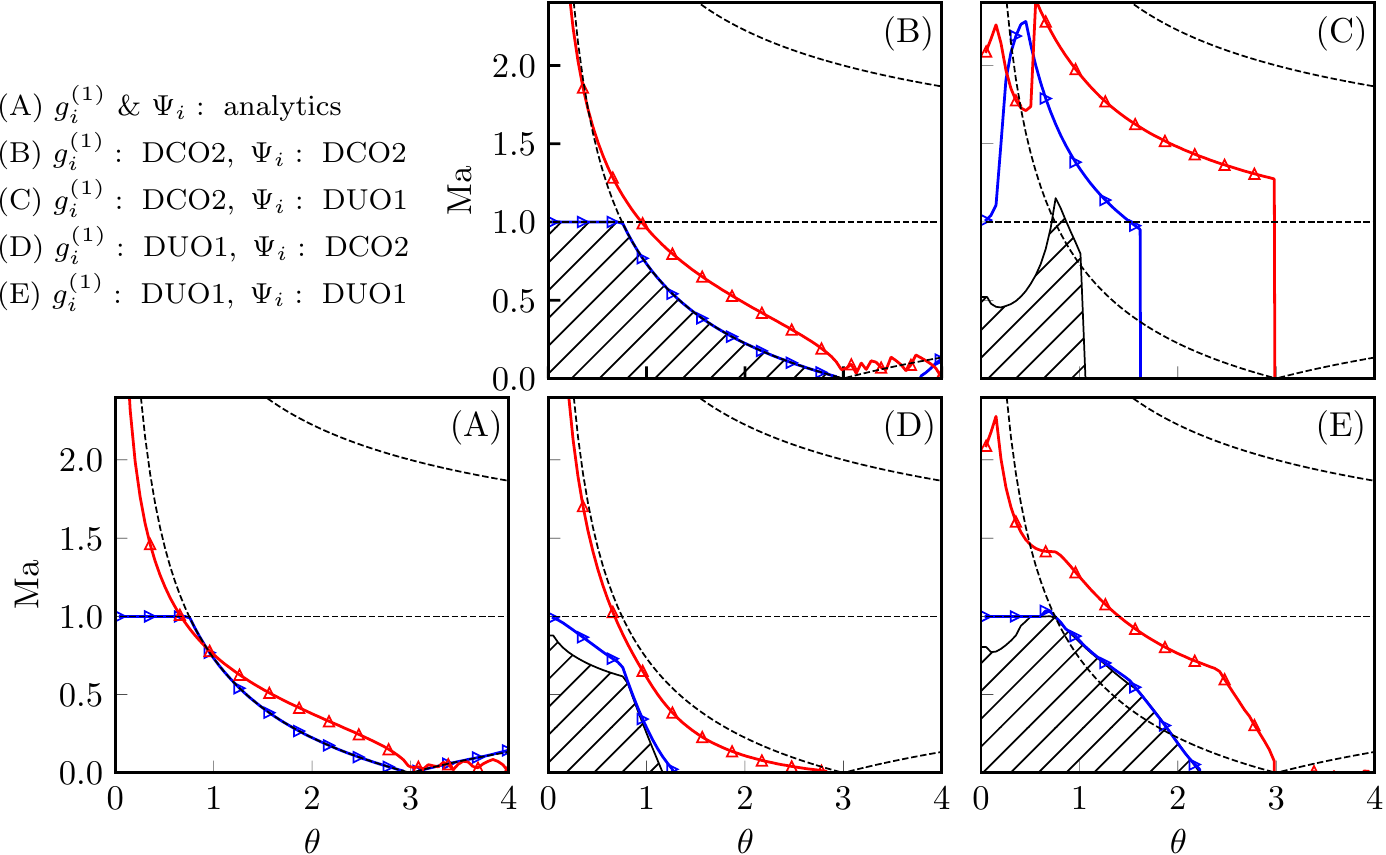}
    \subcaption{$N=3^*$}
    \end{minipage}
    \caption{Impact of the degenerating hyperviscosity on the linear stability of the corrected D2Q9 lattice equipped with the complete regularization, with a Hermite-based equilibrium at order $N$, considering all the possible directions. Contours of a null imaginary eigenvalue $\omega_i$ of Eq.~(\ref{eq:Regul_AR_D1Q3_degen_hypernu_only}) are displayed as: acoustic waves (\bluelinetriangleright), shear wave (\redlinetriangleup). Dash lines represent the theoretical curves of Eq.~(\ref{eq:Regul_AR_theory_Ma_theta}). Hatched area: linear stability region of the LB scheme at $\tau/\Delta t=10^{-5}$.} 
    \label{fig:D2Q9_AR_degen_hypernu}
\end{figure}

\begin{figure}[h!]
    \begin{minipage}[b]{\linewidth}
    \centering
    \includegraphics[scale=1.]{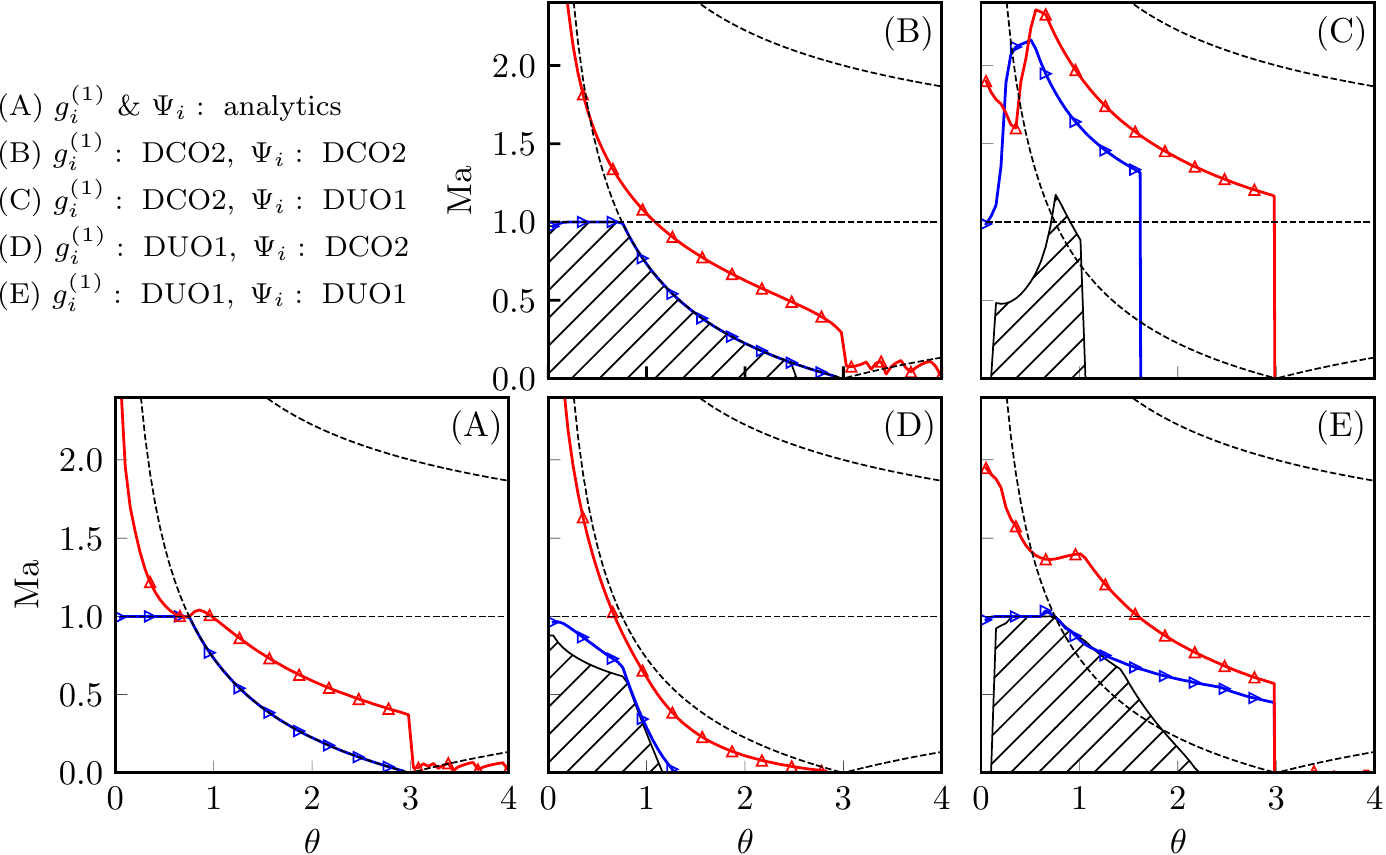}
    \subcaption{$N=4^*$}
    \end{minipage}
    \caption{Impact of the degenerating hyperviscosity on the linear stability of the corrected D2Q9 lattice equipped with the complete regularization, with a Hermite-based equilibrium at order $N$, considering all the possible directions. Contours of a null imaginary eigenvalue $\omega_i$ of Eq.~(\ref{eq:Regul_AR_D1Q3_degen_hypernu_only}) are displayed as: acoustic waves (\bluelinetriangleright), shear wave (\redlinetriangleup). Dash lines represent the theoretical curves of Eq.~(\ref{eq:Regul_AR_theory_Ma_theta}). Hatched area: linear stability region of the LB scheme at $\tau/\Delta t=10^{-5}$.} 
    \label{fig:D2Q9_AR_degen_hypernu_N4}
\end{figure}

Finally, every directions are considered with the D2Q9 lattice, and for the three Hermite-based equilibria, in Figs.~\ref{fig:D2Q9_AR_degen_hypernu}-\ref{fig:D2Q9_AR_degen_hypernu_N4}. To this extent and following previous studies~\cite{Wissocq2020}, the analyses are performed for any orientation of the mean flow $M_\theta \in [0^\circ, 45^\circ]$ and for $k_\theta \in [0^\circ, 180^\circ]$, where $k_\theta=\mathrm{atan2}(k_y, k_x)$. For each value of $\theta \in [0, 4]$, the minimal value of the Mach number ensuring a stabilizing degenerating hyperviscosity is retained. Note that, when performing studies in any direction, downstream and upstream acoustic waves cannot be distinguished from each other, they are simply referred to as acoustic waves. This time, every value of $N$ leads to different results. In any case, no linear stability region can be observed when both $g_i^{(1)}$ and $\Psi_i$ are computed with analytical gradients. This is due to an instability occurring for very large values of $k$, \textit{i.e.} for under-resolved waves, which cannot be captured by the Taylor expansion. Same phenomena explains all the discrepancies obtained between the linear stability areas (hatched areas) and the nullity curves of the hyperviscous degeneracy. Regarding the DCO2 discretization of $g_i^{(1)}$ and $\Psi_i$, similar stability limits as previously are observed in any case: $\mathrm{Ma}<1$ and $\mathrm{CFL}<1$. With $N=2$, using a DUO1 discretization for $\Psi_i$ still allows reaching Mach numbers $\mathrm{Ma}>1$, even though this effect is reduced compared to the D1Q3 case because of a destabilizing hyperviscosity affecting the shear wave. However, with $N=3^*$ and $N=4^*$, the stability region is much reduced, which is not due to the degenerating hyperviscosity but to a under-resolution effect. %One can finally note that, when $g_i^{(1)}$ and $\Psi_i$ are both discretized using a DUO1 scheme, stability regions are surprisingly identical while the numerical hyperviscosity has not the same behavior on the acoustic waves.

\subsection{Summary of the numerical properties of the complete off-equilibrium reconstruction}

This comprehensive study of the shear stress tensor reconstruction (or HRR with $\sigma=0$) highlights important properties shared by several LB schemes. The main interest of this model is that, since the regularized part $g_i^{(1)}$ is known as an explicit function of $\rho$ and $u_\alpha$, deviation terms of the regularized models can be deeply scrutinized and general conclusions can be drawn. Moreover, the analysis of the D1Q3 lattice is made possible, while PR and RR regularizations, as well as MRT models, are reduced to the BGK collision in the D1Q3 case. The following major conclusions can be drawn:
\begin{itemize}
    \item The shear stress reconstruction reconstruction is sufficient to recover a complete consistency with the NS equations. In particular, the Galilean invariance errors in the stress tensor are cured.
    \item In absence of correction term $\Psi_i$, a numerical viscosity is evidenced, consequent of the first-order accuracy of the numerical scheme. Surprisingly, introducing the correction term $\Psi_i$, designed to fix consistency errors, cancels this numerical viscosity. This properties is summarized in Table \ref{tab:nature_visc_BGK_SSTR}.
    \item A degenerating numerical hyperviscosity is at the origin of over-dissipation and severe instabilities occurring for well resolved length scales, especially when the dimensionless relaxation time is low.
    \item With the DCO2 discretization for $g_i^{(1)}$ and $\Psi_i$, theoretical stability limits have been found as $\mathrm{Ma}=1$ and $\mathrm{CFL}=1$.
    \item Using a DUO1 discretization for $\Psi_i$ allows increasing the range of stable Mach numbers, up to $\mathrm{Ma} \approx 2$.
\end{itemize}

\begin{table}[h!]
    \centering
%    {\renewcommand{\arraystretch}{1.7}
    \begin{tabular}{Sc||Sc|Sc}
%    	\hline
         & BGK model & HRR-$\sigma=0$ \\
         \hhline{=||=|=}
         No Mach correction & Consistency error & Numerical error \\
         \hline
         Correction with $\Psi_i$ & Correct viscosity & Correct viscosity \\
    \end{tabular}
    \caption{Summary of the nature of viscous deviations ($Kn^1$-terms) affecting the BGK and HRR-$\sigma=0$ collision models.}
    \label{tab:nature_visc_BGK_SSTR}
\end{table}

Obviously, the present study being restricted to isothermal LB methods, simulations of large Mach numbers would be unphysical. However, the present conclusions are in perfect agreement with previous observations drawn with compressible models based on a hybrid approach~\cite{Renard2020, Renard2020b}, indicating that the numerical properties of their LB part is mainly responsible for the overall behavior. Especially, this confirms the benefits of an upwind estimation of the correction term.

This shear stress tensor reconstruction also provides the interest of filtering out any non-hydrodynamic mode~\cite{Astoul2020, Wissocq2020}. Unfortunately, the aforementioned numerical errors in hyperviscosity yields an over-dissipation that makes it unusable as it. This is why one usually prefer to hybridize it with the more common projected and recursive regularizations. The latter are investigated in the next section.

\section{Projected and recursive regularizations}
\label{sec:Regul_PR_RR}

This section aims at performing the Taylor expansion of two types of regularized models sharing strong similarities: the PR and RR models. Since these models are reduced to the BGK one in the D1Q3 case, only the D2Q9 lattice is considered here. One purpose of the present work is to provide some explanations to their unexpected dissipation properties exhibited in a previous work~\cite{Wissocq2020}. A unified formalism will be derived to simplify the analyses.

\subsection{Unified formalism}
\label{sec:Regul_PR_RR_unified_formalism}

In the projected (PR) and the recursive regularization (RR) including a body-force term, off-equilibrium distribution functions read~\cite{Latt2006, Malaspinas2015, Coreixas2017, Feng2019}
\begin{align}
    & \mathrm{PR}: \quad g_i^{(1)} = \frac{w_i}{2 c_s^4} a_{1, \alpha \beta}^{(2)} \mathcal{H}_{i, \alpha \beta}^{(2)} \quad \mathrm{where} \quad a_{1, \alpha \beta}^{(2)} = \sum_i \mathcal{H}_{i, \alpha \beta}^{(2)} \left(g_i - f_i^{eq} + \mathrm{Kn} \frac{\Delta t}{2 \tau} \Psi_i^d \right), \\
    & \mathrm{RR}: \quad g_i^{(1)} = w_i \sum_{n=2}^{N_r} \frac{1}{n! c_s^{2n}} \boldsymbol{a}_1^{(n)}:\boldsymbol{\mathcal{H}}_i^{(n)} \quad \mathrm{where} \quad a_{1, \alpha_1..\alpha_n}^{(n \geq 3)} = u_{\alpha_n} a_{1, \alpha_1..\alpha_{n-1}}^{(n-1)} + \left[ u_{\alpha_1}..u_{\alpha_{n-2}} a_{1, \alpha_{n-1} \alpha_n}^{(2)} + \mathrm{perm}(\alpha_n)  \right],
\end{align}
where $N_r$ is the order of the recursive regularization, which obeys the same constraints as $N$, and $\mathrm{perm}(\alpha_n)$ stands for all the cyclic permutations of indexes from $\alpha_1$ to $\alpha_{n-1}$. With the D2Q9 lattice, the off-equilibrium coefficients of interest read
\begin{align}
    & a_{1, xxy}^{(3)} = u_y a_{1, xx}^{(2)} + 2u_x a_{1, xy}^{(2)}, \quad a_{1, xyy}^{(3)} = 2u_y a_{1, xy}^{(2)} + u_x a_{1, yy}^{(2)}, \quad a_{1, xxyy}^{(4)} = u_y^2 a_{1, xx}^{(2)} + 4 u_x u_y a_{1, xy}^{(2)} + u_x^2 a_{1, xx}^{(2)}.
\end{align}
In the D2Q9 case, regularized distribution functions can be equivalently written in a general way as
\begin{align}
    g_i^{(1)} = \left[ \mathbf{H}^{-1} \mathbf{P} \mathbf{H} \right]_{ij} \left(g_j - f_j^{eq} + \mathrm{Kn} \frac{\Delta t}{2 \tau} \Psi_j^d \right),
    \label{eq:PR_RR_gi1}
\end{align}
where matrices $\mathbf{H}$ and $\mathbf{P}$ are defined as
\begin{align}
    & \mathbf{H} = \begin{pmatrix}
1 & \dots & 1  \\
e_{0, x} & \dots & e_{8, x} \\
e_{0, y} & \dots & e_{8, y} \\
\mathcal{H}_{0, xx}^{(2)} & \dots & \mathcal{H}_{8, xx}^{(2)} \\
\mathcal{H}_{0, xy}^{(2)} & \dots & \mathcal{H}_{8, xy}^{(2)} \\
\mathcal{H}_{0, xx}^{(2)} & \dots & \mathcal{H}_{8, xx}^{(2)} \\
\mathcal{H}_{0, xxy}^{(3)} & \dots & \mathcal{H}_{8, xxy}^{(3)} \\
\mathcal{H}_{0, xyy}^{(3)} & \dots & \mathcal{H}_{8, xyy}^{(3)} \\
\mathcal{H}_{0, xxyy}^{(4)} & \dots & \mathcal{H}_{8, xxyy}^{(4)} \\
\end{pmatrix}, \qquad 
\mathbf{P} = \begin{pmatrix}
0 & 0 & 0 & 0 & 0 & 0 & 0 & 0 & 0 \\
0 & 0 & 0 & 0 & 0 & 0 & 0 & 0 & 0 \\
0 & 0 & 0 & 0 & 0 & 0 & 0 & 0 & 0 \\
0 & 0 & 0 & 1 & 0 & 0 & 0 & 0 & 0 \\
0 & 0 & 0 & 0 & 1 & 0 & 0 & 0 & 0 \\
0 & 0 & 0 & 0 & 0 & 1 & 0 & 0 & 0 \\
0 & 0 & 0 & \lambda_1 & \lambda_2 & \lambda_3 & 0 & 0 & 0 \\
0 & 0 & 0 & \lambda_4 & \lambda_5 & \lambda_6 & 0 & 0 & 0 \\
0 & 0 & 0 & \lambda_7 & \lambda_8 & \lambda_9 & 0 & 0 & 0 \\
\end{pmatrix}.
\label{eq:Regul_matrix_P}
\end{align}
The coefficients $\lambda_i$ depend on the regularization and are provided in Table~\ref{tab:Regul_P_lambda_i}.
\begin{table}[h]
    \centering
    \begin{tabular}{|c|c|c|c|c|c|c|c|c|c|}
         \hline
         Regularization & $\lambda_1$ & $\lambda_2$ & $\lambda_3$ & $\lambda_4$ & $\lambda_5$ & $\lambda_6$ & $\lambda_7$ & $\lambda_8$ & $\lambda_9$ \\
         \hline
         PR & 0 & $0$ & $0$ & $0$ & $0$ & $0$ & $0$ & $0$ & $0$ \\
         RR,\ $N_r=3^*$ & $u_y$ & $2u_x$ & $0$ & $0$ & $2u_y$ & $u_x$ & $0$ & $0$ & $0$ \\
         RR,\ $N_r=4^*$ & $u_y$ & $2u_x$ & $0$ & $0$ & $2u_y$ & $u_x$ & $u_y^2$ & $4u_x u_y$ & $u_x^2$\\
         \hline
    \end{tabular}
    \caption{Coefficients of the matrix $\mathbf{P}$ of Eq.~(\ref{eq:Regul_matrix_P}) depending on the regularization scheme (PR or RR).}
    \label{tab:Regul_P_lambda_i}
\end{table}

Thanks to this equivalent matrix formulation of the PR and RR models, a straightforward link can be drawn with the MRT family of collision kernels, since
\begin{align}
    g_i^c = g_i + \underbrace{\left[ \left(1- \frac{\Delta t}{\tau + \Delta t/2} \right)\left[ \mathbf{H}^{-1} \mathbf{P} \mathbf{H} \right]_{ij} + \delta_{ij} \right]}_{M_{ij}} \left(g_j - f_j^{eq} + \mathrm{Kn} \frac{\Delta t}{2 \tau} \Psi_j^d \right) + \frac{\Delta t}{\tau} \mathrm{Kn} \Psi_i^d,
\end{align}
and $M_{ij}$ can be interpreted as a matrix for a collision with multiple relaxation times. It can be easily checked that with $M_{ij}=-\Delta t/(\tau + \Delta t/2) \delta_{ij}$, the BGK case of Sec.~\ref{sec:BGK} is recovered. Previous relations between the PR model and Hermite-based MRT models have already been found in the literature~\cite{Latt2007, Wissocq2020}, as well as between the RR scheme with $N=N_r=4^*$ and a MRT model based on centered Hermite moments~\cite{Coreixas2019}. Here, such a unified formalism is interesting since it allows an easy switch between the three models of interest.

\subsection{Variable change}

In order to incorporate this definition of $g_i^{(1)}$ into the generic Taylor expansion of regularized models of Eq.~(\ref{eq:Regul_Taylor_2}), a variable change now has to be performed so that the $f_i$ variables explicitly appear. Combining Eq.~(\ref{eq:PR_RR_gi1}) with the variable change of Eq.~(\ref{eq:Regul_variable_change}) yields
\begin{align}
    g_i^{(1)} = \left[ \mathbf{H}^{-1} \mathbf{P} \mathbf{H} \right]_{ij} \left( \left( 1-\frac{\Delta t}{\tau + \Delta t/2} \right) g_j^{(1)} + \frac{\Delta t}{\tau} (f_j - f_j^{eq}) \right) \\
    \Rightarrow \left[ \mathbf{I} + \left( \frac{\Delta t}{\tau + \Delta t/2} \right) \mathbf{H}^{-1} \mathbf{P} \mathbf{H}  \right]_{ij} g_j^{(1)} = \frac{\Delta t}{\tau} \left[ \mathbf{H}^{-1} \mathbf{P} \mathbf{H} \right]_{ij} \left( f_j - f_j^{eq} \right),
\end{align}
so that, by inverting this system, one finally has
\begin{align}
    g_i^{(1)} = \frac{\Delta t}{\tau} \mathbf{T}_{ij} (f_j-f_j^{eq}) \quad \mathrm{with} \quad \mathbf{T} = \left[  \mathbf{I} + \left( \frac{\Delta t}{\tau + \Delta t/2} \right) \mathbf{H}^{-1} \mathbf{P} \mathbf{H} \right]^{-1} \mathbf{H}^{-1} \mathbf{P} \mathbf{H}.
\end{align}
Regularized distribution functions are then directly related to the off-equilibrium populations $f_i-f_i^{eq}$, which is in the order of magnitude of the Knudsen number. Then, the definition of $\hat{g}_i^{(1)}$ of Eq.~(\ref{eq:Regul_def_hat_gi1}) makes sense in the context of PR and RR models, leading to
\begin{align}
    \hat{g}_i^{(1)} = \frac{1}{\mathrm{Kn}} \frac{\Delta t}{\tau + \Delta t/2} \mathbf{T}_{ij} \left( f_j - f_j^{eq} \right).
\end{align}

\subsection{Computation of error terms in the linear approximation}

Once $\hat{g}_i^{(1)}$ is explicitly known for both models, the deviation terms with regards to the DVBE $E_i^{(n)}$ can be computed  using the Taylor expansion of Eq.~(\ref{eq:Regul_Taylor_2}). This is done following the same procedure as with the complete regularization of Sec.~\ref{sec:Regul_FR_D2Q9_deviations}. A particularity of the PR and RR models is that, given the matrix-shape of the regularized distributions, cross derivatives appear as
\begin{align}
    & D_i D_j = \left( \partial_t + e_{i, \alpha} \partial_\alpha \right) \left( \partial_t + e_{j, \beta} \partial_\beta \right), \quad D_i D_j D_k = \left( \partial_t + e_{i, \alpha} \partial_\alpha \right) \left( \partial_t + e_{j, \beta} \partial_\beta \right) \left( \partial_t + e_{k, \gamma} \partial_\gamma \right), \nonumber \\
    & D_i D_j D_k D_l = \left( \partial_t + e_{i, \alpha} \partial_\alpha \right) \left( \partial_t + e_{j, \beta} \partial_\beta \right) \left( \partial_t + e_{k, \gamma} \partial_\gamma \right) \left( \partial_t + e_{l, \delta} \partial_\delta \right).
\end{align}
Here again, the computer algebra system Maxima~\cite{maxima} is used to compile the equations, providing the $E_i^{(n)}$ terms as well as the macroscopic deviations $E_\rho^{(n)}$ and $E_{\rho u_\alpha}^{(n)}$, following \ref{app:macroscopic_equations}. The latter are provided below in the D2Q9 case for the PR and RR models, for several equilibria and with different implementations of the correction terms. Note that similar deviations terms are obtained in the $x$ direction for $N=3^*$ and $N=4^*$, as well as for $N_r=3^*$ and $N_r=4^*$ in the case of the RR collision model.

Below, only the modified error terms compared to the BGK counterpart are provided, considering derivatives in the $x$-direction for the sake of simplicity.

\subsubsection{Non corrected PR model with $N=2$}

\begin{align}
    %& E^{(1)}_\rho = 0, \qquad E^{(1)}_{\rho u_x} = -\partial^2_{xx} a^{(3)}_{eq, xxx} - 2 \partial^2_{xy} a^{(3)}_{eq, xxy} - \partial^2_{yy} a^{(3)}_{eq, xyy}, \qquad E^{(1)}_{\rho u_y} = -\partial^2_{xx} a^{(3)}_{eq, xxy} - 2 \partial^2_{xy} a^{(3)}_{eq, xyy} - \partial^2_{yy} a^{(3)}_{eq, yyy}, \nonumber \\
	%& E^{(2)}_\rho = \left( \frac{\Delta t}{\tau} \right)^2 \frac{1}{12} \left( -\partial_{xxx}^3 a^{(3)}_{eq, xxx} - 3 \partial^3_{xxy} a^{(3)}_{eq, xxy} - 3\partial_{xyy}^3 a^{(3)}_{eq, xyy} - \partial_{yyy}^3 a^{(3)}_{eq, yyy} + \rho c_s^2 \theta (\partial^2_{xx} S_{xx} + 2\partial^2_{xy} S_{xy} + \partial^2_{yy} S_{yy}) \right), \nonumber\\
	& E_{\rho u_y}^{(2)} = \left( C_5 + 2c_s^2 \frac{\partial^3 (\rho u_x u_y)}{\partial x^3}  \right) - \left( \frac{\Delta t}{\tau} \right) c_s^2 \frac{\partial^3 (\rho u_x u_y)}{\partial x^3} + \left( \frac{\Delta t}{\tau} \right)^2 N_5, \nonumber \\
	& E_{\rho u_y}^{(3)} = C_{11} + \left( \frac{\Delta t}{\tau} \right) N_{27} + \left( \frac{\Delta t}{\tau} \right)^2 \left( N_6 - N_{28} \right) + \left( \frac{\Delta t}{\tau} \right)^3 \frac{1}{4} N_{28}.
\end{align}
Two types of numerical errors arise in the $y$-momentum equation: a first-order error in time, in agreement with a first-order accuracy expected from a previous work~\cite{Wissocq2019_these, Wissocq2020}, and a degenerating numerical hyperviscosity related to the $N_{28}$ term. For a horizontal mean flow, the latter behaves as
\begin{align}
    N_{28}\left|_{(u_y=0)} \right. = \frac{\rho u_x^2}{3} \frac{\partial^4 u_y}{\partial x^4}, 
\end{align}
\textit{i.e.} as a positive hyperviscosity affecting shear flows. Let us recall here that a positive hyperviscosity yields an anti-dissipation. Hence this numerical error is likely to be at the origin of the instability of the shear waves exhibited in a previous work~\cite{Wissocq2020}, which will be further confirmed in Sec.~\ref{sec:Regul_PR_RR_validation}. 

\subsubsection{Non corrected RR model with $N_r=3^*, 4^*$ and $N=2$}

\begin{align}
	& E^{(2)}_{\rho u_y} = C_{13} + \left( \frac{\Delta t}{\tau} \right) N_{30} + \left( \frac{\Delta t}{\tau} \right)^2 N_5, \qquad E_{\rho u_y}^{(3)} = C_{14} +  \left( \frac{\Delta t}{\tau} \right) N_{31} + \left( \frac{\Delta t}{\tau} \right)^2 \left( N_6 + N_{32} \right) + \left( \frac{\Delta t}{\tau} \right)^3 N_{33},
\end{align}
Exactly like the PR model, both first-order error in $\Delta t$ and a degenerating hyperviscosity ($N_{33}$) are introduced. For a horizontal mean flow, the latter behaves as
\begin{align}
    N_{33}\left|_{(u_y=0)} \right. = \frac{\rho u_x^4}{4} \frac{\partial^4 u_y}{\partial x^4}, 
\end{align}
which is a destabilizing hyperviscosity affecting the shear wave only.

\subsubsection{Non corrected PR model with $N=3^*, 4^*$}

\begin{align}
	& E^{(2)}_{\rho u_y} = a_{eq, xxx}^{(3)} \frac{\partial^3 u_y}{\partial x^3} + C_{15} - \left( \frac{\Delta t}{2 \tau} \right) C_{15}  + \left( \frac{\Delta t}{\tau} \right)^2 N_{15}, \nonumber \\
	& E^{(3)}_{\rho u_y} = C_{16} + \left( \frac{\Delta t}{\tau} \right) N_{40} + \left( \frac{\Delta t}{\tau} \right)^2 \left( N_{10} + N_{41} \right) + \left( \frac{\Delta t}{\tau} \right)^3 N_{42}.
\end{align}
Again, a first-order error in $\Delta t$ together with a degenerating hyperviscosity, related to the $N_{42}$ term, occur. For a horizontal mean flow, the latter behaves as
\begin{align}
    N_{42}\left|_{(u_y=0)} \right. = -\frac{1}{8} \rho\left(u_x^4 + (c_s^2 \theta-1)u_x^2 \right) \frac{\partial^4 u_y}{\partial x^4}.
\end{align}

\subsubsection{Non corrected RR model with $N_r=3^*, 4^*$ and $N=3^*, 4^*$}

\begin{align}
	& E^{(2)}_{\rho u_y} = \left( \frac{\Delta t}{2 \tau} \right) a_{eq, xxx}^{(3)} \frac{\partial^3 u_y}{\partial x^3} + \left( \frac{\Delta t}{\tau} \right)^2 N_{15}, \nonumber \\
	& E^{(3)}_{\rho u_y} = -\frac{\rho \theta^2}{9} \frac{\partial^4 u_y}{\partial x^4} -\left( \frac{\Delta t}{6\tau} \right) (3 \theta u_x^2+c_s^2\theta^2-\theta) \rho \frac{\partial^4 u_y}{\partial x^4} + \left( \frac{\Delta t}{\tau} \right)^2 \left( N_{10} + \rho u_x a_{eq, xxx}^{(3)} \frac{\partial^4 u_y}{\partial x^4} \right) + \left( \frac{\Delta t}{\tau} \right)^3 N_{44}.
\end{align}

\subsection{Corrected models}

Deviation terms of corrected PR and RR models are provided in \ref{app:deviations_corrected_PR_RR}. The correction term $\Psi_i$ is either computed analytically, or discretized with a DCO2 and a DUO1 scheme.

\subsection{Hyperviscous degeneracy of PR and RR models}

Expliciting the deviation terms of the PR and RR models allows highlighting three specific features of the regularization. First, the scheme is globally first-order accurate in time, as prospected in a previous work~\cite{Wissocq2020}, but the corresponding errors do not affect the recovery of the NS equations. This first-order accuracy in $\Delta t$ could be guessed from the general Taylor expansion of regularized models of Eq.~(\ref{eq:Regul_Taylor_2}).  It is now clearly exhibited for each model. The fact that the recovery of the NS equations remains second-order accurate is in agreement with past convergence studies~\cite{Latt2007, Malaspinas2015, Montessori2014, Montessori2014b, Mattila2015, Li2019}. Secondly, the regularization step induces a degenerating hyperviscosity in each model. The latter remains uncontrolled as an unexpected consequence of time and space disretization errors. This point could also be guessed from the general form of Eq.~(\ref{eq:Regul_Taylor_2}). This uncontrolled error is as large as the dimensionless relaxation time is low, and causes an unintended dissipation or instability, depending on the sign of the numerical hyperviscosity. Thirdly, while the add of a correction term $\Psi_i$ affects the degenerating hyperviscosity, the way it is discretized (DCO2 or DUO1 schemes) has no effect on it when the $x$ direction is considered only.

\begin{table}[h!]
    \centering
%    {\renewcommand{\arraystretch}{1.7}
    \begin{tabular}{Sc||Sc|Sc}
%    	\hline
         & PR model & RR model with $N_r=3^*, 4^*$ \\
         \hhline{=||=|=}
         \multirow{2}{*}{$N=2$, non corrected} & $\displaystyle \frac{\rho u_x^2}{12} \frac{\partial^4 u_y}{\partial x^4}$ & $\displaystyle \frac{\rho u_x^4}{4} \frac{\partial^4 u_y}{\partial x^4}$ \\
         & (destabilizing) & (destabilizing) \\
         \hline
         \multirow{2}{*}{$N=2$, corrected} & $\displaystyle \frac{\rho u_x^2}{12} \frac{\partial^4 u_y}{\partial x^4}$ & $ \displaystyle (1-\theta)\frac{\rho u_x^2}{12} \frac{\partial^4 u_y}{\partial x^4}$ \\
         & (destabilizing) & (stabilizing when $\theta > 1$) \\
         \hline
         \multirow{2}{*}{$N=3^*, 4^*$, non corrected} & $\displaystyle -\frac{1}{8} \rho\left(u_x^4 + (c_s^2 \theta-1)u_x^2 \right) \frac{\partial^4 u_y}{\partial x^4}$  & $\displaystyle -\frac{1}{8} \rho \left(u_x^4+(\theta-1)u_x^2 \right) \frac{\partial^4 u_y}{\partial x^4} $ \\
         & (stabilizing when $\mathrm{Ma}^2 \geq 1/(c_s^2 \theta) -1$) & (stabilizing when $\mathrm{Ma}^2 \geq (1-\theta)/(c_s^2\theta )$) \\
         \hline
         \multirow{2}{*}{$N=3^*, 4^*$, corrected} & $\displaystyle -\frac{1}{8} \rho\left(u_x^4 + (c_s^2 \theta-1)u_x^2 \right) \frac{\partial^4 u_y}{\partial x^4}$ & $ \displaystyle-\frac{1}{8} \rho \left(u_x^4+(\theta-1)u_x^2 \right) \frac{\partial^4 u_y}{\partial x^4}$ \\
         & (stabilizing when $\mathrm{Ma}^2 \geq 1/(c_s^2 \theta) -1$) & (stabilizing when $\mathrm{Ma}^2 \geq (1-\theta)/(c_s^2\theta )$) \\
%         \hline
    \end{tabular}
    \caption{Summary of the degenerating hyperviscous terms arising in $E_{\rho u_y}^{(3)}$ with the PR and RR models on the D2Q9 lattice, when only $x$ derivatives are considered and with a horizontal mean flow assumption.}
    \label{tab:Regul_degen_hypernu_PR_RR}
\end{table}

A summary of the degenerating hyperviscous terms affecting the PR and RR models is provided in Table~\ref{tab:Regul_degen_hypernu_PR_RR} for different equilibria, corrected or not, when only the $x$-derivatives are considered together with the assumption of a horizontal mean flow. In any case, only a space derivative of the transverse velocity ($u_y$) affects the conservation equation of the $y$-momentum, which means that this numerical error has an impact on the shear waves only. Also note that this error is responsible for instabilities when the numerical hyperviscosity is positive, and for an increasing disssipation when it is negative. These errors terms are provided here along the $x$-direction only for the sake of simplicity, without questioning the generalities of the methodology. When other directions are considered, far more complex phenomena occur which prevent from drawing such simple conclusions. In particular, an anisotropy of both acoustic and shear phenomena is expected in two dimensions~\cite{Wissocq2020}.

\subsection{Validation with linear analyses}
\label{sec:Regul_PR_RR_validation}

As previously, the error terms provided above have been quantitatively validated by comparing the linear analyses of the regularized LB schemes with that of the macroscopic equations obtained by the Taylor expansion, following the methodology detailed in Sec.~\ref{sec:BGK_validation}. In the present section, a more qualitative study is proposed to investigate the impact of the numerical error on the dissipation properties in the linear approximation. 

\begin{figure}
    \begin{minipage}[b]{\linewidth}
    \centering
    \includegraphics[scale=1]{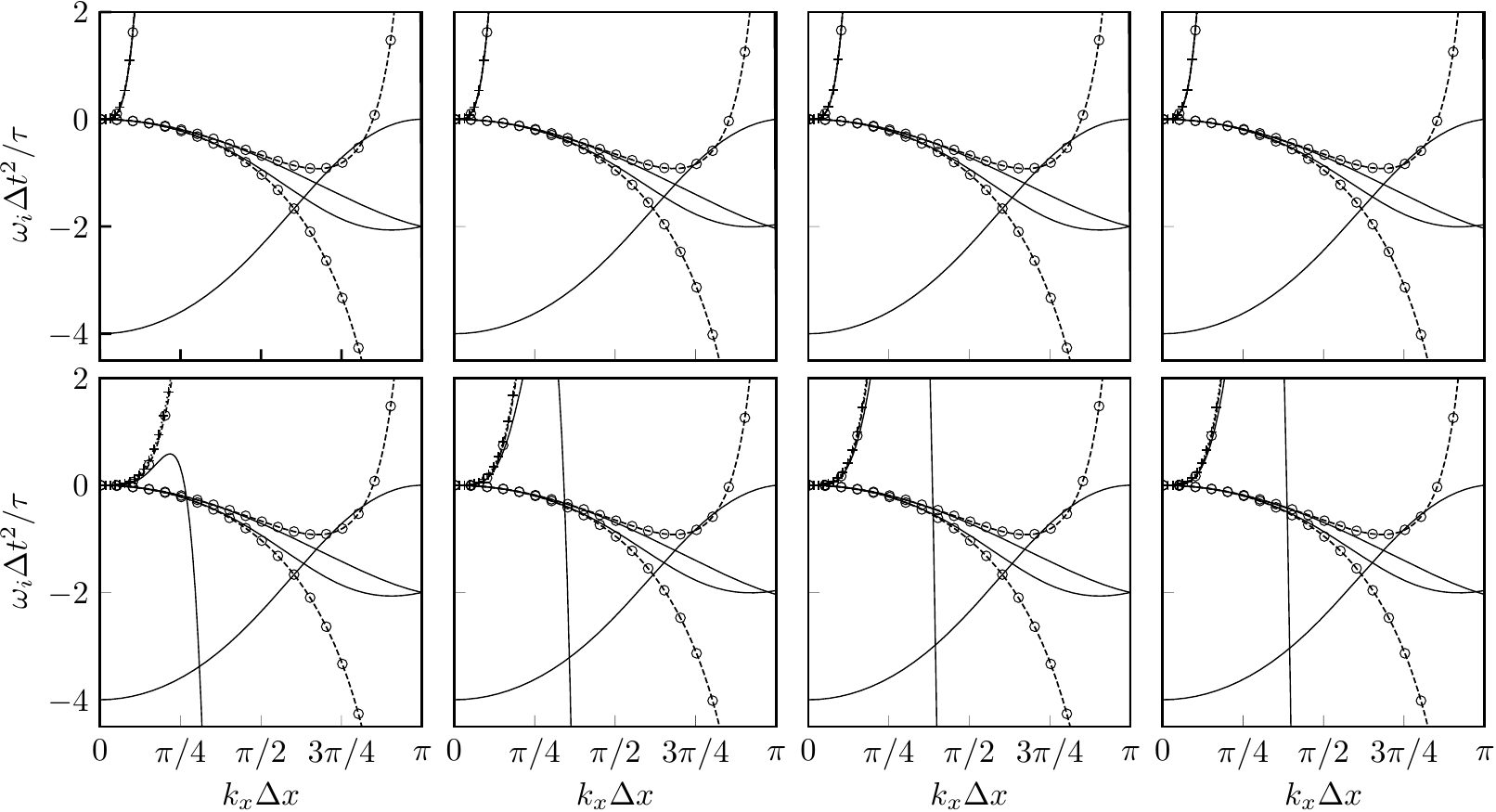}
    \subcaption{$\mathrm{Ma}=0.25$, $\theta=0.9$}
    \end{minipage}\\ \vspace{5mm}
    \begin{minipage}[b]{\linewidth}
    \centering
    \includegraphics[scale=1]{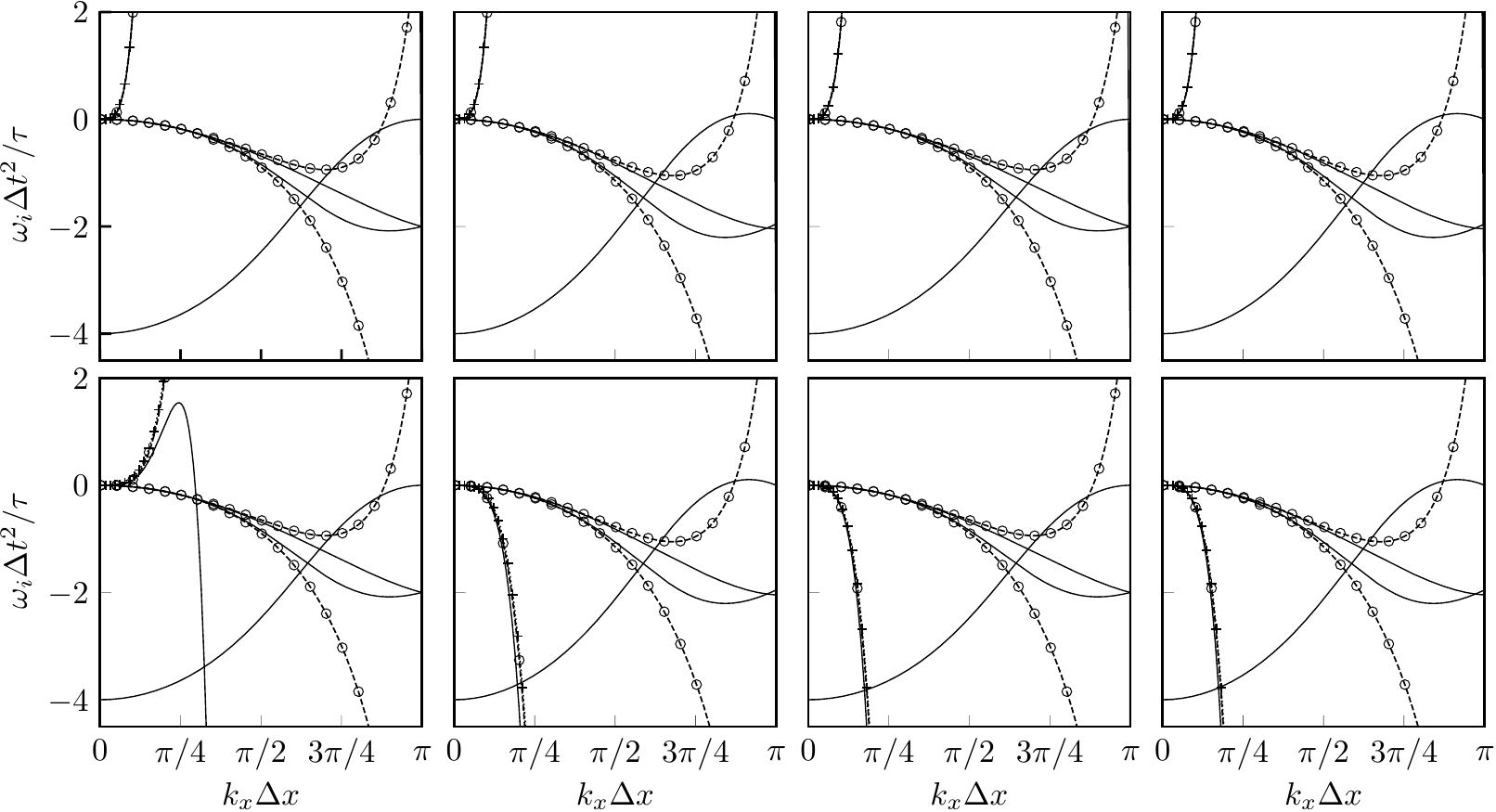}
    \subcaption{$\mathrm{Ma}=0.25$, $\theta=1.1$}
    \label{fig:fig_PR_RR_M025_theta1.1}
    \end{minipage}
    \caption{Dissipation curves of the PR and RR models with the D2Q9 lattice considering horizontal plane waves ($k_y=0$), a horizontal mean flow at $\mathrm{Ma}=0.25$ and $\tau/\Delta t=10^{-5}$. Top: PR model, bottom: RR model with $N_r=3^*$. From left to right: $N=2$ non corrected, $N=2$ with analytical correction, $N=3^*$ non corrected, $N=3^*$ with analytical correction. Three linear analyses are compared: the LB scheme (\blacklineveryfine), the NS equations including deviation terms up to $\mathrm{Kn}^3$ (\dashedlinecircle), and the estimated degenerating hyperviscosity of Table~\ref{tab:Regul_degen_hypernu_PR_RR} (\dottedlinecross).}
    \label{fig:fig_PR_RR_M025}
\end{figure}

Fig.~\ref{fig:fig_PR_RR_M025} displays the dissipation curves obtained from linear analyses of the PR and RR models in the $x$-direction, with a horizontal mean flow at $\mathrm{Ma}=0.25$. In the present case, the discretization effects of the correction term $\Psi_i$ are not considered since they do not affect the hyperviscous degeneracy. These curves are compared with the analyses of the isothermal NS equations including the deviation terms up to the third-order in $\mathrm{Kn}$. The analytical estimation of the degenerating hyperviscosity from Table~\ref{tab:Regul_degen_hypernu_PR_RR} is also presented. For the latter, the dimensionless amplification rate is simply obtained by replacing the fourth-order derivative of $u_y$ by $(k_x \Delta x)^4$, \textit{e.g.} for the non corrected PR model with $N=2$:
\begin{align}
    \omega_i \Delta t = \frac{\rho u_x^2}{12} (k_x \Delta x)^4.
\end{align}
Analyses with a fourth-order equilibrium and a fourth-order regularization show no difference, in the $x$-direction, with the third-order cases. Therefore, only combinations of $N=2, 3^*$ and $N_r=2, 3^*$ are displayed on the figure. 

A very good agreement is obtained between all the curves, except a supplementary non-hydrodynamic wave, whose behavior cannot be predicted by Taylor expansions in $\mathrm{Kn}$. As expected, only one wave (the shear wave) is greatly affected by the regularization in the $x$-direction. Especially, the analytical estimation of the numerical hyperviscosity provides an accurate estimation of its behavior. This indicates that the hyperviscous degeneracy is responsible for the large instabilities and over-dissipation of this wave compared to the theoretically expected curves, for a dimensionless relaxation time $\tau/\Delta t=10^{-5}$. As a consequence, every models are unstable at low Mach number when $\theta < 1$, even though stability can be recovered for large wavenumbers with the RR model, which is due to uncontrolled higher Knudsen effects. For $3 > \theta \geq 1$, only three models can ensure a stable shear wave in the $x$-direction for a non zero Mach number: the corrected RR model with $N=2$ and the RR model with $N=3^*$, corrected or not.

Unfortunately, other studies of the hyperviscous degeneracy performed in any direction (not shown here) indicate that, whatever the considered model (PR or RR), there still exists a combination of directions for which it is destabilizing even at very low Mach number ($\mathrm{Ma}=0.01$). Hence, the degenerating hyperviscosity is globally destabilizing for both the PR and RR models on the D2Q9 lattice. As a consequence, the numerical scheme becomes unstable when the dimensionless relaxation time decreases, even for very low Mach numbers. This explains why the critical Mach numbers obtained by linear stability analyses conducted in previous work tend to zero in the zero-viscosity limit~\cite{Coreixas2018, Wissocq2019_these, Coreixas2020, Wissocq2020}.

\subsection{Summary}

In summary, the Taylor expansions performed on the PR and RR models lead to the following general conclusions:
\begin{itemize}
    \item Even if the scheme is globally first-order accurate in $\Delta t$, no first-order error affects the isothermal NS equations. In particular, these schemes induce no numerical viscosity.
    \item A degenerating hyperviscosity has been exhibited, which is as large as the dimensionless relaxation time is small. When only the $x$-direction is considered, only the shear wave is affected. Its analytical expression could be obtained and validated with linear analyses.
    \item Considering all the possible directions, the degenerating hyperviscosity is destabilizing for both PR and RR models.
\end{itemize}

In addition to the instabilities, the numerical error in hyperviscosity results in a strong anisotropy of the dissipation properties of the PR and RR models, exhibited in a previous work~\cite{Wissocq2020}.

In practice, the instabilities can be of relatively low amplitude, such as with the non corrected RR model with $N=2$ of Fig.~\ref{fig:fig_PR_RR_M025_theta1.1}. But the scheme remains linearly unstable, and the instabilities are all the more pronounced as the dimensionless relaxation time decreases. As a consequence, there is no combination of parameters $(\mathrm{Ma}, \theta)$ for which the hyperviscosity is stabilizing in any direction, unlike the complete reconstruction of Sec.~\ref{sec:Regul_FR}. This is the main reason why these models are sometimes coupled with each other with a parameter $\sigma$, giving birth to the so-called hybrid recursive regularized (HRR) scheme~\cite{Jacob2018}. Unfortunately, its numerical errors are unlikely to be a simple linear combination of that of the RR and the complete regularization. This is due to three reasons: (1) the complex moment cascade, (2) the fact that the variable change of the PR/RR model is also affected by $\sigma$, (3) the fact that, in the original model of Jacob \textit{et al.}~\cite{Jacob2018}, the reccurrence also applies to the complete reconstruction of the shear stress tensor, which has not been considered here. A Taylor expansion of this model could be the purpose of future work.

\section{Application to MRT models}
\label{sec:MRT_models}

Given the strong analogy that can be drawn between PR/RR schemes and MRT models, one can finally wonder how the latter behave in terms of numerical error, more precisely regarding their dissipation properties. In this regard, an increasing numerical error of some athermal MRT models has already been exhibited in the incompressible limit by Dellar~\cite{Dellar2003}. Noticing that reducing the Mach number of a simulation for a fixed Reynolds number comes down to decreasing the dimensionless relaxation time related to the kinematic viscosity, it is likely that a degenerating hyperviscosity is at the origin of such an error. To shed some light on these phenomena, a Taylor expansion in $\mathrm{Kn}$ is proposed here for some MRT models of interest. For the sake of simplicity, the correction term $\Psi_i$ is not considered in this section. As a consequence, $\theta=1$ is adopted here, since a value of $\theta \neq 1$ would yield a very large error (linear in Mach number) in the shear stress tensor. Moreover, only the D2Q9 lattice is considered in this section.

\subsection{Description of the MRT models under consideration}
\label{sec:MRT_descriptions}

Post-collision population of MRT models can generally be written as
\begin{align}
    g_i^c = g_i - \mathbf{C}_{ij} \left( g_j - f_j^{eq} \right),
    \label{eq:MRT_general_formulation}
\end{align}
where $\mathbf{C}$, the collision matrix, is supposed to be diagonal when written in a certain basis of moments, which means that every moments are relaxed independently towards an equilibrium state. The collision matrix then reads
\begin{align}
    \mathbf{C} = \mathbf{M}^{-1} \mathbf{S} \mathbf{M},
\end{align}
where $\mathbf{S}$ is a diagonal matrix of relaxation rates and $\mathbf{M}$ is the matrix of the considered moments. In the present work, four bases of moments are considered: the so-called raw-moments (RM), the central moments (CM), the Hermite moments (HM) and the central Hermite moments (CHM). More information on these bases can be found in~\cite{Coreixas2019}. The MRT model based on central moments is sometimes referred to as \textit{cascaded} model~\cite{Geier2006, Dubois2015}. In the present context, the choice of second-order moments is slightly modified so that the bulk viscosity can be freely adjusted only changing one diagonal value of $\mathbf{S}$. The moments matrices read

\begingroup
\allowdisplaybreaks
\begin{align}
    & \mathbf{M}^{\mathrm{RM}} = \begin{pmatrix}
1 & \dots & 1  \\
e_{0, x} & \dots & e_{8, x} \\
e_{0, y} & \dots & e_{8, y} \\
e_{0, x}^2 - e_{0, y}^2 & \dots & e_{8, x}^2 - e_{8, y}^2 \\
e_{0, x} e_{0, y} & \dots & e_{8, x} e_{8, y} \\
e_{0, x}^2 + e_{0, y}^2 & \dots & e_{8, x}^2 + e_{8, y}^2 \\
e_{0, x}^2 e_{0, y} & \dots & e_{8, x}^2 e_{8, y} \\
e_{0, x} e_{0, y}^2 & \dots & e_{8, x} e_{8, y}^2 \\
e_{0, x}^2 e_{0, y}^2 & \dots & e_{8, x}^2 e_{8, y}^2
\end{pmatrix}, \qquad
\mathbf{M}^{\mathrm{CM}} = \begin{pmatrix}
1 & \dots & 1  \\
c_{0, x} & \dots & c_{8, x} \\
c_{0, y} & \dots & c_{8, y} \\
c_{0, x}^2 - c_{0, y}^2 & \dots & c_{8, x}^2 - c_{8, y}^2 \\
c_{0, x} c_{0, y} & \dots & c_{8, x} c_{8, y} \\
c_{0, x}^2 + c_{0, y}^2 & \dots & c_{8, x}^2 + c_{8, y}^2 \\
c_{0, x}^2 c_{0, y} & \dots & c_{8, x}^2 c_{8, y} \\
c_{0, x} c_{0, y}^2 & \dots & c_{8, x} c_{8, y}^2 \\
c_{0, x}^2 c_{0, y}^2 & \dots & c_{8, x}^2 c_{8, y}^2
\end{pmatrix}, \\
& \mathbf{M}^{\mathrm{HM}} = \begin{pmatrix}
1 & \dots & 1  \\
e_{0, x} & \dots & e_{8, x} \\
e_{0, y} & \dots & e_{8, y} \\
\mathcal{H}_{0, xx}^{(2)} - \mathcal{H}_{0, yy}^{(2)} & \dots & \mathcal{H}_{8, xx}^{(2)} - \mathcal{H}_{8, yy}^{(2)} \\
\mathcal{H}_{0, xy}^{(2)} & \dots & \mathcal{H}_{8, xy}^{(2)} \\
\mathcal{H}_{0, xx}^{(2)} + \mathcal{H}_{0, yy}^{(2)} & \dots & \mathcal{H}_{8, xx}^{(2)} + \mathcal{H}_{8, yy}^{(2)} \\
\mathcal{H}_{0, xxy}^{(3)} & \dots & \mathcal{H}_{8, xxy}^{(3)} \\
\mathcal{H}_{0, xyy}^{(3)} & \dots & \mathcal{H}_{8, xyy}^{(3)} \\
\mathcal{H}_{0, xxyy}^{(4)} & \dots & \mathcal{H}_{8, xxyy}^{(4)} \\
\end{pmatrix}, \qquad 
\mathbf{M}^{\mathrm{CHM}} = \begin{pmatrix}
1 & \dots & 1  \\
c_{0, x} & \dots & c_{8, x} \\
c_{0, y} & \dots & c_{8, y} \\
\mathcal{H}_{0, xx}^{*,(2)} - \mathcal{H}_{0, yy}^{*,(2)} & \dots & \mathcal{H}_{8, xx}^{*,(2)} - \mathcal{H}_{8, yy}^{*,(2)} \\
\mathcal{H}_{0, xy}^{*,(2)} & \dots & \mathcal{H}_{8, xy}^{*,(2)} \\
\mathcal{H}_{0, xx}^{*,(2)} + \mathcal{H}_{0, yy}^{*,(2)} & \dots & \mathcal{H}_{8, xx}^{*,(2)} + \mathcal{H}_{8, yy}^{*,(2)} \\
\mathcal{H}_{0, xxy}^{*,(3)} & \dots & \mathcal{H}_{8, xxy}^{*,(3)} \\
\mathcal{H}_{0, xyy}^{*,(3)} & \dots & \mathcal{H}_{8, xyy}^{*,(3)} \\
\mathcal{H}_{0, xxyy}^{*,(4)} & \dots & \mathcal{H}_{8, xxyy}^{*,(4)} \\
\end{pmatrix},
\end{align}
\endgroup

where $c_{i, \alpha}=e_{i, \alpha} - u_\alpha$ and $\boldsymbol{\mathcal{H}}_i^{*, (n)}$ are centered Hermite polynomials built using $c_{i, \alpha}$ instead of $e_{i, \alpha}$. A fundamental property of $\mathbf{C}$ is that it preserves collision invariants which are the mass density $\rho$ and the momentum $\rho u_x$ and $\rho u_y$. As a consequence, the sub-space spanned by $(1,1,1,1,1,1,1,1,1)^T$, $\boldsymbol{e_x}$ and $\boldsymbol{e_y}$ forms an eigenspace of the collision matrix. Its corresponding eigenvalues have no influence on the collision and can be arbitrarily set to $1$, so that $\mathbf{S}=\mathrm{diag}(1, 1, 1, s_4, s_5, s_6, s_7, s_8, s_9)$. With these definitions, the dimensionless kinematic viscosity $\nu$ and bulk viscosity $\nu_b$ are related to $s_4$, $s_5$ and $s_6$ as:
\begin{align}
    s_4=s_5=\frac{\Delta t}{\tau + \Delta t/2}, \qquad s_6=\frac{\Delta t}{\tau_b + \Delta t/2}, \qquad \nu=k^*\frac{\tau}{\Delta t} c_s^2, \qquad \nu_b=k^* \frac{\tau_b}{\Delta t} c_s^2.
    \label{eq:MRT_relations_s_tau}
\end{align}
Even though several relaxation times appear, a Taylor expansion in Knudsen number is still possible, provided that $\mathrm{Kn}$ is based on the relaxation time $\tau$ related to the shear viscosity $\nu$. Such a work is performed below for a fixed ratio $\tau_b/\tau=\nu_b/\nu$ and for fixed additional relaxation rates $s_7$, $s_8$ and $s_9$.

\subsection{General Taylor expansion of MRT models}

A Taylor expansion of the general expression of MRT models given by Eq.~(\ref{eq:MRT_general_formulation}) yields
\begin{align}
    D_i g_i = -\frac{1}{\mathrm{Kn}} \frac{\tau}{\Delta t} \mathbf{C}_{ij} \left( g_j - f_j^{eq} \right) - \frac{\mathrm{Kn}}{2} \frac{\Delta t}{\tau} D_i^2 g_i - \frac{\mathrm{Kn}^2}{6} \left( \frac{\Delta t}{\tau} \right)^2 D_i^3 g_i - \frac{\mathrm{Kn}^3}{24} \left( \frac{\Delta t}{\tau} \right)^3 D_i^4 g_i + O(\mathrm{Kn}^4).
    \label{eq:MRT_general_Taylor_1}
\end{align}
The following change of variable is then performed:
\begin{align}
    -\frac{1}{\mathrm{Kn}} \frac{\Delta t}{\tau} \mathbf{C}_{ij} \left( g_j - f_j^{eq} \right) = -\frac{1}{\mathrm{Kn}} \left( f_i - f_i^{eq} \right),
\end{align}
which leads to
\begin{align}
    g_i = f_i + \mathbf{A}_{ij} \left( f_j - f_j^{eq} \right), \qquad \mathbf{A} = \frac{\Delta t}{\tau} \mathbf{C}^{-1} - \mathbf{I},
\end{align}
where $\mathbf{I}$ is the Identity matrix. Since $\rho$ and $\rho u_\alpha$ are collision invariants, the equalities $\sum_i g_i = \sum_i f_i = \rho$ and $\sum_i e_{i, \alpha} g_i = \sum_i e_{i, \alpha} f_i = \rho u_\alpha$ are ensured, so that the variable change is admissible. Injecting it in Eq.~(\ref{eq:MRT_general_Taylor_1}) yields
\begin{align}
    D_i f_i &= -\frac{1}{\mathrm{Kn}} \left( f_i - f_i^{eq} \right) - \frac{\mathrm{Kn}}{2} \frac{\Delta t}{\tau} D_i^2 f_i - \frac{\mathrm{Kn}^2}{6} \left( \frac{\Delta t}{\tau} \right)^2 D_i^3 f_i - \frac{\mathrm{Kn}^3}{24} \left( \frac{\Delta t}{\tau} \right)^3 D_i^4 f_i \nonumber \\
    & \qquad - \mathbf{A}_{ij} D_i f_j^{(1)} - \frac{\mathrm{Kn}}{2} \frac{\Delta t}{\tau} \mathbf{A}_{ij} D_i^2 f_j^{(1)} - \frac{\mathrm{Kn}^2}{6} \left( \frac{\Delta t}{\tau} \right)^2 \mathbf{A}_{ij} D_i^3 f_j^{(1)} - \frac{\mathrm{Kn}^3}{24} \left( \frac{\Delta t}{\tau} \right)^3 \mathbf{A}_{ij} D_i^4 f_j^{(1)} + O(\mathrm{Kn}^4),
    \label{eq:MRT_general_Taylor_2}
\end{align}
with $f_i^{(1)} = f_i - f_i^{eq}$. Note that the choice is made here to compare the MRT-LB scheme with the BGK-DVBE. The deviations should not be interpreted as numerical errors only, since the MRT-LB scheme is not supposed to converge towards the BGK-DVBE, but to its \textit{a priori} unknown MRT counterpart. In particular, some deviations between the two equations are physically expected, such as a change of bulk viscosity enabled by $s_6$.

To obtain a general equation in the same form as that of the BGK collision model (\ref{eq:Taylor_BGK_3}) or the regularized ones (\ref{eq:Regul_Taylor_2}), one then needs to express the derivatives of $f_i^{(1)}$ as derivatives of $f_i$ only. To this extent, Eq.~(\ref{eq:MRT_general_Taylor_2}) can be re-written as
\begin{align}
    f_j^{(1)} = -\mathrm{Kn}\, \left( D_j f_j + \mathbf{A}_{jk} D_j f_k^{(1)} \right) - \frac{\mathrm{Kn}^2}{2} \frac{\Delta t}{\tau} \left( D_j^2 f_j + \mathbf{A}_{jk} D_j^2 f_k^{(1)} \right) - \frac{\mathrm{Kn}^3}{6} \left( \frac{\Delta t}{\tau} \right)^2 \left( D_j^3 f_j + \mathbf{A}_{jk} D_j^3 f_k^{(1)} \right) + O(\mathrm{Kn}^4).
\end{align}
%By successive derivations of this equation, the following relations can be obtained:
%\begin{align}
%    & \mathrm{Kn}^3 D_i^4 f_j^{(1)} = O(\mathrm{Kn}^4), \nonumber \\
%    & \mathrm{Kn}^2 D_i^3 f_j^{(1)} = -\mathrm{Kn}^3 D_i^3 D_j f_j + O(\mathrm{Kn}^4), \nonumber \\
%    & \mathrm{Kn}\, D_i^2 f_j^{(1)} = -\mathrm{Kn}^2 D_i^2 D_j f_j + \mathrm{Kn}^3 \left( \mathbf{A}_{jk} D_i^2 D_j D_k f_k - \frac{\Delta t}{2 \tau} D_i^2 D_j^2 f_j \right) + O(\mathrm{Kn}^4), \nonumber \\
%    & D_i f_j^{(1)} = -\mathrm{Kn} \, D_i D_j f_j + \mathrm{Kn}^2 \left( \mathbf{A}_{jk}D_i D_j D_k f_k - \frac{\Delta t}{2 \tau} D_i D_j^2 f_j \right) \nonumber \\
%    & \qquad + \mathrm{Kn}^3 \left( -\frac{1}{6} \left( \frac{\Delta t}{\tau} \right)^2 D_i D_j^3 f_j + \frac{\Delta t}{2\tau} \mathbf{A}_{jk} \left( D_i D_j D_k^2 f_k + D_i D_j^2 D_k f_k \right) - \mathbf{A}_{jk} \mathbf{A}_{kl} D_i D_j D_k D_l f_l \right) + O(\mathrm{Kn}^4).
%\end{align}
After successive derivations of this equation with respect to $D_i$, injecting these expressions into Eq.~(\ref{eq:MRT_general_Taylor_2}) leads to
\begin{align}
    D_i f_i &= -\frac{1}{\mathrm{Kn}} \left( f_i - f_i^{eq} \right) + \mathrm{Kn} \left[ \mathbf{A}_{ij} D_i D_j f_j - \frac{\Delta t}{2 \tau}D_i^2 f_i \right] \nonumber \\
    & + \mathrm{Kn}^2 \left[ -\mathbf{A}_{ij} \mathbf{A}_{jk} D_i D_j D_k f_k + \frac{\Delta t}{2 \tau} \mathbf{A}_{ij} \left( D_i D_j^2 f_j + D_i^2 D_j f_j \right) - \frac{1}{6} \left( \frac{\Delta t}{\tau} \right)^2 D_i^3 f_i \right] \nonumber \\
    & + \mathrm{Kn}^3 \Bigg[ \mathbf{A}_{ij} \mathbf{A}_{jk} \mathbf{A}_{kl} D_i D_j D_k D_l f_l - \frac{\Delta t}{2 \tau} \mathbf{A}_{ij} \mathbf{A}_{jk} \left( D_i D_j D_k^2 f_k + D_i D_j^2 D_k f_k + D_i^2 D_j D_k f_k \right) + \frac{1}{4} \left( \frac{\Delta t}{\tau} \right)^2 \mathbf{A}_{ij} D_i^2 D_j^2 f_j \nonumber \\
    & \qquad + \frac{1}{6} \left( \frac{\Delta t}{\tau} \right)^2 \mathbf{A}_{ij} \left( D_i D_j^3 f_j + D_i^3 D_j f_j \right) - \frac{1}{24} \left( \frac{\Delta t}{\tau} \right)^3 D_i^4 f_i  \Bigg] + O(\mathrm{Kn}^4).
    \label{eq:MRT_general_Taylor_3}
\end{align}
Looking at this equation, several observations can be drawn:
\begin{itemize}
    \item In the case of the BGK collision, the matrix $\mathbf{A}$ is reduced to $\mathbf{A}=\Delta t/(2\tau) \mathbf{I}$. Replacing this expression into Eq.~(\ref{eq:MRT_general_Taylor_3}) allows recovering the general Taylor expansion of the BGK model of Eq.~(\ref{eq:Taylor_BGK_3}), up to the fourth-order in Knudsen number.
    \item Like regularized models, a first-order error in $\mathrm{Kn}$ appear. Its impact on the macroscopic equations in terms of potential numerical viscosity will be investigated.
    \item Similarly, a first-order error in time arises like with regularized models. This is in agreement with a previous demonstration of the first-order accuracy of the PR model based on a MRT formalism~\cite{Wissocq2020}.
    \item A degeneracy of the numerical hyperviscosity is exhibited by the presence of $(\Delta t/\tau)^3$-related terms in the third-order in $\mathrm{Kn}$. Noticing that the matrix $\mathbf{A}$ involves $(\Delta t/\tau)$ itself, the pre-factor of this term cannot be estimated in a general way. Analyses adapted to each MRT model are performed in the next section to interpret the effects of the degeneracy.
\end{itemize}

Starting from Eq.~(\ref{eq:MRT_general_Taylor_3}), the same procedure as with the BGK and regularized models can be followed to obtain the deviation terms $E_i^{(n)}$, which are function of space gradients only. Then, deviation terms with the isothermal NS equations $E_{\Phi}^{(n)}$ can be obtained following \ref{app:macroscopic_equations}. Here again, the computer algebra system Maxima~\cite{maxima} is used to simplify the equations.

\subsection{Error terms in the linear approximation}

In this section, deviation terms with the isothermal NS equations are provided for each MRT model of interest, in the linear approximation. Contrary to Eqs.~(\ref{eq:mass_errors})-(\ref{eq:momentum_errors}), an adjustable bulk viscosity has to be considered here. The adopted formalism for the macroscospic equations is then
\begin{align}
	& \partial_t \rho + \partial_\alpha (\rho u_\alpha) = \mathrm{Kn} E_\rho^{(1)} + \mathrm{Kn}^2 E_\rho^{(2)} + \mathrm{Kn}^3 E_\rho^{(3)} + O(\mathrm{Kn}^4), \\
	& \partial_t (\rho u_\alpha) + \partial_\beta (\rho u_\alpha u_\beta) + c_s^2 \partial_\alpha \rho = \mathrm{Kn} \, \rho c_s^2 \, \partial_\beta \left(S_{\alpha \beta} + \left( \frac{\tau_b}{\tau} - \frac{2}{D}\right) \partial_\gamma u_\gamma \delta_{\alpha \beta} \right) + \mathrm{Kn}E_{\rho u_\alpha}^{(1)} + \mathrm{Kn}^2 E_{\rho u_\alpha}^{(2)} + \mathrm{Kn}^3 E_{\rho u_\alpha}^{(3)} + O(\mathrm{Kn}^4).
\end{align} 
Because of the very large size of the error terms involved in the MRT models (due to the large numbers of free parameters), only two informations will be provided below:
\begin{enumerate}
    \item the first-order deviation terms $E_\rho^{(1)}$ and $E_{\rho u_\alpha}^{(1)}$, including consistency errors and possible numerical errors, in every direction,
    \item the degenerating hyperviscosity, \textit{i.e.} $(\Delta t/\tau)^3$-related terms in $E_\rho^{(3)}$ and $E_{\rho u_\alpha}^{(3)}$, considering the $x$-direction only.
\end{enumerate}
In common with all the models investigated up to now, there is no first-order deviation from the mass equation with MRT models: $E_\rho^{(1)}=0$. The latter will not be recalled below.

\subsubsection{MRT with raw-moments}

With a second-order equilibrium ($N=2$), the first-order deviation terms on the momentum equation read
\begin{align}
        & E_{\rho u_x}^{(1)} = - \left( \frac{\kappa+1}{2} \right) \partial^2_{xx} a^{(3)}_{eq,xxx} - \left( \frac{\kappa-1}{2} \right) \partial^2_{xx} a_{eq, xyy}^{(3)} - \left( \frac{\kappa +3}{2} \right) \partial^2_{xy} a_{eq, xxy}^{(3)} - \left( \frac{\kappa-1}{2} \right) \partial^2_{xy} a_{eq, yyy}^{(3)} - \partial^2_{yy} a_{eq, xyy}^{(3)} , \nonumber \\
    & E_{\rho u_y}^{(1)} = - \partial^2_{xx} a_{eq, xxy}^{(3)} - \left( \frac{\kappa-1}{2} \right) \partial^2_{xy} a_{eq, xxx}^{(3)} - \left( \frac{\kappa +3}{2} \right) \partial^2_{xy} a_{eq, xyy}^{(3)}  - \left( \frac{\kappa-1}{2} \right) \partial^2_{yy} a_{eq, xxy}^{(3)} - \left( \frac{\kappa+1}{2} \right) \partial^2_{yy} a^{(3)}_{eq,yyy},
    \label{eq:MRT_RM_N2_first_order_deviation}
\end{align}
where $\kappa=\tau_b/\tau$. This results in a consistency error only, which is due to the truncation of the equilibrium distribution function to the second-order. Like the PR and the RR model, there is no numerical viscosity even if the scheme is first-order accurate.

%\begin{align}
%    & E_\rho^{(1)} = 0, \qquad E_{\rho u_x}^{(1)} = - \left( \frac{\kappa+1}{2} \right) \partial^2_{xx} a^{(3)}_{eq,xxx} - \left( \frac{\kappa-1}{2} \right) \partial^2_{xx} a_{eq, xyy}^{(3)} - \left( \frac{\kappa +3}{2} \right) \partial^2_{xy} a_{eq, xxy}^{(3)} - \left( \frac{\kappa-1}{2} \right) \partial^2_{xy} a_{eq, yyy}^{(3)} - \partial^2_{yy} a_{eq, xyy}^{(3)} , \nonumber \\
%    & E_{\rho u_y}^{(1)} = - \partial^2_{xx} a_{eq, xxy}^{(3)} - \left( \frac{\kappa-1}{2} \right) \partial^2_{xy} a_{eq, xxx}^{(3)} - \left( \frac{\kappa +3}{2} \right) \partial^2_{xy} a_{eq, xyy}^{(3)}  - \left( \frac{\kappa-1}{2} \right) \partial^2_{yy} a_{eq, xxy}^{(3)} - \left( \frac{\kappa+1}{2} \right) \partial^2_{yy} a^{(3)}_{eq,yyy}, \nonumber \\
%	& E^{(2)}_\rho = \left( \frac{\Delta t}{\tau} \right)^2 \frac{1}{12} \left( -\partial_{xxx}^3 a^{(3)}_{eq, xxx} - 3 \partial^3_{xxy} a^{(3)}_{eq, xxy} - 3\partial_{xyy}^3 a^{(3)}_{eq, xyy} - \partial_{yyy}^3 a^{(3)}_{eq, yyy} + \rho c_s^2 \theta (\partial^2_{xx} S_{xx} + 2\partial^2_{xy} S_{xy} + \partial^2_{yy} S_{yy}) \right), \nonumber \\
%	& E_{\rho u_x}^{(2)} = C_1 + (\kappa-1) C_{20} + \left( \frac{\Delta t}{\tau} \right) \frac{1}{s_8} N_{55} + \left( \frac{\Delta t}{\tau} \right)^2 N_{56}, \qquad E_{\rho u_y}^{(2)} = C_5 + C_{21} + \left( \frac{\Delta t}{\tau} \right) \frac{1}{s_7} N_{57} + \left( \frac{\Delta t}{\tau} \right)^2 N_{58}, \nonumber \\
%	& E_\rho^{(3)} = \left( \frac{\Delta t}{\tau} \right)^2 \frac{1}{12} \partial_x C_1 + (\kappa-1)C_{22}, \qquad E_{\rho u_x}^{(3)} = 
%\end{align}

When only the $x$-direction is considered, there is no hyperviscous degeneracy on the mass equation, nor on the $x$-momentum equation. However, the $y$-momentum equation is affected as
\begin{align}
    E_{\rho u_y}^{(3)} \sim - \left( \frac{\Delta t}{\tau} \right)^3 \frac{s_7-2}{36 s_7} \left( (3 u_x^2+1) u_y \frac{\partial^4 \rho}{\partial x^4} + 6 \rho u_x u_y \frac{\partial^4 u_x}{\partial x^4}  + 3 \rho u_x^2 \frac{\partial^4 u_y}{\partial x^4} \right).
\end{align}

When the equilibrium distribution is enriched with third- and/or fourth-order terms ($N=3^*$ or $N=4^*$), the first-order errors on the momentum equations read
\begin{align}
    & E_{\rho u_x}^{(1)} = - \left( \frac{\kappa+1}{2} \right)\partial^2_{xx} a_{eq, xxx}^{(3)}  - \left( \frac{\kappa-1}{2} \right) \partial^2_{xy} a_{eq,yyy}^{(3)}, \qquad E_{\rho u_y}^{(1)} = - \left( \frac{\kappa+1}{2} \right)\partial^2_{yy} a_{eq, yyy}^{(3)}  - \left( \frac{\kappa-1}{2} \right) \partial^2_{xy} a_{eq,xxx}^{(3)}.
    \label{eq:MRT_RM_N3_first_order_deviation}
\end{align}
All the error terms that were related to $a_{eq,xxy}^{(3)}$ and $a_{eq,xyy}^{(3)}$ have been cancelled thanks to the improved equilibrium. Still with $N=3^*$ and $N=4^*$, only the $y$-momentum equation is affected by a hyperviscous degeneracy when the $x$-direction is considered. It reads
\begin{align}
    E_{\rho u_y}^{(3)} \sim \left( \frac{\Delta t}{\tau} \right)^3 \frac{s_7-2}{72 s_7} \left( (9 u_x^4+9 u_x^2-2) u_y \frac{\partial^4 \rho}{\partial x^4} + (36 u_x^3-6 u_x) \rho u_y \frac{\partial^4 u_x}{\partial x^4} + (9 u_x^4-6 u_x^2) \rho \frac{\partial^4 u_y}{\partial x^4} \right)
\end{align}

\subsubsection{MRT with central moments}

When MRT models based on central moments are considered, the exactly same first-order deviations as the raw-moments formulation are recovered when $N=2$ (Eq.~(\ref{eq:MRT_RM_N2_first_order_deviation})) and with $N=3^*$ or $N=4^*$ (Eq.~(\ref{eq:MRT_RM_N3_first_order_deviation})). This means that, contrary to a common intuition, performing the collision step in the local basis of centered moments does not lead to a correction of the Galilean invariance error in the NS equations: the errors remain unaffected by the moments basis. This observation is in agreement with previous studies of the linearized DVBE with multiple relaxation times~\cite{PAM2019}.

When the $x$-direction is considered only, there is no degenerating hyperviscosity on the mass and $x$-momentum equations, while on the $y$-momentum equations it reads, when $N=2$,
\begin{align}
    E_{\rho u_y}^{(3)} \sim -\left( \frac{\Delta t}{\tau} \right)^3 \frac{s_7-2}{24s_7} \left(  9(u_x^4+u_x^2) u_y \frac{\partial^4 \rho}{\partial x^4} + (30 u_x^3+2 u_x) \rho u_y \frac{\partial^4 u_x}{\partial x^4} + 6 \rho u_x^4 \frac{\partial^4 u_y}{\partial x^4} \right).
\end{align}
For $N=3^*$ and $N=4^*$, it is reduced to
\begin{align}
    E_{\rho u_y}^{(3)} \sim \left( \frac{\Delta t}{\tau} \right)^3 \frac{s_7-2}{8s_7} \rho u_x^4 \frac{\partial^4 u_y}{\partial x^4}.
\end{align}

\subsubsection{MRT with Hermite moments}

With the MRT-HM scheme, the same first-order errors are obtained as with the raw-moment formulation, \textit{i.e.} Eq.~(\ref{eq:MRT_RM_N2_first_order_deviation}) with $N=2$ and Eq.~(\ref{eq:MRT_RM_N3_first_order_deviation}) with $N=3^*$ or $N=4^*$. Regarding the degenerating hyperviscosity, one has, for $N=2$,
\begin{align}
    & E_{\rho u_y}^{(3)} \sim -\left( \frac{\Delta t}{\tau} \right)^3 \frac{s_7-2}{36s_7} \left( (3 u_x^2+1) u_y \frac{\partial^4 \rho}{\partial x^4} + 6 \rho u_x u_y \frac{\partial^4 u_x}{\partial x^4} + 3 \rho u_x^2 \frac{\partial^4 u_y}{\partial x^4}  \right),
\end{align}
which is similar to that of the MRT-RM model. For $N=3^*$ or $N=4^*$, one has
\begin{align}
    & E_{\rho u_y}^{(3)} \sim \left( \frac{\Delta t}{\tau} \right)^3 \frac{s_7-2}{72s_7} \left( (9 u_x^4+9 u_x^2-2) u_y \frac{\partial^4 \rho}{\partial x^4} + (36 u_x^3-6 u_x) \rho u_y \frac{\partial^4 u_x}{\partial x^4} + (9 u_x^4-6 u_x^2) \rho \frac{\partial^4 u_y}{\partial x^4} \right).
\end{align}
Note that when $s_7=1$, the error terms of the projected regularization (PR) are recovered.

\subsubsection{MRT with central Hermite-moments}

Finally, with the MRT-CHM scheme, the first-order errors in Knudsen number are still similar to that of the aforementioned MRT models. In the $x$-direction, only the $y$-momentum equation is affected by a degenerating hyperviscosity, which reads, for $N=2$:
\begin{align}
    & E_{\rho u_y}^{(3)} \sim -\left( \frac{\Delta t}{\tau} \right)^3 \frac{s_7-2}{24s_7} \left( 9( u_x^4+ u_x^2) u_y \frac{\partial^4 \rho}{\partial x^4} + (30 u_x^3+2 u_x) \rho u_y \frac{\partial^4 u_x}{\partial x^4} + 6 \rho u_x^4 \frac{\partial^4 u_y}{\partial x^4} \right),
\end{align}
which is similar to that of the MRT-CM model. For $N=3^*$ or $N=4^*$, one has
\begin{align}
    & E_{\rho u_y}^{(3)} \sim \left( \frac{\Delta t}{\tau} \right)^3 \frac{s_7-2}{8s_7} \rho u_x^4 \frac{\partial^4 u_y}{\partial x^4}
\end{align}
This term is similar to that of the MRT-CM model. It is reduced to the error term of the RR model with $N_R=3^*, 4^*$ when $N=3^*, 4^*$ when $s_7=1$, as prospected by Coreixas \textit{et al.}~\cite{Coreixas2019}.

\subsection{Hyperviscous degeneracy of MRT models}

Looking at the form of the degenerating error in hyperviscosity allows drawing general conclusions for all the MRT models of interest. When only the $x$-direction is considered:
\begin{itemize}
    \item the degenerating hyperviscosity does not involve the bulk viscosity, but only the relaxation rate $s_7$,
    \item when $s_7$=2, the degeneracy is cancelled.
\end{itemize}
These conclusions can even be generalized in every direction, expanding them to the three relaxation rates related to non-hydrodynamic variables: $s_7$, $s_8$ and $s_9$. The particular case $s_7=s_8=s_9=2$ will be further discussed in Sec.~\ref{sec:MRT_degen_hypernu_cancelled}. \newline

Exactly like PR and RR models, the hyperviscous degeneracy only affects the dissipation of the shear wave when the $x$-direction is considered together with a horizontal mean flow ($u_y=0$). A summary of their analytical expression is compiled in Table~\ref{tab:MRT_degen_hypernu}. Let us recall here that a positive hyperviscosity yields a negative dissipation, which can lead to an instability. To this regards, the drastic effects of a simple change of equilibrium distribution function on the linear stability are exhibited. In any case, a change of the dissipation properties is expected at the critical value $s_7=2$.

\begin{table}[ht]
    \centering
%    {\renewcommand{\arraystretch}{1.7}
    \begin{tabular}{Sc||Sc|Sc}
%    	\hline
         & $N=2$ & $N=3^*, 4^*$ \\
         \hhline{=||=|=}
         \multirow{2}{*}{MRT-RM} & $\displaystyle -\left( \frac{s_7-2}{12 s_7} \right) \rho u_x^2 \frac{\partial^4 u_y}{\partial x^4} $ & $\displaystyle \left( \frac{s_7-2}{24 s_7} \right) (3 u_x^2-2) \rho u_x^2 \frac{\partial^4 u_y}{\partial x^4} $ \\
          & (stabilizing when $s_7>2$) & (stabilizing when $s_7>2$ while $\mathrm{Ma}<\sqrt{2}$) \\
         \hline
         \multirow{2}{*}{MRT-CM} & $\displaystyle  - \left( \frac{s_7-2}{4s_7} \right) \rho u_x^4 \frac{\partial^4 u_y}{\partial x^4}$ & $\displaystyle \left( \frac{s_7-2}{8s_7} \right) \rho u_x^4 \frac{\partial^4 u_y}{\partial x^4}$ \\
          & (stabilizing when $s_7>2$) & (stabilizing when $s_7<2$) \\
         \hline
         \multirow{2}{*}{MRT-HM} & $\displaystyle  - \left( \frac{s_7-2}{12s_7} \right) \rho u_x^2 \frac{\partial^4 u_y}{\partial x^4} ,
$  & $\displaystyle  \left( \frac{s_7-2}{24s_7} \right) (3 u_x^2-2) \rho u_x^2 \frac{\partial^4 u_y}{\partial x^4} $ \\
         & (stabilizing when $s_7>2$) & (stabilizing when $s_7>2$ while $\mathrm{Ma}<\sqrt{2}$) \\
         \hline
         \multirow{2}{*}{MRT-CHM} & $\displaystyle - \left( \frac{s_7-2}{4s_7} \right) \rho u_x^4 \frac{\partial^4 u_y}{\partial x^4}$ & $\displaystyle  \left(\frac{s_7-2}{8s_7} \right) \rho u_x^4 \frac{\partial^4 u_y}{\partial x^4}$ \\
         & (stabilizing when $s_7>2$) & (stabilizing when $s_7<2$) \\
%         \hline
    \end{tabular}
    \caption{Summary of the degenerating hyperviscous terms arising in $E_{\rho u_y}^{(3)}$ with the non corrected MRT models on the D2Q9 lattice, when only $x$ derivatives are considered and with a horizontal mean flow assumption.}
    \label{tab:MRT_degen_hypernu}
\end{table}

\subsection{Validation with linear analyses}

The hyperviscous degeneracy exhibited in the above section is now confronted to the linear analyses of the real MRT LB schemes. To this purpose, the eigenvalue problem corresponding to the general MRT scheme is provided in \ref{app:matrices_LSA}.

\begin{figure}
    \begin{minipage}[b]{\linewidth}
    \centering
    \includegraphics[scale=1]{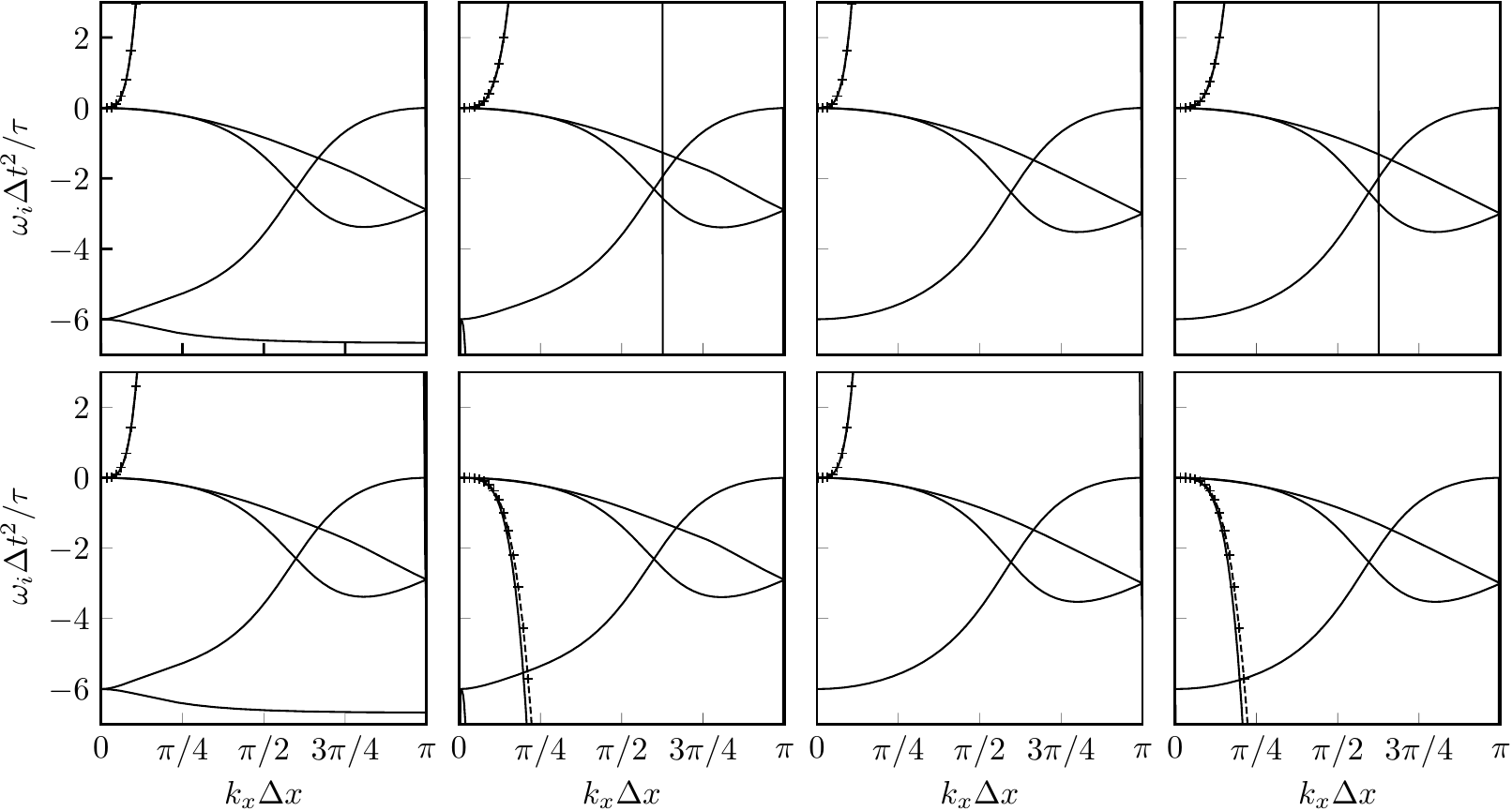}
    \subcaption{$s_7=s_8=s_9=1.5$, $\tau_b=2\tau$}
    \end{minipage}\\ \vspace{5mm}
    \begin{minipage}[b]{\linewidth}
    \centering
    \includegraphics[scale=1]{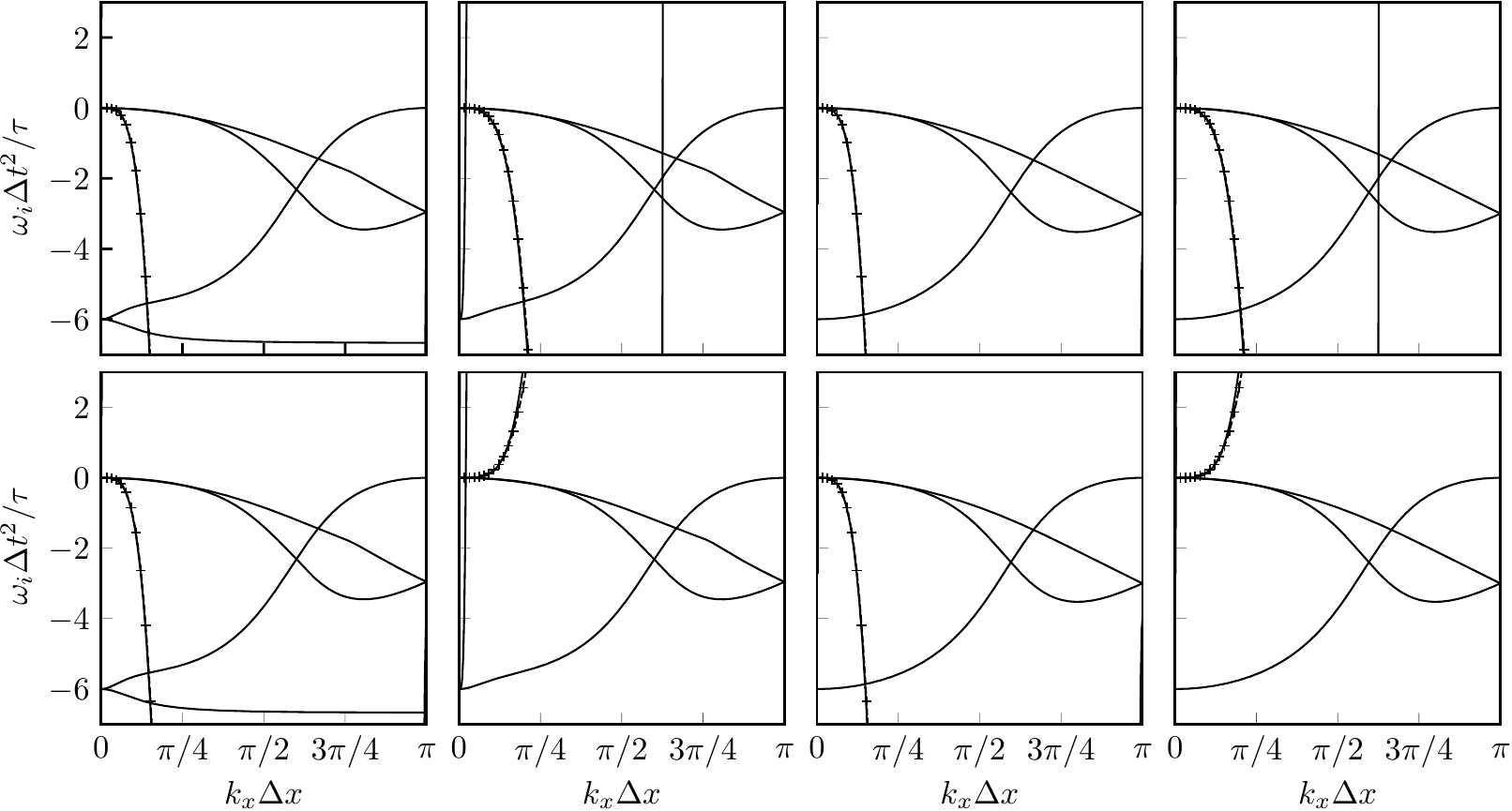}
    \subcaption{$s_7=s_8=s_9=2.5$, $\tau_b=2\tau$}
    \label{fig:D2Q9_MRT_hyperviscous_s7_2.5}
    \end{minipage}
    \caption{Dissipation curves of the MRT models with the D2Q9 lattice considering horizontal plane waves ($k_y=0$), a horizontal mean flow at $\mathrm{Ma}=0.5$ and $\tau/\Delta t=10^{-5}$. Top: $N=2$, bottom: $N=3^*$ (similar results with $N=4^*$). From left to right: MRT-RM, MRT-CM, MRT-HM and MRT-CHM. The linear analyses of the LB scheme (\blacklinefine) are compared to the estimated degenerating hyperviscosity of Table~\ref{tab:MRT_degen_hypernu} (\dashedlinecross).}
    \label{fig:D2Q9_MRT_hyperviscous}
\end{figure}

Fig.~\ref{fig:D2Q9_MRT_hyperviscous} displays the dissipation curves of the four MRT models with $N=2$ and $N=3^*$, in comparison with the analytical estimation of the degenerating hyperviscosity affecting $x$-aligned shear waves of Table \ref{tab:MRT_degen_hypernu}. For instance, for the MRT-RM model with $N=2$, this estimation reads
\begin{align}
    \omega_i \Delta t = -\left( \frac{s_7-2}{12s_7} \right) \rho u_x^2 (k_x \Delta x)^4.
\end{align}
In every case, waves travelling along the $x$-direction are investigated ($k_y=0$) with a horizontal mean flow at Mach number $\mathrm{Ma}=0.5$, together with a dimensionless relaxation time $\tau/\Delta t=10^{-5}$. The bulk relaxation time is set to $\tau_b=2\tau$ to avoid the particular case of the BGK collision model $\tau_b=\tau$, and two parameters are evaluated for the relaxation rate of third- and fourth-order moments: $s_7=s_8=s_9=1.5$ and $s_7=s_8=s_9=2.5$. Note that the second case is above the bounds proposed by Lallemand \& Luo ($0<\tau_N<2$)~\cite{Lallemand2000}.

The numerical prediction of the behavior of the three physical waves provided by the Taylor expansion is in very good agreement with the linear analyses. In particular, two waves, which are actually acoustic waves, seem unaffected by the choice of MRT model and compared to the BGK collision. This is in agreement with the estimation of the degenerating hyperviscosity in the $x$-direction, only affecting the shear wave. Regarding the latter, it is, exactly like the PR and RR models of Sec.~\ref{sec:Regul_PR_RR}, either over-dissipated or anti-dissipated, \textit{i.e.} leading to linear instabilities. The results from Table \ref{tab:MRT_degen_hypernu} allow an accurate estimation of this behavior, indicating that the degenerating hyperviscosity is responsible for these phenomena. Notably, the change of behavior above the critical value $s_7=2$ is recovered for all models.

Looking at these linear analyses is also the opportunity to focus on the behavior of non-hydrodynamic modes, which cannot be predicted by the Taylor expansions in $\mathrm{Kn}$. Even if only two of them are visible in Fig.~\ref{fig:D2Q9_MRT_hyperviscous}, six hydrodynamic modes are actually present. The four remaining modes are much more damped, so that they cannot be observed on the figure. Note that, with every MRT model except the RM formulation, one non-hydrodynamic mode is largely dissipated when $s_7<2$, and amplified for $s_7>2$. The latter is at the origin of a strong instability in the $x$-direction.

\subsection{Cancelling the degeneracy with MRT models}
\label{sec:MRT_degen_hypernu_cancelled}

Two particular cases, for which the degeneracy is cancelled, can be exhibited. Denoting here $s_N=s_7=s_8=s_9$ for a generalization in every direction:
\begin{enumerate}
    \item In the case of the BGK collision model, one has
    \begin{align}
        s_N=\frac{\Delta t}{\tau+\Delta t/2} \quad \Rightarrow \quad \frac{s_N-2}{s_N} = -\frac{2\tau}{\Delta t},
    \end{align}
    so that every $(\Delta t/\tau)^3$-related terms are removed from the numerical hyperviscosity. A result of Sec.~\ref{sec:BGK} is recovered: there is no hyperviscous degeneracy with the BGK collision model. This observation can even be extended to a BGK collision with adaptive bulk viscosity, since the degenerating hyperviscosity does not depend on $\kappa=\tau_b/\tau$.
    \item With $s_N=2$, the degenerating hyperviscosity is cancelled in any direction. Considering fixed relaxation parameters (independent of $\tau/\Delta t$), this is the only value that leads to such a property.
\end{enumerate}
The second case is in line with several linear stability analyses of MRT models aiming at finding optimal sets of free parameters, which often result in some values close to the critical parameter $\tau_N=2$~\cite{Lallemand2000, DHumieres2002, xu2011, xu2012, Gendre2018}. By re-writing this relaxation rate as
\begin{align}
    s_N = \frac{\Delta t}{\tau_N + \Delta t/2},
\end{align}
this particular case is equivalent to $\tau_N=0$, meaning that high-order moments are immediately imposed to their equilibrium value by the collision model, without relaxation. Note that setting $s_N > 2$ is equivalent to a negative relaxation time $\tau_N$, which can explain the large amplification of a non-hydrodynamic mode observed in Fig.~\ref{fig:D2Q9_MRT_hyperviscous_s7_2.5}.\newline

Interestingly, imposing higher-order moments to their equilibrium value is precisely the purpose of the PR model (in the Hermite basis) and the RR one (in the central Hermite basis). This fact seems in disagreement with the above observation. To shed the light on this property, let us re-write the collision step affecting these moments as
\begin{align}
    \boldsymbol{m}^{(n), coll} = \boldsymbol{m}^{(n)} - s_N \left( \boldsymbol{m}^{(n)} - \boldsymbol{m}^{(n)}_{eq} \right) = \boldsymbol{m}^{(n)}_{eq} - (s_N-1) \left( \boldsymbol{m}^{(n)} - \boldsymbol{m}^{(n)}_{eq} \right),
\end{align}
where $\boldsymbol{m}^{(n)}$ and $\boldsymbol{m}^{(n)}_{eq}$ are respectively moments and equilibrium moments of order $n \geq 3$, no matter the definition of moments (RM, CM, HM or CHM). With this notation, the regularization procedure is equivalent to setting $s_N=1$. Hence, by denoting $\boldsymbol{m}^{(n), coll}_{neq} = \boldsymbol{m}^{(n), coll} - \boldsymbol{m}^{(n), coll}_{eq}$, two cases are noticeable:

\begin{flalign}
    \mathrm{MRT\ with\ } \tau_N=0\ (s_N=2):\qquad &\boldsymbol{m}^{(n), coll}_{neq} = -\boldsymbol{m}^{(n)}_{neq}, \label{eq:MRT_overrelaxation} \\
    \mathrm{Regularization}\ (s_N=1):\qquad & \boldsymbol{m}^{(n), coll}_{neq} = 0.
\end{flalign}

The first case has the advantage of cancelling the degenerating hyperviscosity thanks to the particular value $s_N=2$. However, this value enforces an over-relaxation of the corresponding moments resulting in a sign inversion of their off-equilibrium parts, as shown by Eq.~(\ref{eq:MRT_overrelaxation}). It is at the origin of spurious oscillations~\cite{Kruger2017}, also interpreted as an amplitude inversion of the non-hydrodynamic modes at each time step~\cite{Astoul2020}. The second case (regularization) allows filtering out these non-hydrodynamic modes as shown in~\cite{Wissocq2020}, which is done at the cost of the degeneracy.

\begin{figure}
    \centering
    \includegraphics[scale=1]{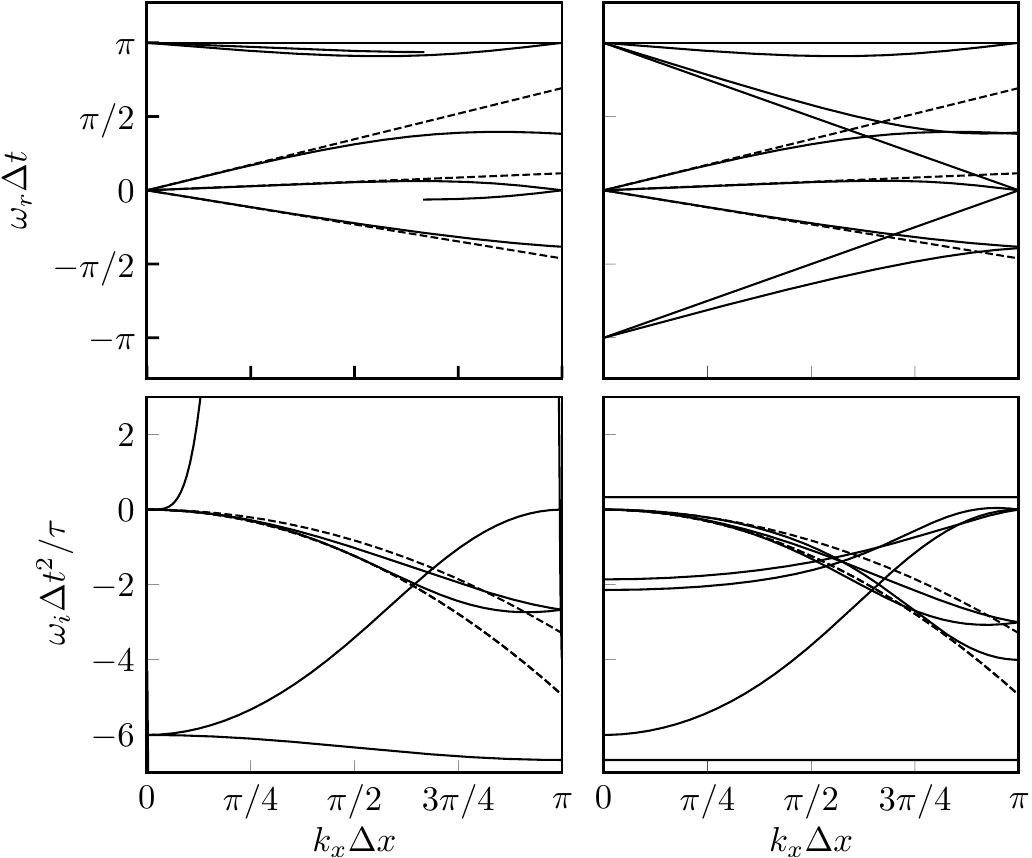}
    \caption{Propagation (top) and dissipation (bottom) curves of the linear analyses performed on the non-corrected MRT-RM scheme with $N=2$ (\blacklinefine) and the isothermal NS equations (\blackdashedline). A horizontal mean flow is considered at $\mathrm{Ma}=0.2$, with $\tau/\Delta t=10^{-5}$ and $\tau_b/\tau=2$. Left: $s_N=1$, right: $s_N=2$.} 
    \label{fig:LSA_MRT_RM_sN1_2}
\end{figure}

These properties are confirmed by Fig.~\ref{fig:LSA_MRT_RM_sN1_2}, displaying dispersion and dissipation curves of the MRT-RM model with $\tau_b/\tau=2$ and the characteristic values $s_N=1$ and $s_N=2$. With $s_N=1$, three modes are filtered out (six modes remain), which can clearly be visible on the dispersion curve. Unfortunately, a physical wave is amplified because of the hyperviscous degeneracy. The behavior of the PR model is well recovered here. With $s_N=2$ the degeneracy is cancelled, so that the expected dissipation of physical waves can be correctly recovered. However, the non-hydrodynamic modes are not filtered any more. A careful study of this model might lead to three kinds of stability issues caused by the particular value $s_N=2$:
\begin{enumerate}
	\item A non-hydrodynamic mode is surprisingly amplified whatever the considered wavenumber. The latter is related to the bulk viscosity and can only be attenuated by setting $\tau_b < \tau$.
	\item Other non-hydrodynamic modes are slightly unstable, even in the $x$-direction, for large values of $k_x \Delta x$.
	\item Due to the large number of non-hydrodynamic modes, eigenvalue collisions are likely to occur in every direction, so that this model is concerned by similar stability issues as the BGK one. 
\end{enumerate}

\subsection{Note on the bulk viscosity and the trace of the shear stress tensor}

The above remark on the values $s_N=1$ (regularization) and $s_N=2$ (no degeneracy) can equivalently be applied to the bulk viscosity regarding the relaxation rate $s_6$. In the present section, the relationship between the bulk viscosity and the trace of $\boldsymbol{\Pi}^{neq}= \boldsymbol{\Pi} - \boldsymbol{\Pi}^{eq}$, which can be obtained thanks to a Chapman-Enskog expansion, is particularly considered. More precisely, the focus is put here on several collision models enforcing a traceless $\boldsymbol{\Pi}^{neq}$ in the aim of cancelling the bulk viscosity~\cite{Farag2020, Farag2021}.

 Without loss of generality, the MRT-RM model described in Sec.~\ref{sec:MRT_descriptions} is considered here. The collision affects the trace of $\boldsymbol{\Pi}^{neq}$ as
\begin{align}
	\mathrm{Tr}\left( \boldsymbol{\Pi}^{neq, coll} \right) = -(s_6-1) \, \mathrm{Tr}\left( \boldsymbol{\Pi}^{neq} \right).
\end{align}
With this formulation, imposing a traceless $\boldsymbol{\Pi}^{neq}$ tensor during the collision step is equivalent to setting $s_6=1$ in the MRT collision model. However, using Eq.~(\ref{eq:MRT_relations_s_tau}), a null bulk viscosity can only be obtained with $s_6=2$. The particular value $s_6=1$ actually leads to
\begin{align}
 	\tau_b = \frac{\Delta t}{2},
\end{align} 
so that $\tau_b/\tau$ may be very large for low values of the dimensionless relaxation time. This non-desired bulk viscosity may be at the origin of an improved stability when enforcing a traceless $\boldsymbol{\Pi}^{neq}$, already noticed in the literature~\cite{Farag2020, Farag2021}.
%This choice has three consequences: (1) a numerical bulk viscosity is effectively introduced, (2) a part of the consistency error related to the bulk viscosity is translated to a numerical error in viscosity, (3) several terms related to $\tau_b$ in the numerical hyperviscosity are likely to degenerate for low values of $\tau/\Delta t$. In particular, considering $x$-derivatives only, a horizontal mean flow and $N=2$, one has with $\tau_b=\Delta t/2$:
%\begin{align}
%	& E_\rho^{(3)} \sim -\left( \frac{\Delta t}{\tau} \right)^3 \frac{1}{432}\left( (27 u_x^4+9u_x^2-2)\frac{\partial^4 \rho}{\partial x^4} + (90 u_x^3-18 u_x) \rho \frac{\partial^4 u_x}{\partial x^4} \right), \nonumber \\
%	& E_{\rho u_x}^{(3)} \sim \left( \frac{\Delta t}{\tau} \right)^3 \frac{1}{864 s_8 s_9}\Bigg( (27 s_8 s_9 u_x^5 + ((15 s_8+54) s_9+36 s_8-72) u_x^3-8 s_8 s_9 u_x)\frac{\partial^4 \rho}{\partial x^4} \nonumber \\
%	& \qquad \qquad + (117 s_8 s_9 u_x^4 +((162 -81 s_8) s_9+108 s_8-216)u_x^2+(14 s_8-36) s_9-24 s_8+48) \rho \frac{\partial^4 u_x}{\partial x^4} \Bigg),
%\end{align}
%while $E_{\rho u_y}^{(3)}$ remain unchanged. The hyperviscous degeneracy now affects both the mass ans momentum equations, which make its effect on the physical waves more difficult to interpret. In particular, both acoustic and shear waves are affected by the degeneracy. This property may be at the origin of an improved stability when enforcing a traceless $\boldsymbol{\Pi}^{neq}$.

Like in the previous section, two cases are then noticeable:
\begin{flalign}
    \mathrm{Traceless}\ \boldsymbol{\Pi}^{neq}\ (s_6=1):\qquad & \mathrm{Tr} \left( \boldsymbol{\Pi}^{neq, coll} \right) = 0, \\
    \mathrm{Zero\ bulk\ viscosity}\ (s_6=2):\qquad & \mathrm{Tr} \left( \boldsymbol{\Pi}^{neq, coll} \right) = - \mathrm{Tr} \left( \boldsymbol{\Pi}^{neq} \right).
\end{flalign}
The first case can be viewed as a new regularization, imposing another moment to its equilibrium value, leading to the filtering of another non-hydrodynamic mode. However, this is done at the cost of an additional numerical bulk viscosity. The second case effectively cancels the bulk viscosity.

\subsection{Particular case of the TRT model}

Studying the linearized error terms of MRT models is the opportunity to investigate the properties of the two-relaxation-time (TRT) scheme, especially regarding the role of the so-called magic parameter $\Lambda$~\cite{Ginzburg2005, DHumieres2009, Ginzburg2010}. The purpose of the TRT collision model is to relax, with two different relaxation parameters $\lambda^+$ and $\lambda^-$, the symmetric and anti-symmetric parts of the distribution functions. Equivalently, it can be viewed as a MRT model relaxing even moments with the parameter $\lambda^+$ and odd ones with $\lambda^-$~\cite{Kruger2017}. Without loss of generality, the MRT-RM formulation is adopted here with
\begin{align}
	s_4=s_5=s_6=s_9 = \lambda^+ = \frac{\Delta t}{\tau + \Delta t/2}, \qquad s_7=s_8 = \lambda^- = \frac{\Delta t}{\tau^- + \Delta t/2}.
\end{align}
Note that with these definitions, the bulk viscosity cannot be tuned any more. The great interest of this model is that only one free parameter remains: $\lambda^-$, or equivalently $\tau^-$. It is usually related to $\lambda^+$ as
\begin{align}
	\Lambda = \lambda^+ \lambda^- = \left( \frac{\Delta t}{\tau + \Delta t/2} \right) \left( \frac{\Delta t}{\tau^- + \Delta t/2} \right) \quad \Rightarrow \quad \lambda^- = \frac{1}{\Lambda} \left( \frac{\tau}{\Delta t} + \frac{1}{2} \right) ,
\end{align}
where $\Lambda$ is referred to as the magic parameter of the TRT model. For a fixed value $\Lambda$ (independent of $\tau/\Delta t$), the relaxation rate of odd-moments behaves as $\lambda^- \sim \tau/\Delta t$. Given the form of the degenerating hyperviscosity exhibited with the MRT-RM model, especially the appearance of $(1/s_7)$-related terms in the degenerating hyperviscosity of Table~\ref{tab:MRT_degen_hypernu}, the degeneracy is expected to be even more prominent with the TRT model with a fixed magic parameter.

When $x$-derivatives only are considered for the sake of simplicity, a computation of the linearized deviation terms of the TRT model yields, for a second-order equilibrium distribution function,
\begin{align}
	& E_\rho^{(3)} = O\left( \left( \Delta t/\tau \right)^2 \right), \qquad E_{\rho u_x}^{(3)} =  O\left( \left( \Delta t/\tau \right)^2 \right), \nonumber \\
	& E_{\rho u_y}^{(3)} = - \left( \frac{\Delta t}{\tau} \right)^4 \frac{\Lambda (\Lambda-1/4)}{9}\left( (6u_x^2+2) u_y \frac{\partial^4 \rho}{\partial x^4} + 12 \rho u_x u_y \frac{\partial^4 u_x}{\partial x^4} + 6 \rho u_x^2 \frac{\partial^4 u_y}{\partial x^4} \right) + O\left( \left( \Delta t/\tau \right)^2 \right).
\end{align}
Exactly like the MRT-RM model with fixed high-order relaxation parameters, no degeneracy is obtained on the mass and $x$-momentum equations. However, a $(\Delta t/\tau)^4$-related term arises in the $y$-momentum equation which, as predicted, increases the degeneracy when the dimensionless relaxation time is small. Also note that no $(\Delta t/\tau)^3$-term arises, which remains true in any direction. It is also noteworthy that this degeneracy is cancelled when $\Lambda=1/4$, which is precisely the optimal parameter found by d'Humi\`{e}res and Ginzburg by derivation of steady recurrence equations~\cite{DHumieres2009}. However, this property is not true when considering all the possible directions.

\begin{figure}[h]
    \centering
    \includegraphics[scale=1]{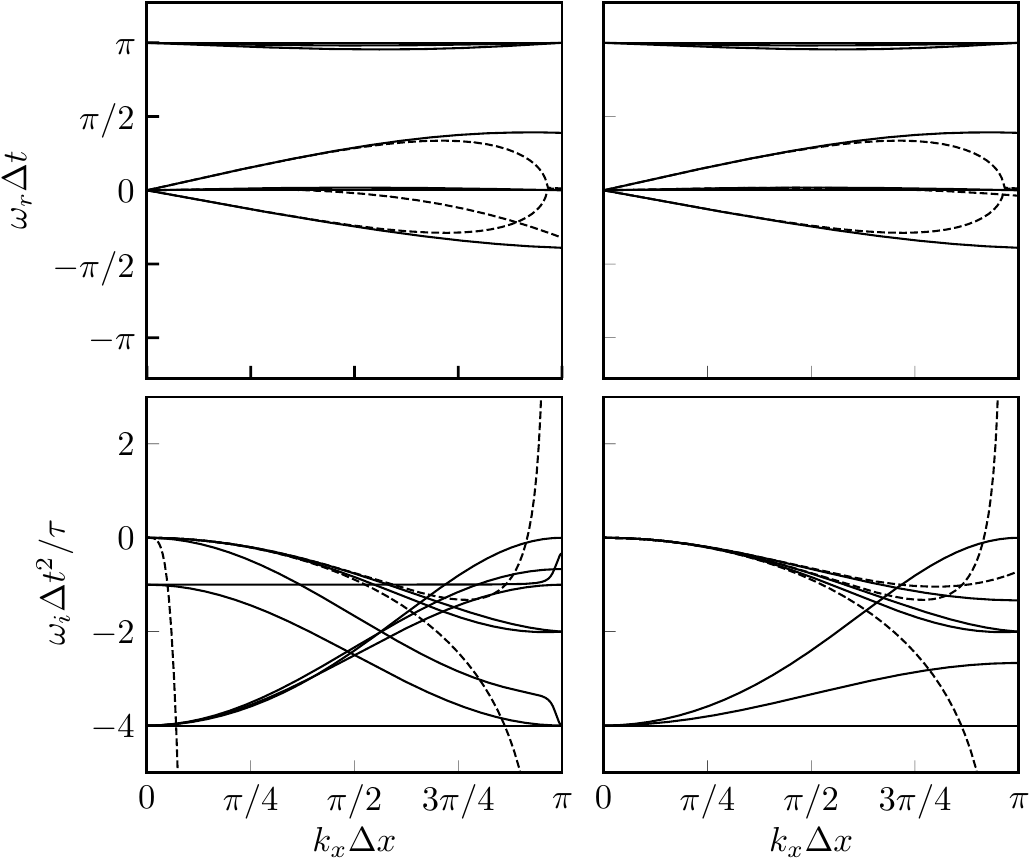}
    \caption{Propagation (top) and dissipation (bottom) curves of the linear analyses performed on the TRT scheme with $N=2$ (\blacklinefine) and the isothermal NS equations including deviation terms up to $\mathrm{Kn}^3$ (\blackdashedline). A horizontal mean flow is considered at $\mathrm{Ma}=0.1$ with $\tau/\Delta t=10^{-3}$. Left: $\Lambda=1$, right: $\Lambda=1/4$.} 
    \label{fig:TRT_spectrum}
\end{figure}

In order to investigate the effects of this degeneracy, linear stability analyses of the TRT scheme are provided in Fig.~\ref{fig:TRT_spectrum}. Dispersion and dissipation curves of two systems are displayed: (1) the TRT-LB scheme following \ref{app:matrices_LSA} and (2) the isothermal NS-equations including deviation terms up to the third-order in Knudsen number. Two values of the magic parameter are considered: $\Lambda=1$ and $\Lambda=1/4$. These analyses are performed along the $x$-direction, with a horizontal mean flow at $\mathrm{Ma}=0.1$, a second-order equilibrium distribution function and with a dimensionless relaxation time $\tau/\Delta t=10^{-3}$. For the sake of completeness, the complete form of the error terms in that case ($x$-aligned mean flow and $x$-derivatives only) is provided below:
\begingroup
\allowdisplaybreaks
\begin{align}
	& E_\rho^{(1)} = 0, \qquad E_{\rho u_x}^{(1)} = -\frac{\partial^2 a_{eq, xxx}^{(3)}}{\partial x^2}, \qquad E_{\rho u_y}^{(1)} = - \rho u_x^2 \frac{\partial^2 u_y}{\partial x^2}, \qquad E_\rho^{(2)} = - \left( \frac{\Delta t}{\tau} \right)^2 \frac{1}{36} \left (3 u_x^3 \frac{\partial^3 \rho}{\partial x^3} + (9 u_x^2-2) \rho \frac{\partial^3 \rho}{\partial x^3} \right), \nonumber \\
	& E_{\rho u_x}^{(2)} = - \frac{1}{9} \left( (27 u_x^4+9 u_x^2-2) \frac{\partial^3 \rho}{\partial x^3} + (90 u_x^3-18 u_x) \rho \frac{\partial^3 u_x}{\partial x^3} \right) + \left( \frac{\Delta t}{\tau} \right)^2 \frac{1}{54} \left( (9 u_x^4+9 u_x^2-2) \frac{\partial^3 \rho}{\partial x^3} + (36 u_x^3-6 u_x) \rho \frac{\partial^3 u_x}{\partial x^3} \right), \nonumber \\
	& E_{\rho u_y}^{(2)} = - 2 u_x(u_x^2-c_s^2) \frac{\partial^3 u_y}{\partial x^3} + \left( \frac{\Delta t}{\tau} \right)^2 \frac{1}{6} (u_x^3-4\Lambda u_x) \rho \frac{\partial^3 u_y}{\partial x^3}, \nonumber \\
	& E_\rho^{(3)} = - \left( \frac{\Delta t}{\tau} \right)^2 \frac{1}{108} \left( (27 u_x^4+9 u_x^2-2) \frac{\partial^4 \rho}{\partial x^4} + (90 u_x^3-18 u_x) \frac{\partial^4 u_x}{\partial x^4} \right), \nonumber \\
	& E_{\rho u_x}^{(3)} = - \frac{1}{9} \left( (108 u_x^5+51 u_x^3-12 u_x)\frac{\partial^4 \rho}{\partial x^4} + (378u_x^4-81u_x^2+2) \rho \frac{\partial^4 u_x}{\partial x^4} \right) \nonumber \\
	& \qquad \qquad + \left( \frac{\Delta t}{\tau} \right)^2 \frac{1}{108} \left( (162u_x^5+129 u_x^3-28 u_x)\frac{\partial^4 \rho}{\partial x^4} + (612u_x^4-99u_x^2-2)\rho \frac{\partial^4 u_x}{\partial x^4} \right), \nonumber \\
	& E_{\rho u_y}^{(3)} = - \frac{1}{9} (45u_x^4-18u_x^2+1) \rho \frac{\partial^4 u_y}{\partial x^4} + \left( \frac{\Delta t}{\tau} \right)^2 \frac{1}{36} (27 u_x^4-72\Lambda u_x^2 + (8\Lambda-1) ) \rho \frac{\partial^4 u_y}{\partial x^4} - \left( \frac{\Delta t}{\tau} \right)^4 \frac{\Lambda (\Lambda-1/4)}{3} 2\rho u_x^2 \frac{\partial^4 u_y}{\partial x^4}.
\end{align}
\endgroup

%\begin{align}
%	& E_{\rho u_y}^{(3)} = -\frac{1}{9}\left( (108 u_x^4+42 u_x^2-4) u_y \frac{\partial^4 \rho}{\partial x^4} + (333 u_x^3-42 u_x) \rho u_y \frac{\partial^4 u_x}{\partial x^4} +  (45 u_x^4-18 u_x^2+1) \rho \frac{\partial^4 u_y}{\partial x^4} \right) \nonumber \\
%	& \qquad - \left( \frac{\Delta t}{\tau} \right)^4 \frac{\Lambda (\Lambda-1/4)}{9}\left( (6u_x^2+2) u_y \frac{\partial^4 \rho}{\partial x^4} + 12 \rho u_x u_y \frac{\partial^4 u_x}{\partial x^4} + 6 \rho u_x^2 \frac{\partial^4 u_y}{\partial x^4} \right) \nonumber \\
%	& \qquad + \left( \frac{\Delta t}{\tau} \right)^3 0 \nonumber \\
%	& \qquad + \left( \frac{\Delta t}{\tau} \right)^2 ( (162 u_x^4+(132-216\Lambda) u_x^2-24\Lambda-4) u_y \frac{\partial^4 \rho}{\partial x^4} + (531 u_x^3+(12-360\Lambda) u_x) \rho u_y \frac{\partial^4 u_x}{\partial x^4} + (81 u_x^4-216\Lambda u_x^2+(24\Lambda-3)) \rho \frac{\partial^4 u_y}{\partial x^4} )/108
%\end{align}

Surprisingly, and unlike all the other models investigated in the present work, the dissipation of the vorticity wave is not accurately predicted by the Taylor expansion when $\Lambda \neq 1/4$. More precisely, the degeneracy predicted by the Taylor expansion expects an over-dissipation of this wave, which is not noticed on the linear analysis. Such a discrepancy is also observed when all the directions are considered. Regarding the acoustic waves, they do not suffer from any large error in dissipation, which is in agreement with the Taylor expansion. It is also noticeable that, looking on the dispersion curves, only four non-hydrodynamic modes have $\omega_r \approx \pi$, and two other ones have a phase velocity close to $0$. Let us recall here that, with the BGK collision model, all the non-hydrodynamic modes have a real pulsation shifted of $\pi$, which is attributed to an over-relaxation of non-hydrodynamic variables in Sec.~\ref{sec:non_hydrodynamic_modes}. With the TRT model, the very large relaxation time applied to the two odd moments ($\tau^-/\Delta t$) leads to an under-relaxation of these quantities instead of an over-relaxation~\cite{Kruger2017}. This is at the origin of the observed phenomena.

\begin{figure}
    \centering
    \begin{subfigure}[b]{0.48\textwidth}
    \includegraphics{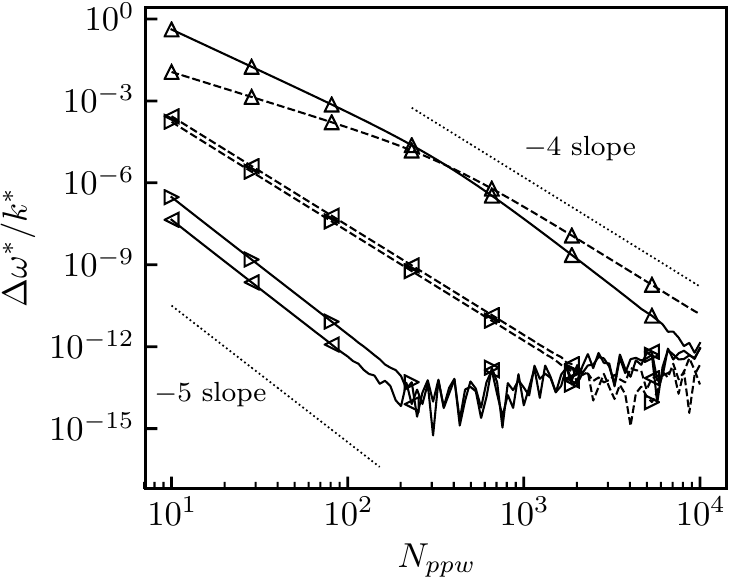}
    \caption{$\Lambda=1$}
    \label{fig:TRT_accuracy_error_Lambda1}
    \end{subfigure}
    \hfill
    \begin{subfigure}[b]{0.48\textwidth}
    \includegraphics{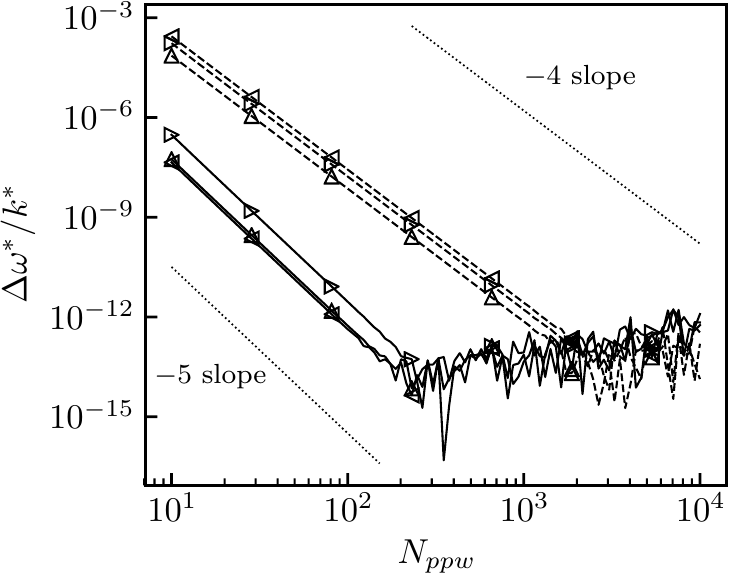}
    \caption{$\Lambda=1/4$}
    \label{fig:TRT_accuracy_error_Lambda025}
    \end{subfigure}
    \caption{$\mathrm{Ma}=0.1$, $\tau/\Delta t=10^{-3}$ \label{fig:TRT_accuracy_error}}
\end{figure}

Given the discrepancies observed in the behavior of the vorticity wave of Fig.~\ref{fig:TRT_spectrum}, and in order to ensure that the Taylor expansions have been properly performed, convergence plots of the linearized error terms are proposed in Fig.~\ref{fig:TRT_accuracy_error}. Exactly like for the validation of the deviation terms of the BGK collision model proposed in Sec.~\ref{sec:BGK_validation}, this figure displays the behavior of $\Delta \omega^*/k^*$ when the resolution of the wave increases, where $ \Delta \omega^*$ is the difference in the eigenvalues of the TRT-LB scheme and that of the isothermal NS equations including the computed error terms. Displayed in log-scale, a $-4$-slope in the asymptotic limit of the dispersion properties (real part of $\Delta \omega^*/k^*$), as well as a $-5$-slope in the dissipation properties (imaginary part of $\Delta \omega^*/k^*$) theoretically validate the accuracy of the Taylor expansion in $\mathrm{Kn}$. With $\Lambda=1$ (Fig.~\ref{fig:TRT_accuracy_error_Lambda1}), such a behavior is correctly recovered on the two acoustic waves travelling in the $x$-direction, up to the machine precision when they are resolved with more than 200 points per wavelength. However, regarding the vorticity, such an agreement is only obtained for very well resolved waves, with $N_{ppw}>200$. With a lower resolution, a $-3$-slope (resp. a $-2$-slope) is observed in the dissipation (resp. dispersion) properties. This indicates that, even though the deviation terms obtained by the Taylor expansion are theoretically validated as suggested by the asymptotic behavior of the remaining error, they cannot accurately predict the numerical properties of waves discretized with less than 200 points. This observation is in agreement with the incorrect prediction provided by the Taylor expansion in Fig.~\ref{fig:TRT_spectrum}. The only explanation for such a behavior is that high-order $\mathrm{Kn}$ deviation terms, that have been neglected in the present work, are subject to a degeneracy that compensates the dissipation error induced by the numerical hyperviscosity. This phenomenon, still not understood by the present authors, might be related to the fact that no $(\Delta t/\tau)^3$-related term arises in the hyperviscosity error. A better understanding of this very surprising property may be the purpose of future work.

Finally note that with $\Lambda=1/4$ (Fig.~\ref{fig:TRT_accuracy_error_Lambda025}), a correct slope is obtained up to the machine precision for all the waves discretized with more than ten points. This is in line with the fact that the degeneracy is cancelled in the $x$-direction for this value of the magic parameter. %As displayed in Fig.~\ref{fig:TRT_spectrum}, the Taylor expansion provides a correct estimation of the dissipation in that case.

\subsection{Summary of the numerical errors of MRT models}

The Taylor expansion performed on several forms of MRT collision models provides important conclusions regarding their numerical errors. Every formulation investigated in the present work depending on the choice of moments (RM, CM, HM, CHM), lead to very similar properties. In particular, setting the non-hydrodynamic relaxation parameters to a fixed value (not related to $\tau/\Delta t$) generally induces a degeneracy in hyperviscosity similar to that of regularized models. On this point, there is no \textit{a priori} advantage of performing the collision step in one basis of moments rather than another. In fact, only the value of the relaxation times of non-hydrodynamic moments seem to have an interesting impact on the degeneracy. The particular value $s_N=2$, which is equivalent to a null relaxation time for high-order moments, allows cancelling the degeneracy. However, unlike the regularization which comes down to setting $s_N=1$, this does not filter out any non-hydrodynamic mode, so that the advantage compared to the BGK collision remains unclear. This is probably why these relaxation parameters are usually prescribed to a value between $1$ and $2$ in the literature, usually tuned by linear analyses or academic test cases~\cite{DHumieres1994, Lallemand2000, DHumieres2002, xu2011, xu2012, Gendre2018}.

The degenerating hyperviscosity is likely to be at the origin of the numerical errors exhibited by Dellar in the incompressible limit with a relaxation parameter $s_N=1$~\cite{Dellar2003}. In this work, simulations of a Kelvin-Helmholtz instability were performed for decreasing Mach numbers with a constant Reynolds number and spatial resolution. This is equivalent to decreasing the dimensionless relaxation time $\tau/\Delta t$ which, according to the Taylor expansions, lead to an increasing numerical error compared to the BGK collision model.

Finally, an investigation of the TRT model exhibits an interesting feature of its numerical errors. Even though, for a fixed magic parameter $\Lambda$, the degeneracy is still present and even amplified, it does not lead to any strong error in dissipation compared to BGK and MRT models with fixed relaxation parameters. Furthermore, the under-relaxation of odd moments may reduce spurious oscillations which can lead to an increased stability.

\section{Conclusion}
\label{sec:conclusion}

In this paper, a thorough asymptotic analysis is proposed to derive the hydrodynamic limits of some isothermal LB schemes in the discrete setting. Since such a Taylor expansion in Knudsen number is not mathematically rigorous because of the presence of discontinuous NH modes, it is then systematically validated by linear analyses. In a sense, the mathematical developments performed in this work are quite close to other Taylor expansions of the literature, such as the work of Holdych \textit{et al.}~\cite{Holdych2004}, Dubois~\cite{Dubois2008} or more recently Fucik and Straka~\cite{Fucik2021}. In these articles, the Taylor expansion is performed assuming that it does not affect the collision step, which comes down to keeping a constant value of $\Delta t/\tau$. Given the two independent dimensionless numbers defined in Sec.~\ref{sec:Methodology}, namely $\mathrm{Kn}$ and $\Delta t/\tau$, these expansions can be interpreted in terms of Knudsen number as well. In this regard, one purpose of the present work is to restore, in a discrete setting, the recovery of macroscopic equations as hydrodynamic limits of the LBM, instead of a consistency limit as it is commonly assumed in the LBM community~\cite{Caiazzo2009}.

An interest of the present methodology is to discriminate consistency errors, inherited from the kinetic modelling of the DVBE, and numerical errors attributed to the time and space discretization. Applied to different collision models and confronted to linear analyses, it leads to the following conclusions:
\begin{itemize}
    \item The low dissipation of the BGK collision model is explained by the appearance of the Bernoulli numbers in the Taylor expansion (\textit{cf.} Sec.~\ref{sec:BGK}).
    \item Both regularized and MRT models with fixed relaxations suffer from a degenerating hyperviscosity, which consists in a resurgence of high-Kn phenomena induced by a large numerical pre-factor. This phenomenon is at the origin of an over-dissipation and instability of hydrodynamic modes, even in well-resolved conditions.
    \item The shear stress reconstruction of the HRR model induces a numerical viscosity which can be addressed by introducing the physical correction $\Psi_i$. In the latter case, the hyperviscosity is stabilizing for $\mathrm{Ma}<1$ and $\mathrm{CFL}<1$. The finite-difference scheme adopted in the computation of space gradients can affect these limits.
\item With the TRT collision model, even though a degeneracy is theoretically present, it surprisingly does not considerably affect the dissipation properties.
\end{itemize}
More generally, an important result of the moment cascade in the discrete setting is to translate consistency errors to numerical ones, by introducing $(\Delta t/\tau)$ pre-factors. This observation may be at the origin of a better numerical behavior when the consistency of the scheme is improved~\cite{Farag2021}.

Even if the analysis has been performed in an isothermal context, several conclusions seem to be directly transposed to a compressible framework. Especially, the $\mathrm{Ma}<1$ and $\mathrm{CFL}<1$ conditions of the HRR model are in line with recent linear stability analyses of hybrid LBM~\cite{Renard2020b}. Further investigations of the hydrodynamic limits and numerical errors of compressible LBM may be the purpose of future work. Furthermore, to complete the analysis, a particular attention may be paid to the behavior of non-hydrodynamic modes which have been neglected in the present work. Last but not least, error terms being explicitly known at each Knudsen order, a benefit of the present work is to pave the way for possible numerical corrections.

 \section*{Acknowledgments}

Enlightening discussions with P.-A. Masset, F. Renard, T. Astoul and G. Farag are greatly acknowledged. This work was supported by ANR Industrial Chair ALBUMS (grant ANR-18-CHIN-0003-01).

\appendix

\section{Macroscopic equations}
\label{app:macroscopic_equations}

Let us start with the following DVBE with deviation terms,
\begin{align}
    \partial_t f_i + e_{i, \alpha} \partial_\alpha f_i = -\frac{1}{\mathrm{Kn}} \left( f_i - f_i^{eq} - \mathrm{Kn} \Psi_i \right) + \sum_{n \geq 1} \mathrm{Kn}^n E_i^{(n)},
\end{align}
where it is supposed that every deviation term $E_i^{(n)}$ is explicitly expressed as space-gradients of macroscopic quantities $\rho$ and $u_\alpha$. Taking the first five moments of this equation yields

\begingroup
\allowdisplaybreaks
\begin{align}
    & \partial_t \rho + \partial_\alpha (\rho u_\alpha) = \sum_i \Psi_i + \mathrm{Kn} \sum_i E_i^{(1)} + \mathrm{Kn}^2 \sum_i E_i^{(2)} + \mathrm{Kn}^3 \sum_i E_i^{(3)} + O(\mathrm{Kn}^4), \label{eq:app_mass} \\
    & \partial_t (\rho u_\alpha) + \partial_\beta \Pi_{\alpha \beta} = \sum_i e_{i, \alpha} \Psi_i + \mathrm{Kn} \sum_i e_{i, \alpha} E_i^{(1)} + \mathrm{Kn}^2 \sum_i e_{i, \alpha} E_i^{(2)} + \mathrm{Kn}^3 \sum_i e_{i, \alpha} E_i^{(3)} + O(\mathrm{Kn}^4), \label{eq:app_momentum} \\
    & \partial_t \Pi_{\alpha \beta} + \partial_\gamma Q_{\alpha \beta \gamma} = -\frac{1}{\mathrm{Kn}} \left( \Pi_{\alpha \beta} - \Pi^{eq}_{\alpha \beta} \right) + \sum_i e_{i, \alpha} e_{i, \beta} \Psi_i + \mathrm{Kn} \sum_i e_{i, \alpha} e_{i, \beta} E_i^{(1)} + \mathrm{Kn}^2 \sum_i e_{i, \alpha} e_{i, \beta} E_i^{(2)} + O(\mathrm{Kn}^3), \label{eq:app_Pi}  \\
    & \partial_t Q_{\alpha \beta \gamma} + \partial_\delta R_{\alpha \beta \gamma \delta} = -\frac{1}{\mathrm{Kn}} \left( Q_{\alpha \beta \gamma} - Q^{eq}_{\alpha \beta \gamma } \right) + \sum_i e_{i, \alpha} e_{i, \beta} e_{i, \gamma} \Psi_i + \mathrm{Kn} \sum_i e_{i, \alpha} e_{i, \beta} e_{i, \gamma} E_i^{(1)} + O(\mathrm{Kn}^2), \label{eq:app_Q} \\
    & \partial_t R_{\alpha \beta \gamma \delta} + \partial_\epsilon S_{\alpha \beta \gamma \delta \epsilon} = -\frac{1}{\mathrm{Kn}} \left( R_{\alpha \beta \gamma \delta} - R^{eq}_{\alpha \beta \gamma \delta} \right) + \sum_i e_{i, \alpha} e_{i, \beta} e_{i, \gamma} e_{i, \delta} \Psi_i + O(\mathrm{Kn}),
\end{align}
\endgroup
where
\begin{align}
    & \Pi_{\alpha \beta} = \sum_i e_{i, \alpha} e_{i, \beta} f_i, \quad Q_{\alpha \beta \gamma} = \sum_i e_{i, \alpha} e_{i, \beta} e_{i, \gamma} f_i, \quad R_{\alpha \beta \gamma \delta} = \sum_i e_{i, \alpha} e_{i, \beta} e_{i, \gamma} e_{i, \delta} f_i, \quad S_{\alpha \beta \gamma \delta \epsilon} = \sum_i e_{i, \alpha} e_{i, \beta} e_{i, \gamma} e_{i, \delta} e_{i, \epsilon} f_i,
    %& \Pi^{eq}_{\alpha \beta} = \sum_i e_{i, \alpha} e_{i, \beta} f_i^{eq}, \quad Q^{eq}_{\alpha \beta \gamma} = \sum_i e_{i, \alpha} e_{i, \beta} e_{i, \gamma} f_i^{eq}, \quad R^{eq}_{\alpha \beta \gamma \delta} = \sum_i e_{i, \alpha} e_{i, \beta} e_{i, \gamma} e_{i, \delta} f_i^{eq}.
\end{align}
and $\Pi_{\alpha \beta}^{eq}$, $Q_{\alpha \beta \gamma}^{eq}$, $R_{\alpha \beta \gamma \delta}^{eq}$ and $S_{\alpha \beta \gamma \delta \epsilon}^{eq}$ are their equilibrium counterpart. Note that an aliasing error can occur on these moments because of the lattice closure, so that they may not be independent from each other~\cite{Karlin2010}. For example, with the D1Q3 and D2Q9 lattices, one has $Q_{xxx}=Q_{xxx}^{eq}=\rho u_x$. The stress tensor $\boldsymbol{\Pi}$ appearing in the momentum equation (\ref{eq:app_momentum}) can be expressed using Eq.~(\ref{eq:app_Pi}):
\begin{align}
    \Pi_{\alpha \beta} = \Pi^{eq}_{\alpha \beta} - \mathrm{Kn} \left( \partial_t \Pi_{\alpha \beta} + \partial_\gamma Q_{\alpha \beta \gamma} - \sum_i e_{i, \alpha} e_{i, \beta} \Psi_i \right) + \mathrm{Kn}^2 \sum_i e_{i, \alpha} e_{i, \beta} E_i^{(1)} + \mathrm{Kn}^3 \sum_i e_{i, \alpha} e_{i, \beta} E_i^{(2)} + O(\mathrm{Kn}^4). \label{eq:app_Pi_2}
\end{align}
In the above equation, two terms remain to be expressed at $O(\mathrm{Kn}^3)$: $\partial_t \Pi_{\alpha \beta}$ and $\partial_\gamma Q_{\alpha \beta \gamma}$. For the first one, by successive derivations of Eqs.~(\ref{eq:app_Pi_2})-(\ref{eq:app_Q}) one has
\begin{align}
    \partial_t \Pi_{\alpha \beta} & = \partial_t \Pi^{eq}_{\alpha \beta} - \mathrm{Kn} \left( \partial^2_{tt} \Pi_{\alpha \beta} + \partial^2_{\gamma t} Q_{\alpha \beta \gamma} - \sum_i e_{i, \alpha} e_{i, \beta} \partial_t \Psi_i \right) + \mathrm{Kn}^2 \sum_i e_{i, \alpha} e_{i, \beta} \partial_t E_i^{(1)} + O(\mathrm{Kn}^3) \\
    %&= \partial_t \Pi^{eq}_{\alpha \beta}  - \mathrm{Kn} \Bigg[ \partial^2_{tt} \left( \Pi_{\alpha \beta}^{eq} - \mathrm{Kn} \left( \partial_t \Pi^{eq}_{\alpha \beta} + \partial_\gamma Q^{eq}_{\alpha \beta \gamma} - \sum_i e_{i, \alpha} e_{i, \beta} \Psi_i \right) \right) \nonumber \\
    %& + \partial^2_{\gamma t} \left( Q^{eq}_{\alpha \beta \gamma} - \mathrm{Kn} \left( \partial_t Q^{eq}_{\alpha \beta \gamma} + \partial_\delta R^{eq}_{\alpha \beta \gamma \delta} - \sum_i e_{i, \alpha} e_{i, \beta} e_{i, \gamma} \Psi_i \right) \right) - \sum_i e_{i, \alpha} e_{i, \beta} \partial_t \Psi_i \Bigg] + \mathrm{Kn}^2 \sum_i e_{i, \alpha} e_{i, \beta} \partial_t E_i^{(1)} + O(\mathrm{Kn}^3) \\
    & = \partial_t \Pi^{eq}_{\alpha \beta} - \mathrm{Kn} \left( \partial^2_{tt} \Pi_{\alpha \beta}^{eq} + \partial^2_{\gamma t} Q^{eq}_{\alpha \beta \gamma} - \sum_i e_{i, \alpha} e_{i, \beta} \partial_t \Psi_i \right) + \mathrm{Kn}^2 \Bigg( \partial^3_{ttt} \Pi^{eq}_{\alpha \beta} + 2 \partial^3_{tt\gamma} Q^{eq}_{\alpha \beta \gamma} + \partial^3_{t\gamma \delta} R^{eq}_{\alpha \beta \gamma \delta} \nonumber \\
    & - \sum_i e_{i, \alpha} e_{i, \beta} \partial^2_{tt} \Psi_i - \sum_i e_{i, \alpha} e_{i, \beta} e_{i, \gamma} \partial^2_{\gamma t} \Psi_i + \sum_i e_{i, \alpha} e_{i, \beta} \partial_t E_i^{(1)} \Bigg) + O(\mathrm{Kn}^3).
\end{align}
Similarly, the second term can be evaluated as
\begin{align}
    \partial_\gamma Q_{\alpha \beta \gamma} & = \partial_\gamma Q^{eq}_{\alpha \beta \gamma} - \mathrm{Kn} \left( \partial^2_{\gamma t} Q_{\alpha \beta \gamma} + \partial^2_{\gamma \delta} R_{\alpha \beta \gamma \delta} - \sum_i e_{i, \alpha} e_{i, \beta} e_{i, \gamma} \partial_\gamma \Psi_i \right) + \mathrm{Kn}^2 \sum_i e_{i, \alpha} e_{i, \beta} e_{i, \gamma}\partial_\gamma E_i^{(1)} + O(\mathrm{Kn}^3) \\
    %& = \partial_\gamma Q^{eq}_{\alpha \beta \gamma} - \mathrm{Kn} \Bigg( \partial^2_{\gamma t} \left( Q^{eq}_{\alpha \beta \gamma} - \mathrm{Kn} \left( \partial_t Q^{eq}_{\alpha \beta \gamma} + \partial_\delta R^{eq}_{\alpha \beta \gamma \delta} - \sum_i e_{i, \alpha} e_{i, \beta} e_{i, \gamma} \Psi_i \right) \right) \\ 
    %& + \partial^2_{\gamma \delta} \left( R^{eq}_{\alpha \beta \gamma \delta} - \mathrm{Kn} \left( \partial_t R^{eq}_{\alpha \beta \gamma \delta} + \partial_\epsilon S^{eq}_{\alpha \beta \gamma \delta \epsilon} - \sum_i e_{i, \alpha} e_{i, \beta} e_{i, \gamma} e_{i, \delta} e_{i, \epsilon} \Psi_i \right) \right) - \sum_i e_{i, \alpha} e_{i, \beta} e_{i, \gamma} \partial_\gamma \Psi_i \Bigg) + \mathrm{Kn}^2 \sum_i e_{i, \alpha} e_{i, \beta} e_{i, \gamma}\partial_\gamma E_i^{(1)} + O(\mathrm{Kn}^3) \\
    & = \partial_\gamma Q^{eq}_{\alpha \beta \gamma} - \mathrm{Kn} \left( \partial^2_{\gamma t} Q^{eq}_{\alpha \beta \gamma} + \partial^2_{\gamma \delta} R^{eq}_{\alpha \beta \gamma \delta} -  \sum_i e_{i, \alpha} e_{i, \beta} e_{i, \gamma} \partial_\gamma \Psi_i \right) + \mathrm{Kn}^2 \Bigg( \partial^3_{\gamma tt} Q^{eq}_{\alpha \beta \gamma} + 2 \partial^3_{\gamma \delta t} R^{eq}_{\alpha \beta \gamma \delta} + \partial^3_{\gamma \delta \epsilon} S^{eq}_{\alpha \beta \gamma \delta \epsilon} \nonumber \\
    & - \sum_i e_{i, \alpha} e_{i, \beta} e_{i, \gamma} \partial^2_{\gamma t} \Psi_i - \sum_i e_{i, \alpha} e_{i, \beta} e_{i, \gamma} e_{i, \delta} \partial^2_{\gamma \delta} \Psi_i + \sum_i e_{i, \alpha} e_{i, \beta} e_{i, \gamma}\partial_\gamma E_i^{(1)} \Bigg) + O(\mathrm{Kn}^3). 
\end{align}
The flux in the momentum equation (\ref{eq:app_momentum}) can then be written as
\begingroup
\allowdisplaybreaks
\begin{align}
    \partial_\beta \Pi_{\alpha \beta} & = \partial_\beta \Pi^{eq}_{\alpha \beta} - \mathrm{Kn} \left( \partial^2_{\beta t} \Pi^{eq}_{\alpha \beta} + \partial^2_{\beta \gamma} Q^{eq}_{\alpha \beta \gamma} - \sum_i e_{i, \alpha} e_{i, \beta} \partial_\beta \Psi_i \right) + \mathrm{Kn}^2 \Bigg( \partial^3_{\beta tt} \Pi_{\alpha \beta}^{eq} + 2\partial^3_{\beta \gamma t} Q^{eq}_{\alpha \beta \gamma} + \partial^3_{\beta \gamma \delta} R^{eq}_{\alpha \beta \gamma \delta} - \sum_i e_{i, \alpha} e_{i, \beta} \partial^2_{\beta t} \Psi_i \nonumber \\
    & - \sum_i e_{i, \alpha} e_{i, \beta} e_{i, \gamma} \partial^2_{\beta \gamma} \Psi_i + \sum_i e_{i, \alpha} e_{i, \beta} \partial_\beta E_i^{(1)} \Bigg) - \mathrm{Kn}^3 \Bigg( \partial^4_{\beta ttt} \Pi^{eq}_{\alpha \beta} + 3 \partial^4_{\beta \gamma tt} Q^{eq}_{\alpha \beta \gamma} + 3 \partial^4_{\beta \gamma \delta t} R^{eq}_{\alpha \beta \gamma \delta} + \partial^4_{\beta \gamma \delta \epsilon} S^{eq}_{\alpha \beta \gamma \delta \epsilon} \nonumber \\
    & - \sum_i e_{i, \alpha} e_{i, \beta} \partial^3_{\beta tt} \Psi_i - 2\sum_i e_{i, \alpha} e_{i, \beta} e_{i, \gamma} \partial^3_{\beta \gamma t} \Psi_i - \sum_i e_{i, \alpha} e_{i, \beta} e_{i, \gamma} e_{i, \delta} \partial^3_{\beta \gamma \delta} \Psi_i \nonumber \\
    & + \sum_i e_{i, \alpha} e_{i, \beta} \partial^2_{\beta t} E_i^{(1)} + \sum_i e_{i, \alpha} e_{i, \beta} e_{i, \gamma}\partial^2_{\beta \gamma} E_i^{(1)} - \sum_i e_{i, \alpha} e_{i, \beta} \partial_\beta E_i^{(2)} \Bigg) + O(\mathrm{Kn}^4).
\end{align}
\endgroup
In this expression, any $n$th-order derivative can be expressed in the linear approximation with the chain rule
\begin{align}
    \partial^n_{\alpha_1..\alpha_n} \Phi = \frac{\partial \Phi}{\partial \rho} \partial^n_{\alpha_1..\alpha_n} \rho + \frac{\partial \Phi}{\partial (\rho u_\alpha) } \partial^n_{\alpha_1..\alpha_n} (\rho u_\alpha), \qquad \alpha_i \in \{x, y, t\}, 
\end{align}
which can be used to express the derivatives of $\Pi^{eq}_{\alpha \beta}$, $Q^{eq}_{\alpha \beta \gamma}$, $R^{eq}_{\alpha \beta \gamma \delta}$, $S^{eq}_{\alpha \beta \gamma \delta \epsilon}$, $\Psi_i$, $E_i^{(1)}$ and $E_i^{(2)}$ as function of derivatives of the macroscopic quantities $\rho$ and $\rho u_\alpha$ only. Their time-derivatives can then be disposed using the mass and momentum equations (\ref{eq:app_mass})-(\ref{eq:app_momentum}) at the correct order in $\mathrm{Kn}$. Applying this technique iteratively to dispose of the second-order time-derivatives and above, the following closed set of equations are obtained
\begin{align}
	& \partial_t \rho + \partial_\alpha (\rho u_\alpha) = \mathrm{Kn} E_\rho^{(1)} + \mathrm{Kn}^2 E_\rho^{(2)} + \mathrm{Kn}^3 E_\rho^{(3)} + O(\mathrm{Kn}^4),
	\label{eq:app_mass_errors} \\
	& \partial_t (\rho u_\alpha) + \partial_\beta (\rho u_\alpha u_\beta) + c_s^2 \theta \partial_\alpha \rho = \mathrm{Kn} \left( \rho c_s^2 \theta \partial_\beta S_{\alpha \beta} \right) + \mathrm{Kn}E_{\rho u_\alpha}^{(1)} + \mathrm{Kn}^2 E_{\rho u_\alpha}^{(2)} + \mathrm{Kn}^3 E_{\rho u_\alpha}^{(3)} + O(\mathrm{Kn}^4),
	\label{eq:app_momentum_errors}
\end{align} 
with $S_{\alpha \beta} = \partial_\alpha u_\beta + \partial_\beta u_\alpha$ and where the fact that $\Pi^{eq}_{\alpha \beta} = \rho u_\alpha u_\beta + \rho c_s^2 \theta \delta_{\alpha \beta}$ has been used. $E^{(n)}_\rho$ and $E^{(n)}_{\rho u_\alpha}$ are deviation terms from the mass and momentum equations of the isothermal Navier-Stokes equations, respectively. They both include: 
\begin{itemize}
    \item the effects of the moments cascade of the DVBE, which can be viewed as errors in consistency with the isothermal NS equations,
    \item the macroscopic effects of the deviations from the DVBE $E_i^{(n)}$, which include the numerical errors of the model. 
\end{itemize}
Explicit expressions of these deviation terms are obtained using the computer algebra system Maxima~\cite{maxima}. Numerical errors can then be clearly identified as $\Delta t/\tau$-dependent terms, while consistency errors are the remaining terms when $\Delta t/\tau \rightarrow 0$.

\section{Discretized correction terms}
\label{app:discretized_correction_terms}

In the case of corrected LB models, the body-force term $\Psi_i$ is usually discretized thanks to finite differences. The aim of this appendix is to focus on the numerical errors performed during their discretization in the context of Taylor expansions in Knudsen number $\mathrm{Kn}$. Starting from the general expression of Eq.~(\ref{eq:correction_D2Q9_N2}), it can be re-written in its dimensionless form using a chain rule as:
\begin{align}
	\Psi_i = -\frac{w_i}{2c_s^4} \mathcal{H}_{i, \alpha \beta}^{(2)} \left( \frac{\partial a^{(3)}_{eq, \alpha \beta \gamma}}{\partial \rho} \partial_\gamma \rho + \frac{\partial a_{eq, \alpha \beta \gamma}^{(3)}}{\partial (\rho u_\delta)} \partial_\gamma (\rho u_\delta) \right).
\end{align}
The discrete form of the correction term $\Psi_i^d$ is then obtained by using a finite-difference scheme for the evaluation of space gradients $\partial_\gamma$. Note that the choice of variables used for this discretization, \textit{e.g.} $(\rho, \boldsymbol{u})$ or $(\rho, \rho \boldsymbol{u})$, has no impact in the linear approximation, and the second one is arbitrarily adopted here. Two kinds of discretizations are considered: a first-order upwind scheme (DUO1) and a second-order centered scheme (DCO2):
\begin{align}
	& \mathrm{DCO2}: \qquad (\partial_\gamma \Phi)^{\mathrm{DCO2}} = \frac{\Phi(\boldsymbol{x}+k\Delta x \boldsymbol{e_\gamma})-\Phi(\boldsymbol{x}-k\Delta x \boldsymbol{e_\gamma})}{2k\Delta x}, \\
	& \mathrm{DUO1}: \qquad (\partial_\gamma \Phi)^{\mathrm{DUO1}} = \frac{\Phi(\boldsymbol{x}+d^+ k\Delta x \boldsymbol{e_\gamma})-\Phi(\boldsymbol{x}-d^-k\Delta x \boldsymbol{e_\gamma})}{k\Delta x} \quad \mathrm{with} \quad    
    \begin{cases}
      d^+ = |\mathrm{sgn}(u_\gamma)-1|/2  \\
      d^-=|\mathrm{sgn}(u_\gamma)+1|/2  \\
    \end{cases},
\end{align}
for any function $\Phi$, where $\boldsymbol{e_\gamma}$ is the unit vector in the direction $\gamma$. A Taylor series expansion of these expressions yields
\begin{align}
	& \mathrm{DCO2}: \quad (\partial_\gamma \Phi)^{\mathrm{DCO2}} = \partial_\gamma \Phi + \frac{\mathrm{Kn}^2}{6} \left( \frac{\Delta t}{\tau} \right)^2 \partial^3_{\gamma \gamma \gamma} \Phi + O(\mathrm{Kn}^4), \\
	& \mathrm{DUO1}: \quad (\partial_\gamma \Phi)^{\mathrm{DUO1}} = \partial_\gamma \Phi - \mathrm{sgn}(u_\gamma) \frac{\mathrm{Kn}}{2} \frac{\Delta t}{\tau} \partial^2_{\gamma \gamma} \Phi + \frac{\mathrm{Kn}^2}{6} \left( \frac{\Delta t}{\tau} \right)^2 \partial^3_{\gamma \gamma \gamma} \Phi - \mathrm{sgn}(u_\gamma) \frac{\mathrm{Kn}^3}{24} \left( \frac{\Delta t}{\tau} \right)^3 \partial^4_{\gamma \gamma \gamma \gamma} \Phi + O(\mathrm{Kn}^4).
\end{align}
This leads to the following Taylor series expansions for the linearized $\Psi_i^d$ with a DCO2 discretization:
\begin{align}
	\Psi_i^{\mathrm{DCO2}} = \Psi_i - \frac{w_i}{2c_s^4} \mathcal{H}_{i, \alpha \beta}^{(2)} \frac{\mathrm{Kn}^2}{6} \left( \frac{\Delta t}{\tau} \right)^2 \left( \frac{\partial a_{eq, \alpha \beta \gamma}^{(3)}}{\partial \rho} \partial^3_{\gamma \gamma \gamma} \rho + \frac{\partial a_{eq, \alpha \beta \gamma}^{(3)}}{\partial (\rho u_\delta)} \partial^3_{\gamma \gamma \gamma} (\rho u_\delta)  \right) + O(\mathrm{Kn}^4),
\end{align}
and with a DUO1 discretization:
\begin{align}
	\Psi_i^{\mathrm{DUO1}} = \Psi_i^{\mathrm{DCO2}} + \mathrm{sgn}(u_\gamma) \frac{w_i}{2c_s^4} \mathcal{H}_{i, \alpha \beta}^{(2)} \Bigg( \frac{\mathrm{Kn}}{2} \frac{\Delta t}{\tau} & \left( \frac{\partial a_{eq, \alpha \beta \gamma}^{(3)}}{\partial \rho} \partial^2_{\gamma \gamma} \rho + \frac{\partial a_{eq, \alpha \beta \gamma}^{(3)}}{\partial (\rho u_\delta)} \partial^2_{\gamma \gamma} (\rho u_\delta) \right) \nonumber \\
	& + \frac{\mathrm{Kn}^3}{24} \left( \frac{\Delta t}{\tau} \right)^3 \left( \frac{\partial a_{eq, \alpha \beta \gamma}^{(3)}}{\partial \rho} \partial^4_{\gamma \gamma \gamma \gamma} \rho + \frac{\partial a_{eq, \alpha \beta \gamma}^{(3)}}{\partial (\rho u_\delta)} \partial^4_{\gamma \gamma \gamma \gamma} (\rho u_\delta) \right) \Bigg).
\end{align}

\section{Matrices of the linear analyses}
\label{app:matrices_LSA}

\subsection{Macroscopic equations with deviation terms}

To perform analyses of Eqs.~(\ref{eq:mass_errors})-(\ref{eq:momentum_errors}), the linearized deviation terms $E_\Phi^{(n)}$ can be written under the following general form:
\begin{align}
	E_\Phi^{(n)} = A_{\Phi,\alpha_1..\alpha_{n+1}}^{(n+1)} \frac{\partial^{n+1} \rho}{\partial \alpha_1 .. \partial \alpha_{n+1}} + B_{\Phi,\alpha_1..\alpha_{n+1}}^{(n+1)} \frac{\partial^{n+1} (\rho u_x)}{\partial {\alpha_1} .. \partial {\alpha_{n+1}}} + C_{\Phi,\alpha_1..\alpha_{n+1}}^{(n+1)} \frac{\partial^{n+1} (\rho u_y)}{\partial {\alpha_1} .. \partial {\alpha_{n+1}}}, \qquad \Phi \in \{\rho, \rho u_\alpha \},
	\label{eq:deviation_terms_linearized}
\end{align} 
where implicit summations are performed on indices $ \alpha_1,..,\alpha_{n+1} \in \{x,y\} $ and $\boldsymbol{A}_\Phi^{(n)}$, $\boldsymbol{B}_\Phi^{(n)}$ and $\boldsymbol{C}_\Phi^{(n)}$ are constant coefficients involving the mean base (space- and time-averaged) density and velocity. In 1D, every index is equal to $x$ and $\boldsymbol{C}_\Phi^{(n)}=0$ so that, for instance, the first-order deviation of the mass equation $E_\rho^{(1)}$ can be written as
\begin{align}
	E_\rho^{(1)} = A_{\rho,xx}^{(2)} \frac{\partial^2 \rho}{\partial x^2} + B_{\rho,xx}^{(2)} \frac{\partial^2 (\rho u_x)}{\partial x^2}.
\end{align}
A linear analysis of Eqs.~(\ref{eq:mass_errors})-(\ref{eq:momentum_errors}) yields the following system of equations
\begin{align}
    & \omega^* \hat{\rho} = k^*_\alpha \hat{j_\alpha} -\sum_{n \geq 1} \mathrm{i}^n \left( \frac{\tau}{\Delta t} \right)^n  k^*_{\alpha_1}..k^*_{\alpha_{n+1}} \left( A^{(n+1)}_{\rho, \alpha_1..\alpha_{n+1}} \hat{\rho} + B^{(n+1)}_{\rho, \alpha_1..\alpha_{n+1}} \hat{j_x} + C^{(n+1)}_{\rho, \alpha_1..\alpha_{n+1}} \hat{j_y} \right), \\
    & \omega^* \hat{j_\alpha} = k_\beta^* \left( \overline{u_\beta} \hat{j_\alpha} + \overline{u_\alpha} \hat{j_\beta} - \overline{u_\alpha} \overline{u_\beta} \hat{\rho} \right) + k^*_\alpha c_s^2 \theta \hat{\rho} -\mathrm{i} \frac{\tau}{\Delta t}\, c_s^2 \theta k^*_\beta \left( k^*_\alpha \hat{j_\beta} + k^*_\beta \hat{j_\alpha} - \overline{u_\beta} k^*_\alpha \hat{\rho} - \overline{u_\alpha} k^*_\beta \hat{\rho}  \right) \nonumber \\
    & \qquad \qquad \qquad - \sum_{n \geq 1} \mathrm{i}^n \left(\frac{\tau}{\Delta t} \right)^{n} k^*_{\alpha_1}..k^*_{\alpha_{n+1}} \left( A^{(n+1)}_{\rho u_\alpha, \alpha_1..\alpha_{n+1}} \hat{\rho} + B^{(n+1)}_{\rho u_\alpha, \alpha_1..\alpha_{n+1}} \hat{j_x} + C^{(n+1)}_{\rho u_\alpha, \alpha_1..\alpha_{n+1}} \hat{j_y} \right),
\end{align}
where $j_\alpha = \rho u_\alpha$. The eigenvalue problem of Eq.~(\ref{eq:LSA_macros_Errors}) is then recovered with
\begingroup
\allowdisplaybreaks
\begin{align}
    M_{00}^{(p)} &= -\sum_{n=1}^p \mathrm{i}^n \left( \frac{\tau}{\Delta t} \right)^n  k^*_{\alpha_1}..k^*_{\alpha_{n+1}} A^{(n+1)}_{\rho, \alpha_1..\alpha_{n+1}}, \qquad M_{01}^{(p)} = k^*_x -\sum_{n=1}^p \mathrm{i}^n \left( \frac{\tau}{\Delta t} \right)^n  k^*_{\alpha_1}..k^*_{\alpha_{n+1}} B^{(n+1)}_{\rho, \alpha_1..\alpha_{n+1}}, \\
    M_{02}^{(p)} &= k^*_y -\sum_{n=1}^p \mathrm{i}^n \left( \frac{\tau}{\Delta t} \right)^n  k^*_{\alpha_1}..k^*_{\alpha_{n+1}} C^{(n+1)}_{\rho, \alpha_1..\alpha_{n+1}}, \\
    M_{10}^{(p)} &= -k_x^* \overline{u_x}^2 - k_y^* \overline{u_x} \overline{u_y} + k_x^* c_s^2 \theta + \mathrm{i} \frac{\tau}{\Delta t} c_s^2 \theta \left( 2{k_x^*}^2\overline{u_x} + k_x^* k_y^* \overline{u_y} + {k_y^*}^2 \overline{u_x} \right) - \sum_{n=1}^p \mathrm{i}^n \left( \frac{\tau}{\Delta t} \right)^n  k^*_{\alpha_1}..k^*_{\alpha_{n+1}} A^{(n+1)}_{\rho u_x, \alpha_1..\alpha_{n+1}}, \\
    M_{11}^{(p)} &= 2k_x^* \overline{u_x} + k_y^* \overline{u_y} - \mathrm{i} \frac{\tau}{\Delta t} c_s^2 \theta \left( 2 {k_x^*}^2 + {k_y^*}^2 \right) - \sum_{n=1}^p \mathrm{i}^n \left( \frac{\tau}{\Delta t} \right)^n  k^*_{\alpha_1}..k^*_{\alpha_{n+1}} B^{(n+1)}_{\rho u_x, \alpha_1..\alpha_{n+1}}, \\
    M_{12}^{(p)} &= k_y^* \overline{u_x} - \mathrm{i} \frac{\tau}{\Delta t} c_s^2 \theta \left( k_x^* k_y^* \right), \\
    M_{20}^{(p)} &= -k_x^* \overline{u_x} \overline{u_y} - k_y \overline{u_y}^2 + k_y c_s^2 \theta + \mathrm{i} \frac{\tau}{\Delta t} c_s^2 \theta \left( k_x^* k_y^* \overline{u_x} + {k_x^*}^2 \overline{u_y} + 2 {k_y^*}^2 \overline{u_y} \right) - \sum_{n=1}^p \mathrm{i}^n \left( \frac{\tau}{\Delta t} \right)^n  k^*_{\alpha_1}..k^*_{\alpha_{n+1}} A^{(n+1)}_{\rho u_y, \alpha_1..\alpha_{n+1}}, \\
    M_{21}^{(p)} &= k_x^* \overline{u_y} - \mathrm{i} \frac{\tau}{\Delta t} c_s^2 \theta \left( k_x^* k_y^* \right) - \sum_{n=1}^p \mathrm{i}^n \left( \frac{\tau}{\Delta t} \right)^n  k^*_{\alpha_1}..k^*_{\alpha_{n+1}} B^{(n+1)}_{\rho u_y, \alpha_1..\alpha_{n+1}}, \\
    M_{22}^{(p)} &= k_x^* \overline{u_x} + 2 k_y^* \overline{u_y} - \mathrm{i} \frac{\tau}{\Delta t} c_s^2 \theta \left( {k_x^*}^2 + 2 {k_y^*}^2 \right) - \sum_{n=1}^p \mathrm{i}^n \left( \frac{\tau}{\Delta t} \right)^n  k^*_{\alpha_1}..k^*_{\alpha_{n+1}} C^{(n+1)}_{\rho u_y, \alpha_1..\alpha_{n+1}},
\end{align}
\endgroup
where $p$ is the order of truncation of the Taylor expansion in $\mathrm{Kn}$. Note that the error done on the eigenvalue problem is of the order $O\left({k^*}^{p+2} \right)$ for a given value of the dimensionless relaxation time $\tau/\Delta t$. This is an important observation for validating the deviation terms by comparison with the eigenvalues of the linearized LB scheme.

\subsection{LBM with BGK collision}

The linear stability analysis of the LB scheme with the BGK collision model can be found in previous references~\cite{Sterling1996, Worthing1997, Lallemand2000, Dellar2002, Marie2009, Wissocq2019, Hosseini2019a, Coreixas2020, Wissocq2020}. The eigenvalue problem of Eq.~(\ref{eq:LSA_LBM}) is recovered with 
\begin{align}
     \mathbf{M}^{\mathrm{LBM}}_{ij} = e^{-\mathrm{i}k^*_\alpha e_{i, \alpha}} \left( \delta_{ij} - \frac{\Delta t}{\tau+\Delta t/2} \left( \delta_{ij} - \mathbf{J}^{eq, N}_{ij} \right) + \frac{\tau}{\tau + \Delta t/2} \left( D^{\mathrm{FD}}_\gamma \psi_{ij, \gamma} \right) \right),
     \label{eq:app_LSA_MLBM}
\end{align}
where $\mathbf{J}^{eq, N}$ is the Jacobian matrix of the equilibrium distribution, which can be analytically computed as~\cite{Wissocq2020}
\begin{align}
    \mathbf{J}_{ij}^{eq, N} = \sum_{n=0}^N \frac{w_i}{n! c_s^{2n}} \boldsymbol{\Lambda}^{(n)}_{eq, j}: \boldsymbol{\mathcal{H}}_i^{(n)},
\end{align}
where 
\begin{align}
    & \Lambda^{(0)}_{eq, j} = 1, \qquad \left( \boldsymbol{\Lambda}^{(1)}_{eq, j} \right)_\alpha = e_{j, \alpha}, \qquad \left( \boldsymbol{\Lambda}^{(2)}_{eq, j} \right)_{\alpha \beta} = \overline{u_\alpha} e_{j, \beta} + \overline{u_\beta} e_{j, \alpha} - \overline{u_\alpha} \overline{u_\beta}, \\
    & \left( \boldsymbol{\Lambda}^{(3)}_{eq, j} \right)_{\alpha \beta \gamma} = \overline{u_\alpha} \overline{u_\beta} e_{j, \gamma} + \overline{u_\alpha} \overline{u_\gamma} e_{j, \beta} + \overline{u_\beta} \overline{u_\gamma} e_{j, \alpha} - 2 \overline{u_\alpha} \overline{u_\beta} \overline{u_\gamma}, \\
    & \left( \boldsymbol{\Lambda}^{(4)}_{eq, j} \right)_{\alpha \beta \gamma \delta} = \overline{u_\alpha} \overline{u_\beta} \overline{u_\gamma} e_{j, \delta} + \overline{u_\alpha} \overline{u_\beta} \overline{u_\delta} e_{j, \gamma} + \overline{u_\alpha} \overline{u_\gamma} \overline{u_\delta} e_{j, \beta} + \overline{u_\beta} \overline{u_\gamma} \overline{u_\delta} e_{j, \alpha} - 3 \overline{u_\alpha} \overline{u_\beta} \overline{u_\gamma} \overline{u_\delta}.
\end{align} 
The last right-hand-side term of Eq.~(\ref{eq:app_LSA_MLBM}) aims at considering the body-force term $\Psi_i$ of Eq.~(\ref{eq:correction_D2Q9_N2}) in the linear analysis,
with 
\begin{align}
    \psi_{ij, \gamma} = -\frac{w_i}{2c_s^4} \mathcal{H}_{i, \alpha \beta}^{(2)} \left( \boldsymbol{\Lambda}^{(3)}_{eq, j} \right)_{\alpha \beta \gamma}, 
\end{align}
and where $D_\gamma^{FD}$ accounts for the linearized space gradients in the direction $\gamma$, depending on the way they are discretized:
\begin{align}
    & \mathrm{Analytical\ gradients}: \qquad D_\gamma^{\mathrm{FD}} = \mathrm{i} k^*_\gamma, \\
    & \mathrm{DCO2}: \qquad D_\gamma^{\mathrm{FD}} = \mathrm{i} \sin(k^*_\gamma), \\
    & \mathrm{DUO1}: \qquad D_\gamma^{\mathrm{FD}} = \mathrm{sgn}(\overline{u_\gamma}) \left( 1-\exp(-\mathrm{sgn}(\overline{u_\gamma}) \mathrm{i} k^*_\gamma) \right)
\end{align}

\subsection{LBM with regularized collision}

Linear stability analyses of the regularized LBM can be found in~\cite{Wissocq2020, Renard2020b}. The eigenvalue problem of Eq.~(\ref{eq:LSA_LBM}) is generally recovered with
\begin{align}
    \mathbf{M}_{ij}^{\mathrm{LBM}} = e^{-\mathrm{i} k_\alpha^* e_{i, \alpha}} \left( \mathbf{J}_{ij}^{eq, N} + \left( 1-\frac{\Delta t}{\tau + \Delta t/2} \right)\mathbf{J}_{ij}^{(1)} + \frac{1}{2} D_\gamma^{\mathrm{FD}} \psi_{ij, \gamma} \right),
\end{align}
where $\mathbf{J}_{ij}^{(1)}$ is the Jacobian matrix of the off-equilibrium distribution functions. It depends on the regularized scheme:
\begin{align}
    & \mathrm{Complete\ reconstruction}: \mathbf{J}_{ij}^{(1)}= -\frac{w_i}{c_s^2} \left( \frac{\tau}{\Delta t} + \frac{1}{2} \right) \left( e_{i, \alpha} e_{i, \beta} - c_s^2 \delta_{\alpha \beta} \right) \theta \left( D_\alpha^{\mathrm{FD}}(e_{j, \beta}-\overline{u_\beta}) + D_\beta^{\mathrm{FD}}(e_{j, \alpha}-\overline{u_\alpha}) \right), \\
    & \mathrm{PR/RR}: \mathbf{J}_{ij}^{(1)}=\left[\mathbf{H}^{-1} \mathbf{P} \mathbf{H}\right]_{ik} \left[ \delta_{kj} - \mathbf{J}_{kj}^{eq, N} + \frac{1}{2} D_\gamma^{\mathrm{FD}} \psi_{ij, \gamma} \right],
\end{align}
where $\mathbf{H}$ and $\mathbf{P}$ are defined in Sec.~\ref{sec:Regul_PR_RR_unified_formalism}.

\subsection{MRT models}

For MRT models in absence of body-force term as considered in Sec.~\ref{sec:MRT_models}, the eigenvalue problem of Eq.~(\ref{eq:LSA_LBM}) is recovered with 
\begin{align}
     \mathbf{M}^{\mathrm{LBM}}_{ij} = e^{-\mathrm{i}k^*_\alpha e_{i, \alpha}} \left( \delta_{ij} - \overline{\mathbf{C}}_{ik} \left( \delta_{kj} - \mathbf{J}^{eq, N}_{kj} \right) \right),
     \label{eq:app_LSA_MMRT}
\end{align}
where $\overline{\mathbf{C}}$ is the collision matrix evaluated at the mean base flow.

\section{Relations with Genocchi and Bernoulli numbers for the BGK collision model}
\label{app:Genocchi_Bernoulli}

Let's consider the coefficients $c_n$ defined by the following recurrence relation
\begin{align}
    c_0=2, \qquad \forall n \geq 1,\ c_n = - \sum_{k=0}^n \binom{n}{k} c_k.
    \label{eq:reccurence_cn}
\end{align}
In order to find out any interesting property in the Taylor expansion of $f_i^{neq}$ with the BGK collision model, let us focus on the generating function of $c_n$:
\begin{align}
    f(x) = \sum_{n \geq 0} c_n \frac{x^n}{n!}.
\end{align}
Multiplying Eq.~(\ref{eq:reccurence_cn}) by $x^n/n!$ and doing a summation over $n\geq 1$ yield
\begin{align}
    \sum_{n \geq 1} c_n \frac{x^n}{n!} = - \sum_{n \geq 1} \sum_{k=0}^n \binom{n}{k}c_k \frac{x^n}{n!} \qquad & \Rightarrow \sum_{n\geq 0} c_n \frac{x^n}{n!} - c_0 = - \sum_{n \geq 0} \sum_{k=0}^n \binom{n}{k}c_k \frac{x^n}{n!} + c_0 \\
    & \Rightarrow \underbrace{\sum_{n\geq 0} c_n \frac{x^n}{n!}}_{f(x)} - 2 = - \underbrace{\left( \sum_{n \geq 0} c_n \frac{x^n}{n!} \right)}_{f(x)} \underbrace{\left( \sum_{n \geq 0} \frac{x^n}{n!} \right)}_{\exp(x)} + 2.
\end{align}
Hence, the generating function of $c_n$ is given by
\begin{align}
    f(x) = \frac{4}{\exp(x)+1}.
\end{align}
Then, knowing that $f(x)+f(-x)=4$, one has
\begin{align}
    \sum_{n \geq 0} c_n \frac{x^n}{n!} (1+(-1)^n) = 4,
\end{align}
so that any even coefficient is null:
\begin{align}
    \forall p \geq 1, \ c_{2p} = 0.
\end{align}
Note that, given the generating function $f$ of $c_n$, these coefficients can be related to the Genocchi numbers $G_n$ defined by their generating function $g$~\cite{Comtet}:
\begin{align}
    g(x) = \sum_{n \geq 1} G_n \frac{x^n}{n!} = \frac{2x}{\exp(x)+1} = \frac{x}{2} f(x).
\end{align}
Hence,
\begin{align}
    \forall n \geq 0,\ c_n = 2 \frac{G_{n+1}}{n+1}.
\end{align}
Genocchi nummbers can alsso be related to the Bernoulli numbers $B_n^-$ as~\cite{Arfken, Comtet}
\begin{align}
    \forall n \geq 1,\ G_n = 2(1-2^n) B_n^-,
\end{align}
so that
\begin{align}
    \forall n \geq 0,\ c_n = 4(1-2^{n+1}) \frac{B_{n+1}^-}{n+1}.
\end{align}

\section{Deviation terms of the BGK model with the D2Q9 lattice}
\label{app:deviations_BGK_D2Q9}

This appendix provides the deviation terms $E_\Phi^{(n)}$ of the BGK-D2Q9 model, from the isothermal NS equations. For the sake of simplicity, only the $x$-derivatives will be provided in the deviation terms $E_{\rho u_x}^{(2)}$, $E_{\rho u_y}^{(2)}$, $E_\rho^{(3)}$, $E_{\rho u_x}^{(3)}$ and $E_{\rho u_y}^{(3)}$. The expression of the other terms ($E_\rho^{(1)}$, $E_{\rho u_x}^{(1)}$, $E_{\rho u_x}^{(2)}$ and $E_\rho^{(2)}$) is much simpler, they will be provided in their complete form. Three equilibrium are considered: $N=2$, $N=3^*$ and $N=4^*$, with and without correction term $\Psi_i$.

Note that deviation terms are completely equivalent in the $x$-direction with the partial third-order ($N=3^*$) and the partial fourth-order ($N=4^*$) equilibria, they are therefore only provided once.

\subsection{Second-order equilibrium (N=2)}

%\subsubsection{Absence of correction}
No correction :
\begingroup
\allowdisplaybreaks
\begin{align}
    & E^{(1)}_\rho = 0, \qquad E^{(1)}_{\rho u_x} = -\partial^2_{xx} a^{(3)}_{eq, xxx} - 2 \partial^2_{xy} a^{(3)}_{eq, xxy} - \partial^2_{yy} a^{(3)}_{eq, xyy}, \qquad E^{(1)}_{\rho u_y} = -\partial^2_{xx} a^{(3)}_{eq, xxy} - 2 \partial^2_{xy} a^{(3)}_{eq, xyy} - \partial^2_{yy} a^{(3)}_{eq, yyy}, \nonumber \\
	& E^{(2)}_\rho = \left( \frac{\Delta t}{\tau} \right)^2 \frac{1}{12} \left( -\partial_{xxx}^3 a^{(3)}_{eq, xxx} - 3 \partial^3_{xxy} a^{(3)}_{eq, xxy} - 3\partial_{xyy}^3 a^{(3)}_{eq, xyy} - \partial_{yyy}^3 a^{(3)}_{eq, yyy} + \rho c_s^2 \theta (\partial^2_{xx} S_{xx} + 2\partial^2_{xy} S_{xy} + \partial^2_{yy} S_{yy}) \right), \nonumber \\
	& E^{(2)}_{\rho u_x} = C_1 + \left( \frac{\Delta t}{\tau} \right)^2 N_1, \qquad  E^{(2)}_{\rho u_y} = C_5 + \left( \frac{\Delta t}{\tau} \right)^2 N_5, \qquad E^{(3)}_\rho = \left( \frac{\Delta t}{\tau} \right)^2 \frac{1}{12} \partial_x C_1, \nonumber  \\
	& E_{\rho u_x}^{(3)} = C_2 + \left( \frac{\Delta t}{\tau} \right)^2 N_2, \qquad E_{\rho u_y}^{(3)} = C_6 + \left( \frac{\Delta t}{\tau} \right)^2 N_6.
\end{align}
\endgroup

%\subsubsection{Analytically computed correction}
Including now a correction term $\Psi_i$ without considering its discretization errors, one has
\begingroup
\allowdisplaybreaks
\begin{align}
    & E^{(1)}_\rho = 0, \quad \qquad E^{(1)}_{\rho u_x} = 0, \quad \qquad E^{(1)}_{\rho u_y} = 0, \nonumber \\
	& E^{(2)}_\rho = \left( \frac{\Delta t}{\tau} \right)^2 \frac{1}{12} \left( -\partial_{xxx}^3 a^{(3)}_{eq, xxx} - 3 \partial^3_{xxy} a^{(3)}_{eq, xxy} - 3\partial_{xyy}^3 a^{(3)}_{eq, xyy} - \partial_{yyy}^3 a^{(3)}_{eq, yyy} + \rho c_s^2 \theta (\partial^2_{xx} S_{xx} + 2\partial^2_{xy} S_{xy} + \partial^2_{yy} S_{yy}) \right), \nonumber \\
	& E_{\rho u_x}^{(2)} = C_3 + \left( \frac{\Delta t}{\tau} \right)^2 N_1, \qquad E_{\rho u_y}^{(2)} = C_7 + \left( \frac{\Delta t}{\tau} \right)^2 N_5, \qquad E_\rho^{(3)} = \left( \frac{\Delta t}{\tau} \right)^2 \frac{1}{12} \partial_x C_3, \nonumber \\
	& E^{(3)}_{\rho u_x} = C_4 + \left( \frac{\Delta t}{\tau} \right)^2 N_3, \qquad E^{(3)}_{\rho u_y} = C_8 + \left( \frac{\Delta t}{\tau} \right)^2 N_7.
\end{align}
\endgroup

%\subsubsection{Correction with the DCO2 scheme}
Including the discretization effects of the DCO2 scheme, only third-order errors in the momentum equation are affected:
\begin{align}
    & E^{(3)}_{\rho u_x} = C_4 + \left(\frac{\Delta t}{\tau} \right)^2 \left( N_3 + \frac{1}{6} \frac{\partial^4 a^{(3)}_{eq, xxx}}{\partial x^4} \right), \qquad E^{(3)}_{\rho u_y} = C_8 + \left(\frac{\Delta t}{\tau} \right)^2 \left( N_7 + \frac{1}{6} \frac{\partial^4 a_{eq, xxy}^{(3)}}{\partial x^4} \right).
\end{align}

%\subsubsection{Correction with the DUO1 scheme}
Finally, with the DUO1 scheme, the numerical errors of the momentum equations are altered as
\begin{align}
    & E^{(2)}_{\rho u_x} = C_3 + \left( \frac{\Delta t}{\tau} \right)^2 N_1 - \left( \frac{\Delta t}{2\tau} \right) \mathrm{sgn}(u_x) \frac{\partial^3 a^{(3)}_{eq, xxx}}{\partial x^3}, \qquad E^{(2)}_{\rho u_y} = C_7 + \left( \frac{\Delta t}{\tau} \right)^2 N_5 - \left( \frac{\Delta t}{2\tau} \right) \mathrm{sgn}(u_x) \frac{\partial^3 a^{(3)}_{eq, xxy}}{\partial x^3}, \nonumber \\
    & E^{(3)}_{\rho u_x} = C_4 +  \left( \frac{\Delta t}{\tau} \right)^2 \left( N_3 + \frac{1}{6} \frac{\partial^4 a^{(3)}_{eq, xxx}}{\partial x^4} \right) + \left( \frac{\Delta t}{2\tau} \right) \mathrm{sgn}(u_x) N_4, \qquad E_{\rho u_y}^{(3)} = C_8 +  \left( \frac{\Delta t}{\tau} \right)^2 \left( N_7 + \frac{1}{6} \frac{\partial^4 a^{(3)}_{eq, xxy}}{\partial x^4} \right) + \left( \frac{\Delta t}{2\tau} \right) \mathrm{sgn}(u_x) N_{9}.
\end{align}

\subsection{Partial third- and fourth-order equilibria ($N=3^*, 4^*$)}

%\subsubsection{Absence of correction}
No correction:
\begingroup
\allowdisplaybreaks
\begin{align}
	& E^{(1)}_\rho = 0, \quad \qquad E^{(1)}_{\rho u_x} = -\frac{\partial^2 a^{(3)}_{eq, xxx}}{\partial x^2}, \quad \qquad E^{(1)}_{\rho u_y} = -\frac{\partial^2 a^{(3)}_{eq, yyy}}{\partial y^2}, \nonumber \\
	& E^{(2)}_\rho = \left( \frac{\Delta t}{\tau} \right)^2 \frac{1}{12} \left( -\frac{\partial^3 a^{(3)}_{eq, xxx}}{\partial x^3} - \frac{\partial^3 a^{(3)}_{eq, yyy}}{\partial y^3} + \rho c_s^2 \theta \left( \frac{\partial^2 S_{xx}}{\partial x^2} + 2\frac{\partial^2 S_{xy}}{\partial x \partial y} + \frac{\partial^2 S_{yy}}{\partial y^2} \right) \right), \nonumber \\
	& E^{(2)}_{\rho u_x} = C_1 + \left( \frac{\Delta t}{\tau} \right)^2 N_1, \qquad E^{(2)}_{\rho u_y} = a_{eq, xxx}^{(3)} \frac{\partial^3 u_y}{\partial x^3} +  \left( \frac{\Delta t}{\tau} \right)^2 N_{15}, \nonumber \\
	& E^{(3)}_\rho = \left( \frac{\Delta t}{\tau} \right)^2 \frac{1}{12} \partial_x C_1, \qquad E^{(3)}_{\rho u_x} = C_2 + \left( \frac{\Delta t}{\tau} \right)^2 N_2, \qquad  E^{(3)}_{\rho u_y} = C_9 + \left( \frac{\Delta t}{\tau} \right)^2 N_{10}.
\end{align}
\endgroup

%\subsubsection{Analytically computed correction}
Analytically computed correction:
\begingroup
\allowdisplaybreaks
\begin{align}
    & E_\rho^{(1)} = 0, \qquad E_{\rho u_x}^{(1)} = 0, \qquad E_{\rho u_y}^{(1)} = 0, \nonumber \\
	& E^{(2)}_\rho = \left( \frac{\Delta t}{\tau} \right)^2 \frac{1}{12} \left( -\frac{\partial^3 a^{(3)}_{eq, xxx}}{\partial x^3} - \frac{\partial^3 a^{(3)}_{eq, yyy}}{\partial y^3} + \rho c_s^2 \theta \left( \frac{\partial^2 S_{xx}}{\partial x^2} + 2\frac{\partial^2 S_{xy}}{\partial x \partial y} + \frac{\partial^2 S_{yy}}{\partial y^2} \right) \right), \nonumber \\
    & E_{\rho u_x}^{(2)} = C_3 + \left( \frac{\Delta t}{\tau} \right)^2 N_1, \qquad E_{\rho u_y}^{(2)} = \frac{\partial^3}{\partial x^3} (u_y a_{eq, xxx}^{(3)}) + \left( \frac{\Delta t}{\tau} \right)^2 N_{15}, \nonumber \\
    & E_\rho^{(3)} = \left( \frac{\Delta t}{\tau} \right)^2 \frac{1}{12} \partial_x C_3, \qquad E_{\rho u_x}^{(3)} = C_4 + \left( \frac{\Delta t}{\tau} \right)^2 N_3, \qquad E^{(3)}_{\rho u_y} = C_{10} + \left( \frac{\Delta t}{\tau} \right)^2 N_{8}.
\end{align}
\endgroup

%\subsubsection{Correction with the DCO2 scheme}
Including the discretization effects of the DCO2 scheme, only the third-order error in the $x$-momentum equation is affected:
\begin{align}
    & E^{(3)}_{\rho u_x} = C_4 + \left(\frac{\Delta t}{\tau} \right)^2 \left( N_3 + \frac{1}{6} \frac{\partial^4 a^{(3)}_{eq, xxx}}{\partial x^4} \right).%, \qquad E^{(3)}_{\rho u_y} = C_{10} + \left(\frac{\Delta t}{\tau} \right)^2  N_7.
\end{align}

%\subsubsection{Correction with the DUO1 scheme}
Finally, with the DUO1 scheme, the numerical errors of the momentum equations are altered as
\begin{align}
    & E^{(2)}_{\rho u_x} = C_3 + \left( \frac{\Delta t}{\tau} \right)^2 N_1 - \left( \frac{\Delta t}{2\tau} \right) \mathrm{sgn}(u_x) \frac{\partial^3 a^{(3)}_{eq, xxx}}{\partial x^3},  % \qquad E_{\rho u_y}^{(2)} = C_{11} + \left( \frac{\Delta t}{\tau} \right)^2 N_{14}, \\
    \qquad E^{(3)}_{\rho u_x} = C_4 + \left(\frac{\Delta t}{\tau} \right)^2 \left( N_3 + \frac{1}{6} \frac{\partial^4 a^{(3)}_{eq, xxx}}{\partial x^4} \right) + \left( \frac{\Delta t}{2\tau} \right) \mathrm{sgn}(u_x) N_4, \nonumber \\
    & E_{\rho u_y}^{(3)} = C_{10} + \left( \frac{\Delta t}{\tau} \right)^2 N_{8} - \left( \frac{\Delta t}{2 \tau} \right) \mathrm{sgn}(u_x) u_y \frac{\partial^4 a^{(3)}_{eq, xxx}}{\partial x^4}.
\end{align}

\section{Deviation terms of the shear stress reconstruction (HRR-$\sigma=0$) with the D2Q9 lattice}
\label{app:deviations_AR_D2Q9}

This appendix provides the deviation terms (with the athermal NS equations) of the D2Q9 lattice with the shear stress reconstruction of Sec.~\ref{sec:Regul_FR}, also referred to as HRR-$\sigma=0$, in the linear approximation. Only the equilibrium at second-order ($N=2$) is considered here, noticing that changing the order of the Hermite-based equilibrium has very few impact on the general conclusions. Furthermore, for the sake of simplicity, only $x-$derivatives are provided for $E^{(2)}_{\rho u_\gamma}$, $E^{(3)}_\rho$ and $E^{(3)}_{\rho u_\gamma}$.

\subsection{Non corrected D2Q9 lattice, analytical reconstruction of $g_i^{(1)}$}
\begingroup
\allowdisplaybreaks
\begin{align}
    & E^{(1)}_\rho = 0, \quad E^{(1)}_{\rho u_x} = - \left( \frac{\Delta t}{2 \tau} \right) \left( \partial^2_{xx} a^{(3)}_{eq, xxx} + 2 \partial^2_{xy} a^{(3)}_{eq, xxy} + \partial^2_{yy} a^{(3)}_{eq, xyy} \right), \quad E^{(1)}_{\rho u_y} = - \left( \frac{\Delta t}{2 \tau} \right) \left( \partial^2_{xx} a^{(3)}_{eq, xxy} + 2 \partial^2_{xy} a^{(3)}_{eq, xyy} + \partial^2_{yy} a^{(3)}_{eq, yyy} \right), \nonumber \\
    & E^{(2)}_\rho = \left( \frac{\Delta t}{\tau} \right)^2 \frac{1}{12} \left( -\partial_{xxx}^3 a^{(3)}_{eq, xxx} - 3 \partial^3_{xxy} a^{(3)}_{eq, xxy} - 3\partial_{xyy}^3 a^{(3)}_{eq, xyy} - \partial_{yyy}^3 a^{(3)}_{eq, yyy} + \rho c_s^2 \theta (\partial^2_{xx} S_{xx} + 2\partial^2_{xy} S_{xy} + \partial^2_{yy} S_{yy}) \right), \nonumber \\
    & E^{(2)}_{\rho u_x} = \left( \frac{\Delta t}{\tau} \right) \frac{C_3}{2} + \left( \frac{\Delta t}{\tau} \right)^2 N_{11}, \qquad E^{(2)}_{\rho u_y} = \left( \frac{\Delta t}{\tau} \right) N_{19} + \left( \frac{\Delta t}{\tau} \right)^2 N_{20}, \qquad E^{(3)}_\rho = \frac{1}{12} \left( \frac{\Delta t}{\tau} \right)^2 \partial_x C_3 + \left( \frac{\Delta t}{\tau} \right)^3 N_{12}, \nonumber \\
    & E^{(3)}_{\rho u_x} =  \left( \frac{\Delta t}{\tau} \right) N_{13} + \left( \frac{\Delta t }{\tau} \right)^2 N_{14} + \left( \frac{\Delta t}{\tau} \right)^3 N_{16}, \qquad E^{(3)}_{\rho u_y} = -\left( \frac{\Delta t}{\tau} \right) \frac{\rho \theta^2}{18} \frac{\partial^4 u_y}{\partial x^4} + \left( \frac{\Delta t}{\tau} \right)^2 N_{21} + \left( \frac{\Delta t}{\tau} \right)^3 N_{22}.
\end{align}
\endgroup 

\subsection{$N=2$, analytically corrected D2Q9 lattice, analytical reconstruction of $g_i^{(1)}$}

\begingroup
\allowdisplaybreaks
\begin{align}
    & E^{(1)}_\rho = 0, \quad E^{(1)}_{\rho u_x} =0, \quad E^{(1)}_{\rho u_y} = 0, \nonumber \\
    & E^{(2)}_\rho = \left( \frac{\Delta t}{\tau} \right)^2 \frac{1}{12} \left( -\partial_{xxx}^3 a^{(3)}_{eq, xxx} - 3 \partial^3_{xxy} a^{(3)}_{eq, xxy} - 3\partial_{xyy}^3 a^{(3)}_{eq, xyy} - \partial_{yyy}^3 a^{(3)}_{eq, yyy} + \rho c_s^2 \theta (\partial^2_{xx} S_{xx} + 2\partial^2_{xy} S_{xy} + \partial^2_{yy} S_{yy}) \right), \nonumber \\
    & E^{(2)}_{\rho u_x} = \left( \frac{\Delta t}{2\tau} \right) C_3 + \left( \frac{\Delta t}{\tau} \right)^2 N_1, \qquad E^{(2)}_{\rho u_y} = \left( \frac{\Delta t}{\tau} \right) N_{19} + \left( \frac{\Delta t}{\tau} \right)^2 N_{23}, \qquad E^{(3)}_\rho = \frac{1}{12} \left( \frac{\Delta t}{\tau} \right)^2 \partial_x C_3, \nonumber \\
    & E^{(3)}_{\rho u_x} = \left( \frac{\Delta t}{\tau} \right) N_{13} + \left( \frac{\Delta t}{\tau} \right)^2 N_{17} + \left( \frac{\Delta t}{\tau} \right)^3 N_{18}, \qquad E^{(3)}_{\rho u_y} = -\left( \frac{\Delta t}{\tau} \right) \frac{\rho \theta^2}{18} \frac{\partial^4 u_y}{\partial x^4} + \left( \frac{\Delta t}{\tau} \right)^2 N_{24} + \left( \frac{\Delta t}{\tau} \right)^3 N_{25}.
    \label{eq:Regul_errors_D2Q9_N2_AR_fneqAnalytical_PsiAnalytical}
\end{align}
\endgroup

\subsection{$N=2$, correction with a DCO2 scheme, reconstruction of $g_i^{(1)}$ with a DCO2 scheme}

Compared to Eq.~(\ref{eq:Regul_errors_D2Q9_N2_AR_fneqAnalytical_PsiAnalytical}), only the numerical errors in the momentum equation are affected as
\begingroup
\allowdisplaybreaks
\begin{align}
   & E^{(3)}_{\rho u_x} = \left( \frac{\Delta t}{\tau} \right) N_{13} + \left( \frac{\Delta t}{\tau} \right)^2 \left( N_{17} + \frac{\rho \theta}{9} \frac{\partial^4 u_x}{\partial x^4} \right) + \left( \frac{\Delta t}{\tau} \right)^3 \left( N_{18} + \frac{1}{12}\frac{\partial^4 a^{(3)}_{eq, xxx}}{\partial x^4} - \frac{\rho \theta}{18} \frac{\partial^4 u_x}{\partial x^4} \right), \nonumber \\
   & E^{(3)}_{\rho u_y} = -\left( \frac{\Delta t}{\tau} \right) \frac{\rho \theta^2}{18} \frac{\partial^4 u_y}{\partial x^4} + \left( \frac{\Delta t}{\tau} \right)^2 \left( N_{24} + \frac{\rho \theta}{18} \frac{\partial^4 u_y}{\partial x^4} \right) + \left( \frac{\Delta t}{\tau} \right)^3 \left( N_{25} + \frac{1}{12}\frac{\partial^4 a^{(3)}_{eq, xxy}}{\partial x^4} - \frac{\rho \theta}{36} \frac{\partial^4 u_y}{\partial x^4} \right).
\end{align}
\endgroup

\subsection{$N=2$, correction with a DCO2 scheme, reconstruction of $g_i^{(1)}$ with a DUO1 scheme}

Compared to Eq.~(\ref{eq:Regul_errors_D2Q9_N2_AR_fneqAnalytical_PsiAnalytical}), only the numerical errors in the momentum equation are affected as
\begingroup
\allowdisplaybreaks
\begin{align}
    & E^{(2)}_{\rho u_x} = \left( \frac{\Delta t}{\tau} \right) \left( \frac{C_3}{2} - \mathrm{sgn}(u_x) \rho c_s^2 \theta \frac{\partial^3 u_x}{\partial x^3} \right) + \left( \frac{\Delta t}{\tau} \right)^2 \left( N_1 + \mathrm{sgn}(u_x) \frac{\rho \theta}{6} \frac{\partial^3 u_x}{\partial x^3} \right), \nonumber \\
   & E^{(2)}_{\rho u_y} = \left( \frac{\Delta t}{\tau} \right) \left( N_{19} - \mathrm{sgn}(u_x) \frac{\rho \theta}{6} \frac{\partial^3 u_y}{\partial x^3} \right) + \left( \frac{\Delta t}{\tau} \right)^2 \left( N_{23} + \mathrm{sgn}(u_x) \frac{\rho \theta}{12} \frac{\partial^3 u_y}{\partial x^3} \right), \nonumber \\ 
   & E^{(3)}_{\rho u_x} = \left( \frac{\Delta t}{\tau} \right) N_{13} + \left( \frac{\Delta t}{\tau} \right)^2 \left( N_{17} + \frac{\rho \theta}{9} \frac{\partial^4 u_x}{\partial x^4} - \mathrm{sgn}(u_x) \frac{\partial_x C_3}{4} \right) + \left( \frac{\Delta t}{\tau} \right)^3 \left( N_{18} + \frac{1}{12}\frac{\partial^4 a^{(3)}_{eq, xxx}}{\partial x^4} - \frac{\rho \theta}{18} \frac{\partial^4 u_x}{\partial x^4} + \mathrm{sgn}(u_x) \frac{\partial_x C_3}{8} \right), \nonumber \\
   & E^{(3)}_{\rho u_y} = -\left( \frac{\Delta t}{\tau} \right) \frac{\rho \theta^2}{18} \frac{\partial^4 u_y}{\partial x^4} + \left( \frac{\Delta t}{\tau} \right)^2 \left( N_{24} + \frac{\rho \theta}{18} \frac{\partial^4 u_y}{\partial x^4} - \mathrm{sgn}(u_x) \frac{N_{19}}{2} \right) + \left( \frac{\Delta t}{\tau} \right)^3 \left( N_{25} + \frac{1}{12}\frac{\partial^4 a^{(3)}_{eq, xxy}}{\partial x^4} - \frac{\rho \theta}{36} \frac{\partial^4 u_y}{\partial x^4} + \mathrm{sgn}(u_x) \frac{N_{19}}{4} \right).
\end{align}
\endgroup

\subsection{$N=2$, correction with a DUO1 scheme, reconstruction of $g_i^{(1)}$ with a DCO2 scheme}

Compared to Eq.~(\ref{eq:Regul_errors_D2Q9_N2_AR_fneqAnalytical_PsiAnalytical}), only the numerical errors in the momentum equation are affected as
\begingroup
\allowdisplaybreaks
\begin{align}
    & E^{(2)}_{\rho u_x} = \left( \frac{\Delta t}{2\tau} \right) C_3 + \left( \frac{\Delta t}{\tau} \right)^2 \left( N_1 - \frac{\mathrm{sgn}(u_x)}{4} \frac{\partial^3 a^{(3)}_{eq, xxx}}{\partial x^3} \right), \qquad E^{(2)}_{\rho u_y} = \left( \frac{\Delta t}{\tau} \right) N_{19} + \left( \frac{\Delta t}{\tau} \right)^2 \left( N_{23} - \frac{\mathrm{sgn}(u_x)}{4}\frac{\partial^3 a^{(3)}_{eq, xxy}}{\partial x^3} \right) , \nonumber \\
    & E^{(3)}_{\rho u_x} = \left( \frac{\Delta t}{\tau} \right) N_{13} + \left( \frac{\Delta t}{\tau} \right)^2 \left( N_{17} + \frac{\rho \theta}{9} \frac{\partial^4 u_x}{\partial x^4} \right) + \left( \frac{\Delta t}{\tau} \right)^3 \left( N_{18} + \frac{1}{12}\frac{\partial^4 a^{(3)}_{eq, xxx}}{\partial x^4} - \frac{\rho \theta}{18} \frac{\partial^4 u_x}{\partial x^4} + \mathrm{sgn}(u_x) \frac{N_4}{8} \right), \nonumber \\
    & E^{(3)}_{\rho u_y} = -\left( \frac{\Delta t}{\tau} \right) \frac{\rho \theta^2}{18} \frac{\partial^4 u_y}{\partial x^4} + \left( \frac{\Delta t}{\tau} \right)^2 \left( N_{24} + \frac{\rho \theta}{18} \frac{\partial^4 u_y}{\partial x^4} \right) + \left( \frac{\Delta t}{\tau} \right)^3 \left( N_{25} + \frac{1}{12}\frac{\partial^4 a^{(3)}_{eq, xxy}}{\partial x^4} - \frac{\rho \theta}{36} \frac{\partial^4 u_y}{\partial x^4} + \mathrm{sgn}(u_x) N_{26} \right).
\end{align}
\endgroup

\subsection{$N=2$, correction with a DUO1 scheme, reconstruction of $g_i^{(1)}$ with a DUO1 scheme}

Compared to Eq.~(\ref{eq:Regul_errors_D2Q9_N2_AR_fneqAnalytical_PsiAnalytical}), only the numerical errors in the momentum equation are affected as
\begingroup
\allowdisplaybreaks
\begin{align}
    & E^{(2)}_{\rho u_x} = \left( \frac{\Delta t}{\tau} \right) \left(\frac{C_3}{2} - \mathrm{sgn}(u_x) \rho c_s^2 \theta \frac{\partial^3 u_x}{\partial x^3} \right) + \left( \frac{\Delta t}{\tau} \right)^2 \left( N_1 - \frac{\mathrm{sgn}(u_x)}{4} \frac{\partial^3 a^{(3)}_{eq, xxx}}{\partial x^3} + \mathrm{sgn}(u_x) \frac{\rho \theta}{6} \frac{\partial^3 u_x}{\partial x^3} \right), \nonumber \\
    & E^{(2)}_{\rho u_y} = \left( \frac{\Delta t}{\tau} \right) \left( N_{19} - \mathrm{sgn}(u_x) \frac{\rho \theta}{6} \frac{\partial^3 u_y}{\partial x^3} \right) + \left( \frac{\Delta t}{\tau} \right)^2 \left( N_{23} + \mathrm{sgn}(u_x) \frac{\rho \theta}{12} \frac{\partial^3 u_y}{\partial x^3} -  \frac{\mathrm{sgn}(u_x)}{4} \frac{\partial^3 a^{(3)}_{eq, xxy}}{\partial x^3} \right), \nonumber \\ 
    & E^{(3)}_{\rho u_x} = \left( \frac{\Delta t}{\tau} \right) N_{13} + \left( \frac{\Delta t}{\tau} \right)^2 \left( N_{17} + \frac{\rho \theta}{9} \frac{\partial^4 u_x}{\partial x^4} - \mathrm{sgn}(u_x)\frac{\partial_x C_3}{4} \right) + \left( \frac{\Delta t}{\tau} \right)^3 \left( N_{18} + \frac{1}{12}\frac{\partial^4 a^{(3)}_{eq, xxx}}{\partial x^4} - \frac{\rho \theta}{18} \frac{\partial^4 u_x}{\partial x^4} + \frac{\mathrm{sgn}(u_x)}{8} \left( N_4 + \partial_x C_3 \right) \right), \nonumber \\
    %& E^{(3)}_{\rho u_y} = -\left( \frac{\Delta t}{\tau} \right) \frac{\rho \theta^2}{18} \frac{\partial^4 u_y}{\partial x^4} + \left( \frac{\Delta t}{\tau} \right)^2 \left( N_{24} + \frac{\rho \theta}{18} \frac{\partial^4 u_y}{\partial x^4} - \mathrm{sgn}(u_x) \frac{\rho \theta}{6} \left( u_y \frac{\partial^4 u_x}{\partial x^4} + u_x \frac{\partial^4 u_y}{\partial x^4}  \right) \right) \nonumber \\
    %& \qquad + \left( \frac{\Delta t}{\tau} \right)^3 \Bigg( N_{25} + \frac{1}{12}\frac{\partial^4 a^{(3)}_{eq, xxy}}{\partial x^4} - \frac{\rho \theta}{36} \frac{\partial^4 u_y}{\partial x^4} + \mathrm{sgn}(u_x) \frac{\rho \theta}{12} \left( u_y \frac{\partial^4 u_x}{\partial x^4} + u_x \frac{\partial^4 u_y}{\partial x^4} \right) \nonumber \\
    %& \qquad - \frac{\mathrm{sgn}(u_x)}{24} \left( (9u_x^2 + (7\theta-5) u_x)u_y \frac{\partial^4 \rho}{\partial x^4} + (24u_x^2 + 4(\theta-1))\rho u_y \frac{\partial^4 u_x}{\partial x^4}  + (6u_x^3 + 2(\theta-1)u_x) \rho \frac{\partial^4 u_y}{\partial x^4} \right) \Bigg). \\
    & E^{(3)}_{\rho u_y} = -\left( \frac{\Delta t}{\tau} \right) \frac{\rho \theta^2}{18} \frac{\partial^4 u_y}{\partial x^4} + \left( \frac{\Delta t}{\tau} \right)^2 \left( N_{24} + \frac{\rho \theta}{18} \frac{\partial^4 u_y}{\partial x^4} - \mathrm{sgn}(u_x) \frac{N_{19}}{2} \right) \nonumber \\
    & \qquad \qquad + \left( \frac{\Delta t}{\tau} \right)^3 \left( N_{25} + \frac{1}{12}\frac{\partial^4 a^{(3)}_{eq, xxy}}{\partial x^4} - \frac{\rho \theta}{36} \frac{\partial^4 u_y}{\partial x^4} + \mathrm{sgn}(u_x) \left( \frac{N_{19}}{4} + N_{26} \right) \right).
\end{align}
\endgroup

\section{Deviation terms of corrected PR and RR models}
\label{app:deviations_corrected_PR_RR}

This appendix details the deviation terms of corrected PR and RR models. For the sake of simplicity, only modifications from their BGK counterpart are provided with $x$-derivatives only.

\subsection{PR model with $N=2$}
Analytical correction:
\begin{align}
	& E_{\rho u_y}^{(2)} = \left( C_7 + 2c_s^2 \frac{\partial^3 (\rho u_x u_y)}{\partial x^3} \right) - \left( \frac{\Delta t}{\tau} \right) c_s^2 \frac{\partial^3 (\rho u_x u_y)}{\partial x^3} + \left( \frac{\Delta t}{\tau} \right)^2 N_5, \nonumber \\
	& E^{(3)}_{\rho u_y} = C_{12} + \left( \frac{\Delta t}{\tau} \right) N_{29} +  \left( \frac{\Delta t}{\tau} \right)^2 \left( N_7 - N_{28} \right) + \left( \frac{\Delta t}{\tau} \right)^3 \frac{1}{4} N_{28}.
\end{align}
%Note that the degenerating hyperviscosity, still related to $N_{28}$, is not altered by the correction.

%\subsection{PR model with $N=2$}
Correction with a DCO2 scheme:
\begin{align}
	& E_{\rho u_y}^{(2)} = \left( C_7 + 2c_s^2 \frac{\partial^3 (\rho u_x u_y)}{\partial x^3} \right) - \left( \frac{\Delta t}{\tau} \right) c_s^2 \frac{\partial^3 (\rho u_x u_y)}{\partial x^3} + \left( \frac{\Delta t}{\tau} \right)^2 N_5, \nonumber \\
	& E^{(3)}_{\rho u_y} = C_{12} + \left( \frac{\Delta t}{\tau} \right) N_{29} + \left(\frac{\Delta t}{\tau} \right)^2 \left( N_7 + \frac{1}{6} \frac{\partial^4 a_{eq, xxy}^{(3)}}{\partial x^4} - N_{28} \right) + \left( \frac{\Delta t}{\tau} \right)^3 \frac{1}{4} N_{28}.
\end{align}
%Here again, the degenerating hyperviscosity is not affected by the (discretized) correction.

%\subsection{PR model with $N=2$: correction with a DUO1 scheme}
Correction with a DUO1 scheme:
\begin{align}
    & E^{(2)}_{\rho u_y} =  \left( C_7 + 2c_s^2 \frac{\partial^3 (\rho u_x u_y)}{\partial x^3} \right) - \left( \frac{\Delta t}{\tau} \right) c_s^2 \frac{\partial^3 (\rho u_x u_y)}{\partial x^3} + \left( \frac{\Delta t}{\tau} \right)^2 N_5 - \left( \frac{\Delta t}{2\tau} \right) \mathrm{sgn}(u_x) \frac{\partial^3 a^{(3)}_{eq, xxy}}{\partial x^3}, \nonumber \\
    & E^{(3)}_{\rho u_y} = C_{12} + \left( \frac{\Delta t}{\tau} \right) N_{29} + \left( \frac{\Delta t}{\tau} \right)^2 \left( N_7 + \frac{1}{6} \frac{\partial^4 a_{eq, xxy}^{(3)}}{\partial x^4} - N_{28} \right) + \left( \frac{\Delta t}{2\tau} \right) \mathrm{sgn}(u_x) N_{9} + \left( \frac{\Delta t}{\tau} \right)^3 \frac{1}{4} N_{28}.
\end{align}
%Here again, the degenerating hyperviscosity is not affected by the (discretized) correction.

\subsection{RR model with $N_r=3^*, 4^*$ and $N=2$}

Analytical correction:
\begin{align}
    E_{\rho u_y}^{(2)} = \left( \frac{\Delta t}{3 \tau} \right) N_{34} +  \left( \frac{\Delta t}{\tau}\right)^2 N_{35}, \qquad E_{\rho u_y}^{(3)} = -\frac{\rho \theta^2}{9} \frac{\partial^4 u_y}{\partial x^4} + \left( \frac{\Delta t}{\tau} \right) N_{36} + \left( \frac{\Delta t}{\tau} \right)^2 \left( N_7 - N_{37} \right) + \left( \frac{\Delta t}{\tau} \right)^3 N_{38},
\end{align}
where
%The difference in consistency with the NS equations is greatly reduced in the $y$-momentum equation. This observation can be attributed to the role of the recursive regularization, which aims at reconstructing off-equilibrium populations using the $\boldsymbol{a}_1^{(n)}$ coefficients based on the Chapman-Enskog expansion leading to NS. Thanks to the body-force correction, $\boldsymbol{a}_1^{(2)}$ is in agreement with the NS equations, and thanks to the recurisve regularization, some $\boldsymbol{a}_1^{(3)}$ terms are in agreement with the CE expansion. Furthermore, as previously, both first-order errors in $\Delta t$ and a degenerating hyperviscosity ($N_{38}$) are introduced. For a horizontal mean flow, the latter behaves as
\begin{align}
    N_{38}\left|_{(u_y=0)} \right. = (1-\theta)\frac{\rho u_x^2}{12} \frac{\partial^4 u_y}{\partial x^4}. 
\end{align}
%which affects the shear wave only: it is dissipative for $\theta>1$, destabilizing for $\theta < 1$ and only cancels for $\theta=1$.

%\subsection{RR model with $N_r=3^*, 4^*$ and $N=2$: correction with a DCO2 scheme}
Correction with a DCO2 scheme:
\begin{align}
    E_{\rho u_y}^{(2)} = \left( \frac{\Delta t}{3 \tau} \right) N_{34} +  \left( \frac{\Delta t}{\tau}\right)^2 N_{35}, \qquad E_{\rho u_y}^{(3)} = -\frac{\rho \theta^2}{9} \frac{\partial^4 u_y}{\partial x^4} + \left( \frac{\Delta t}{\tau} \right) N_{36} + \left( \frac{\Delta t}{\tau} \right)^2 \left( N_7 + \frac{1}{6} \frac{\partial^4 a_{eq, xxy}^{(3)}}{\partial x^4} - N_{37} \right) + \left( \frac{\Delta t}{\tau} \right)^3 N_{38}.
\end{align}
%The DCO2 discretization of $\Psi_i$ does not affect the degenerating hyperviscosity (in the $x$-direction).

%\subsection{RR model with $N_r=3^*, 4^*$ and $N=2$: correction with a DUO1 scheme}
Correction with a DUO1 scheme:
\begin{align}
    & E_{\rho u_y}^{(2)} = \left( \frac{\Delta t}{3 \tau} \right) N_{34} +  \left( \frac{\Delta t}{\tau}\right)^2 N_{35} - \left( \frac{\Delta t}{2\tau} \right) \mathrm{sgn}(u_x) \frac{\partial^3 a^{(3)}_{eq, xxy}}{\partial x^3}, \nonumber \\
    & E_{\rho u_y}^{(3)} = -\frac{\rho \theta^2}{9} \frac{\partial^4 u_y}{\partial x^4} + \left( \frac{\Delta t}{\tau} \right) \left( N_{36} + \mathrm{sgn}(u_x) \partial_x C_{13} \right) + \left( \frac{\Delta t}{\tau} \right)^2  \left( N_7 + \frac{1}{6} \frac{\partial^4 a_{eq, xxy}^{(3)}}{\partial x^4} - N_{37} + \mathrm{sgn}(u_x) N_{39} \right) + \left( \frac{\Delta t}{\tau} \right)^3 N_{38}.
\end{align}

\subsection{PR model with $N=3^*, 4^*$}

Analytical correction:
\begin{align}
	& E_{\rho u_y}^{(2)} = 2 \rho c_s^2 \theta \frac{\partial^3 (u_x u_y)}{\partial x^3} - \left( \frac{\Delta t}{2 \tau} \right) C_{15} + \left( \frac{\Delta t}{\tau} \right)^2 N_{15}, \qquad E^{(3)}_{\rho u_y} = C_{17} + \left( \frac{\Delta t}{\tau}\right) N_{43} + \left( \frac{\Delta t}{\tau} \right)^2 \left( N_{8} + N_{41} \right) + \left( \frac{\Delta t}{\tau} \right)^3 N_{42}.
\end{align}
%The degenerating hyperviscosity is unchanged compared to the non corrected case.

%\subsection{PR model with $N=3^*, 4^*$: correction with a DCO2 scheme}

Correction with a DCO2 scheme:
\begin{align}
	& E_{\rho u_y}^{(2)} = 2 \rho c_s^2 \theta \frac{\partial^3 (u_x u_y)}{\partial x^3} - \left( \frac{\Delta t}{2 \tau} \right) C_{15} + \left( \frac{\Delta t}{\tau} \right)^2 N_{15}, \quad E^{(3)}_{\rho u_y} = C_{17} + \left( \frac{\Delta t}{\tau} \right) N_{43} + \left( \frac{\Delta t}{\tau} \right)^2 \left( N_{8} + N_{41} \right) + \left( \frac{\Delta t}{\tau} \right)^3 N_{42}.
\end{align}

%\subsection{PR model with $N=3^*, 4^*$: correction with a DUO1 scheme}
Correction with a DUO1 scheme:
\begin{align}
	& E_{\rho u_y}^{(2)} = 2 \rho c_s^2 \theta \frac{\partial^3 (u_x u_y)}{\partial x^3} - \left( \frac{\Delta t}{2 \tau} \right) C_{15} + \left( \frac{\Delta t}{\tau} \right)^2 N_{15},\nonumber \\
	& E^{(3)}_{\rho u_y} = C_{17} + \left( \frac{\Delta t}{\tau} \right) N_{43} - \left( \frac{\Delta t}{2 \tau} \right) \mathrm{sgn}(u_x) u_y \frac{\partial^4 a_{eq, xxx}^{(3)}}{\partial x^4} + \left( \frac{\Delta t}{\tau} \right)^2 \left( N_{8} + N_{41} \right) + \left( \frac{\Delta t}{\tau} \right)^3 N_{42}.
\end{align}

\subsection{RR model with $N_r=3^*, 4^*$ and $N=3^*, 4^*$}

Analytical correction:
\begin{align}
    & E_{\rho u_y}^{(2)} = \left( \frac{\Delta t}{2 \tau} \right) \frac{\partial^3 (u_y a_{eq, xxx}^{(3)})}{\partial x^3} + \left( \frac{\Delta t}{\tau} \right)^2 N_{15}, \qquad E^{(3)}_{\rho u_y} = -\frac{\rho \theta^2}{9} \frac{\partial^4 u_y}{\partial x^4} + \left( \frac{\Delta t}{\tau} \right) N_{45} +\left( \frac{\Delta t}{\tau} \right)^2 \left( N_{8} + N_{46} \right) + \left( \frac{\Delta t}{\tau} \right)^3 N_{47}.
\end{align}
For a horizontal mean flow, the degenerating hyperviscosity behaves as
\begin{align}
    N_{47}\left|_{(u_y=0)} \right. = N_{44} = -\frac{1}{8} \rho (u_x^4+(\theta-1)u_x^2) \frac{\partial^4 u_y}{\partial x^4}.
\end{align}

%\subsection{RR model with $N_r=3^*, 4^*$ and $N=3^*, 4^*$: correction with a DCO2 scheme}
Correction with a DCO2 scheme:
\begin{align}
    & E_{\rho u_y}^{(2)} = \left( \frac{\Delta t}{2 \tau} \right) \frac{\partial^3 (u_y a_{eq, xxx}^{(3)})}{\partial x^3} + \left( \frac{\Delta t}{\tau} \right)^2 N_{15}, \qquad E^{(3)}_{\rho u_y} = -\frac{\rho \theta^2}{9} \frac{\partial^4 u_y}{\partial x^4} + \left( \frac{\Delta t}{\tau} \right) N_{45} +\left( \frac{\Delta t}{\tau} \right)^2 \left( N_{8} + N_{46} \right) + \left( \frac{\Delta t}{\tau} \right)^3 N_{47}.
\end{align}

%\subsection{RR model with $N_r=3^*, 4^*$ and $N=3^*, 4^*$: correction with a DUO1 scheme}
Correction with a DUO1 scheme:
\begin{align}
    & E_{\rho u_y}^{(2)} = \left( \frac{\Delta t}{2 \tau} \right) \frac{\partial^3 (u_y a_{eq, xxx}^{(3)})}{\partial x^3} + \left( \frac{\Delta t}{\tau} \right)^2 N_{15}, \nonumber \\
    & E^{(3)}_{\rho u_y} = -\frac{\rho \theta^2}{9} \frac{\partial^4 u_y}{\partial x^4} + \left( \frac{\Delta t}{\tau} \right) N_{45} +\left( \frac{\Delta t}{\tau} \right)^2 \left( N_{8} + N_{46} \right) - \mathrm{sgn}(u_x) \left( \frac{\Delta t}{\tau} \right)^2 \frac{u_y}{4} \frac{\partial^4 a_{eq, xxx}^{(3)}}{\partial x^4} + \left( \frac{\Delta t}{\tau} \right)^3 N_{47}.
\end{align}

\section{Deviation terms}
\label{app:deviation_terms}

\subsection{Consistency errors with the isothermal NS equations}

\begingroup
\allowdisplaybreaks
\begin{align*}
    & C_1 = - \left( 3u_x^4 + (4\theta-3)u_x^2 + \theta^2/9 - \theta/3 \right) \frac{\partial^3 \rho}{\partial x^3} - \left( 10\rho u_x^3 + (2\theta-4)\rho u_x \right) \frac{\partial^3 u_x}{\partial x^3}, \\
    & C_2 = -\frac{1}{3} \left( 36u_x^5 + (56\theta-39)u_x^3 + (4\theta^2-11\theta + 3)u_x \right) \frac{\partial^4 \rho}{\partial x^4} -\frac{1}{9}\left( 378u_x^4+(108\theta-189)u_x^2+2\theta^2-9\theta+9 \right) \rho \, \frac{\partial^4 u_x}{\partial x^4}, \\
    & C_3 =  \frac{2 \theta^2}{9} \frac{\partial^3 \rho}{\partial x^3} + 2 \rho \theta u_x \frac{\partial^3 u_x}{\partial x^3}, \qquad C_4 = \frac{4}{3} \theta^2 u_x \frac{\partial^4 \rho}{\partial x^4} + \frac{1}{9} \left(54 \theta u_x^2-2 \theta^2 \right) \rho \frac{\partial^4 u_x}{\partial x^4}, \\
    & C_5 = -\left(3 u_x^3+(7 c_s^2\theta-1) u_x \right) u_y \frac{\partial^3 \rho}{\partial x^3} - \left(8u_x^2+2c_s^2 (\theta-1) \right) \rho u_y \frac{\partial^3 u_x}{\partial x^3} - 2\rho u_x^3 \frac{\partial^3 u_y}{\partial x^3}, \\
    & C_6 = -\frac{1}{9}\left(108 u_x^4+(123 \theta-57) u_x^2+3 \theta^2-2 \theta-1 \right) u_y \frac{\partial^4 \rho}{\partial x^4} - \frac{1}{3}\left(111 u_x^3+(19 \theta-19) u_x \right) \rho u_y \frac{\partial^4 u_x}{\partial x^4} - \frac{1}{9} \left(45u_x^4+6 u_x^2-1 \right) \rho \frac{\partial^4 u_y}{\partial x^4}, \\
    & C_7 = \frac{2}{3} \left(-u_x u_y \frac{\partial^3 \rho}{\partial x^3} + \rho(\theta-1) \left( u_y \frac{\partial^3 u_x}{\partial x^3} + u_x \frac{\partial^3 u_y}{\partial x^3} \right) \right), \\
    & C_8 = -\frac{1}{9} \left( 18 u_x^2-4 \theta^2+4 \theta \right) u_y \frac{\partial^4 \rho}{\partial x^4} + 10 \rho c_s^2(\theta-1) u_x u_y \frac{\partial^4 u_x}{\partial x^4} + \frac{1}{9} \left((12 \theta-18) u_x^2- \theta^2+2 \theta \right) \rho \frac{\partial^4 u_y}{\partial x^4}, \\
    & C_9 =\left(3 u_x^4+(2 \theta-3) u_x^2-2c_s^4 \theta^2+ c_s^2 \theta \right) \rho \frac{\partial^4 u_y}{\partial x^4} , \\
    & C_{10} = \frac{1}{3}\left( 9u_x^4+(15\theta-9)u_x^2+2\theta^2-2\theta\right)u_y \frac{\partial^4 \rho}{\partial x^4} + (11 u_x^3+5(\theta-1)u_x)\rho u_y \frac{\partial^4 u_x}{\partial x^4} + \frac{1}{9} \left(27u_x^4+(18\theta-27)u_x^2-2\theta^2+3\theta \right) \rho \frac{\partial^4 u_y}{\partial x^4}, \\ %-\frac{1}{9} \left(18 u_x^2-4 \theta^2+4 \theta \right) u_y \frac{\partial^4 \rho}{\partial x^4} + \frac{10}{3} (\theta-1) \rho u_x u_y \frac{\partial^4 u_x}{\partial x^4} + \frac{1}{9} \left( (12\theta-18) u_x^2-\theta^2+2 \theta \right) \rho \frac{\partial^4 u_y}{\partial x^4} , \\
    & C_{11} = C_6 + \frac{1}{9} \left( (24u_x^2 + 6\theta-2) u_y \frac{\partial^4 \rho}{\partial x^4} +  42 \rho u_x u_y \frac{\partial^4 u_x}{\partial x^4} + (24u_x^2-2) \rho \frac{\partial^4 u_y}{\partial x^4} \right), \\
    & C_{12} = C_8 + \frac{1}{9} \left( (18u_x^2+4\theta)u_y \frac{\partial^4 \rho}{\partial x^4} +30 \rho u_x u_y \frac{\partial^4 u_x}{\partial x^4} + (18 u_x^2-2 \theta) \rho \frac{\partial^4 u_y}{\partial x^4} \right), \\
    & C_{13} = -2c_s^2 \theta u_x u_y \frac{\partial^3 \rho}{\partial x^3} - (u_x^2+ c_s^2(\theta-1)) \rho u_y \frac{\partial^3 u_x}{\partial x^3}, \\
    & C_{14} = - \frac{1}{9} (27 u_x^4+(24\theta-24) u_x^2+\theta^2-2\theta+1) u_y \frac{\partial^4 \rho}{\partial x^4} - \frac{1}{3} (24u_x^3+(4 \theta-8)u_x) \rho u_y \frac{\partial^4 u_x}{\partial x^4} - \frac{1}{9} (9 u_x^4-6 u_x^2+1) \rho \frac{\partial^4 u_y}{\partial x^4}, \\
	& C_{15} = -(u_x^3+(\theta-1) u_x) u_y \frac{\partial^3 \rho}{\partial x^3} -  (3 u_x^2+ c_s^2 \theta-1) \rho u_y \frac{\partial^3 u_x}{\partial x^3} - ( u_x^3 + (c_s^2 \theta-1) u_x) \rho \frac{\partial^3 u_y}{\partial x^3}, \nonumber \\
	& C_{16} = \frac{1}{3} \left( (-15 u_x^4+(15-21\theta)u_x^2-2c_s^2\theta^2+2\theta) u_y \frac{\partial^4 \rho}{\partial x^4} + ((21-11 \theta) u_x - 51 u_x^3) \rho u_y \frac{\partial^4 u_x}{\partial x^4} + (4 \theta u_x^2- c_s^2 \theta^2) \rho \frac{\partial^4 u_y}{\partial x^4} \right), \\
	& C_{17} = \frac{4}{9} \theta^2 u_y \frac{\partial^4 \rho}{\partial x^4} + 10 \rho c_s^2 \theta u_x u_y \frac{\partial^4 u_x}{\partial x^4} + \frac{1}{9}(12 \theta u_x^2- \theta^2) \rho \frac{\partial^4 u_y}{\partial x^4},  \\
\end{align*}
\endgroup

\subsection{Numerical errors}

\begingroup
\allowdisplaybreaks
\begin{align*}
    & N_1 = \frac{1}{54} \left( 9u_x^4 + (18\theta-9)u_x^2 + \theta^2-3\theta \right) \frac{\partial^3 \rho}{\partial x^3} + \frac{1}{9}\left( 6 u_x^3 + (2\theta-3)u_x \right) \rho \frac{\partial^3 u_x}{\partial x^3}, \\
    & N_2 = \frac{1}{108} \Bigg( \left(162u_x^5+(300\theta-171)u_x^3+(26\theta^2-63\theta+9)u_x \right) \frac{\partial^4 \rho}{\partial x^4} + \left(612u_x^4+(216\theta-315)u_x^2+4\theta^2-15\theta+9\right) \rho \frac{\partial^4 u_x}{\partial x^4} \Bigg), \\
    & N_3 = \frac{1}{108} \left( \left(54 u_x^5+(108\theta-45)u_x^3+(-10\theta^2-9\theta-9) u_x \right) \frac{\partial^4 \rho}{\partial x^4} + \left(216 u_x^4-81 u_x^2-8\theta^2+15 \theta-9 \right)\rho \frac{\partial^4 u_x}{\partial x^4} \right), \\
    & N_4 = \left( -3 u_x^4+(3-4 \theta)u_x^2-\theta^2/3+ \theta/3 \right) \frac{\partial^4 \rho}{\partial x^4} + \left(-10 u_x^3+(4 -4 \theta) u_x \right) \rho \frac{\partial^3 u_x}{\partial x^3}, \\
    & N_5 = \frac{1}{6} \left( (u_x^3+\theta u_x) u_y \frac{\partial^3 \rho}{\partial x^3} +  (3 u_x^2+ c_s^2 \theta) \rho u_y \frac{\partial^3 u_x}{\partial x^3} + \rho u_x^3 \frac{\partial^3 u_y}{\partial x^3} \right), \\
    & N_6 = \frac{1}{36} \Bigg( \left(54 u_x^4+(71 \theta-21) u_x^2+7 c_s^2\theta^2-2c_s^2 \theta-1 \right) u_y \frac{\partial^4 \rho}{\partial x^4} + \left(177u_x^3+(37 \theta-21) u_x\right) \rho u_y \frac{\partial^4 u_x}{\partial x^4} + \left(27u_x^4+6  u_x^2-1 \right) \rho \frac{\partial^4 u_y}{\partial x^4} \Bigg), \\
    & N_7 = \frac{1}{108} \left( 54 u_x^4+(81 \theta+9) u_x^2-\theta^2+6 \theta-3 \right) u_y \frac{\partial^4 \rho}{\partial x^4} + \frac{1}{12} \left(21 u_x^3+(\theta+3) u_x \right) \rho u_y \frac{\partial^4 u_x}{\partial x^4} + \frac{1}{36} \left(9 u_x^4+(12-6 \theta) u_x^2-1 \right) \rho \frac{\partial^4 u_y}{\partial x^4}, \\
    & N_{8} = -\frac{1}{108} \left(27u_x^4+(54\theta-27) u_x^2+7\theta^2-9\theta \right) u_y \frac{\partial^4 \rho}{\partial x^4} -\frac{1}{6} \left(6u_x^3+(2\theta-3)u_x\right)\rho u_y \frac{\partial^4 u_x}{\partial x^4} \\
    & \qquad \qquad \qquad -\frac{1}{36} \left(18u_x^4+(12\theta-18)u_x^2-\theta^2+2\theta \right) \rho \frac{\partial^4 u_y}{\partial x^4},\\
    & N_{9} = - \frac{1}{3} \left( (9 u_x^3+(7\theta-5) u_x) u_y \frac{\partial^4 \rho}{\partial x^4} + (24 u_x^2+4 \theta-4) \rho u_y \frac{\partial^4 u_x}{\partial x^4} + (6 u_x^3+2 (\theta-1)u_x) \rho \frac{\partial^4 u_y}{\partial x^4} \right), \\
    & N_{10} = -\frac{1}{108} \left(27 u_x^4+(36 \theta-27) u_x^2+\theta^2-3 \theta \right) u_y \frac{\partial^4 \rho}{\partial x^4} - \frac{1}{6} \left(5 u_x^3+(\theta-2) u_x \right) \rho u_y \frac{\partial^4 u_x}{\partial x^4}  \\
    & \qquad \qquad \qquad - \frac{1}{36} \left(18 u_x^4+(12\theta-18) u_x^2- \theta^2+2 \theta \right) \rho \frac{\partial^4 u_y}{\partial x^4}, \\
    & N_{11} = -\frac{1}{108} \left(63 u_x^4+(72 \theta-63) u_x^2+7\theta^2-3 \theta \right) \frac{\partial^3 \rho}{\partial x^3} -\frac{1}{18} \left(33 u_x^3+(14 \theta-12) u_x \right) \rho \frac{\partial^3 u_x}{\partial x^3}, \\
    & N_{12} = - \frac{1}{72} \left( (9 u_x^4+(12 \theta-9) u_x^2+\theta^2-\theta) \frac{\partial^4 \rho}{\partial x^4} + (30 u_x^3+(12 \theta-12) u_x) \rho \frac{\partial^4 u_x}{\partial x^4} \right), \\
    & N_{13} = -\frac{2}{9} \rho \theta^2 \frac{\partial^4 u_x}{\partial x^4}, \qquad N_{14} = \frac{1}{54} \left( \left(9 \theta u_x^2 + 19 \theta^2 - 9 \theta \right)u_x \frac{\partial^4 \rho}{\partial x^4} + \left( 99 \theta u_x^2 +17\theta^2 - 12 \theta \right) \rho \frac{\partial^4 u_x}{\partial x^4}  \right), \\
    & N_{15} = -\frac{1}{36} \left( (3 u_x^3+3(\theta-1) u_x) u_y \frac{\partial^3 \rho}{\partial x^3} + (9u_x^2+\theta-3) \rho u_y \frac{\partial^3 u_x}{\partial x^3} + (3u_x^3+3(\theta-1) u_x) \rho \frac{\partial^3 u_y}{\partial x^3} \right),\\
    & N_{16} = -\frac{1}{36} \left( (27 u_x^5+(37\theta-30) u_x^3+(8\theta^2-9\theta+3) u_x)\frac{\partial^4 \rho}{\partial x^4} + + (87 u_x^4+(51\theta-42) u_x^2+4\theta^2-6\theta+3) \rho \frac{\partial^4 u_x}{\partial x^4} \right) , \\
    & N_{17} = \frac{1}{54} \left( 10 \theta^2 u_x \frac{\partial^4 \rho}{\partial x^4} + \left( 45\theta u_x^2-\theta^2+6\theta \right) \rho \frac{\partial^4 u_x}{\partial x^4} \right), \\
    & N_{18} = \frac{1}{72} \left( (18 u_x^5+(36 \theta-15) u_x^3+(2 \theta^2-3 \theta-3) u_x)\frac{\partial^4 \rho}{\partial x^4} + \left(72 u_x^4+(24 \theta-27) u_x^2+\theta-3 \right) \rho \frac{\partial^4 u_x}{\partial x^4} \right), \\
    %& N_{19} = \left( \frac{\theta^2}{18} \frac{\partial^4 \rho}{\partial x^4} + \frac{\rho \theta u_x}{2} \frac{\partial^4 u_x}{\partial x^4} \right), \qquad 
    & N_{19} = \rho c_s^2 \theta \frac{\partial^3 (u_x u_y)}{\partial x^3}, \\
    & N_{20} = -\frac{1}{12}(7 u_x^3+(5 \theta-3) u_x) u_y \frac{\partial^3 \rho}{\partial x^3} - \frac{1}{18} (27 u_x^2+5 \theta-3) \rho u_y \frac{\partial^3 u_x}{\partial x^3} - \frac{1}{6}(2 u_x^3+ \theta u_x) \rho \frac{\partial^3 u_y}{\partial x^3}, \\
    & N_{21} =  \frac{1}{108}(9\theta u_x^2 + 11 \theta^2 - 3\theta) u_y \frac{\partial^4 \rho}{\partial x^4} + \rho \theta u_x u_y \frac{\partial^4 u_x}{\partial x^4} + \frac{1}{18} (6 \theta u_x^2 + \theta^2) \rho \frac{\partial^4 u_y}{\partial x^4}, \\
    & N_{22} = -\frac{1}{216} (162 u_x^4+(165 \theta-108) u_x^2+13 \theta^2-7 \theta) u_y \frac{\partial^4 \rho}{\partial x^4} - \frac{1}{18} (39 u_x^3+(14 \theta-9) u_x) \rho u_y \frac{\partial^4 u_x}{\partial x^4} - \frac{1}{72} (18 u_x^4+12 \theta u_x^2+\theta^2) \rho \frac{\partial^4 u_y}{\partial x^4}, \\
    & N_{23} = \frac{1}{6} (u_x^3+(\theta-1) u_x) u_y \frac{\partial^3 \rho}{\partial x^3} + \frac{1}{18} (9 u_x^2+\theta-3) \rho u_y  \frac{\partial^3 u_x}{\partial x^3} + \frac{1}{6} (u_x^3- u_x) \rho \frac{\partial^3 u_y}{\partial x^3}, \\
    & N_{24} =  \frac{2}{27}\theta^2 u_y \frac{\partial^4 \rho}{\partial x^4} + \frac{\rho \theta}{2} u_x u_y \frac{\partial^4 u_x}{\partial x^4} + \frac{\rho \theta}{18} (3 u_x^2 + 1) \frac{\partial^4 u_y}{\partial x^4}, \\
    & N_{25} = \frac{1}{72} \left( (18 u_x^4+(27\theta-15) u_x^2+\theta^2-2\theta-1) u_y \frac{\partial^4 \rho}{\partial x^4} + (63 u_x^3+(15 \theta-21) u_x) \rho u_y \frac{\partial^4 u_x}{\partial x^4} + (9u_x^4-6 u_x^2-1) \rho \frac{\partial^4 u_y}{\partial x^4} \right), \\
    & N_{26} = -\frac{1}{24} \left( (9 u_x^3+(7\theta-5) u_x) u_y \frac{\partial^4 \rho}{\partial x^4} + (24 u_x^2+4(\theta-1)) \rho u_y \frac{\partial^4 u_x}{\partial x^4} + (6u_x^3+2(\theta-1)u_x) \rho \frac{\partial^4 u_y}{\partial x^4} \right), \\
    & N_{27} =  -\frac{1}{9} \left( (9u_x^2+2\theta-1) u_y \frac{\partial^4 \rho}{\partial x^4} + 15\rho u_x u_y \frac{\partial^4 u_x}{\partial x^4} + (9 u_x^2-1) \rho \frac{\partial^4 u_y}{\partial x^4} \right), \\
    & N_{28} = \frac{1}{9} \left( (3u_x^2+\theta)u_y \frac{\partial^4 \rho}{\partial x^4} + 6\rho u_x u_y \frac{\partial^4 u_x}{\partial x^4}c + 3\rho u_x^2 \frac{\partial^4 u_y}{\partial x^4} \right), \quad N_{29} =  - \frac{1}{9} \left( (6 u_x^2+\theta) u_y \frac{\partial^4 \rho}{\partial x^4} + 9\rho u_x u_y \frac{\partial^4 u_x}{\partial x^4} + (6 u_x^2-\theta) \rho \frac{\partial^4 u_y}{\partial x^4} \right), \\
    & N_{30} = -\frac{1}{6}  \left( (9 u_x^3+(5\theta-3) u_x) u_y\frac{\partial^3 \rho}{\partial x^3}  +(21 u_x^2+\theta-1 ) \rho u_y \frac{\partial^3 u_x}{\partial x^3} + 6\rho u_x^3 \frac{\partial^3 u_y}{\partial x^3} \right), \\
    & N_{31} = \frac{1}{18} \left( -  ((33\theta-6)u_x^2+\theta^2+\theta-2) u_y \frac{\partial^4 \rho}{\partial x^4} - ( 45 u_x^3+(21\theta-15)u_x) \rho u_y \frac{\partial^4 u_x}{\partial x^4} + (18 u_x^4-12 u_x^2+2) \rho \frac{\partial^4 u_y}{\partial x^4} \right), \\
	& N_{32} = \frac{1}{18} \left( (-54 u_x^4+(-51\theta+18) u_x^2-\theta^2+\theta) u_y \frac{\partial^4 \rho}{\partial x^4} + (-153 u_x^3+(-21\theta+15) u_x) \rho u_y \frac{\partial^4 u_x}{\partial x^4} - 36\rho u_x^4 \frac{\partial^4 u_y}{\partial x^4} \right), \\
    & N_{33} = \frac{1}{72}\left( (27 u_x^4+(36\theta-9)u_x^2+\theta^2-\theta) u_y \frac{\partial^4 \rho}{\partial x^4} + (90 u_x^3+(18\theta-12) u_x)\rho u_y \frac{\partial^4 u_x}{\partial x^4} + 18 \rho u_x^4 \frac{\partial^4 u_y}{\partial x^4} \right), \\
    & N_{34} =  - u_x u_y \frac{\partial^3 \rho}{\partial x^3} + (\theta-1) \rho u_y \frac{\partial^3 u_x}{\partial x^3} + (\theta-1) \rho u_x \frac{\partial^3 u_y}{\partial x^3}, \\
    &  N_{35} = \frac{1}{6} \left( (u_x^3+\theta u_x) u_y \frac{\partial^3 \rho}{\partial x^3} +  (3 u_x^2+ c_s^2 \theta) \rho u_y \frac{\partial^3 u_x}{\partial x^3} + \rho u_x^3 \frac{\partial^3 u_y}{\partial x^3} \right), \\
    & N_{36} =  \frac{1}{9} (\theta^2-\theta) u_y \frac{\partial^4 \rho}{\partial x^4} + \frac{1}{3} ( 2 \theta-1) \rho u_x u_y \frac{\partial^4 u_x}{\partial x^4} + \frac{1}{9} (1-3 u_x^2) \rho \theta \frac{\partial^4 u_y}{\partial x^4}, \\
	& N_{37} = \frac{1}{9}(6 u_x^2-\theta^2 + \theta) u_y \frac{\partial^4 \rho}{\partial x^4} + \frac{1}{3}(-2\theta+3) \rho u_x u_y \frac{\partial^4 u_x}{\partial x^4}  +  \frac{2}{3}(1-\theta) \rho u_x^2 \frac{\partial^4 u_y}{\partial x^4}, \\
    & N_{38} = \frac{1}{36} (3 u_x^2-\theta^2+\theta) u_y \frac{\partial^4 \rho}{\partial x^4}  + \frac{1}{12} (2-\theta) \rho u_x u_y \frac{\partial^4 u_x}{\partial x^4} + (1-\theta) \frac{\rho u_x^2}{12} \frac{\partial^4 u_y}{\partial x^4}, \\
	& N_{39} = - \frac{1}{12} \left( (9 u_x^3+(5\theta-5) u_x) u_y \frac{\partial^4 \rho}{\partial x^4} +  (21 u_x^2+3(\theta-1)) \rho u_y \frac{\partial^4 u_x}{\partial x^4} +  (6 u_x^3 + 2(\theta-1) u_x) \rho \frac{\partial^4 u_y}{\partial x^4} \right), \\
	& N_{40} = \frac{1}{6} \left( (12 u_x^4+(15\theta-12) u_x^2+c_s^2\theta^2-\theta) u_y \frac{\partial^4 \rho}{\partial x^4} + (39 u_x^3 + (7 \theta-15)u_x) \rho u_y \frac{\partial^4 u_x}{\partial x^4} + (6 u_x^4 + (\theta-6) u_x^2- c_s^2 \theta^2+\theta) \rho \frac{\partial^4 u_y}{\partial x^4} \right), \\
	& N_{41} = \frac{1}{18} \left( (9 u_x^4 + (18\theta-9) u_x^2+\theta^2-3\theta) u_y \frac{\partial^4 \rho}{\partial x^4} + (36 u_x^3 + (12\theta-18) u_x) \rho u_y \frac{\partial^4 u_x}{\partial x^4} +  (9 u_x^4 +(3\theta-9) u_x^2) \rho \frac{\partial^4 u_y}{\partial x^4} \right), \\
	& N_{42} = \frac{1}{24} \left( (-3 u_x^4 +(-6\theta+3) u_x^2-c_s^2\theta^2+\theta) u_y \frac{\partial^4 \rho}{\partial x^4} + (-12 u_x^3 + (-4 \theta+6) u_x) \rho u_y \frac{\partial^4 u_x}{\partial x^4} +  (-3 u_x^4 +(3-\theta) u_x^2)\rho \frac{\partial^4 u_y}{\partial x^4} \right), \\
	& N_{43} = \frac{1}{18} \left( \frac{\partial^4 \rho}{\partial x^4} (18 u_x^4 + (27\theta-18) u_x^2+\theta^2-3\theta) u_y + (63 u_x^3 +(3 \theta-27) u_x) \rho u_y \frac{\partial^4 u_x}{\partial x^4} + (18 u_x^4 + (3 \theta-18) u_x^2-\theta^2 + 3\theta)\rho \frac{\partial^4 u_y}{\partial x^4} \right), \\
	&  N_{44} = -\frac{1}{8} \rho (u_x^4+(\theta-1)u_x^2) \frac{\partial^4 u_y}{\partial x^4}, \\
	& N_{45} = \frac{1}{6} \left( (3\theta u_x^2+\theta^2 - \theta) u_y \frac{\partial^4 \rho}{\partial x^4} + (3u_x^3 + 3(\theta-1)u_x) \rho u_y \frac{\partial^4 u_x}{\partial x^4} + (-3\theta u_x^2- c_s^2\theta^2+\theta) \rho \frac{\partial^4 u_y}{\partial x^4} \right), \nonumber \\
	& N_{46} = \frac{1}{2} \left( (2 u_x^4 + (3\theta-2) u_x^2+c_s^2\theta^2 - c_s^2\theta) u_y \frac{\partial^4 \rho}{\partial x^4} + (7u_x^3 + 3(\theta-1)u_x) u_y \frac{\partial^4 u_x}{\partial x^4} + (2 u_x^4 + 2(\theta-1) u_x^2) \rho \frac{\partial^4 u_y}{\partial x^4} \right), \nonumber \\
	& N_{47} = \frac{1}{8} \left( (-u_x^4 + (-2\theta+1) u_x^2-c_s^2\theta^2 + c_s^2 \theta) u_y \frac{\partial^4 \rho}{\partial x^4} + (-4 u_x^3 + (-2 \theta+2)u_x) \rho u_y \frac{\partial^4 u_x}{\partial x^4} - ( u_x^4 +(\theta-1) u_x^2) \rho \frac{\partial^4 u_y}{\partial x^4} \right). 
\end{align*}
\endgroup

\bibliographystyle{acm}

\end{document}